\documentclass[]{scrreprt}

\usepackage{feynmp-auto}
\usepackage[square,numbers]{natbib}
\usepackage{amsthm,mathrsfs}
\usepackage[vcentermath]{youngtab}
\usepackage{framed}
\usepackage{tensor}
\usepackage[colorlinks=true,
linkcolor=blue,
filecolor=black,      
urlcolor=blue,
citecolor=magenta]{hyperref}
\usepackage{amsmath,mathtools}
\usepackage{amsfonts}
\usepackage{amssymb}
\usepackage{color}
\usepackage{slashed}

\usepackage{newpxtext} \usepackage[euler-digits]{eulervm}

\def\Ddot{D\cdot}
\def\it{\textit}

\def\3s{{s \choose 3}}
\def\4s{{s \choose 4}}
\def\5s{{s \choose 5}}
\def\6s{{s \choose 6}}

\def\12{\dfrac{1}{2}}
\def\fr{\frac}
\def\ft{\footnote}

\def\nn{\nonumber}

\def\2{\ell_2}

\def\pr{\partial}

\def\scri{\mathscr{I}}
\def\scrim{\mathscr{I}^-}
\def\scrip{\mathscr{I}^+}

\def\be{\begin{equation}}
\def\ee{\end{equation}}
\def\bea{\begin{eqnarray}}
\def\eea{\end{eqnarray}}
\def\beali{\begin{equation}\begin{aligned}}
\def\eeali{\end{aligned}\end{equation}}
\def\ba{\begin{array}}
	\def\ea{\end{array}}
\def\bec{\begin{center}}
	\def\ec{\end{center}}
\newcommand{\beal}{\begin{aligned}}

\def\g{\gamma} 

\def\d{\delta} 
\def\D{\Delta}
\def\e{\epsilon}

\def\r{\rho}

\def\vf{\varphi}
\def\O{\Omega}

\def\cF{{\cal F}}
\def\cP{{\cal G}}

\def\cK{{\cal K}}
\def\cL{{\cal L}}
\def\cM{{\cal M}}
\def\cN{{\cal N}}
\def\cO{{\cal O}}
\def\cP{{\cal P}}
\def\cR{{\cal R}}

\def\cT{{\cal T}}

\numberwithin{equation}{section}

\title{Asymptotic Symmetries of Yang-Mills, Gravity, Two-Forms and Higher-Spins}
\author{Carlo Heissenberg}
\date{}

\begin{document}
%\maketitle
%--------------------------------------------------------------
\begin{titlepage}
	\centering
	\includegraphics[width=.3\textwidth]{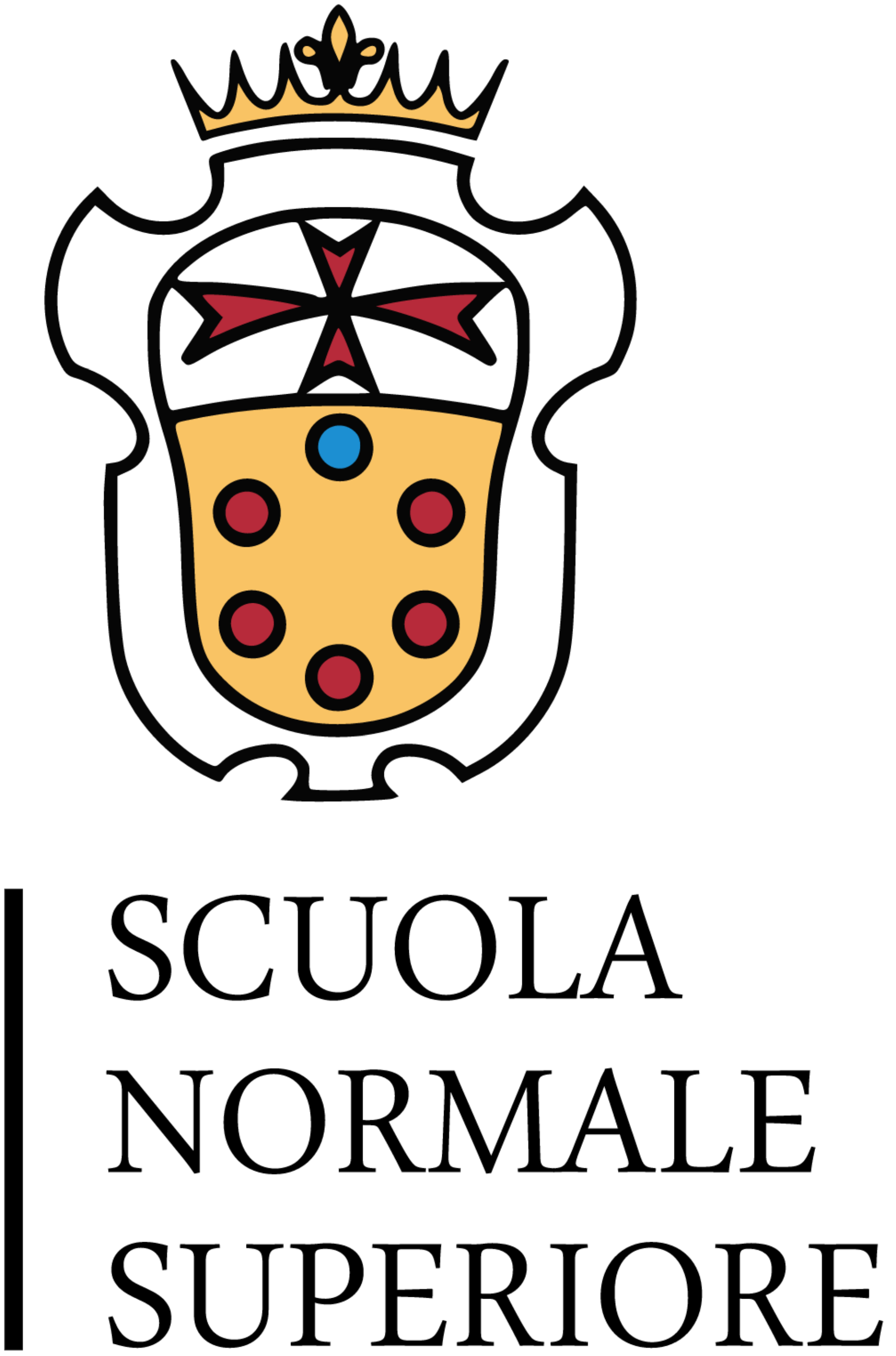}\par\vspace{.5cm}
	{\scshape\LARGE Classe di Scienze\par}
	\vspace{-.2cm}
	\hrulefill
	\vspace{.3cm}
	{\par\Large Corso di Perfezionamento in Fisica\par
		%\par
	}
	
	\vspace{1cm}
	{\Large \scshape 
		tesi di dottorato\par}
	\vspace{1cm}
	{\fontfamily{lmss}\selectfont \huge\bfseries Topics in Asymptotic Symmetries and Infrared Effects\par}
	\vspace{2cm}
	
	\begin{minipage}[t]{5cm}
		{\large{\normalfont\scshape relatore: 
				\vspace{2mm}\\
				\itshape{\large Dr. Dario Francia}
		}}
	\end{minipage}
	\hfill
	\begin{minipage}[t]{3.6cm}%\raggedleft 
		{\large{\normalfont\scshape candidato:
				\vspace{2mm}\\
				\large{\itshape Carlo Heissenberg}
		}}
	\end{minipage}

	\vfill
	
	% Bottom of the page
	%\vhrulefill{.5pt}\par \vspace{-.43cm}
	%\vhrulefill{1pt}
	{\large
		\par
		XXXII Ciclo
		\par }
	\vspace{5pt}
	{\large
		\par
		2016-2019
		\par}\vspace{5pt}
	{\large
		\par
		Pisa
		\par}
\end{titlepage}
%--------------------------------------------------------------

\tableofcontents

%%%%%%%%%%%%%%%%%%%%%%%%%%%%%%%%%%%%%%
\chapter*{Introduction}
\addcontentsline{toc}{chapter}{Introduction}

How much physical information can be extracted from local symmetries? This question surely bears relevance to our understanding of theoretical physics, due to the prominent examples of well-established physical theories taking the form of gauge field theories: Maxwell's electrodynamics and general relativity at the classical level, the standard model of elementary particles at the quantum level.

By their very nature, local gauge transformations leave all observable quantities invariant, so that they can be defined only by introducing nonobservable fields, namely gauge fields acting as force mediators and charged matter fields, that transform nontrivially under the local gauge group. The Hamiltonian and the equations of motion are invariant under such transformations.
Therefore, due to the intrinsic redundancy introduced in the dynamical problem by the gauge symmetry, the time evolution of gauge fields is not well defined, already at the classical level, unless one introduces a gauge fixing condition that breaks local gauge invariance to a residual subset, which may or may not coincide with the set of global transformations depending on the strength of such a condition.
This picture may thus superficially leave the impression that local gauge transformations are only a convenient artifact that permits to write down interesting theories, but which does not carry any intrinsic physical relevance.

On the contrary, the physical meaning of the global counterparts of local gauge symmetries is rather well-understood. They give rise to very important conserved observable quantities, at the classical level: the electric charge $Q$ for electrodynamics, the color charge $Q^A$ for Yang-Mills theory, energy-momentum $P_\mu$ and angular momentum $M_{\mu\nu}$ for general relativity in the case of asymptotically flat spacetimes. In quantum theories, unbroken global symmetries arising from local gauge symmetries induce a decomposition of the Hilbert space into superselection sectors and allow one to derive selection rules in scattering processes. 

However, motivated by recently-established connections with  observable phenomena, in particular with soft factorization theorems for scattering amplitudes and with memory effects, a larger class of local transformations appears worthy of attention as far as direct physical information is concerned: asymptotic symmetries.

Such symmetries were originally introduced in the sixties by Bondi Metzner, Van der Burg and Sachs \cite{BMS,Sachs_Waves,Sachs_Symmetries} in the context of general relativity, within the study of asymptotically flat spacetimes, namely of the solutions of the Einstein equations that describe weakly radiating mass distributions---for instance, a planet orbiting around the sun, a black-hole merger or a pair of coalescing neutron stars. As such, these systems possess a metric tensor that tends to the flat one as one follows the outflowing gravitational waves and goes ``very far'' from the sources, thus approaching null infinity. More precisely, after specifying properly-defined coordinates, the components of the metric tensor approach those of the Minkowski one as the radial coordinate $r$ tends to infinity for fixed retarded time $u$, up to corrections that become negligibly small in this limit. The delicate point in the definition of asymptotically flat spacetimes, and the important result of the above authors, is precisely the identification of the proper assignment of ``falloff conditions'' on these corrections, specifying how a nontrivial geometry becomes Minkowski as $r\to\infty$; in four spacetime dimensions, this falloff is $\mathcal O(1/r)$ for the normalized corrections to the angular components.

Asymptotic symmetries of asymptotically flat spacetimes arise as those diffeomorphisms that map the set of asymptotically flat spacetimes to itself and transform a given solution of the Einstein equations to a physically inequivalent one. Familiar examples of these symmetries are given by the ordinary translations, rotations and boosts, namely the isometries of flat spacetime, which indeed preserve (asymptotic) flatness, while still altering physically relevant quantities, \emph{e.g.} the angular momentum and the energy-momentum. 
In fact, in the sixties, it was expected that Poincar\'e symmetries could be uniquely selected as the asymptotic symmetries of asymptotically flat spacetimes. However, this is not the case. In four spacetime dimensions, one cannot adopt falloff conditions so stringent as to reduce the asymptotic symmetries of gravity to the Poincar\'e group without giving up on gravitational waves, and the asymptotic symmetries of gravity take the form of an infinite-dimensional enhancement of Poincar\'e, termed $BMS$ group.

It was later realized that the concept of asymptotic symmetry can be in principle extended to any gauge theory, in the following way. One first specifies, after fixing the gauge (the analog of choosing suitable coordinates), a physically relevant set of solutions to the equations of motion by assigning falloff conditions on the gauge fields, \emph{i.e.} by specifying the manner in which they reduce to a trivial (flat) configuration as one goes very far from matter sources. Asymptotic symmetries are then defined as those local symmetries that preserve the set of solutions under consideration and that map a given solution to a different, inequivalent one. A prominent feature of these symmetries is that they often comprise an infinite-dimensional enhancement of standard global symmetries \cite{Strominger_YM, Einstein-YM.Barnich, Strominger_QED, StromingerKac}, again in analogy with the gravitational setup. 

The concept of asymptotic symmetry appears thus to be both quite flexible, as it may find applicability in many interesting theories, and very rich, at least from a formal point of view.
Actually, as we anticipated, the interest in asymptotic symmetries is not merely formal, since these symmetries have been shown to imply observable consequences: memory effects in classical gauge theories and soft theorems in scattering amplitudes. Mainly thanks to the works of Strominger and collaborators, the interplay among these three features of gauge theories, sometimes termed the \emph{infrared triangle}, triggered a significant trend of original research, the main aspects of which are reviewed in \cite{Strominger_rev}.

More specifically, it has been observed that the passage of a wave packet near a test charge can leave a permanent observable imprint on the physical properties of the probe, \emph{i.e.} a memory effect, that can be understood in terms of the underlying action of an asymptotic symmetry. The detection of a memory effect is understood as a concrete manifestation of the fact that the underlying gauge field underwent a vacuum transition: the passage of radiation induced a sharp transformation from a given radiative vacuum to an inequivalent vacuum, connected to the previous one by an asymptotic symmetry. In this respect, one can clearly see that asymptotic symmetries are akin to \emph{spontaneously broken} global symmetries.

The first instance of memory effect again dates back to studies in general relativity, in particular \cite{Zeldovich, ChristodoulouMem}, where memory was identified in the form of a permanent relative displacement induced by the passage of gravitational radiation on a system of detectors. The underlying connection with $BMS$ symmetry was realized only much later in \cite{memory}. An electromagnetic analog of gravitational wave memory was instead put forward in \cite{Bieri:2013hqa}, where it occurs as a velocity kick, and the corresponding asymptotic-symmetry interpretation followed in \cite{MagicaSabrina}. Another interesting type of memory arising in the electromagnetic theory is the one encoded in the phases of suitably-placed superconducting nodes \cite{Susskind}, while a Yang-Mills counterpart of this phase memory is given by color memory \cite{StromingerColor}, which occurs as a  color rotation in the Hilbert space of two test quarks. Many other types of memories have been discussed, see \emph{e.g.} \cite{Mao_em,Mao_note,Hamada:2017atr,Afshar:2018sbq,Chumemory}, and occur both in linearized theories, with massive or massless background sources, and in nonlinear theories. More precisely, memory effects induced by radiation emitted by massive sources are identified as \emph{linear} or \emph{ordinary} and can be regarded as a picture of the movements of bulk sources that is stored in the properties of faraway probes, while memory effects associated to massless sources or nonlinear wave-like perturbations give rise to the so-called \emph{nonlinear} or \emph{null} memory.

At the level of scattering amplitudes, asymptotic symmetries bear relevance in connection with factorization results that are valid when an external massless particle is evaluated in the low-energy regime: soft theorems. These identities have been remarked to be equivalent to semiclassical Ward identities stemming from asymptotic symmetries, so that soft theorems can be actually seen as an expression of the invariance of the $S$ matrix under these infinite-dimensional asymptotic symmetries. More specifically, Weinberg's soft photon and soft graviton theorems \cite{Weinberg_64, Weinberg_65} have been shown to be equivalent to the invariance of the $S$ matrix under asymptotic $U(1)$ symmetries \cite{Strominger_QED, Campiglia} and under $BMS$ symmetries \cite{Strominger_Invariance,Strominger_Weinberg}, respectively. This link has been extended to many different contexts and has been  generalized to encompass subleading corrections, in the soft frequency, to Weinberg's result \cite{soft-subleading,Bern:2014vva,soft_QED_Strominger, Sen:2017nim,LaddhaSen:2017ygw}. Moreover, the presence of these symmetries appears to be at the basis of the universality of certain soft theorems, whose validity does not depend on the specific interactions under consideration but rather only on the gauge symmetry of the theory. Compatibly with the spontaneous breaking of asymptotic symmetries, soft photons and gravitons have been interpreted as the massless bosons whose existence is then ensured by the Goldstone theorem.
Since they unveil a new aspect of the infrared physics associated to gauge mediators, asymptotic symmetries also gave rise to a resurgence of interest in the definition of asymptotic states in QED, gravity and QCD \cite{Sever,PerryStates,HiraiSugiDressed,Tristan} and in the infrared problem, which lies at the heart of a possible nonperturbative definition of the $S$ matrix that does not  rely on inclusive processes (see also \cite{Strocchi-Erice} for a detailed review).

Asymptotic symmetries possess another piece of appeal, not unrelated to the previous aspects which we have touched upon, in relation with the so-called black-hole information paradox. It has been observed that a proper description of black-hole solutions may require taking into account additional symmetries, \emph{i.e.} asymptotic symmetries acting at null infinity and near the horizon, and hence additional conserved charges \cite{Barnich-Brandt} compared to the one at the basis of the proof of the no-hair theorem. However, the presence of additional soft charges associated to asymptotic symmetries in itself may be insufficient to the purpose of fully resolving the information paradox \cite{Perry,PorratiMir,PerrySuperrot,PorratiWig,StromingerInfo, PerryEntropy}.  

Up to now, we have been considering the case of gauge theories defined in four spacetime dimensions.
Indeed, the attention of the literature focused at first on the arena of four-dimensional (asymptotically) Minkowski spacetime, where certain issues concerning the asymptotic behavior of the fields and the calculation of asymptotic charges simplify, and attempts to carry out the analysis in arbitrary dimensions have faced a number of difficulties since their inception. 

Indeed, considering for the moment the gravitational case, a first  aspect is that, in dimensions $D>4$, it is indeed possible to impose falloff conditions, termed radiation falloffs, that select Poincar\'e as the asymptotic symmetry group of asymptotically flat spacetimes, without spoiling the description of gravitational waves \cite{gravity_evenD_1,Hollands_SI, angular-momentum, No_ST}. Thus, infinite-dimensional asymptotic symmetries of higher-dimensional gravity appear to be less fundamental than in $D=4$ as they can be safely trivialized without losing interesting solutions. Actually, from the point of view of general relativity itself, it may appear rather natural to do so, also because radiation falloffs are instrumental in ensuring the finiteness of energy fluxes and of other physical quantities as $r\to\infty$ , while being compatible with the finiteness of Poincar\'e charges. Quantitatively, the radiation falloff for the angular components of the normalized metric fluctuation is $\mathcal O(1/r^{\frac{D-2}{2}})$.
These features are shared by electromagnetism, where, for a time, it appeared impossible to retrieve an infinite-dimensional asymptotic symmetry group in any spacetime dimension, without derogating from the falloff conditions that ensure the finiteness of fluxes/charges at infinity. 

A different approach was then put forward, for the case of even-dimensional spacetimes, in \cite{alternative-bnd,StromingerQEDevenD}, where such falloff conditions were relaxed, from $\mathcal O(1/r^{\frac{D-2}{2}})$ to $\mathcal O(1/r)$, while at the same time introducing constraints on gauge-invariant quantities that ensure the finiteness of physical observables, in order to allow for the presence of infinite-dimensional asymptotic symmetries. 
Indeed, the absence of the latter in higher dimensions would be at odds with the validity of soft theorems, which hold independently of the dimension of spacetime. Therefore, from this point of view, it appears more natural to adopt (if possible) falloff conditions that do not select standard global symmetries only, \emph{even if} this can be in principle achieved without renouncing the description of radiation. Recent achievements on the connection between soft theorems and asymptotic symmetries in spacetimes of arbitrary dimensions, either even or odd, have been presented in \cite{He-Mitra-photond+2, He-Mitra-gauged+2, He-Mitra-magneticd+2} for the case of Maxwell and Yang-Mills theories, by analyzing the asymptotic behavior of field strengths rather than the gauge fields themselves, while the canonical realization of the asymptotic symmetries for the Maxwell theory in any dimension was given in \cite{HenneauxQEDanyD}.

Another partially unsatisfactory feature of the early investigations of the asymptotic structure of gauge theories in higher dimensions was the purported absence of memory effects \cite{AltroWald,BMS-memory,Garfinkle:2017fre,WaldOddD}. The elusive nature of memory in higher dimensions, which was later elucidated in \cite{Mao_EvenD,StromingerGravmemevD,SatishchandranWald} in the case of even-dimensional spacetimes, can be traced back to the fact that, despite being imprints left by the passage of radiation, memory effects do not scale asymptotically in the same manner as radiation itself, $\mathcal O(1/r^{\frac{D-2}{2}})$. Instead, they possess a much faster falloff, $\mathcal O(1/r^{D-3})$, subleading in $1/r$ for $D>4$, which is associated to stationary solutions and, in general, to static or ``DC'' effects; this is also the asymptotic behavior corresponding to fields giving rise to finite and nontrivial global charges and is therefore termed Coulombic falloff. The interpretation of the gravitational memory effect experienced by geodesic detectors in terms of an underlying symmetry acting at large $r$ was in particular clarified in \cite{StromingerGravmemevD}. Contrary to the four-dimensional case, the corresponding symmetry responsible for the vacuum transition underwent by the gravitational field in higher, even dimensions is not a true asymptotic symmetry, since in particular it does not give rise to a nonvanishing soft charge; nevertheless, it has been noted that a tight link exists between this effect and the corresponding soft theorem \cite{Mao_EvenD}. 

Difficulties associated to the discussion of memory effects in odd dimensions, instead, can be essentially ascribed to the phenomenon of dispersion that characterizes the propagation of waves in those spacetimes, \emph{i.e.} the failure of the Huygens principle, due to which an idealized point-like perturbation gives rise to disturbances that persist after the passage of the first wavefront. This peculiarity causes a blurring of ordinary memory effects in odd dimensions, as also noted in \cite{SatishchandranWald}, but may be resolved, as we shall see, by analyzing null memory, where null sources effectively reach the region near the probe and bypass the problem introduced by dispersion phenomena.

An interesting feature of the higher-dimensional context, as already exhibited by the example of memory effects, is thus the interplay between radiation and Coulombic terms, $\mathcal O(1/r^{\frac{D-2}{2}})$ and $\mathcal O(1/r^{D-3})$ respectively, which indeed become separated into two different orders of $1/r$ only for $D>4$ and whose identification would be therefore precluded in a purely four-dimensional approach. Another issue pertaining to this interplay is related to the possibility of defining nonzero and finite asymptotic charges in the presence of radiation. Indeed, asymptotic charges are generically finite as $r\to\infty$ only when sourced by Coulombic fields, while in the presence of radiation fields, which are leading in $1/r$, they exhibit seemingly divergent contributions. This problem is particularly interesting in the case of asymptotic symmetries that do not reduce to the standard global symmetries and in nonlinear theories. In these cases, no general results prevent the charges from diverging in the limit $r\to\infty$ in the presence of radiation contributions. Furthermore, waves in nonlinear theories are actually charged under the global group (for instance they carry energy-momentum in general relativity); therefore, they will induce a leak of global charges, as they reach null infinity, and a dependence of the latter on retarded time $u$.

As we have seen, the analysis of asymptotic symmetries, charges and soft theorems focused at first on the case of four spacetime dimensions, but a lot of mileage has been covered since then in the context of higher dimensions. The extension of this program to the case of spins greater than two is instead a much less-explored subject. However, this research direction is worth pursuing essentially for the very same reason that motivates the higher-dimensional investigations: soft theorems are valid not only for an arbitrary dimension but also for an arbitrary spin. Therefore one may wonder whether or not they can be understood in terms of  underlying asymptotic symmetries also in the case of higher-spin theories.

Another physical motivation for the extension to higher spins is provided by the role that such symmetries play in the understanding of asymptotic states and of the infrared problem in lower-spin gauge theories. In view of both the impossibility of long-range higher-spin forces already
implied by Weinberg's results, and of the established lore against interacting higher-spin theories in
flat space \cite{BBS}, one may envision that unraveling the asymptotic structure of higher-spin theories could lead to some clarifications on their infrared structure. The relevance of the role played by the infrared limit in connection with the puzzling issues concerning higher-spin interactions is also clearly indicated by the two main frameworks where the most direct obstructions are seemingly evaded. 
On the one hand, as originally proposed by Fradkin and Vasiliev in their seminal paper \cite{Vasi}, a nonvanishing, negative
cosmological constant can indeed act as an infrared regulator for some of the most immediate singularities met by massless higher spins in interaction with gravity on flat backgrounds. This suggestion paved the way to a complete on-shell description of gauge-invariant, nonlinear higher-spin dynamics, encoded in the Vasiliev equations, up to the more recent developments in the context of AdS/CFT (see \emph{e.g.} \cite{BCIValgebras,HOLO,Did}).
On the other hand, another context in which higher-spin interactions can be naturally formulated is string theory, where higher spins are lifted to being massive, the string tension playing the role in
this context of the needed infrared regulator (see \emph{e.g.} \cite{GSW}). A long-standing conjecture concerns the interpretation of massive higher spins as the manifestation of some underlying higher-spin gauge symmetry breaking mechanism \cite{Gross, Sagnotti}, and one may speculate that a deeper understanding of the infrared physics of higher-spin interactions should be ultimately related to this picture.

Let us also mention that an underlying driving force for the study of asymptotic symmetries on asymptotically flat spacetimes in any dimension and for gauge theories of any spin is the possibility to find a suitable generalization of the AdS/CFT duality to the case of Minkowski spacetime, the so-called \emph{flat space holography}, which could be exploited to gain insight into the structure of bulk theories, such as quantum gravity or higher spins, by studying their dual boundary theory at null infinity. At the gravitational level in four dimensions this program is currently being pursued \cite{FSHolo1, FSHolo2, FSHolo3, FSHolo4} and the infinitesimal symmetries of the boundary theory take the form of a semidirect sum of supertranslations, an infinite-dimensional enhancement of Poincar\'e translations occurring within the $BMS$ group, and superrotations, Virasoro symmetries that locally generalize ordinary rotations and boosts \cite{Barnich_Revisited,Barnich_BMS/CFT,Barnich_Superrotations, Strominger-semiVir,StromingerKac}.

\section*{Results}
In this Ph.D. thesis we aim, first, to review the main aspects of the above-described connection between asymptotic symmetries and observable effects in the context of gravity, electromagnetism and Yang-Mills theory in four dimensions and, second, to present original results concerning the extension of this program to the case of arbitrary-dimensional spacetimes and to higher-spin gauge theories.

Motivated by the appeal possessed by memory effects in connection with the asymptotic structure of gauge theories, we present a collection of results concerning memory effects in higher-dimensional scalar, electromagnetic and non-Abelian theories and on their interpretation in terms of residual symmetries acting at large distances in even dimensions \cite{Memory-io}. 
In this context, we derive explicit formulas for kick memory effects due to the interaction of test particles with massless scalar fields and electromagnetic fields emitted by specific background sources in any dimensions, either even or odd; we observe that null memory can be employed as a tool for obtaining nontrivial memory kicks in odd spacetime dimensions. We then analyze the significance that these memory effects bear with respect to the residual symmetries of Maxwell and Yang-Mills theories in the Lorenz gauge, including a discussion of (Abelian) phase memory and (non-Abelian) color memory, in arbitrary even dimensions. We obtain an interpretation of spin-one memories in terms of underlying symmetries acting at large $r$, which, as in the case of gravitational memories, act at Coulombic order and do not comprise \emph{bona fide} asymptotic symmetries for even $D>4$ because their charge tends to zero as $r\to\infty$.

Concerning the discussion of asymptotic observables at null infinity, we provide an explicit evaluation of the global color charge, energy flux and color flux for Yang-Mills theory in any spacetime dimension \cite{Cariche-io}. We also propose a strategy for the definition of surface charges associated to infinite-dimensional asymptotic symmetries at null infinity that overcomes the above-mentioned difficulties \cite{Memory-io}. The main idea of the proposal is to evaluate these charges by taking the limit $r\to\infty$ in a region where radiation is absent, so that no divergence may arise in this step, and then define their evolution with respect to retarded time $u$ by means of the asymptotic expansion of the equations of motion in powers of $1/r$, even in the presence of radiation.

Another issue which we address is the following seeming paradox arising in the case of a massless scalar force mediator. Theories containing a massless scalar have been shown to possess nontrivial asymptotic charges, which are linked with soft scalar factorization theorems \cite{CampigliaCoitoMizera, Campiglia_scalars}. However, no gauge symmetry is obviously present in this context, and hence no asymptotic symmetry is available in order to explain the presence of such charges. As a possible resolution of this purported contradiction, in the case of four dimensions, we propose a matching with the asymptotic analysis of a two-form theory, to which a free on-shell scalar is linked by a duality transformation \cite{Campigliascoop,twoform-io}. 

Finally, we discuss higher-spin asymptotic symmetries, whose analysis was initiated in \cite{carlo_tesi}.
A first relevant feature is that, in four spacetime dimensions, these symmetries turn out to comprise an infinite-dimensional family and to underlie Weinberg's leading soft theorem for higher-spin soft emissions \cite{super-io}. 

As in the case of lower spins, the imposition of radiation falloffs, while instrumental to the purpose of ensuring the finiteness of fluxes at infinity, actually trivializes the asymptotic symmetries in dimensions $D>4$ by reducing them to the exact Killing tensors \cite{Cariche-io}. 
The realization of infinite-dimensional asymptotic symmetries is indeed possible, but it requires to relax these falloffs to match the corresponding four-dimensional ones in any $D$, while at the same time ensuring the finiteness of relevant physical observables by means of suitable constraints involving gauge-invariant quantities. 

A possible way to carry out this program is offered by the following strategy. One may first consider the solution space characterized by radiation falloffs, whose physical consistency is under control, and then act upon this space by means of gauge transformations that leave the form of the equations of motion invariant, and hence map solutions to solutions, and have finite charges, but that do not preserve the falloff conditions. In this way, one generates a wider solution space, different from the previous one, that by construction exhibits infinite-dimensional asymptotic symmetries, while still ensuring the finiteness of observables \cite{highspinsrel-io}.

A promising aspect of the higher-spin asymptotic symmetries uncovered in our approach is that they display a structure closely resembling an infinite-dimensional enhancement of (a suitable quotient of) the enveloping algebra of Poincar\'e, which one may interpret as the possible remnant of an underlying higher-spin algebra \cite{flat-algebras_1,flat-algebras_2}. Indeed, focusing on spin three, we find three families of symmetries corresponding to (traceless projections of symmetrized) products of two Poincar\'e generators, namely generalized supertranslations associated to $P_{(\mu}P_{\nu)}$ and superrotations, $M_{\mu(\nu}M_{\rho)\sigma}$, together with a third family corresponding to $P_{(\mu}M_{\nu)\rho}$. More concrete steps towards this identification could be performed by discussing the non-Abelian deformation of our higher-spin asymptotic symmetries in the presence of cubic vertices, which could lead to an understanding of their commutation relations.

\subsection*{Structure of this thesis}
The material is organized in two parts, the first being devoted to the analysis of asymptotic symmetries and of their implications in spin-one theories, Maxwell and Yang-Mills, and spin-two theories, linearized gravity and general relativity, while the second part deals with extensions to the case of more exotic spins: the scalar and higher spins, $s\ge3$. 

We begin by reviewing the setup in which asymptotic symmetries first made their appearance, \emph{i.e.} the study of four-dimensional asymptotically flat spacetimes, in Chapter \ref{chap:basics}, while also providing an overview of the calculation of the associated charges and furnishing a brief account of the analogous discussions in the case of Maxwell and Yang-Mills theory.
Chapter \ref{chap:obs} offers then a review of the connection between the symmetries highlighted in the previous chapter and observable effects in four dimensions: Weinberg's soft photon and soft graviton theorem, and various types of memory effects.

The remaining three chapters contain instead the original material of this thesis. In Chapter \ref{chap:spin1higherD}, we present the extension of the calculation of asymptotic charges and of the discussion of memory effects in connection with residual symmetries to the case of spin-one theories in higher spacetime dimensions. Chapter \ref{chap:scalar2form} deals with the puzzling case of scalar asymptotic charges in four dimensions, which are tightly linked with scalar soft theorems but seemingly lack an underlying gauge symmetry. Finally, Chapter \ref{chap:HSP} collects the results on the asymptotic structure of higher-spin theories in four-dimensional spacetime and a discussion of their generalization to higher dimensions.

\begin{center}
{\large	$\ast$  $\ast$  $\ast$}
\end{center}

The research work I carried out during my Ph.D., which is reviewed in this thesis, appeared in the following papers \cite{super-io,Cariche-io,proceedings-io,twoform-io,Memory-io}.

I was additionally involved in different projects, whose content is not detailed in the present thesis, mainly focusing on foundational aspects of spontaneous symmetry breaking in quantum systems:
\begin{itemize}
	\item[{[\hypertarget{refa}{a}]}] C.~Heissenberg and F.~Strocchi, ``{Gauge invariance and symmetry breaking by topology and energy gap}'', \textit{Mathematics} \textbf{3} (2015) 984--1000, arXiv: \href{https://arxiv.org/abs/1511.01757}{1511.01757 [math-ph]}.
	\item[{[\hypertarget{refb}{b}]}] C.~Heissenberg and F.~Strocchi, ``{Existence of quantum time crystals}'', 2016, arXiv: \href{https://arxiv.org/abs/1605.04188}{1605.04188 [quant-ph]}.
	\item[{[\hypertarget{refc}{c}]}] C.~M.~Bender and C.~Heissenberg, ``{Convergent and Divergent Series in Physics}'', \textit{22th Saalburg Summer School on Foundations and New Methods in Theoretical Physics (2016), Wolfersdorf, Germany}, 2017, arXiv: \href{https://arxiv.org/abs/1703.05164}{1703.05164 [math-ph]}.
	\item[{[\hypertarget{refd}{d}]}] C.~Heissenberg and F.~Strocchi, ``{Generalized criteria of symmetry breaking. A strategy for quantum time crystals}'', 2019, arXiv: \href{https://arxiv.org/abs/1906.12293}{1906.12293 [cond-mat.stat-mech]}.
\end{itemize} 
In particular, in [\hyperlink{refa}{a}], we discuss a mechanism of spontaneous symmetry breaking in which the topology of the algebra of observables allows for the presence of a gap in the energy spectrum, thus evading the main consequence of Goldstone's theorem.
The main aim of [\hyperlink{refb}{b}, \hyperlink{refd}{d}]  is instead to propose a generalized criterion of spontaneous symmetry breaking that does not rely on the properties of the ground state, thereby allowing for the breaking of time translation symmetry, together with the discussion of simple models with time-independent Hamiltonian which realize this feature. The study of such systems was motivated by a pioneering proposal, due to Wilczek \cite{Wilczek}, of quantum time crystals, namely putative systems exhibiting the spontaneous breaking of continuous time translation symmetry down to a discrete subgroup, in analogy with the situation occurring in ordinary space crystals. 

%\acknowledgements

%%%%%%%%%%%%%%%%%%%%%%%%%%%%%%%%%%%%%%

\part{Lower Spins}

%%%%%%%%%%%%%%%%%%%%%%%%%%%%%%%%%%%%%%
\chapter{Asymptotic Symmetries of Lower Spins in Four Dimensions}\label{chap:basics}

An interesting class of gravitational systems is the one
characterized by the presence of an ideally isolated or weakly radiating mass distribution: for instance a black hole, a system of planets orbiting around a star or a pair of coalescing neutron stars. Intuitively, as one goes very far from the sources under consideration, the gravitational field induced by their presence becomes very weak, \emph{i.e.} the underlying spacetime metric reduces to the flat, Minkowski metric. A precise description for this class of systems is provided by the definition of  asymptotically flat spacetimes, first investigated in the sixties by Bondi, Metzner, van der Burg and Sachs \cite{BMS, Sachs_Waves}, and later reformulated in a more geometrical fashion by Penrose and Carter \cite{Penrose1, Penrose2, Carter}.

A natural question that arises in this context concerns the characterization of those diffeomorphisms that leave the set of asymptotically flat spacetimes invariant, \emph{i.e.} that map any isolated or weakly radiating system to another one. These transformations are, by definition, the asymptotic symmetries of asymptotically flat spacetimes and form the $BMS$ group, which comprises an infinite-dimensional enhancement of the Poincar\'e group.

As is usually the case when dealing with symmetry transformations, one can then wonder whether asymptotic symmetries give rise to nontrivial conserved quantities, \emph{i.e.} asymptotic charges.

In this chapter we begin by reviewing in detail the characterization of asymptotically flat spacetimes and the structure of the $BMS$ group in Section \ref{sec:asymptflat}. Section \ref{sec:charges} then offers an overview of the problem of associating conserved charges to asymptotic symmetries.

Although asymptotic symmetries were first introduced in the gravitational context, the problem of characterizing residual local gauge transformations that preserve assigned asymptotics is actually well-posed in any gauge theory. Indeed, as we will detail in Sections \ref{sec:EMD=4} and \ref{sec:YMD=4}, this asymptotic analysis already furnishes nontrivial results in spin-one gauge theories, namely Maxwell and Yang-Mills.

While the main focus of the discussion in this chapter is on the case of four spacetime dimensions, we will also include comments and explicit calculations concerning its higher-dimensional generalization, also in view of Chapters \ref{chap:spin1higherD} and \ref{chap:HSP}.

\section{Asymptotically Flat Spacetimes}\label{sec:asymptflat}
In any four-dimensional spacetime, it is always possible \cite{Sachs_Waves} to define locally a retarded time $u$, a radial coordinate $r$ and two angular coordinates $x^i=(\theta, \phi)$, with $i=1$, $2$, such that the spacetime metric can be written in the following form
\be\label{Sachsform}
ds^2 = e^{2\beta}(\tfrac{V}{r}du^2-2du\,dr)+ r^2l_{ij}(dx^i -\mathcal U^i du)(dx^j-\mathcal U^j du)\,,
\ee
where $\beta$, $V$ and $U^i$ are independent spacetime functions, while $l_{ij}$ is subject to $\sqrt{\mathrm{det}(l_{ij})}=\sin\theta$.
In total, the expression \eqref{Sachsform}, which as we shall see can be reached independently of the Einstein equations, is specified by six functions, so that the definition of the coordinates $u$, $r$, $\theta$ and $\phi$, discussed below, has the effect of eliminating four redundant functions that are present in a generic metric, \emph{i.e.} of performing a gauge-fixing. 

For instance, in the case of Minkowski space \eqref{Sachsform} is achieved globally by letting $t=u+r$ and $\mathbf x = r\,\mathbf n$, where $\mathbf n(\theta,\phi)=(\sin\theta\cos\phi, \sin\theta\sin\phi,\cos\theta)$ is the standard parametrization of the points on the unit sphere, in terms of which $ds^2 = -du^2-2du\,dr+r^2 (d\theta^2+\sin^2\theta d\phi^2)$. Namely, $\beta=0=\mathcal U^i$ and $V=-r$, while $l_{ij}=\gamma_{ij}$ is the metric on the Euclidean two-sphere. 

To see that \eqref{Sachsform} can be reached in a generic situation, consider a spacetime with a given metric $g_{ab}$ in some local coordinates $x^a$. There exists locally a function $u(x)$ such that its gradient $k_a=\partial_a u$ satisfies $k^ak_a=0$, \emph{i.e.} the hypersurfaces of constant $u$ are null. The integral curves of $k^a$, called \emph{rays}, are also geodesic, since $\nabla_a k_b=\nabla_a\nabla_b u=\nabla_b \nabla_a u=\nabla_b k_a$ and hence
$
k^b \nabla_b k^a = k^b \nabla^a k_b = \frac{1}{2}\nabla^a(k^b k_b)=0
$.
Introduce functions $\theta(x)$ and $\phi(x)$ by the condition that they be constant along rays, namely $k^a\partial_a\theta=0$ and $k^a\partial_a\phi=0$. 
Choosing $u$, $\theta$ and $\phi$, together with a fourth function $r$, yet to be determined, as coordinates, the components of the inverse metric satisfy $g^{uu}=k^ak_a=0$, $g^{u\theta}=k^a\partial_a\theta=0$  and $g^{u\phi}=k^a\partial_a\phi=0$, in such coordinates, and hence the determinant of the inverse metric reads
\be\label{determinanturtf}
-(g^{ur})^2[g^{\theta\theta}g^{\phi\phi}-(g^{\theta\phi})^2]= -(g^{ur})^2\mathrm{det}(g^{ij})\,.
\ee
Since the the Jacobian of the transformation $x^a=x^a(u,r,\theta,\phi)$ vanishes if and only if \eqref{determinanturtf} is zero, we require $\mathrm{det}(g^{ij})\neq0$ together with $g^{ur}\neq0$. We can then define $r$ by setting $\sqrt{\mathrm{det}(g^{ij})}=(r^2\sin\theta)^{-1}$. Adopting the notation
\be\label{inverseSachs}
g^{ab}=\left(
\begin{matrix}
0 & g^{ur} & 0\\
g^{ur} & g^{rr} & g^{ri}\\
0 & g^{ri} & g^{ij}
\end{matrix}
\right)
=\left(
\begin{matrix}
	0 & -e^{-2\beta} & 0\\
	-e^{-2\beta} & -\frac{V}{r}e^{-2\beta} & -\mathcal U^je^{-2\beta}\\
	0 & -\mathcal U^ie^{-2\beta} & \frac{1}{r^2}l^{ij}
\end{matrix}
\right)\,,
\ee 
one can invert $g^{ab}$ and solve for the metric components $g_{ab}$ obtaining \eqref{Sachsform}, where $l_{ij}$ denotes the inverse of $l^{ij}$.

To summarize, the above coordinates are characterized by the following properties:
\begin{itemize}
\item the hypersurfaces of constant $u$ are tangent to the local light-cone and the integral curves of $k^a = g^{ab}\partial_bu$ are null geodesics (light rays);
\item the coordinates $\theta$ and $\phi$ are constant along each ray and can be interpreted as optical angles,
\item the radial coordinate $r$ is the luminosity distance, which means that a spherical wavefront $u=\ $constant has a total measure of $4\pi r^2$, in view of the condition $\sqrt{\mathrm{det}(l_{ij})}=\sin\theta$, and hence its luminosity (or total power) $L$ is related to its mean flux (or intensity) $F$ by $L =4\pi r^2 F$.
\end{itemize}
Such coordinates are well-suited to describing gravitational radiation emitted by sources localized within a region of total radius $r_0$. A faraway detector placed at a fixed distance $r>r_0$ will characterize the properties of outflowing radiation by specifying at what (retarded) time $u$ it receives a signal and the direction $\theta$, $\phi$ from which it arrived.

Although the discussion of the present chapter will mostly focus on the case of four spacetime dimensions, let us note that the above definition of retarded coordinates can be straightforwardly extended to arbitrary dimension $D$. This can be achieved by specifying $D-2$ angular coordinates $x^i$ with $i=1$, $2$, $\ldots$ , $D-2$, which generalize the ordinary azimuthal and polar angles, $\theta$ and $\phi$. This leads to \eqref{Sachsform} provided one adopts the constraint
$ 
\sqrt{\mathrm{det}(l_{ij})}=\sqrt{\mathrm{det}(\gamma_{ij})}
$,
implicitly providing the definition of the luminosity-distance $r$, where $\gamma_{ij}$ is the metric on the Euclidean $(D-2)$-unit sphere. Again quoting the example of flat space, letting 
\be\label{retBondicoord}
t=u+r\,,\qquad  
x^I=r\, n^I\,,
\ee 
for $I=1,2,\ldots,D-1$, where $\mathbf n$ is a parametrization of the unit sphere in terms of the angular coordinates $x^i$, obtains
\be\label{MetricMinkD}
ds^2 = -du^2 - 2du\, dr + r^2 \gamma_{ij} \, dx^i\,dx^j\,,
\ee
with $\gamma_{ij}=\partial_i\mathbf n \cdot \partial_j \mathbf n$ the Euclidean metric on the sphere.
For future reference, we also list the Christoffel symbols for the flat connection in this coordinate system (more details are provided in Appendix \ref{app:Laplaciano})
\be \label{christoffel}
\Gamma{^i_{jr}}=\frac{1}{r}\,\delta{^i_j}\,,\qquad
\Gamma{^u_{ij}}=r\, \gamma_{ij}=-\,\Gamma{^r_{ij}}\,,\qquad
\Gamma{^k_{ij}}=\frac{1}{2}\, \gamma^{kl}\left(\partial_i \gamma_{lj}+\partial_j \gamma_{il}-\partial_l \gamma_{ij} \right)\, .
\ee

In a series of pioneering papers \cite{BMS, Sachs_Waves} (see \emph{e.g.} \cite{Barnich_BMS/CFT} for a more recent approach), employing these coordinates, the authors have studied the space of solutions to Einstein's equations that describes physical situations of the type under scrutiny, namely ideally isolated or weakly radiating gravitational systems, in four dimensions. 

The characterization of this type of solutions requires a specification of the manner in which the gravitational field becomes weak as one follows the outflowing radiation, \emph{i.e.}  $r\to\infty$ for $u$ fixed, as one approaches so-called \emph{future null infinity}. In other words one must clarify \emph{how} $g_{ab}$ given by \eqref{Sachsform} reduces to the Minkowski metric $ds^2 = -du^2-2du\,dr+r^2 \gamma_{ij}\,dx^idx^j$ in this limit and the spacetime becomes asymptotically flat. The most restrictive way of assigning the \emph{falloff conditions} characterizing asymptotically flat spacetimes that still allows for the presence of gravitational radiation is given, in four dimensions, by 
\be\begin{aligned}\label{BMSFalloffs}
\frac{V}{r}&=-1+\frac{2m_B}{r}+\mathcal O(r^{-2})\,,\\
\beta&=\frac{\bar \beta}{r^2}+\mathcal O(r^{-3})\,,\\
l_{ij}&=\gamma_{ij}+\frac{C_{ij}}{r}+\mathcal O(r^{-2})\,,\\
\mathcal U^i &= \frac{U^i}{r^2}+\mathcal O(r^{-3})\,,
\end{aligned}\ee 
where $m_B$, $\bar\beta$, $U^i$, $C_{ij}$ are functions of $u$, $\theta$ and $\phi$, and in addition $\gamma^{ij}C_{ij}=0$ in view of the requirement that $\sqrt{\mathrm{det}(l_{ij})}=\sin\theta$. The function $m_B(u,\theta,\phi)$ is called the Bondi mass aspect and, as we shall see later, is related to the notion of energy at a given retarded time. The definition of the angular momentum requires the discussion of a function appearing to subleading order in \eqref{BMSFalloffs}, the angular momentum aspect, and is here omitted for simplicity. 

In particular, the Einstein equations are satisfied to leading order provided that
\be\label{Bondieq}
U^i = -\frac{1}{2}D_jC^{ij}\,,\qquad
\partial_{u}m_B = - \frac{1}{8}N_{ij}N^{ij}+\frac{1}{4}D_iD_jN^{ij}\,,
\qquad
\bar\beta = - \frac{1}{32} C_{ij}C^{ij}\,,
\ee 
where $N_{ij}=\partial_u C_{ij}$ is called the Bondi news tensor and the angular indices are raised using $\gamma^{ij}$. This identifies the two independent components of the traceless tensor $C_{ij}$ as the free radiative boundary data, namely as the two independent graviton polarizations that propagate to infinity. This interpretation is further strengthened by the fact that the time dependence of $C_{ij}$ gives rise to the Bondi news and hence to energy flux carried by radiation.
Indeed, a linearized perturbation on Minkowski spacetime, $h_{ab}=g_{ab}-\eta_{ab}$, satisfying the above falloffs carries a nontrivial energy flux across a sphere of retarded time $u$ and radius $r$, proportional to
\be
\oint N_{ij}N^{ij} d\Omega\,,
\ee
where $d\Omega$ is the measure element on the Euclidean unit sphere. This result can be obtained by considering the hypersurface of constant $r$ and retarded time smaller than $u$, integrating $T_{ab}n^a t^b$ on such hypersurface (where $T_{ab}$ is the stress-energy tensor for such perturbation, $n_a=\partial_a r$ is its unit normal and $t^b=(\partial_u)^b$ is the background Killing vector associated to energy) and taking the derivative with respect to $u$. 

Consider now a hypersurface of fixed radius $r$. Such a manifold is parametrized by $u$ and the two angles on the sphere, and can be pictured as a set of static measuring devices covering the surface of a sphere at a fixed radius $r$. Intuitively, as $r$ becomes large,   the detectors are able to measure the properties of outgoing radiation very far away from the sources, thus approaching the future null ``boundary'' of the spacetime. This notion of a boundary, which at this level is not precise, can be made rigorous and covariant by means of a geometric definition of \emph{future null infinity}, $\mathscr I^+$, due to Penrose and Carter (see \cite{Geroch_Lectures, Ashtekar_Lectures} and \cite[Chapter 11]{Wald} for excellent reviews; a handy account is also available in \cite{carlo_tesi}), involving a conformal mapping that brings infinity to a finite distance. 

To illustrate the idea behind this construction, let us consider  Minkowski spacetime in $D$ dimensions, with metric \eqref{MetricMinkD}. One defines the coordinate $\Omega= 1/r$, in a region $r>r_0$, so that
\be
ds^2 = -du^2 + \frac{1}{\Omega^2}(2du\,d\Omega+\gamma_{ij}\,dx^i dx^j)\,,
\ee
and rescales the metric by letting $d\tilde s^2 = \Omega^2 ds^2$, namely 
\be
d\tilde s^2 = - \Omega^2 du^2 + 2du\,d\Omega+\gamma_{ij}\,dx^i dx^j\,.
\ee
Future null infinity $\mathscr I^+$ is then defined as the surface $\Omega = 0$ after the conformal rescaling, which can be therefore parametrized by $u, x^i$, has a degenerate metric
$
0\cdot du^2 + \gamma_{ij}\,dx^idx^j\,,
$
and represents the surface where all null trajectories of Minkowski space have their future endpoints. Following to this strategy one can in fact \emph{define} a four-dimensional spacetime $M$ to be asymptotically flat at null infinity if, roughly speaking, there exists a conformal isometry of $M$ to a spacetime $\tilde M$ whose boundary ``resembles'' Minkowskian $\mathscr I^+$ (we refer again to \cite{Geroch_Lectures, Ashtekar_Lectures,Wald} for a more precise and complete discussion). It can be shown that this definition reduces to the one we have given above in terms of falloff conditions \eqref{BMSFalloffs}, provided one adopts a suitable set of retarded coordinates.

Let us also mention that, while it has been successfully extended to spacetimes with arbitrary even dimensions \cite{Hollands_SI}, this geometric, covariant definition of null infinity is actually precluded in the case of odd-dimensional spacetimes \cite{Wald_NO}, in the presence of gravitational radiation, while the less formal approach adopted for instance in \cite{No_ST}, based on taking the limit $r\to\infty$ in the given coordinate system, appears to be more flexible in this respect, although not covariant, and better suited to highlighting the similarities between gravity and other gauge theories of lower or higher spin, when it comes to discussing asymptotic symmetries.

In the following, we will rely on the latter noncovariant notion of null infinity, still denoted $\mathscr I^+$, which suffices for many practical applications. According to the intuitive picture, future null infinity is therefore a hypersurface parametrized by $u$, and $x^i$ whose sections at constant $u$ are obtained by considering a sphere $S_{u,r}$ at fixed retarded time and radius $r$ and sending $r\to\infty$. The induced metric thereon is $- du^2+r^2 \gamma_{ij}dx^idx^j$ while the volume element is $r^{D-2}du\,d\Omega$,  where $d\Omega$ is the measure on the unit sphere, with the limit $r\to\infty$ to be taken at the very end of the calculations of physical quantities.

\subsection{The BMS group}
Asymptotic symmetries are defined as the large diffeomorphisms that preserve \eqref{Sachsform}  and \eqref{BMSFalloffs}, and hence map the space of asymptotically flat spacetimes to itself. More precisely, they are those diffeomorphisms that respect the coordinate conditions and the falloff conditions ensuring asymptotic flatness, while still acting nontrivially on the leading field components $C_{ij}$, $U^i$, $m_B$. Furthermore, two given asymptotic symmetries should be identified if their action only differs by a small diffeomorphism or, more formally, if they induce the same transformation on $\mathscr I^+$. 

Anticipating a bit the discussion of charges, we may state that large diffeomorphisms are characterized by the fact that they possess a finite and nonzero charge, while small ones are those that give rise to a vanishing charge.

Such symmetries are determined by solving the asymptotic Killing equation, namely imposing that the infinitesimal transformations
\be
\delta_\xi g_{ab}=\mathcal L_\xi g_{ab}\,,
\ee
respect the falloffs. 

In our case, we can start by imposing that $g_{rr}=g_{ri}=0$ and $g^{ij}g_{ij}=0$, which follow from \eqref{Sachsform}, namely $\mathcal L_\xi g_{rr}=\mathcal L_\xi g_{ri}=\mathcal L_\xi (g_{ij}g^{ij})=0$, which explicitly read
\be\begin{aligned}\label{akeprimetre}
g_{ar}\partial_r\xi^{a}&=0\\
g_{ar}\partial_i \xi^{a}+g_{ai}\partial_r\xi^a&=0\\
g^{ij}(\xi^a\partial_a g_{ij}+g_{a(i}\partial_{j)}\xi^a)&=0\,,
\end{aligned}\ee
where round brackets denote symmetrization of the indices. From the first equation, we read off $\partial_r\xi^u=0$, namely 
\be\label{xiu}
\xi^u=f(u,\theta,\phi)\,.
\ee 
The second equation in \eqref{akeprimetre} reads
$
\partial_r\xi^i = e^{2\beta} g^{ij}\partial_j f
$
so that
\be\label{xii}
\xi^i = - \partial_j f \int_r^{\infty}e^{2\beta} g^{ij}dr' + Y^i\,,
\ee
with $Y^i=Y^{i}(u,\theta, \phi)$. Using now the identities $g^{ij}\partial_r g_{ij}=4r^{-1}$, $g^{ij}\partial_ug_{ij}=0$ and $g^{ij}\partial_k g_{ij}=\gamma^{ij}\partial_k\gamma_{ij}$, which follow from the condition $\mathrm{det}(g_{ij})=r^4 \mathrm{det}(\gamma_{ij})$, the third equation in \eqref{akeprimetre} gives
\be\label{xir}
\xi^r=\frac{r}{2} \mathcal U^i \partial_i\xi^u-\frac{r}{2}D_i\xi^i\,,
\ee
where $D_i$ is the covariant derivative on the unit sphere associated with the Euclidean metric $\gamma_{ij}$. Substituting the falloff conditions \eqref{BMSFalloffs} into \eqref{xiu}, \eqref{xii} and \eqref{xir} we have
\be\begin{aligned}
\xi^u &= f(u,\theta,\phi)\,,\\
\xi^r &= - \frac{r}{2}D_iY^i(u,\theta,\phi)+\frac{1}{2}\Delta f(u,\theta,\phi)+\mathcal O(r^{-1})\,,\\
\xi^i &= Y^i(u,\theta,\phi)-\frac{1}{r} D^i f(u,\theta,\phi)+\mathcal O(r^{-2})\,,
\end{aligned}\ee
where $\Delta=D^iD_i$ and $D^i=\gamma^{ij}D_j$.
Imposing that $\mathcal L_\xi g_{ur}=\mathcal O(r^{-2})$ and $\mathcal L_\xi g_{ui}=\mathcal O(1)$ gives
\be
\partial_u f = \frac{1}{2} D_i Y^i\,,\qquad
\partial_u Y^i = 0\,,
\ee
so that $Y^i(u,\theta,\phi)=Y^i(\theta,\phi)$ is actually a vector on the sphere, while 
\be
\xi^u = T(\theta,\phi)+\frac{u}{2}D_i Y^i(\theta, \phi)+\mathcal O(r^{-1})\,.
\ee
Furthermore, $\mathcal L_\xi g_{ij}=\mathcal O(r)$ implies that $Y^i$ satisfies the conformal Killing equation on the two-sphere
\be\label{confKillingeq2}
D_{(i}Y_{j)}= \gamma_{ij} D\cdot Y\,,
\ee
where the dot denotes contraction with $\gamma_{ij}$. Finally, the last remaining equation $\mathcal L_{\xi}g_{uu}=\mathcal O(r^{-1})$ yields the constraint
\be\label{consequenceofKilling}
(\Delta+2)D\cdot Y =0\,.
\ee
The latter equation is actually identically true, since the Ricci tensor satisfies $R_{ij}=\gamma_{ij}$ on the Euclidean two-sphere and hence, taking two divergences of \eqref{confKillingeq2}, we have $\Delta D\cdot Y=D^i D^j(D_i Y_j + D_j Y_i)=2\Delta D\cdot Y + 2 D\cdot Y$.

To summarize, asymptotic symmetries of four-dimensional asymptotically flat spacetimes are given by 
\be\begin{aligned}\label{BMSor}
\xi^u&=T(\theta, \phi)+ \frac{u}{2}D\cdot Y(\theta, \phi)\,,\\
\xi^r&=\frac{1}{2}\Delta T(\theta, \phi)-\frac{u+r}{2}D\cdot Y(\theta, \phi) + \mathcal O(r^{-1})\,,\\
\xi^i &= Y^i(\theta, \phi)-\frac{1}{r}D^i T(\theta, \phi)-\frac{u}{2r}D^iD\cdot Y(\theta, \phi)+ \mathcal O(r^{-2})\,,
\end{aligned}\ee
where $T(\theta, \phi)$ is an arbitrary function of the angles, while $Y^{i}$ satisfies the conformal Killing equation \eqref{confKillingeq2} on the Euclidean unit sphere. 

Since the spacetime becomes asymptotically Minkowski, it is expectable that such a set of asymptotic symmetries contain the ordinary isometries of flat space: translations, rotations and boosts. This is indeed the case. For instance, the exact Killing vector associated to a translation in Minkowski space, $a^\mu= (a^0, \mathbf a)$ in Cartesian coordinates $x^\mu=(x^0,\mathbf x^I)=(t, \mathbf x)$ with $I=1$, $2$, $3$, reads in covariant retarded components $a_u=a_0=-a^0$, $a_r = a_0 + \mathbf n \cdot \mathbf a$, $a_i=r \partial_i\mathbf n \cdot \mathbf a$. Switching to contravariant components 
\be
a^u = a^0-\mathbf n \cdot \mathbf a\,,\qquad
a^r = \mathbf n \cdot \mathbf a\,,\qquad
a^i = \frac{1}{r}\partial^i\mathbf n \cdot \mathbf a\,,
\ee
so that, by comparison with \eqref{BMSor}, 
\be\label{translations}
T= a^0-\mathbf n \cdot \mathbf a\,,\qquad Y^i = 0\,,
\ee  
once we recall that, on the Euclidean two-sphere (see Appendix \ref{app:Laplaciano}), $\Delta \mathbf n = -2\mathbf n$. 
For an infinitesimal rotation/boost given by $l_\mu = \omega_{\mu\nu}x^\nu$ in Cartesian coordinates, with $\omega_{\mu\nu}=-\omega_{\nu\mu}$, we have $l_u=l_0=\omega_{0I}r\, n^I$ $l_r= n^{I}\omega_{I0}u$, $l_i=r \partial_i n^{I}\omega_{I0}(u+r)+r^2\partial_i n^I \omega_{IJ} n^J$, so
\be
l^u=u n^{I}\omega_{0I}\,,\qquad
l^r=-(u+r) n^{I}\omega_{0I}\,,\qquad
l^i=-\partial^i n^{I}\omega_{0I}+\partial^i n^I \omega_{IJ} n^J-\frac{u}{r}\partial^i n^{I}\omega_{0I}\,,
\ee
where we recognize 
\be\label{rotations/boosts}
T=0\,,\qquad
Y^i=-\partial^i n^{I}\omega_{0I}+\partial^i n^I \omega_{IJ} n^J\,.
\ee To check that $Y^i$ is indeed a conformal Killing vector for the Euclidean two-sphere, one need recall that (see Appendix \ref{app:Laplaciano}) $D_{i}D_{j}\mathbf n=-\gamma_{ij}\mathbf n$.

We have checked that the set asymptotic symmetries \eqref{BMSor} contains the Poincar\'e transformations. However, it is in fact much bigger. First, it contains indeed transformations parametrized by an arbitrary function $T(\theta,\phi)$, which give an infinite-dimensional analog of translations \eqref{translations}, termed \emph{supertranslations}. Second, one may note that, locally, the conformal Killing equation admits two infinite-dimensional families of solutions, generalizing the ordinary rotations and boosts \eqref{rotations/boosts}, called \emph{superrotations} \cite{Barnich_BMS/CFT, Barnich_Revisited, Barnich_Superrotations}. These are conveniently described by introducing the stereographic coordinates 
\be\label{stereo}
z=e^{i\phi}\cot(\theta/2)\,,\qquad \bar z=e^{-i\phi}\cot(\theta/2)\,,
\ee 
in terms of which the Euclidean metric on the sphere takes the form $ds^2=2\gamma_{z\bar z}dz d\bar z$, with $\gamma_{z\bar z}=2/(1+z\bar z)^2$, and the conformal Killing equation reads
\be\label{holoantiholo}
\partial_z Y^{\bar z}=0\,,\qquad
\partial_{\bar z} Y^z=0\,,
\ee
so that for any holomorphic (resp. antiholomorphic) function $F(z)$ (resp. $\tilde F(\bar z)$), local solutions are given by $Y^z(z,\bar z)=F(z)$ and $Y^{\bar z}(z,\bar z)=\tilde F(\bar z)$. Globally well-defined solutions are selected by requiring that the Laurent expansion
\be
F(z)=\sum_{n\in\mathbb Z} c_n z^n
\ee
be nonsingular near $z=0$ and $w=\frac{1}{z}=0$, and similarly for $\tilde F(\bar z)$. This leaves us with 
\be
Y^z=c_0 + c_1 z + c_2 z^2\,,\qquad
Y^{\bar z}=\tilde c_0 + \tilde c_1 \bar z + \tilde c_2 \bar z^2\,,
\ee
which can be recast in the form \eqref{rotations/boosts} by noting that $\mathbf n=(z+\bar z, -i(z- \bar z), z\bar z-1)/(1+z\bar z)$, with 
\be\begin{aligned}
2c_0&=\omega_{01}+i\omega_{02}-(\omega_{13}-i\omega_{23})\,,\\
c_1&=\omega_{03}+i\omega_{12}\,,\\
2c_2&=-(\omega_{01}+i\omega_{02})+\omega_{13}-i\omega_{23}
\end{aligned}\ee 
and $\tilde c_0$, $\tilde c_1$, $\tilde c_2$ given by complex conjugation,
and correspond to those infinitesimal superrotations that can be exponentiated to global rotations and boosts. Indeed, the algebra of globally well-defined conformal transformations of the Euclidean sphere is isomorphic to the Lorentz algebra $so(3,1)$. The analogous statement for the associated groups is that the group of global conformal transformations, isomorphic to $SL(2,\mathbb C)/\mathbb Z_2$, is also isomorphic to the proper, orthochronous Lorentz group $SO(3,1)$.  
 
The asymptotic symmetries induced on $\mathscr I^+$ can be obtained by projecting \eqref{BMSor} on a surface of fixed $r$\footnote{More precisely, by considering its action on $r$-independent functions.} and taking the limit $r\to\infty$, for fixed $u$, and are given by the vectors
\be\label{BMSonScri}
	\xi^u=T+ \frac{u}{2}D\cdot Y\,,\qquad
	\xi^i= Y^i\,.
\ee
Such vectors are closed under Lie bracket and give rise to the following algebra:
\be
[\xi_1, \xi_2]^u = Y_{[1}\cdot D T_{2]}+ \frac{1}{2} T_{[1}D\cdot Y_{2]}+\frac{u}{2}D\cdot [Y_1,Y_2]\,,\qquad
[\xi_1,\xi_2]^i = [Y_1, Y_2]^i\,,
\ee
so that, schematically,
\be\label{commBMS}
[(T_1, Y_1),(T_2, Y_2)]=\Big(Y_{[1}\cdot D T_{2]}+ \frac{1}{2} T_{[1}D\cdot Y_{2]} , [Y_1, Y_2]\Big)\,.
\ee
The vector fields \eqref{BMSonScri}, parametrized by angular functions $T$ and by conformal Killing vectors $Y^i$ on the sphere, together with the commutation relations \eqref{commBMS} define the Bondi-Metzner-Sachs ($BMS$) algebra. A key structural feature of this algebra is the existence of an infinite-dimensional Abelian subalgebra given by supertranslations $\xi = T(\theta, \phi)\partial_u$, which is in fact a Lie ideal since by \eqref{commBMS} the commutator of a supertranslation with any transformation is again a supertranslation. At the local level, the subalgebra given by superrotations is also infinite-dimensional and is in fact given by the direct sum of two Virasoro algebras (one for the holomorphic sector and one for the antiholomorphic sector \eqref{holoantiholo}). Globally well-defined superrotations \eqref{rotations/boosts}, as we have seen, instead give rise to the Lorentz algebra.

At the group level, supertranslations give rise to an infinite-dimensional Abelian normal subgroup $ST$ of the full $BMS$ group. Supertranslations contain ordinary translations \eqref{translations} as a four-dimensional subgroup and enter the full asymptotic symmetry group as follows:
\be
BMS = ST \rtimes SO(3,1)\,.
\ee 
Denoting by $x^i=x^i(x')$ a global conformal transformation, in stereographic coordinates $x^1=z$, $x^2=\bar z$, 
\be
z = \frac{\alpha z' + \beta }{\gamma z' + \delta}
\ee
where $\alpha$, $\beta$, $\gamma$ and $\delta$ are complex numbers satisfying $\alpha\delta-\beta\gamma=1$, such that
\be
 \frac{\partial x^k}{\partial x'^i}\gamma_{kl}\frac{\partial x^l}{\partial x'^j}=F(x')^2\gamma_{ij}\,,
\ee 
we can express a global $BMS$ transformation as \cite{Schmidt_Walker_Sommers}
\be
u=u'F(x')-T(x')\,,\qquad
x^i=x^i(x')\,.
\ee 
As a final remark, let us mention that, while the translation subgroup can be characterized  as the unique four-dimensional normal subgroup of the $BMS$ group, the homogeneous Lorentz transformations are not similarly unique, since any two Lorentz subgroups of $BMS$ that differ by a supertranslations are isomorphic \cite{Sachs_Symmetries}.

\subsection{A linearized approach}

From the perspective of linearized gravity (\emph{i.e.} of a generic spin-2 massless field) the form \eqref{Sachsform} for the metric tensor and the falloffs \eqref{BMSFalloffs} are equivalent to the following choice of boundary conditions: to leading order in $1/r$,
\be\label{definizione_Bondi_gaugerip}
h_{ab}dx^a dx^b = \frac{2m_B}{r}du^2 - 2U_i du dx^i+ r C_{ij} dx^i dx^j+\cdots\,,
\ee
with $\gamma^{ij}C_{ij}=0$. In particular $h_{rr}=h_{ri}=0$, while $h_{ur}=\mathcal O(r^{-2})$ and
\be\label{Bondigaugelin}
h_{uu}=\frac{2m_B}{r}+\mathcal O(r^{-2})\,,\qquad
h_{ui}=-U_i+\mathcal O(r^{-1})\,,\qquad
h_{ij}=r C_{ij}+\mathcal O(1)\,.
\ee
In stereographic coordinates, the trace condition takes a particularly simple form: $\gamma^{ij}h_{ij}=2h_{z\bar z}/\gamma_{z\bar z}=0$, so $h_{z\bar z}=0$.
We would like to recover the $BMS$ algebra from the linearized gauge transformations $\delta_\xi h_{ab}=\nabla_{(a}\xi_{b)}$, where $\nabla_a$ denotes the background (flat) connection (see \eqref{christoffel}), that preserve \eqref{Bondigaugelin}. Actually, we shall do so while keeping the dimension $D$ of the spacetime formally arbitrary, as this only requires a minimal modification of the calculation.
 
From $\nabla_{(r}\xi_{r)}=0$ we obtain $\partial_r \xi_r = 0$, hence
\be\label{xiruindep}
\xi_r = \xi_r(u, x^i),
\ee
while from $\nabla_{(u}\xi_{r)}=\mathcal O(r^{-2})$ we have $\partial_r \xi_u + \partial_u \xi_r =\mathcal O(r^{-2})$ and thus
\be\label{ruuur}
\partial_r \partial_u \xi_u + \partial_u^2 \xi_r =\mathcal O(r^{-2})\,.
\ee
However, $\nabla_{(u}\xi_{u)}=\mathcal O(r^{-1})$ also requires $\partial_u \xi_u = \mathcal O(r^{-1})$ and hence
\be\label{uxiu}
\partial_r\partial_u \xi_u = \mathcal O(r^{-2}),
\ee
which, together with \eqref{ruuur}, implies
\be
\partial^2_u \xi_r = \mathcal O(r^{-2})\,.
\ee
Since $\xi_r$ is actually $r$-independent \eqref{xiruindep}, we then have 
\be
\xi_r = - T(x^i) - u F(x^i)\,.
\ee
Integrating the equation $\partial_r \xi_u + \partial_u \xi_r=\mathcal O(r^{-2})$ with respect $r$ we get 
\be
\xi_u = r F(x^i)-S(x^i) + \mathcal O(r^{-1}),
\ee
where $S$ does depend on $u$, by $\partial_u\xi_u=\mathcal O(r^{-1})$.
Then the equation $\nabla_{(r}\xi_{i)}=0$ takes the form
\be
\partial_r \xi_i + \partial_i \xi_r - \frac{2}{r}\xi_i =0
\ee
and substituting the above solutions reads
\be
\partial_r \xi_i- \frac{2}{r}\xi_i - D_i T - u D_i F = 0.
\ee
Looking for a solution in the form of a power series in $r$, one readily sees that 
\be
\xi_i = -r D_i T - ru D_i F + r^2 Y_i(u,z,\bar z).
\ee
for some vector field $Y^i(u,x^j)$ (and $Y_i = \gamma_{ij}Y^j$).

From the equation $\nabla_{(u}\xi_{i)}=\mathcal O(1)$, one immediately obtains
\be
\partial_u Y_i = 0\,,
\ee
so $Y_i = Y_i(x^j)$. 
Now we are left with the conditions $\nabla_{(i}\xi_{j)}=\mathcal O(r)$ and $\gamma^{ij}\nabla_i\xi_j=0$. The traceless projection of the former gives
\be\label{Killingeq}
D_{(i}Y_{j)}-\frac{2}{D-2}\gamma_{ij}D\cdot Y=0\,,
\ee
which is the conformal Killing equation, while the latter instead implies
\be\label{Ddotxi}
\gamma^{ij}D_i\xi_j - (D-2)r(\xi_u-\xi_r)=0\,.
\ee

Up to now we have
\be\begin{aligned}
	\xi_r &= -T - u F\\
	\xi_u &= r F-S+\mathcal O(r^{-1})\\
	\xi_i &= -r D_iT - ur D_iF + r^2 Y_i\,,
\end{aligned}\ee
where $T$, $S$ and $F$ are arbitrary functions on the sphere, while $Y^i$ is a conformal Killing vector. Upon substituting into \eqref{Ddotxi}, we get
\be
r[-\Delta T + (D-2)(S-T)]-ur(\Delta+D-2)F+r^2[D\cdot Y-(D-2)F]=\mathcal O(1)\,.
\ee
This equation can be satisfied only if the coefficient of each independent monomial $r$, $r^2$ and $ur$ is zero: this requires
\be
S = \frac{1}{D-2} (\Delta+D-2) T
\,,\qquad
F = \frac{1}{D-2} D\cdot Y,
\ee
while the last remaining condition, $(\Delta+D-2)D\cdot Y=0$, is identically satisfied as a consequence of the conformal Killing equation, as can be verified by taking two divergences of \eqref{Killingeq} and recalling $[D_i, D_j]v^k=R\indices{^k_{lij}} v^l$, with $R_{ijkl}=\gamma_{ik}\gamma_{jl}-\gamma_{il}\gamma_{jk}$ (see also eq. \eqref{Ridentities}).

 To sum up, the residual gauge freedom is parametrized by
\be \begin{aligned}
	\xi_r &= - T - \frac{u}{D-2}D\cdot Y\,,\\
	\xi_u &= \frac{r}{D-2} D\cdot Y - \frac{1}{D-2} (\Delta+D-2)T + \mathcal O(r^{-1})\,,\\
	\xi_i &= - r D_i T  - \frac{ur}{D-2} D_i D\cdot Y + r^2 Y_i\,.
\end{aligned} \ee
Equivalently, in contravariant components 
\be\begin{aligned}\label{BMSlinonScri}
	\xi^u &= T + \frac{u}{D-2} D\cdot Y \,,\\
	\xi^r &= \frac{1}{D-2}\Delta T - \frac{u+r}{D-2} D\cdot Y + \mathcal O(r^{-1}),\\
	\xi^i &= -\frac{1}{r}D^iT + Y^i - \frac{u}{(D-2)r}D^i D\cdot Y. 
\end{aligned}\ee
In particular, these vector fields induce the same asymptotic symmetries on $\mathscr I^+$, in the relevant case $D=4$, as in \eqref{BMSonScri} (compare also with \eqref{BMSor}).

Let us note that, as far as the above calculation of the asymptotic symmetries in the linearized theory is concerned, it would not have been too restrictive to also impose the condition $h_{ur}=0$. With this choice, equation \eqref{BMSlinonScri} would simply hold sharply and not up to $\mathcal O(r^{-1})$. One might be tempted to set it to zero or conclude that it can be set to zero by means of a small gauge transformation, and indeed inspection of the equations of motion \eqref{Bondieq} implies that the leading $\mathcal O(r^{-2})$ component $\bar\beta$ of $h_{ur}$ is actually zero in the linearized theory in four dimensions.

Although the linearized theory suffices for the purpose of establishing the falloff conditions and deriving the asymptotic symmetries, by its very nature it is insensitive to self-interaction effects, such as the energy flux to null infinity, which are instead captured by the nonlinear description as we shall see explicitly in the next section.

\section{Bondi Energy and Charges}\label{sec:charges}
Having introduced the concept of asymptotic symmetry in an asymptotically flat spacetime, it is natural to ask whether it is possible to find conserved quantities associated to these symmetries. This question, and more generally the subtle issue of defining conserved charges in general relativity and gauge theories, has received an extensive attention in the literature \cite{Barnich-Brandt,Barnich_Charge, Ashtekar_Lectures, Ashtekar-Streubel}, also in connection with black hole entropy \cite{Iyer:1994ys} and  in the context of covariant phase space methods \cite{Wald-Zoupas}. Here, we will follow an approach based on the application of Noether's theorem \cite{Avery_Schwab}, commenting on the subtleties that arise in this context along the way.

\subsection{The Noether two-form}
One can in principle associate a conserved charge to {any} infinitesimal diffeomorphism according to the Noether theorem \cite{Avery_Schwab}. Our starting point is the Einstein-Hilbert action on a generic $D$-dimensional spacetime $M$,
\be
S = \frac{1}{2k^2_{D}}\int_M R\, \omega\,,
\ee 
where $k^2_D=(D-2)\Omega_{D-2}G$, with $\Omega_{D-2}$ the area of the Euclidean $(D-2)$-sphere (\emph{i.e.} the solid angle in $D-1$ space dimensions), and $\omega$ denotes the spacetime volume form, in local coordinates $\omega=\sqrt{-g}\,dx^0\wedge dx^1 \wedge \cdots \wedge dx^{D-1}$. This action is obviously invariant under the symmetry variation $\delta_\xi g_{ab}=\nabla_{(a}\xi_{b)}$ for any vector $\xi$, since
\be\label{EHinv}
\delta_\xi(R\,\omega)=\mathcal L_{\xi}(R\,\omega) = d(R \,\imath_\xi\omega)=\nabla_a (R\xi^a)\omega\,.
\ee 
On the other hand, under a generic variation of the (inverse) metric, we have
\be\begin{aligned}\label{Palatiniinv}
\delta (R\,\omega) &=\Big(R_{ab}-\frac{1}{2}g_{ab}R\Big)\omega\,\delta g^{ab}+\nabla_a\theta^a\omega\,,\\
\theta^a &= \nabla_b\delta g^{ab}-g_{cd}\nabla^a\delta g^{cd}\,,
\end{aligned}\ee
with $\theta^a$ the Palatini surface term.
Comparing \eqref{EHinv} and \eqref{Palatiniinv}, we obtain the current
\be
j^a_\xi =\frac{1}{2k^2_D} \sqrt{-g}(\theta^a_\xi - \xi^a R)\,,
\ee
which is conserved on shell, $\partial_a j^a_\xi=0$. This can be further simplified by rewriting
\be\begin{aligned}
\theta^a_\xi &= \nabla_b\delta_\xi g^{ab}-g_{cd}\nabla^a\delta_\xi g^{cd}\\
&=-\nabla_b\nabla^{(a}\xi^{b)}+2\nabla^a\nabla \cdot \xi\\
&=\nabla_b \nabla^{[b}\xi^{a]} + 2 R^{ab}\xi_b\,,
\end{aligned}\ee
so that
\be
j_\xi^a = \sqrt{-g}
\left[\nabla_b \nabla^{[b}\xi^{a]}+2\Big(R^{ab}-\frac{1}{2}g^{ab}R\Big)\xi_b
\right]
\ee
and, employing again the Einstein equations in the vacuum,
\be\label{Komarform}
j_\xi^a=\partial_b \kappa_\xi^{ab}\,,\qquad
\kappa_\xi^{ab}= \frac{1}{2k^2_D}\sqrt{-g}\nabla^{[b}\xi^{a]}\,,
\ee
where one explicitly sees that, in accordance with Noether's second theorem, the conserved current associated to a local symmetry is equal, on shell, to the divergence of an antisymmetric rank-two tensor. The continuity equation $\partial_aj_\xi^a=0$ becomes then trivial.

Equation \eqref{Komarform} also holds if one takes into account the introduction of the Gibbons-Hawking-York term needed in order to make the Einstein-Hilbert variational problem well-posed:
\be
S =\frac{1}{2k^2_D} \int_M R\, \omega + \frac{1}{k^2_D} \int_{\partial M} K\, \bar\omega\,,
\ee  
where $\bar\omega$ is the induced volume form on $\partial M$ and $K$ is the mean extrinsic curvature thereon (see \cite[Appendix E]{Wald}). Indeed, the introduction of the boundary term modifies the Lagrangian only by a total derivative (or an exact $D$-form) $R\,\omega\mapsto R\,\omega+d\alpha$ and hence does not alter the Noether current:
\be\begin{aligned}
\delta_\xi(R\,\omega+d\alpha)&=d\imath_\xi(R\,\omega+d\alpha)=\nabla_a(R\xi^a)\omega+d\imath_\xi d\alpha\,,\\
\delta(R\,\omega+d\alpha)&=\Big( R_{ab}-\frac{1}{2}g_{ab}R\Big)\delta g^{ab}\omega+\nabla_a\theta^a\omega+d\delta\alpha\,,
\end{aligned}\ee
but $d\delta_\xi\alpha=d\mathcal L_\xi\alpha= d\imath_\xi d\alpha$, by Cartan's formula, and hence the contribution due to $d\alpha$ cancels out in the calculation of $j_\xi^a$.

The expression \eqref{Komarform} is also correct if one adds to the Einstein-Hilbert action a cosmological constant term. For a scalar field $\phi$ coupled to gravity,
\be
S = \int_M \left[
\Big(\frac{1}{2k^2_D}-\frac{\lambda}{2}\phi^2\Big)R
-\frac{1}{2}\nabla_a\phi\nabla^a\phi-\frac{m^2}{2}\phi^2
\right]\omega\,,
\ee 
a calculation similar to the one presented above yields
\be
\kappa_\xi^{ab}=\sqrt{-g}\left[\Big(\frac{1}{2k^2_D}-\frac{\lambda}{2}\phi^2\Big)\nabla^{[b}\xi^{a]}+\lambda(\xi^a\nabla^b\phi^2-\xi^b\nabla^a\phi^2)\right]\,,
\ee
which reduces to \eqref{Komarform} for minimal coupling $\lambda=0$ (the case $\lambda= \tfrac{D-2}{4(D-1)}$, $m=0$ is also of interest due to its conformal invariance).
In the case of Einstein-Maxwell theory, with Lagrangian density $R-\frac{1}{4}F_{ab}F^{ab}$, instead
\be
\kappa_\xi^{ab}=\sqrt{-g}\Big(\frac{1}{2k^2_D}\nabla^{[b}\xi^{a]}+\xi^cA_c F^{ba}\Big)\,.
\ee 

Formally, one may thus define a conserved quantity associated to any infinitesimal diffeomorphism $\xi$ by integrating either $j_\xi^a$ on a Cauchy hypersurface $\Sigma$ or, equivalently, $\kappa_\xi^{ab}$ on its boundary $\partial\Sigma$:
\be\label{Noethercharge}
\mathcal Q_\xi = \int_{\Sigma} j_\xi^a d\Sigma_a = \oint_{\partial\Sigma} \kappa_\xi^{ab} d\sigma_{ab}\,.
\ee
The evaluation of the integral over $\Sigma$, or equivalently of the limit implicitly involved in the definition of $\partial\Sigma$, is in general a nontrivial issue. For instance, it is clear that $\mathcal Q_\xi$ is zero if $\xi$ is a \emph{small} diffeomorphism whose action is localized in a compact region, since in that case the \emph{surface charge} 
\be\label{surfacecharge}
\mathcal Q_\xi[\sigma]=\oint_{\sigma} \kappa_\xi^{ab} d\sigma_{ab}\,,
\ee 
obtained integrating over a closed $(D-2)$-surface $\sigma$, vanishes as soon as $\sigma$ is taken outside that region. On the other hand, \eqref{Noethercharge} is obtained as the limit of \eqref{surfacecharge} where $\sigma$ is taken towards the boundary of the spacetime.
 
\subsection{Global charges} 
Suppose $\xi$ is instead an exact Killing vector for a given solution of interest, $\nabla_{(a}\xi_{b)}=0$, namely a \emph{global} symmetry. This implies $\nabla\cdot\xi=0$ and $\Box\xi^a - R_{ab} \xi^b=0$, therefore 
\be
\partial_b \kappa_\xi^{ab} = \frac{1}{2k^2_D} \sqrt{-g}\,\nabla_b \nabla^{[b}\xi^{a]} =
\frac{1}{k^2_D}\sqrt{-g}\,R^{ab}\xi_b\,.
\ee
By the Einstein equations with matter,
\be
R_{ab}=k^2_D\Big(T_{ab}+\frac{1}{2-D}g_{ab}T\Big)\,,
\ee
so that the charge \eqref{Noethercharge} reads
\be\label{Qstress}
Q_\xi = \int_{\Sigma} \sqrt{-g} \left(T_{ab}+\frac{1}{2-D}g_{ab}T\right)\xi^a d\Sigma^b\,,
\ee
with  $T=g^{ab}T_{ab}$.
In the vacuum, the integrand on the right-hand side of the previous equation vanishes, implying that the charge integral \eqref{Noethercharge} of $j^a_\xi$ over the Cauchy hypersurface $\Sigma$ only receives contributions from the regions of $\Sigma$ where stress-energy is present. Equivalently, the surface charge \eqref{surfacecharge} is actually independent of the specific closed surface on which it is performed as long as $\sigma$ is deformed without crossing any source. 

The simplest example where this situation occurs is provided by the Schwarzschild solution,
\be
ds^2 = -f(r)dt^2+\frac{dr^2}{f(r)}+r^2\gamma_{ij}dx^idx^j\,,\qquad
f(r)=1-\frac{C}{r^{D-3}}\,,
\ee
where $t^a=(\partial_t)^a$ is an exact time translation isometry. Integrating \eqref{Komarform} on a sphere $S_{t,r}$ at fixed time $t$ and radius $r>C$, we have
\be
\mathcal Q_t =
\int_{S_{t,r}}\kappa^{tr}r^{D-2}d\Omega= \frac{1}{2k^2_D}\int_{S_{t,r}}(\tfrac{1}{f}\Gamma^{r}_{tt}+f\Gamma^t_{rt})r^{D-2}d\Omega
\ee
but $\Gamma^{r}_{tt}=\tfrac{1}{2}f\partial_rf$ and $\Gamma^t_{rt}=\tfrac{1}{2f}\partial_rf$, which gives
\be\label{MassBH}
\mathcal Q_t = \frac{C(D-3)}{2G(D-2)}\,.
\ee
The constant $C$ is determined by matching with the Newtonian potential in a regime of weak gravitational fields; a faraway test particle, $\frac{C}{r^{D-3}}\ll1$, moving radially in such a field at nonrelativistic speed will obey, by the geodesic equation, $\ddot r + \Gamma^{r}_{tt}=0$, hence
\be
\ddot r = - \partial_r\left(-\frac{C}{2r^{D-3}}\right)\,.
\ee 
The nonrelativistic interaction potential between a source of mass $M$ and a test mass must match Newton's formula $-\frac{GM}{r^{D-3}}$ and hence $C={2GM}$. Going back to \eqref{MassBH}, we see that the Noether charge equals 
\be\label{BHoff}
\mathcal Q_t = \frac{D-3}{D-2}\,M\,.
\ee

We may additionally cross-check the rather awkward normalization factor by performing the static Newtonian limit directly in \eqref{Noethercharge}. In this limit, the only nonvanishing component of the stress-energy tensor is $T_{00}=\rho$ in Cartesian coordinates, $\rho$ denoting the mass density. Then, \eqref{Qstress} evaluated on a slice of constant time reads 
\be\label{normoff}
\mathcal Q_t=\frac{D-3}{D-2}\int \rho\, d\mathbf x\,,
\ee 
as expected. Equivalently, for a static linearized fluctuation $g_{00}=-1+h_{00}$, the Einstein equations give $\Delta_{\mathbb R^3} h_{00}= 2k^2_D \frac{D-3}{D-2} \rho$,  and hence $h_{00}=-\frac{2GM}{r^{D-3}}$, twice the Newton potential, in the case of a particle of mass $M$ sitting in the origin $\rho(\mathbf x)=M\delta(\mathbf x)$.

More generally, for any stationary spacetime, where there exists a time-like killing vector $\xi^a$, one can define the energy content as the integral
\be\label{Komartrue}
M[\sigma]=\frac{1}{2(D-3)G\Omega_{D-2}}\int_\sigma \nabla^{[b}\xi^{a]} d\sigma_{ab}
=\frac{D-2}{D-3}\int_\sigma \kappa_\xi^{ab}d\sigma_{ab}\,,
\ee 
evaluated on any $(D-2)$-surface $\sigma$ enclosing all the sources. This quantity, whose definition takes into account and solves the above normalization issues by introducing the correct prefactor, is called the Komar mass of the spacetime. In a static spacetime, it can be understood in terms of the total force that must be exerted on a unit surface mass density distributed over a sphere enclosing all sources in order to hold it in place.

Taking $\sigma=S_{u,r}$ to be a sphere of fixed retarded time $u$ and large radius $r$ in retarded coordinates, the above discussion ensures that the Komar mass evaluated on $S_{u,r}$ is independent of $r$ and of $u$. Thus, the energy can actually be calculated on a section of $\mathscr I^+$, by taking the limit $r\to\infty$, and will be observed to be constant in $u$ by asymptotic measurements on $\mathscr I^+$, as should be the case since in a stationary spacetime energy should be strictly conserved by the absence of radiation.

\subsection{Asymptotic charges}

The situation is potentially more interesting in the case of \emph{asymptotic} symmetries, \emph{i.e.} the solutions of the asymptotic Killing equation. First, being large, these diffeomorphisms are candidates to yielding nontrivial surface charges \eqref{surfacecharge} at infinity. Second, the independence on $r$ and $u$ of the corresponding surface charges is not guaranteed in general (in contrast with the case of exact isometries). This on the one hand requires to check the convergence of the limit $r\to\infty$ case by case, and on the other hand  leaves open the possibility of describing charge leak due to radiation, \emph{i.e.} a dependence of the charges on retarded time.

In the case of the asymptotic symmetries of four dimensional asymptotically flat spacetimes associated to global Poincar\'e transformations \eqref{translations}, \eqref{rotations/boosts} one is able to verify that the energy and angular momentum surface charges admit a finite and nonvanishing limit $r\to\infty$ and that they display a nontrivial dependence on retarded time $u$ due to the presence of radiation, namely due to the energy and angular momentum fluxes carried by outgoing gravitational waves.

For a generic vector $\xi^a=\xi^u(\partial_u)^a+\xi^r(\partial_r)^a+\xi^i(\partial_i)^a$, the surface charge defined by the integral of $\kappa_\xi^{ur}$ \eqref{Komarform} on a sphere $S_{u,r}$ of fixed $u$ and $r$ reads, recalling \eqref{inverseSachs},
\be\begin{aligned}\label{Chargexigen}
\mathcal Q_{\xi}[S_{u,r}]
= \frac{r^2}{16\pi G}\int_{S_{u,r}}
e^{-2\beta}
\Big[
&
\left(-\partial_u-\tfrac{V}{r}\partial_r-\mathcal U^i D_i\right)\xi^u
+
\left(\Gamma^r_{ru}-\Gamma^u_{uu}-\mathcal U^i\Gamma^u_{ui}\right)\xi^u\\
&
+\partial_r\xi^r
+\Gamma^{r}_{rr}\xi^r +
\left(\Gamma^{r}_{ri}-\Gamma^u_{ui}-\mathcal U^j\Gamma^u_{ij}\right)
\xi^i
\Big]d\Omega\,,
\end{aligned}\ee
where we have used that the Christoffel \cite{Barnich_BMS/CFT} satisfy $\Gamma^{u}_{ar}=0$.

Restricting to the case of time translation asymptotic symmetry, given by $T=1$, $Y^i=0$ in \eqref{BMSonScri}, namely $\xi^a=(\partial_u)^a$, this surface charge gives
\be
\mathcal Q_\xi[S_{u,r}]= 
\frac{r^{2}}{16 \pi G} \int_{S_{u,r}} e^{-2\beta} \left(\Gamma^r_{ru}-\Gamma^u_{uu}-\mathcal U^i \Gamma^u_{iu}\right)d\Omega\,.
\ee
Evaluating the Christoffel symbols \cite{Barnich_BMS/CFT} and taking into account the falloffs \eqref{BMSFalloffs}, we see that
\be\label{Christoffelfalloff}
\Gamma^r_{ru}=\frac{m_B}{r^2}+\mathcal O(r^{-3})\,,\qquad
\Gamma^u_{uu}=-\frac{m_B}{r^2}+2\partial_u\bar \beta+\mathcal O(r^{-3})\,,\qquad
\Gamma^u_{iu}=\mathcal O(r^{-3})
\ee
and obtain the following expression 
\be\label{Bondi+spur}
\mathcal Q_{t}[S_{u,r}] = \frac{1}{8\pi G}\int_{S_{u,r}} (m_B-\partial_u\bar\beta)\,d\Omega + \mathcal O(r^{-1})\,,
\ee
where we may use \eqref{Bondieq} to solve for $\partial_u\bar \beta=-\frac{1}{32}\partial_u(C_{ij}C^{ij})$.
The limit of this quantity as $r\to\infty$ is finite and nonzero
\be\label{Bondiin}
8\pi G\, \mathcal Q_{t}[S_{u}] = \int_{S_{u}} m_B\,d\Omega + \frac{1}{32}\partial_u \int_{S_{u}} C_{ij}C^{ij} d\Omega\,,
\ee 
where $S_u$ is the spherical section of $\mathscr I^+$ at fixed $u$.
The linear piece of \eqref{Bondiin} has been given the interpretation \cite{BMS} of the total energy present in the spacetime at a given retarded time, also called \emph{Bondi mass}. The proper normalization is given by the stationary limit, as in \eqref{Komartrue}, which reduces in this case to a factor of two \cite{Iyer:1994ys}, yielding
\be\label{Bondimass}
M_B(u) = \frac{1}{4\pi G} \int_{S_{u}} m_B\,d\Omega\,.
\ee

The nonlinear and time-dependent term in \eqref{Bondiin} can be eliminated by modifying the two-form $\kappa_{\xi}^{ab}$ as proposed by Tamburino and Winicour \cite{Tamburino_Winicour}:
\be\label{improvementTW}
\kappa_{\xi}^{ab}\mapsto \frac{1}{16\pi G} \sqrt{-g}\, (\nabla^{[b} \xi^{a]}+\nabla_c \xi^c\,m^{[b}n^{a]})
\ee
where $n^a$, to leading order, $n^a=(\partial_r)^a$ and $m^a = (\partial_u)^a-\frac{1}{2}(\partial_r)^a$. 
In fact, in the case of the time translation vector field $\xi^a=(\partial_u)^a$, noting that $\nabla_a\xi^a=\partial_u\log\sqrt{-g}$ and using \eqref{inverseSachs}, this additional term yields
\be
\frac{r^2}{16\pi G} \int_{S_{u,r}} \partial_u\log\sqrt{-g} \,d\Omega = \frac{1}{8\pi G} \int_{S_{u,r}} \partial_u\bar \beta\,d\Omega+\mathcal O(r^{-1})\,,
\ee
and hence cancels the second term appearing in \eqref{Bondi+spur}. 

The same result is obtained by the following procedure advocated by Geroch and Winicour \cite{Geroch_Winicour}. In general, the independence of the surface charge on the specific representative vector that reduces to a given  asymptotic symmetry vector on $\mathscr I^+$ is not guaranteed. For instance, by inspection of equation \eqref{Chargexigen}, adding $f(\partial_r)^a$ to a vector $\xi^a$ will give the same transformation on $\mathscr I^+$ (since it projects to zero thereon) but gives rise to a difference in the surface charge given by $r^2(16\pi G)^{-1}\int e^{-2\beta}(\partial_r+\Gamma^r_{rr})fd\Omega$. Therefore, an additional condition is needed in order to resolve this ambiguity: \be\label{Geroch-Wini}\nabla_a\xi^a=0\,,\ee to be imposed on the extension of the infinitesimal symmetry vector field in the interior of the spacetime. In our case, we may extend the time translation asymptotic symmetry by introducing a $\xi^r$ component as in 
\be
\xi^a=(\partial_u)^a-\frac{2\partial_u\bar\beta}{r}(\partial_r)^a+\cdots, 
\ee
so that the additional condition \eqref{Geroch-Wini} is upheld, while the projection of $\xi^a$ on a surface of constant $r$ is still $(\partial_u)^{a}$, as required by \eqref{translations}. With this modification, taking into account $\Gamma^r_{rr}=2\partial_r\beta=\mathcal O(r^{-3})$, the surface charge \eqref{Komarform} yields half the Komar mass \eqref{Bondimass} with no additional terms.

Notice that, in the case of exact isometries as for instance for time translations in stationary spacetimes, the Tamburino-Winicour improvement \eqref{improvementTW} is identically zero and the Geroch-Winicour condition \eqref{Geroch-Wini} is identically satisfied, since $\nabla_a\xi^a=0$ follows from the Killing equation.

Employing the second equation of motion \eqref{Bondieq}, and noting that $D_iD_jN^{ij}$ integrates to zero on the sphere, we also have
\be\label{massleakD4}
\partial_u M_B(u) = -\frac{1}{32\pi G}\int_{S_u} N_{ij} N^{ij}d\Omega\,,
\ee
which is the mass loss formula. Gravitational radiation, responsible for a nontrivial News tensor $N_{ij}=\partial_u C_{ij}$, gives rise to the time-dependence of the Bondi mass, which in fact always decreases as $u$ increases since the right-hand side of the previous equation is negative definite. This is interpreted as the fact that gravitational waves always carry positive amounts of energy, which leak to the null boundary of the spacetime.

Under suitable regularity assumptions, it has been shown that the Bondi mass is always positive \cite{Reula_Tod} and that its upper bound is given by the Arnowitt-Deser-Misner \cite{ADM} mass $M_\mathrm{ADM}$, namely  the total energy content of the spacetime: 
\be
\lim_{u\to-\infty}M_B(u)=M_\mathrm{ADM}\,,
\ee 
the limit being reached from below \cite{Schoen-Yau,Witten_+}. These results consolidate the following picture: a gravitational system ``initially'' (\emph{i.e.} for $u\to-\infty$) possesses a total energy $M_\mathrm{ADM}$, a portion of which leaks to null infinity due to the emission of gravitational waves \eqref{massleakD4}; this leaves behind an energy $M_B(u)$,  which thus represents the amount energy \emph{left} in the spacetime at a given retarded time $u$.

Similar calculations to the one performed above are available in the case of spatial translations, $\xi^a = \mathbf n \cdot \mathbf a (\partial_u)^a$ and, taking in due account subleading terms in the expansion \eqref{BMSFalloffs}, of rotations and boosts \eqref{rotations/boosts}.

Turning our attention to a generic supertranslation $\xi^a=T(\theta,\phi)(\partial_u)^a$, as in \eqref{BMSonScri} with $Y^i=0$ and an arbitrary $T(\theta,\phi)$, we have, 
\be
\mathcal Q_\xi[S_{u,r}]
=\frac{r^2}{16\pi G}\int_{S_{u,r}}e^{-2\beta}\Big[
-\mathcal U^iD_iT + 
\left(\Gamma^r_{ru}-\Gamma^u_{uu}-\mathcal U^i \Gamma^u_{iu}\right)
\Big]d\Omega\,.
\ee
Substituting \eqref{Christoffelfalloff} and \eqref{BMSFalloffs}, and using \eqref{Bondieq}, the limit $r\to\infty$ gives
\be
8\pi G \mathcal Q_\xi[S_u]= \int_{S_u} m_B T\, d\Omega + \frac{1}{4} \int D^i C_{ij}D^j T\,d\Omega + \frac{1}{32}\partial_u \int_{S_u} C_{ij}C^{ij}T\,d\Omega\,.
\ee
As in the case of the global charge, this surface charge is finite and nonvanishing. Its linear part, which can be selected by resorting to the improvement \eqref{improvementTW} or by enforcing the condition \eqref{Geroch-Wini} as was done in the case of asymptotic time translations, and which must be corrected by a factor of two according to \eqref{Komartrue}, reads
\be\begin{aligned}\label{BMSgencharge}
\mathcal Q_T(u) &= \frac{1}{4\pi G}\int_{S_u} m_B T\, d\Omega - \frac{1}{16 \pi G} \int_{S_u} D^i D^j C_{ij}\, T d\Omega\,,\\
\partial_u \mathcal Q_T(u) &=
-\frac{1}{32\pi G}\int_{S_u} N_{ij}N^{ij}\, Td\Omega\,.
\end{aligned}\ee
The term linear in $C_{ij}$ vanishes when $T$ satisfies
\be\label{DiDj1/2}
\left(D_{i}D_{j}-\tfrac{1}{2} \gamma_{ij}\Delta \right)T=0\,,
\ee 
since $C_{ij}$ is traceless. This selects asymptotic translations, because, using that $[\Delta,D_i]T=D_i T$, the divergence of \eqref{DiDj1/2} yields $D_i(\Delta+2)T=0$ and hence either $T$ is a constant or it satisfies $(\Delta+2)T=0$, whose solutions are (linear combinations of) the spherical harmonics $\cos\theta$, $\sin\theta\cos\phi$, $\sin\theta\sin\phi$  (see Appendix \ref{app:Laplaciano} for more details).

As we shall see in the next chapter, expressions of the type \eqref{BMSgencharge} lie at the heart of the connection between $BMS$ symmetries and Weinberg's soft graviton theorem.

Let us conclude this section on asymptotic charges by a word of warning. As anticipated, we have chosen to present the calculation of surface charges without detailing subtleties that must be actually faced in order to give a general and precise definition of conserved quantities in gauge theories. Some of these issues have actually surfaced while discussing the improvement \eqref{improvementTW} and are related to the ambiguities inherent to the definition of the canonical Noether current $j_\xi^a$ starting from a given Lagrangian. Another issue, which arises when trying to attach to conserved quantities the meaning of Hamiltonian generators is that of integrability: in this context one only calculates the \emph{formal} variation $\slashed\delta Q_\xi$ of a given charge in field space and needs to establish whether or not this can be interpreted as an exact variation $\delta Q_\xi$. Nonintegrability of the $BMS$ charges \cite{Barnich_Charge}, for instance, is associated with the presence of nonlinear radiation.
  
We have glossed over such aspects in order to keep the presentation as concise as possible without renouncing on relevant formulas and physical ideas, and refer the interested reader to \cite{Wald-Zoupas,Iyer:1994ys} for detailed discussions in the context of covariant phase space methods (see also \cite{Hajian, Seraj}) and to the general results presented in \cite{Barnich-Brandt}, where in particular the issue of ambiguities is resolved by appealing to the falloffs of the equations of motion. However, we will return to some of the issues raised here in Chapter \ref{chap:spin1higherD}, in the context of Yang-Mills theory, where they are easier to present in a self-contained and concise manner.

\section{Large Gauge Symmetries in Electromagnetism}\label{sec:EMD=4}
Up to this point in the discussion, we have been concerned with the symmetries of weakly radiating gravitational systems in the limit $r\to\infty$. A similar analysis can be performed in the simpler case of electromagnetism \cite{Strominger_QED}, where once again asymptotic symmetries are identified as those residual symmetries of the gauge-fixed theory that preserve the falloff conditions assigned to the field components, while still acting nontrivially near $\mathscr I^+$.

The classical action for electromagnetism coupled to a locally conserved current density $\mathcal J^a$,
\be
S = - \frac{1}{4}\int \mathcal F_{ab} \mathcal F^{ab}\,\omega - \int \mathcal A_a  \mathcal J^a \,\omega,
\ee
with $\mathcal F_{ab}=\partial_a\mathcal A_b-\partial_b\mathcal A_a$, is invariant under the gauge transformation $\delta_\epsilon \mathcal A_a =\partial_a \epsilon$, since the symmetry variation of the Lagrangian is equal to the total divergence term $-\nabla_a( \mathcal  J^a \epsilon )\omega$, while on the other hand for a generic variation one obtains
\be
(\nabla^b \mathcal F_{ba}-\mathcal J_a)\delta A^a+\nabla^b(\mathcal F_{ab}\delta \mathcal A^a)\,.
\ee
This gives, on shell, the canonical current
\be\label{j_s=11}{
	j_\epsilon^a = \sqrt{-g}(\mathcal F^{ba}\partial_b\epsilon + \mathcal J^a\epsilon)=\partial_b\kappa_\epsilon^{ab}\,,\qquad
	\kappa_\epsilon^{ab} = \sqrt{-g}(\mathcal F^{ba}\epsilon)\,.
}
\ee
In retarded Bondi coordinates, we choose the radial gauge condition
\be
\mathcal A_r=0\,,
\ee
which is the spin-one analog of the choice \eqref{Sachsform} we made in the gravitational setting. 
The falloff conditions to be assigned to currents $J^a$ depend on the type of charged matter that on is assuming to include.
In particular, accounting for the possible presence of massless charged particles implies that $J^a$ behaves like
\be
\mathcal J_u= \frac{J_u}{r^2}+\mathcal O(r^{-3})\,,\qquad
\mathcal J_r= \frac{J_r}{r^2}+\mathcal O(r^{-3})\,,\qquad
\mathcal J_i = \frac{J_i}{r}+\mathcal O(r^{-2})\,,
\ee 
as $r\to\infty$ for fixed $u$.
Correspondingly, one finds that the equations of motion are satisfied to leading order provided one adopts the falloff conditions
\be\begin{aligned}
	\mathcal A_u = A_u(u, z, \bar z)/r + \mathcal O(r^{-2})\,,\qquad
	\mathcal A_i = A_i(u, z, \bar z) + \mathcal O(r^{-1})\,,
\end{aligned}\ee
and the field components satisfy the relation
\be\label{chargedensityevM}
\partial_u A_u = \partial_u D^i A_i + J_u
\ee
(compare with \eqref{definizione_Bondi_gaugerip} and \eqref{Bondieq} respectively). 

Residual gauge transformations are therefore given by those gauge parameters that only depend on angular coordinates
\be
\epsilon = T(z, \bar z)\,,
\ee
which thus exhibit a close analogy to supertranslations.
The surface charge, \emph{i.e.} the analog of \eqref{surfacecharge}, associated to this residual gauge freedom is given by
\be
Q_\epsilon[S_u]= \lim_{r\to\infty} r^2\int_{S_{u,r}}\mathcal F_{ur}T d\Omega = \int_{S_u} A_u T d\Omega\,.
\ee
In analogy with the previous case, the global surface charge, \emph{i.e.} the electric charge expressed as the Gauss integral, is in general independent of the specific two-sphere we choose, provided we do not cross charges flowing to $\mathscr I^+$. To see this, setting $\epsilon$ to unity, note that $\partial_b \kappa_1^{ab}=\sqrt{-g}\,\nabla_b \mathcal F^{ba}=\sqrt{-g}\,\mathcal J^a$, so that
\be
\partial_u \mathcal Q_1[S_u]=\int_{S_u} J_u d\Omega\,,
\ee 
which vanishes away from the charges. This equation is the analog, for massless electrodynamics, of the mass leak formula we encountered in the gravitational case. Note in particular that this effect, which is due to the presence of self-interactions in the case of gravity, is here caused by charges moving at the speed of light. We will encounter a similar phenomenon while discussing ordinary and null memory effects in the next chapter.

For a generic function $T(z, \bar z)$, instead, the soft surface charge exhibits a nontrivial $u$-dependence.
In particular,
\be
\partial_u\mathcal Q_\epsilon[S_u]=\int_{S_u} \partial_u A_u T d\Omega=\int_{S_u} (\partial_u D^i A_i + J_u) T d\Omega\,.
\ee
This infinite-dimensional family of $u$-dependent charges plays a crucial role in the connection between asymptotic symmetries of massless electrodynamics and the soft photon theorem.

We conclude by remarking that, if we want to compare the retarded radial gauge to the advanced radial gauge, we need consider that the transformation from retarded to advanced coordinates $v=u+2r$ gives
\be
(\mathcal A'_v, \mathcal A'_r, \mathcal A'_i)=
(\mathcal A_u, 2 \mathcal A_u+\mathcal A_r, \mathcal A_i)\,,
\ee 
so that $\mathcal A_r$ and $\mathcal A'_r$ cannot vanish simultaneously unless $\mathcal A_u$ is identically zero. This indicates that the radial gauge cannot be continued smoothly to the bulk of spacetime \cite{Sever}. This issue can be resolved, for instance, by performing the analysis in the Lorenz gauge; we will return on this point at the end of the next chapter.

\section{Kac-Moody Symmetry of Yang-Mills Theory}\label{sec:YMD=4}
 It is possible to perform an analysis akin to that of asymptotically flat gravitational systems also in classical Yang-Mills theory \cite{Strominger_YM}. Other than being of interest in its own right, this provides a convenient toy version of the corresponding gravitational case, while still encompassing genuinely nonlinear effects, in contrast with the Maxwell case. 

We may start from the classical action for pure Yang-Mills,
\be
S = \frac{1}{4}\int \mathrm{tr}\big(\mathcal F_{ab} \mathcal F^{ab}\big)\,\omega\,,
\ee
where $\mathcal F_{ab}=\partial_a\mathcal A_b-\partial_b\mathcal A_a+[\mathcal A_a, \mathcal A_b]$. We conventionally work with anti-Hermitian fields $A_a = A_a^A X^A$ with $[X^A, X^B]=f^{ABC}X^C$, where $X^A$ and $f^{ABC}$ are the $su(N)$ generators and structure constants respectively. We normalize them according to $\mathrm{tr}(X^AX^B)=-\delta^{AB}$. 

This action is invariant under the gauge transformation $\delta_\epsilon \mathcal A_a =\partial_a \epsilon+[\mathcal A_a, \epsilon]$. On the other hand for a generic variation one obtains
\be
\delta\mathcal L=-\mathrm{tr}\big(\nabla^b \mathcal F_{ba}\,\delta A^a\big)+\nabla^b\mathrm{tr}\big(\mathcal F_{ba}\delta \mathcal A^a\big)\,.
\ee
On shell, this gives the canonical current
\be\label{j_s=12}{
	j_\epsilon^a = \sqrt{-g}\,\mathrm{tr}\big(\mathcal F^{ab}(\partial_b\epsilon + [\mathcal A_b, \epsilon]) \big)=\partial_b\kappa_\epsilon^{ab}\,,\qquad
	\kappa_\epsilon^{ab} = \sqrt{-g}\,\mathrm{tr}\big(\mathcal F^{ab}\epsilon\big)\,.
}
\ee
Adopting the retarded radial gauge condition
\be
\mathcal A_r=0\,,
\ee
one then sees that the equations of motion are compatible with the falloffs
\be\begin{aligned}
	\mathcal A_u = A_u(u, z, \bar z)/r + \mathcal O(r^{-2})\,,\qquad
	\mathcal A_i = A_i(u, z, \bar z) + \mathcal O(r^{-1})\,,
\end{aligned}\ee
provided
\be
\partial_u A_u = \partial_u D^i A_i + \gamma^{ij}[A_i, \partial_u A_j]\,.
\ee
Comparing with \eqref{chargedensityevM}, we see that, asymptotically, the nonlinearities play the same role that was played by massless charges in the previous section. This is inherent to the twofold nature of waves in nonlinear theories, where they play the role on the one hand of propagating perturbations and on the other hand of massless sources. 

Residual gauge transformations are therefore
\be
\epsilon = T(z, \bar z) = T^A(z,\bar z)X^A
\ee
in close analogy to supertranslations.
The surface charge is thus given by
\be
Q_T[S_u]= -\lim_{r\to\infty} r^2\int_{S_{u,r}}\mathcal \mathrm{tr}\big(F_{ur}T\big) d\Omega = -\int_{S_u} \mathrm{tr}\big(A_u T\big) d\Omega\,.
\ee
The global surface charge, \emph{i.e.} the color charge, depends on retarded time due to the nonlinearities: setting $\epsilon=X^A$,
\be
\partial_u \mathcal Q_A[S_u]=\int_{S_u} \gamma^{ij} [A_i, \partial_u A_j]^A d\Omega\,.
\ee 
This equation is the analog, for classical Yang-Mills, of the mass leak formula we encountered in the gravitational case. 
For a generic parameter $T(z, \bar z)$, instead, the soft surface charge exhibits an additional nontrivial $u$-dependence that is observed to be linear in the field.
In particular,
\be
\partial_u\mathcal Q_T[S_u]=-\int_{S_u} \mathrm{tr}(\partial_u D^i A_i + \gamma^{ij}[A_i, \partial_u A_j] ) T d\Omega\,.
\ee
The charges have a nontrivial algebra: 
\be
[\mathcal Q_T, \mathcal Q_{T'}]=\delta_{T}\mathcal Q_{T'}=
-\int_{S_u} \mathrm{tr}\big([A_u,T]T'\big) d\Omega=
-\int_{S_u} \mathrm{tr}\big(A_u[T,T']\big) d\Omega=\mathcal Q_{[T,T']}\,,
\ee
namely a zero-level Kac-Moody algebra.

%%%%%%%%%%%%%%%%%%%%%%%%%%%%%%%%%%%%%%
\chapter{Observable Effects in Four Dimensions} \label{chap:obs}
%%%%%%%%%%%%%%%%%%%%%%%%%%%%%%%%%%%%%%

As we have seen, asymptotic symmetries give rise to an infinite number of independent conserved charges potentially amenable to be connected to observable quantities. Indeed, at least two broad classes of phenomena have been identified as observable effects that afford an explanation in terms of an underlying asymptotic symmetry principle: soft theorems for scattering amplitudes and memory effects \cite{Strominger_rev}. 

This chapter is devoted to reviewing these effects, together with their connection with the underlying asymptotic symmetries discussed in the previous chapter, in the well-studied cases of electromagnetic and gravity theories in four spacetime dimensions. This will also serve as preparation for our exploration of their higher-dimensional counterparts in Chapter \ref{chap:spin1higherD}, in the case of Maxwell and Yang-Mills theory, and of our proposal for higher-spin asymptotic symmetries and charges, and their relation to soft theorems, in Chapter \ref{chap:HSP}.

\section{Soft Theorems in Particle Physics}

Soft theorems are identities relating scattering amplitudes that differ by the emission or absorption of massless, or very light, particles with low energy \cite{GellMannPhysRev.96.1433,Low,Kazes,Yennie:1961ad,Weinberg_64,Weinberg_65, Burnett}. They have received renewed interest in the literature (see \emph{e.g.} \cite{LaddhaSen:2017ygw, Sahoo:2018lxl,InfiniteLi:2018gnc} for some recent advances in this respect, together with the extensive list of references in \cite{Strominger_rev}) in the context of asymptotic symmetries, of which  they have been identified as physical consequences. 

Indeed, although soft theorems can be directly seen to hold at the level of $S$ matrix and Feynman diagrams, they can be often derived as a consequence of the invariance of the theory under a symmetry, typically a spontaneously broken one. 
A textbook example of this situation is that of the spontaneously broken (approximate) symmetry  $SU(2)_L\times SU(2)_R \rightarrow SU(2)_V$ of quantum chromodynamics, which lies at the heart of soft-pion techniques \cite{WeinbergBook2, Ferrari_Picasso_1}, dating back to the sixties, where the pions are interpreted as the (pseudo-)Goldstone bosons of said spontaneous breaking. More generally, the understanding that the dynamics of Goldstone bosons affords a universal description at low energies, dictated solely by symmetry requirements, paved the way to the formulation of effective Lagrangians and nonlinearly realized symmetries \cite{CCWZ1, CCWZ2} (for a more recent approach to soft theorems in the context of effective field theories, see \cite{Larkoski2015,Elvang:2016qvq}). 

The basic mechanism that links symmetries to scattering amplitudes can be summarized as follows. A continuous symmetry of the dynamics of a given theory is locally generated, in a canonical setup, by a conserved current $j^\mu(x)$ according to Noether's theorem. Namely, for any local operator $A$ (which may be thought of, for instance, as $\mathcal O_1(x_1) \mathcal O_2(x_2)\cdots \mathcal O_n(x_n)$), the infinitesimal symmetry variation $\delta$ is given by
\be\label{localgen}
\delta A = i \lim_{R\to\infty}  \int f_R(\mathbf x) [j^0(x),A] d\mathbf x\,,
\ee
with a smooth cutoff function $f_R(\mathbf x)$ that equals $1$ if $|\mathbf x|<R$ and vanishes if $|\mathbf x|>R+\varepsilon$ (for a small $\varepsilon>0$). In relativistic local field theories, the right-hand side converges by the requirement that the commutator vanish when $x$ becomes spacelike with respect to the region where $A$ is localized. Furthermore, it is also independent of $t$, as a consequence of the conservation of $j^\mu$: using the Gauss theorem,
\be
\lim_{R\to\infty}  \int f_R(\mathbf x) [\partial_tj^0(x),A] d\mathbf x
=
\lim_{R\to\infty} \int \nabla f_R(\mathbf x) \cdot [\mathbf j(x),A] d\mathbf x =0\,,
\ee 
because $\nabla f_R(\mathbf x)$ is nonzero only for $|\mathbf x|\simeq R$, a region that becomes spacelike with respect to any point as $R\to\infty$ for fixed $x^0=t$.
 
Taking the vacuum expectation of \eqref{localgen} gives rise to Ward identities of the type
\be\label{motherWard}
\langle 0 | \delta A |0 \rangle =i \lim_{R\to\infty}  \int f_R(\mathbf x) \langle 0 | [j^0(x),A] |0\rangle d\mathbf x = i \lim_{\mathbf k\to0} \langle 0 | [\tilde j^0(\mathbf k,t),A] |0\rangle\,,
\ee 
where $\tilde j^0$ denotes the Fourier transform of the charge density.
Now, if the symmetry is unbroken, namely it admits a unitary operator $U$ that implements the symmetry in the Hilbert space of the theory, then it actually admits a canonical charge
\be
Q=\lim_{R\to\infty}\int f_{R}(\mathbf x) j^0(x) d\mathbf x\,,
\ee 
such that $U=e^{iQ}$
and, restricting for simplicity to the case of internal symmetries,
\be
Q|0\rangle = 0\,,\qquad U|0\rangle = |0\rangle\,,
\ee
since the vacuum $|0\rangle$ is the only translation-invariant state,
so that \eqref{motherWard} simply reduces to 
\be
\langle 0 | \delta A |0\rangle=0\,.
\ee
If instead the symmetry is spontaneously broken, no generator $Q$ exists and the vacuum state is not invariant under the symmetry, namely \eqref{motherWard} imposes
\be\label{GBbasic}
0\neq \langle 0 | \delta A |0\rangle
=
i\lim_{\mathbf k\to0} \langle 0 | [\tilde j^0(\mathbf k,t),A] |0\rangle\,,
\ee 
which relates the vacuum expectation value of the symmetry transformation to the insertion of a \emph{soft} operator. In fact, the intermediate states saturating the right-hand side must clearly have a gap-less dispersion relation, $\omega(\mathbf k)\to 0 $ as $\mathbf k\to0$, in order for it to be time-independent, as required by $\partial_\mu j^\mu=0$, and further inspection shows that they must be (massless) one-particle states: the Goldstone bosons. 

In order to make contact with scattering amplitudes, one takes the operator $A$ to be of the type $\mathcal T(\phi_1(x_1)\phi_2(x_2)\cdots \phi_n(x_n))$, where $\phi_j(x_j)$ are suitable local operators of the theory and $\mathcal T$ denotes time-ordering. The next step is to apply the LSZ reduction formulas \cite{LSZ1, LSZ2} (see \emph{e.g.} \cite{Schweber} for a textbook presentation), which amount to performing a Fourier transform with respect to the $x_j$ and taking the on-shell limit after amputating the propagators that arise from external particles. Restricting for simplicity to the case of a Hermitian scalar field $\varphi$ of mass $m$, with asymptotic Fock oscillators $a_\mathrm{in/out}(\mathbf q)$, the LSZ formula can be expressed as
\be\begin{aligned}
&\langle 0 | a_\mathrm{out}(\mathbf q'_1)\cdots a_\mathrm{out}(\mathbf q'_n) a^\ast_\mathrm{in}(\mathbf q_1)\cdots a^\ast_\mathrm{in}(\mathbf q_m)
|0\rangle
=\int dx'_1\cdots dx'_n dx_1 \cdots dx_m\\
&e^{i \sum_{j=1}^n x'_j \cdot q'_j-i \sum_{k=1}^m x_k \cdot q_k} 
(-\Box_{x'_1}+m^2)\cdots (-\Box_{x'_n}+m^2)
(-\Box_{x_1}+m^2)\cdots (-\Box_{x_m}+m^2)\\
&
\langle 0 |
\mathcal T(
\varphi(x'_1)\cdots \varphi(x'_n) \varphi(x_1) \cdots \varphi(x_m) 
)
|0\rangle\,.
\end{aligned}\ee 
To illustrate the above strategy in one of its simplest applications, we may derive the consequence of the (unbroken) global $U(1)$ symmetry in QED, namely the additive conservation of electric charge in all scattering processes. The Ward identities read in this case
\be
(n-m)
\langle 0 |
\mathcal T(
\psi(x'_1)\cdots \psi(x'_n) \bar\psi(x_1) \cdots \bar\psi(x_m) A_{\mu_1}(y_1)\cdots A_{\mu_l}(y_l) 
)
|0\rangle
=
0\,,
\ee
so that $\langle 0 |
\mathcal T(
\psi(x'_1)\cdots \psi(x'_n) \bar\psi(x_1) \cdots \bar\psi(x_m) A_{\mu_1}(y_1)\cdots A_{\mu_l}(y_l) 
)
|0\rangle=0$ unless  $n=m$. On the other hand, by the LSZ reduction formula, we see that, in terms of the numbers $N$ (resp. $N'$) of incoming (outgoing) positrons and $M$ ($M'$) of electrons, we must have $n=M+N'$ and $m=M'+N$. By comparison, this implies $M-N = M'-N'$, \emph{i.e.} that the \emph{in}  state must have the same electric charge as the \emph{out} state.

In the case of spontaneous breaking, the so-obtained identities instead relate different amplitudes with and without the insertion of soft Goldstone bosons, according to \eqref{GBbasic}, and hence give rise, upon LSZ reduction, to soft theorems. 
 
In fact, this strategy has been applied to a wide range of different models and symmetries (\emph{e.g.} \cite{Ferrari_Picasso_1,Hamada:2017atr}), including also non-internal symmetries such as scaling and conformal symmetry \cite{DiVecchia:2015jaq, DiVecchia:2017uqn} and residual symmetries of QED, Yang-Mills theory, gravity \cite{Ferrari_Picasso_2, Hamada:2018vrw}. Taking successive variations and commutators also allows one to analyze \emph{double} soft limits, schematically
\be
\langle 0 |\delta_1 \delta_2 A |0\rangle = - 
\lim_{\mathbf k_1\to0}\lim_{\mathbf k_2\to0} 
\langle 0 |
[\tilde j_{(1)}^0(\mathbf k_1,t),
[\tilde j_{(2)}^0(\mathbf k_2,t),
A]] 
|0\rangle\,,
\ee
discussing, for instance, the dependence on the orders in which such limits are taken and its relation to the symmetry algebra \cite{Distler:2018rwu, H:2018ktv}.
 
Although the above considerations are in principle nonperturbative, since they only make reference to general properties of symmetries and of the spectrum of the theory, one must in practice face the issue of taking into account possible corrections due to loop effects in order to extend their validity to the full quantum level. A possible way out is furnished by nonrenormalization theorems that protect the commutators involving conserved currents from corrections due to ultraviolet singularities \cite{Symanzik}. Indeed, let us mention that some of these symmetry considerations have been extended to the quantum level \cite{Campiglia:2019wxe}, and that soft theorems have been investigated at loop level, displaying a certain degree of stability under renormalization (see \emph{e.g.} \cite{HeLoop2014,Sen:2017nim}).
Another important point that must be addressed in the discussion of loop-level results is the presence of infrared divergences. In four--dimensional QED, for instance, such infinities arise due to the masslessness of the photon and were observed to cancel out when considering cross sections that sum over emissions/absorptions of photons below a certain energy threshold, so that \textit{soft} photons actually play a major role in the discussion of the infrared problem  \cite{Strocchi-Erice,Weinberg_65,PerryStates}. 

At the level of semiclassical analysis, it has proven useful to investigate the relation between soft theorems in gauge theories and asymptotic or large gauge symmetries, in particular those at null infinity. This permitted in particular to interpret Weinberg's soft photon and graviton theorems \cite{Weinberg_64, Weinberg_65}, which we are going to review in the next section, in terms of underlying large $U(1)$ symmetries and $BMS$ supertranslation symmetry. An analogous program has been carried out also for Yang-Mills theory \cite{StromingerColor} and has been extended to encompass subleading corrections to the Weinberg result \cite{soft_QED_Strominger,Bern:2014vva, Conde:2016csj}. 

The technical steps needed in order to perform this connection are typically a slight variation of those outlined above in the case of a standard symmetry of the quantum theory. One first identifies the asymptotic symmetry and calculates the corresponding (classical) asymptotic charge $Q$. As a consequence of Noether's theorems, this charge takes the form of an integral over a Cauchy surface $\Sigma$ or, equivalently, over its boundary $\partial \Sigma$,
\be
Q = \int_\Sigma j^\mu d\Sigma_{\mu} = \int_{\partial \Sigma} \kappa^{\mu\nu} d\sigma_{\mu\nu}\,,
\ee 
where $j^\mu = \partial_\nu \kappa^{\mu\nu}$ and $\kappa^{\mu\nu}=-\kappa^{\nu\mu}$, and it is actually independent of the specific Cauchy surface $\Sigma$. One then expresses the fact that the symmetry is also a symmetry of the $S$ matrix as
\be\label{Q+SSQ1}
0=[Q,S]=Q^+S-S Q^-\,,
\ee 
where $Q^\pm$ denotes the charges evaluated on the Cauchy surfaces $\mathscr I^\pm \cup i^\pm$. Taking matrix elements of the above identity between \textit{in} and \textit{out} states furnishes Ward identities that lead to the soft theorems, where the piece of $Q$ that is responsible for the soft insertion is termed \emph{soft} charge, $Q_S$, to be contrasted with the \emph{hard} part, $Q_H$, involving the matter fields. An equivalent route, more similar to the steps presented above, is to calculate 
\be\label{Q+SSQ2}
\langle0| \delta A |0\rangle = i\langle0| [Q,A] |0\rangle = i \langle0| (Q^+ A - A Q^-) |0\rangle\,,
\ee
where $A$ stands for a generic product of fields, and then take the LSZ reduction of this identity in order to derive consequences at the level of scattering amplitudes. In the latter approach, which we are going to illustrate in detail in the case of Weinberg's soft photon and graviton theorems, the hard part of the charge can usually be neglected since, in the right-hand side of the above identity, $Q$ acts on the vacuum where no stable matter is present, $Q_H|0\rangle=0$.

Let us stress that the operators $Q^+$ and $Q^-$ appearing in equations \eqref{Q+SSQ1} and \eqref{Q+SSQ2} are not independent; after all, they must be the asymptotic charges corresponding to the same symmetry. Therefore, $Q^+$ and $Q^-$ must be evaluated by taking the limits of the same bulk symmetry transformation to the far future and far past: we will see a concrete example of this procedure in the case of the Maxwell theory in  the Lorenz gauge. The general, $CPT$ and Lorentz-invariant way of specifying this correspondence between quantities evaluated on $\mathscr I^-$ and $\mathscr I^+$ is the antipodal map on the celestial sphere, which is obtained by the identification 
\be
\mathcal O^+(\mathbf n)\big|_{\mathscr I^+_-} = \mathcal O^-(-\mathbf n)\big|_{\mathscr I^-_+}
\ee
across spatial infinity. The antipodal identification is also instrumental in the definition of a meaningful scattering problem from $\mathscr I^-$ to $\mathscr I^+$. The process of trivial scattering is characterized by the fact that the \emph{out} states are obtained as the antipodal images of the \emph{in} states, as can be seen for instance by following the trajectory of a free massless particle in Minkowski space, and therefore the antipodal map provides the natural definition of ``identity'' (the trivial intertwining operator) between \textit{in} and \textit{out} Fock spaces. In the following, this identification is assumed throughout in the discussion of scattering amplitudes at null infinity.

We turn now to the illustration of Weinberg's soft theorems and to their connection with asymptotic symmetries of electromagnetism and gravity.
%universality?

\subsection{Weinberg's theorems}
In his celebrated $1964$ paper \cite{Weinberg_64}, Weinberg showed that, using only the Lorentz invariance and the pole structure of the $S$ matrix, together with masslessness and spins of the photon and of the graviton, it is possible to derive the conservation of electric charge and the equality of gravitational and inertial mass. On the same grounds, he gave a possible explanation as to why we observe no macroscopic fields corresponding to massless particles of spin $3$ or higher.

In particular, exploiting the (assumed) $S$-matrix pole structure and Lorentz covariance, he could prove the following two properties:
\begin{itemize}
	\item[(1)] The $S$ matrix for the emission of a photon or a graviton can be written as the product of a polarization ``vector'' $\varepsilon^\mu$ or ``tensor'' $\varepsilon^\mu \varepsilon^\nu$ with a covariant vector or tensor amplitude, and it vanishes if any of the $\varepsilon^\mu$ is replaced by the photon or graviton momentum $q^\mu$;
	\item[(2)] Electric charge, defined dynamically by the strength of soft-photon interactions, is additively conserved in all reactions. Gravitational mass, defined by the strength of soft graviton interactions, is equal to inertial mass (in the nonrelativistic limit).
\end{itemize}

We will now review the derivation of these results.
For a few technical statements, related to the implications of covariance on the $S$-matrix structure, which are here assumed to hold, we refer to the appendices of Weinberg's paper \cite{Weinberg_64}.

Let us consider a process in which a massless particle is emitted with momentum $\mathbf q$ and helicity $\pm s$, limiting ourselves to integer spins. The transformation rules for $S$-matrix elements under Lorenz transformations can be inferred from the transformation law for one-particle states; using $p$ as a shorthand for a set of $p_i$, with $i=1,\ldots, n$, one has
\be\label{Strasflaw}
S_{\pm s}(\mathbf q, p) = \left(\frac{|\Lambda \mathbf q|}{|\mathbf q|}\right)^{1/2}e^{\pm i s \Theta (\mathbf q, \Lambda)}S_{\pm s}(\Lambda \mathbf q, \Lambda p),
\ee
where $\Theta$ is a function of the massless particle momentum $\mathbf q$ and of the Lorentz transformation $\Lambda$. It is always possible to write $S_{\pm s}$ as a product of a polarization ``tensor'' and an $M$-function in the following way:
\be\label{Spmjmathbf}
S_{\pm s}(\mathbf q, p) = \frac{1}{\sqrt{2\mathbf q}}\varepsilon_{\pm}^{\mu_1\ast}(\mathbf q) \ldots \varepsilon_{\pm}^{\mu_s\ast}(\mathbf q)M_{\pm\mu_1\ldots\mu_s}(\mathbf q, p),
\ee
where $M$ is a symmetric Lorentz tensor. The polarization ``vector'' $\varepsilon_{\pm}^\nu$ obeys the transformation rule
\be
\left( \Lambda\indices{_\nu^\mu}-\frac{q^\mu}{|\mathbf q|} \Lambda\indices{_\nu^0}\right) \varepsilon_{\pm}^\nu(\Lambda q) = e^{\pm i \Theta(\mathbf q, \Lambda)}\varepsilon_{\pm}^\mu(q)\,,
\ee
so that, at a closer scrutiny, $\varepsilon_{\pm}^\mu$ is not a \emph{bona fide} Lorenz vector. An auxiliary condition will be needed, in order to make sure that $S_{\pm s}$ satisfies Lorentz invariance despite this inhomogeneous transformation rule for $\varepsilon_\pm^\mu$. The $S$-matrix transformation law \eqref{Strasflaw} then reads
\begin{equation}\begin{aligned}\relax
		S_{\pm s}(\mathbf q, p) =& \frac{1}{\sqrt{2\mathbf q}}e^{\pm i s \Theta(\mathbf q, \Lambda)}
		\left[ \varepsilon_{\pm}^{\mu_1}(\Lambda q)- \frac{(\Lambda q)\indices{^{\mu_1}}}{|\mathbf q|}\Lambda\indices{_\nu^0}\varepsilon_{\pm}^\nu(\Lambda q) \right]\ldots \\
		&\times\left[ \varepsilon_{\pm}^{\mu_s}(\Lambda q)- \frac{(\Lambda q)\indices{^{\mu_s}}}{|\mathbf q|}\Lambda\indices{_\nu^0}\varepsilon_{\pm}^\nu(\Lambda q) \right]
		M_{\pm \mu_1 \ldots \mu_s}(\Lambda \mathbf q, \Lambda p).
\end{aligned}\end{equation}
For an infinitesimal Lorentz transformation $\Lambda\indices{^\mu_\nu}=\delta\indices{^\mu_\nu}+\omega\indices{^\mu_\nu}$, we can use \eqref{Spmjmathbf} and the symmetry of $M$ to put the previous equation in the form
\begin{equation}\begin{aligned}\relax
		S_{\pm s}(\mathbf q, p) = & \left(\frac{|\Lambda \mathbf q|}{|\mathbf q|}\right)^{1/2} e^{\pm i s \Theta(\mathbf q, \Lambda)} S_{\pm s}(\Lambda \mathbf q, \Lambda p)\\
		&- \frac{s}{\sqrt{2|q|^3}} \left[\omega\indices{_\nu^0}\varepsilon_{\pm}^{\nu\ast}(q)\right]
		q^{\mu_1}\varepsilon_{\pm}^{\mu_2\ast}(q)\ldots \varepsilon_{\pm}^{\mu_s\ast}(q) 
		M_{\pm\mu_1\ldots\mu_s}(\mathbf q, p)\,.
\end{aligned}\end{equation}
Hence the necessary and sufficient condition for this transformation law not to contradict the first one is that $S_{\pm}$ vanishes when one of the $\varepsilon_{\pm}^\mu$ is replaced with $q^\mu$:
\be
q^{\mu_1}\varepsilon_{\pm}^{\mu_2\ast}(q)\ldots \varepsilon_{\pm}^{\mu_s\ast}(q) 
M_{\pm\mu_1\ldots\mu_s}(\mathbf q, p)=0.
\ee
This can be recognized, thinking in terms of fields, as a requirement of gauge invariance of the amplitude.

\subsection[Conservation of the electric charge, universality of the\newline gravitational coupling and higher spins]{Conservation of the electric charge, universality of the gravitational coupling and higher spins}
Considering the vertex amplitude for a very-low-energy massless particle of integer helicity $\pm s$, emitted by a particle of spin $0$ and mass $m$ (perhaps zero), and momentum $p^\mu = (\mathbf p, E)$, the only tensor which can be used to form $M_{\pm}^{\mu_1\ldots \mu_s}$ is $p^{\mu_1}\ldots p^{\mu_s}$, since terms involving $g^{\mu\mu'}$ do not contribute to the $S$ matrix because of $\varepsilon_{\pm}^\mu \varepsilon_{\pm\mu}=0$\footnote{
	This follows from the fact that external particles are on-shell and the corresponding tensors live in traceless representations of the stability group of $p^\mu$.
}. On the other hand, terms involving the soft momentum $q^\mu$ itself would either vanish by $q_\mu \varepsilon_\pm^\mu(q)=0$ or contribute to subleading orders. Therefore, the vertex amplitude must be of the form
\be\label{da_giustificare}
\frac{1}{2E(\mathbf p)\sqrt{2|\mathbf q|}} p_{\mu_1}\ldots p_{\mu_s}\varepsilon_{\pm}^{\mu_1\ast}(q)\ldots \varepsilon_{\pm}^{\mu_s\ast}(q).
\ee
As we shall argue below, even for emitting particles with spin greater than $0$, the $S$-matrix elements will still be given by this expression, times $\delta_{\sigma\sigma'}$ where $\sigma$ and $\sigma'$ are respectively the initial and final helicity of the emitting particle.

We define the soft photon coupling constant $e$ by the statement that the $s=1$ vertex amplitude is 
\be
\frac{2ie(2\pi)^4\delta_{\sigma\sigma'}p_\mu \varepsilon_{\pm}^{\mu\ast}(q)}{(2\pi)^{9/2}2E(\mathbf q) \sqrt{2|\mathbf q|}},
\ee
and similarly for the ``gravitational charge'' $f$ we state that the $s=2$ vertex amplitude is
\be
\frac{2if(8\pi )^{1/2}(2\pi)^4\delta_{\sigma\sigma'}\left(p_\mu \varepsilon_{\pm}^{\mu\ast}(q)\right)^2}{(2\pi)^{9/2}2E(\mathbf q) \sqrt{2|\mathbf q|}}.
\ee

Let $S_{\beta\alpha}$ be the $S$ matrix for some process $\alpha \rightarrow \beta$, the states $\alpha$ and $\beta$ consisting of various charged and uncharged particles, perhaps including gravitons and photons. The same process can occur with emission of a very soft extra photon or graviton of momentum $\mathbf q$ and helicity $\pm 1$, or $\pm 2$, and we will denote the corresponding $S$-matrix element as $S_{\beta\alpha}^{\pm1}(\mathbf q)$ or $S_{\beta\alpha}^{\pm2}(\mathbf q)$. 

As illustrated in the figure below, which represents graphically the amplitude $S^\pm_{\beta\alpha}(\mathbf q)$,
\begin{fmffile}{MyDiagram}
\begin{align*}
\begin{gathered}
\begin{fmfgraph*}(90, 65)
\fmfleft{i1,d1,i2,i3}
\fmfv{label={\color{blue}\vdots}, label.dist=-4mm}{d1}
\fmfright{o1,d2,o2,o4,o3}
\fmfv{label={\color{blue}\vdots}, label.dist=-4mm}{d2}
\fmf{fermion, fore=blue}{i1,v1}
\fmf{fermion, fore=blue}{i2,v1}
\fmf{fermion, fore=blue}{i3,v1}
\fmf{fermion, fore=blue}{v1,o1}
\fmf{fermion, fore=blue}{v1,o2}
\fmf{fermion, fore=blue}{v1,o3}
\fmf{photon, fore=red, tension=0}{v1,o4}
%\fmfblob{.35w}{v1}
\fmfv{d.sh=circle,d.f=empty,d.si=.25w,b=(.5,,0,,1)}{v1}
\end{fmfgraph*}
\end{gathered}\ 
=
\
\begin{gathered}
\begin{fmfgraph*}(90, 65)
\fmfleft{i1,d1,i2,i3}
\fmfv{label={\color{blue}\vdots}, label.dist=-4mm}{d1}
\fmfright{o1,d2,o2,u3,o3}
\fmfv{label={\color{blue}\vdots}, label.dist=-4mm}{d2}
\fmf{fermion, fore=blue}{i1,v1}
\fmf{fermion, fore=blue}{i2,v1}
\fmf{fermion, fore=blue}{i3,v1}
\fmf{fermion, fore=blue}{v1,o1}
\fmf{fermion, fore=blue}{v1,u2}
\fmf{photon, fore=red, tension=0}{u2,u3}
\fmf{fermion, fore=blue}{u2,o2}
\fmf{fermion, fore=blue}{v1,o3}
%\fmfblob{.35w}{v1}
\fmfv{d.sh=circle,d.f=empty,d.si=.25w,b=(.5,,0,,1)}{v1}
\end{fmfgraph*}
\end{gathered}
\ +\ \cdots
\end{align*}
\end{fmffile}these emission matrix elements will have poles at $\mathbf q=0$, corresponding to the Feynman diagrams in which the extra photon or graviton is emitted by one of the incoming or outgoing external particles in states $\alpha$ or $\beta$, since then the $n$-th outgoing, respectively incoming, particle of mass $m_n$ and momentum $p_n$ gives rise to a term of the form
\be
\frac{1}{(p_n\pm q)^2+m_n^2} = \pm \frac{1}{2p_n\cdot q}.
\ee
These poles give the dominant contribution in this limit, while diagrams in which the soft propagator is attached to an internal line will give rise to subleading corrections.

In the limit $\mathbf q \to 0$ we thus get,
\begin{equation}\begin{aligned}\label{WeinbergsWeinberg}\relax
		S_{\beta \alpha}^{\pm1}(\mathbf q) &\approx \frac{1}{(2\pi)^{3/2}\sqrt{2|\mathbf q|}}\left[ \sum_n \eta_n e_n \frac{p_n \cdot \varepsilon_{\pm}^\ast(q)}{p_n\cdot q}\right]S_{\beta\alpha}\\
		S_{\beta \alpha}^{\pm2}(\mathbf q) &\approx \frac{(8\pi )^{1/2}}{(2\pi)^{3/2}\sqrt{2|\mathbf q|}}\left[ \sum_n \eta_n f_n \frac{(p_n \cdot \varepsilon_{\pm}^\ast(q))^2}{p_n\cdot q}\right]S_{\beta\alpha},
\end{aligned}\end{equation}
$\eta_n$ being $+1$ or $-1$ according to whether the particle $n$ is outgoing or incoming.

As we have discussed, Lorentz invariance requires the vanishing of the vertex amplitude when a polarization is substituted with the corresponding four-momentum. This yields for $s=1$
\be
\sum_n \eta_n e_n=0
\ee
and for $s=2$
\be
\sum_{n} \eta_n f_n p_n^\mu=0.
\ee
The first one is precisely the conservation of the electric charge, whereas the second, when compared with the equation of momentum conservation $\sum \eta_n p_n=0$, yields the universality of the gravitational coupling constant $f_n=1$, for all $n$.

From these calculations, Weinberg also argues that the choice of $p_{\mu_1}\ldots p_{\mu_s}$ in \eqref{da_giustificare} gives in fact the only possible form of the $M$-function, also for emitting particles with spin $1$ or higher, since any other helicity-dependent vertex amplitude could never give rise to cancellations between different poles needed to satisfy the Lorentz invariance condition.

For higher helicities $s=3,4,\ldots$ one still has a factorization of the form
\be
S_{\beta \alpha}^{\pm s}(\mathbf q) \approx \frac{1}{(2\pi)^{3/2}\sqrt{2|\mathbf q|}}\left[ \sum_n \eta_n g_n^{(s)} \frac{(p_n \cdot \varepsilon_{\pm}^\ast(q))^s}{p_n\cdot q}\right]S_{\beta\alpha},
\ee
and the requirement
\be\label{sort_of_conservazione}
\sum_n \eta_n g_n^{(s)} \left[ p_n \cdot \varepsilon_\pm^\ast(q)\right]^{s-1}=0,
\ee
which contradicts momentum conservation unless $g_n^{(s)}=0$. This tells us that the low-energy interaction for higher spins is trivial or, in other words, that massless higher-spin particles cannot propagate long-range forces.
On the Lagrangian side, this implies that higher-spin interactions should be of multipolar type, \emph{i.e.} the vertices should contain enough derivatives so that they vanish in the soft limit.

\section{Soft Theorems and Asymptotic Symmetries}
We will now make the link between asymptotic symmetries and soft theorems explicit by providing the details of the derivation of the soft photon and soft graviton theorems from the underlying large $U(1)$ and $BMS$ supertranslation symmetry.

For this purpose, it is useful to recast Weinberg's result in terms of retarded coordinates and stereographic coordinates on the sphere \eqref{stereo}.
To this end, consider a wave packet for a massless particle with spatial momentum centered around $\mathbf{q}$. At large $r$ (and large times $t=u+r$ for fixed retarded time $u$), this wave packet becomes localized on the sphere at null infinity near the point 
\be
\mathbf q = {\omega}\,\widehat{\mathbf x} = \frac{\omega}{1+z\bar z}\,(z + \bar z, -i(z-\bar z), 1-z\bar z ) \, ,
\ee
so that the momentum of massless particles may be equivalently characterized by $q^\mu$ or $(\omega, z , \bar z)$.
The polarization  vectors can be chosen as follows \cite{Choi_Shim_Song}
\be\begin{aligned}\relax
	\varepsilon^+(\mathbf q) &= \frac{1}{\sqrt 2}\left( \bar z, 1, -i, -\bar z\right) ,\\
	\varepsilon^-(\mathbf q) &= \frac{1}{\sqrt 2}\left(  z, 1, i, - z\right) = \overline{\varepsilon^+(\mathbf q) } \, ,
\end{aligned}
\ee
thus allowing to rewrite Weinberg's soft theorem, \emph{e.g.} for a positive-helicity emission, from the momentum space form \eqref{wfact} to its position-space counterpart
\be\label{WeinbergPSPACE}
\lim_{\omega\to0^+}\omega\, S_{\beta\alpha}^{+s}  = (-1)^{s}\,{2^{\frac{s}{2}-1}}
(1+z\bar z)\left[\, \sum_i \eta_i^{\phantom{(s)}}\!\!\! g_i^{(s)} 
\frac{(E_{i})^{s-1}(\bar z-\bar{z}_i)^{s-1}}{(z-z_i)(1+z_i\bar z_i)^{s-1}}
\,\right]S_{\beta\alpha} \, ,
\ee
where, for simplicity, we have assumed that all the particles taking part in the scattering process are massless, the energy $E_i$ and the angular coordinates $(z_i,\bar z_i)$ characterizing their asymptotic states at null infinity.

\subsection{Large $U(1)$ symmetries and the soft photon theorem}
The action for electromagnetism coupled to a locally conserved current $\mathcal J^\mu$,
\be
S = - \frac{1}{4}\int \mathcal F_{\mu\nu} \mathcal F^{\mu\nu} d^D\!x - \int \mathcal A_\mu \mathcal J^\mu d^D\!x,
\ee
being invariant under $\delta \mathcal A_\mu =\partial_\mu \epsilon$ up to the boundary term,
possesses the canonical current
\be\label{j_s=1}{
	j^\mu = \mathcal F^{\nu\mu}\partial_\nu \epsilon + \mathcal J^\mu \epsilon.
}
\ee
In Bondi coordinates, near $\scrip$, in the case $\mathcal J=0$,
\be
Q^+ = \int_{\scrip} j^r \gamma_{z \bar z} r^2 du d^2\!z. 
\ee
As we discussed in Section \ref{sec:EMD=4}, choosing retarded radial gauge $\mathcal A_r = 0$ together with the falloff conditions
\be\begin{aligned}
	\mathcal A_u =& A_u(u, z, \bar z)/r + \mathcal O(r^{-2})\\
	\mathcal A_z =& A_z(u, z, \bar z) + \mathcal O(r^{-1})\\
	\mathcal A_{\bar z} =& A_{\bar z} (u, z, \bar z) + + \mathcal O(r^{-1})\,,
\end{aligned}\ee
and using \eqref{j_s=1}, the charge associated to the residual gauge freedom given by angular functions $T(z, \bar z)$ computed at $\scrip$ reads
\be{
	Q^+ = \int_{\scrip} T(z, \bar z) \left[ \partial_u (D^z  A_{z} +  D^{\bar z}  A_{\bar z} ) + \mathcal J\right] \gamma_{z\bar z} du d^2\!z},
\ee
where 
\be
\mathcal J(u, z, \bar z) \equiv \lim_{r\to\infty}r^2\mathcal J^r(u, r, z, \bar z).
\ee
Since this charge acts on matter fields by $\delta \Phi(x) = i[Q, \Phi(x)] = i e \epsilon(x) \Phi(x)$, any correlation function will satisfy
\be\begin{aligned}
	\langle \delta \prod_{n=1}^N \Phi_n(x_n) \rangle =& i \langle 0 | \left( Q^+ \prod_{n=1}^N \Phi_n(x_n) - \prod_{n=1}^N \Phi_n(x_n) Q^-\right) |0\rangle\\
	=& i \sum_{n=1}^N e_n \varepsilon(x_n) \langle \prod_{n=1}^N \Phi_n(x_n)\rangle.
\end{aligned}\ee
Performing LSZ reduction of the previous formula yields the Ward identity
\be
\langle\text{out}| (Q^+ S - S Q^-) |\text{in}\rangle = \sum_{n=1}^N \eta_{n} e_n T (z_n, \bar z_n) \langle\text{out}| S |\text{in}\rangle.
\ee
This derivation is given in \cite{Avery_Schwab}, where it was also observed that the antipodal identification commonly employed in the literature essentially consists in choosing the same gauge transformation for $\scrip$ and $\scrim$. This means that, in order to write down the correct Ward identity, one should take $T(z, \bar z)$ and its counterpart on $\mathscr I^-$ as limits to $\scri^\pm$ of the same bulk gauge transformation; an example of this will be given in Section \ref{ssec:memoryas} in the Lorenz gauge. Furthermore, the authors also note that since the charge is computed on a surface approximating $\scri^\pm \cup i^\pm$ which necessarily cuts through time-like trajectories, the results also hold for massive fields.

Using the auxiliary boundary condition $\partial_z A_{\bar z} = \partial_{\bar z} A_z$ at $\scrip_{\pm}$, which amounts to imposing the absence of long-range magnetic fields on $\mathscr I^+$, choosing $T(z, \bar z) = \frac{1}{w-z}$, where $w$ is a fixed complex parameter, and exploiting $\partial_{\bar z} \frac{1}{z-w} = 2\pi \delta^2(z-w)$ gives
\be\label{4piout_S}
4\pi \langle\text{out}|\left[\left(\int du \partial_u A_z\right) S - S \left(\int dv \partial_v A_z\right)\right] |\text{in}\rangle =
\sum_{n=1}^N \frac{\eta_{n} e_n}{z-z_n}\langle\text{out}| S |\text{in}\rangle, 
\ee
where we have used that $\mathcal J$ annihilates the vacuum, since the global $U(1)$ symmetry is unbroken. Using the free mode expansion for $A_z$ near $\scri$ and the stationary phase approximation, we obtain
\be
\int du e^{i\omega u} \partial_u A_z=-\frac{i}{8\pi^2}\frac{\sqrt 2}{1+z\bar z} \int_0^\infty d\omega_{\mathbf q} \left[ a^{\text{out}}_+(\omega_{q}\hat x) e^{-i\omega_{\mathbf q}u}
- a^{\text{out}\dagger}_-(\omega_{q}\hat x) e^{i\omega_{\mathbf q}u}\right]
\ee
so that
\be
\int_{-\infty}^{+\infty} du \partial_u A_z = - \frac{1}{8\pi} \frac{\sqrt 2}{1+z\bar z} \lim_{\omega\to 0^+}\left[ \omega a^{\text{out}}_+(\omega \hat x) + \omega a^{\text{out}\dagger}_-(\omega \hat x)\right].
\ee
Substituting this result into \eqref{4piout_S}, together with the analogous one for $\scrim$, and using crossing symmetry yields
\be{
	\lim_{\omega \to 0}\left[ \omega \langle \text{out}|a^{\text{out}}_+S|\text{in}\rangle \right]
	=
	-\frac{1+z\bar z}{\sqrt 2} \sum_n \frac{\eta_n e_n}{z-z_n} \langle\text{out}| S |\text{in}\rangle,
}\ee
which is Weinberg's theorem \eqref{WeinbergPSPACE}.

\subsection{BMS symmetry and the soft graviton theorem:\newline a linearized perspective}
The action for a massless Fierz-Pauli field $h_{\mu\nu}$, describing a linear perturbation of the Minkowski metric tensor, is
\be\label{S_spin2}
S = \frac{1}{2}\int \mathcal E^{\mu\nu}h_{\mu\nu}d^D\!x - \int J^{\mu\nu}h_{\mu\nu}d^D\!x,
\ee
where $\mathcal E^{\mu\nu}$ is the linearized Einstein tensor
\be
\mathcal E_{\mu\nu} = \Box h_{\mu\nu} - \partial_{(\mu} \partial\cdot h_{\nu)} - \partial_\mu\partial_\nu h' + \eta_{\mu\nu} (\partial\cdot\partial\cdot h - \Box h'),
\ee
and $J^{\mu\nu}$ is a conserved ``energy-momentum tensor'', $\partial_\mu J^{\mu\nu}=0$.
The action \eqref{S_spin2} is invariant under $\delta h_{\mu\nu} = \partial_{(\mu}\xi_{\nu)}$ up to the boundary term
\be
\int \partial_\mu \left[\left( \mathcal E^{\mu\nu}-2J^{\mu\nu}\right)\xi_\nu\right] d^D\!x,
\ee
since $\mathcal E^{\mu\nu}$ satisfies the linearized Bianchi identity $\partial\cdot \mathcal E^\nu = 0$ and $J^{\mu\nu}$ is conserved. The equations of motion are $\mathcal E^{\mu\nu} = J^{\mu\nu}$.

The symmetrized derivatives needed for the computation of the current are
\be\begin{aligned}
	\frac{\delta  S}{\delta h_{\alpha\beta,\mu\nu}} =& \frac{1}{2}\left[ 
	\eta^{\alpha\beta}h^{\mu\nu}+\eta^{\mu\nu}h^{\alpha\beta} \right.\\ 
	&\left. - \frac{1}{2}\left(\eta^{\mu\beta} h^{\nu\alpha} + \eta^{\mu\alpha}h^{\nu\beta} - \eta^{\nu\alpha}  h^{\mu\beta} -\eta^{\nu\beta}h^{\alpha\mu}\right) \right.\\
	&\left. -\left(\eta^{\alpha \beta} \eta^{\mu \nu} - \frac{1}{2}(\eta^{\mu\beta}\eta^{\nu\alpha} + \eta^{\mu\alpha} \eta^{\nu\beta}) \right)h'
	\right]\\
	=& \frac{1}{2}\left[ \frac{1}{2} H^{\mu\alpha\nu\beta} + \frac{1}{2} H^{\mu\beta\nu\alpha}\right],
\end{aligned}\ee
where $H^{\mu\alpha\nu\beta}$ is defined by
\be
H^{\mu\alpha\nu\beta} \equiv \eta^{\mu\nu}h^{\alpha\beta} + \eta^{\alpha \beta} h^{\mu\nu} - \eta^{\mu\beta} h^{\nu\alpha} - \eta^{\nu\alpha} h^{\mu\beta} - (\eta^{\mu\nu} \eta^{\alpha\beta} - \eta^{\mu\beta} \eta^{\nu\alpha})h'.
\ee
This tensor has the same symmetries of $R^{\mu\alpha\nu\beta}$,
\be
H^{\mu\alpha\nu\beta} = -H^{\alpha\mu\nu\beta} = - H^{\mu\alpha\beta\nu} = H^{\nu\beta\mu\alpha},
\ee
satisfies the cyclic identity,
\be
H^{\mu\alpha\nu\beta} + H^{\mu\nu\beta\alpha} + H^{\mu\beta\alpha\nu} =0,
\ee
and acts as a superpotential for the linearized Einstein tensor, meaning
\be\label{superpotential}
\mathcal E^{\mu\nu} = \partial_\alpha \partial_\beta H^{\mu\alpha\nu\beta}.
\ee
Defining the trace-reversed tensor $\bar h^{\mu\nu} = h^{\mu\nu} - \frac{1}{2}\eta^{\mu\nu} h'$, satisfying $\bar h'=- h'$, one gets the simpler form
\be
H^{\mu\alpha\nu\beta} = \eta^{\mu\nu}\bar h^{\alpha\beta} + \eta^{\alpha \beta} \bar h^{\mu\nu} - \eta^{\mu\beta} \bar h^{\nu\alpha} - \eta^{\nu\alpha} \bar h^{\mu\beta}. 
\ee
The canonical Noether current is given by
\be
j^\mu = \frac{\delta S}{\delta h_{\alpha\beta,\mu\nu}}\delta h_{\alpha\beta,\nu} - \partial_\nu \frac{\delta S}{\delta h_{\alpha\beta,\mu\nu}} \delta h_{\alpha\beta} + J^{\mu\nu}\xi_{\nu},
\ee
where $\delta h_{\alpha\beta} = \partial_{(\alpha}\xi_{\beta)}$.
The contribution $J^{\mu\nu}\xi_{\nu}$ is given by the boundary term in the variation of the action. Thus
\be\label{jmu_spin2}
j^\mu =
\frac{1}{2}\left( H^{\mu\alpha\nu\beta} \partial_\nu \partial_\alpha \xi_\beta - \partial_\nu H^{\mu\alpha\nu\beta} (\partial_\alpha \xi_\beta + \partial_\beta \xi_\alpha) \right) + J^{\mu\nu}\xi_{\nu},
\ee 
where we have used the antisymmetry of $H^{\mu\alpha\nu\beta}$ in $\nu\beta$ and the symmetry of $\partial_\nu\partial_\beta$ (or analogous considerations for similar contributions) for the first term and symmetrized in $\alpha\beta$ the second term.

Let us now recover the Noether tensor $\kappa^{\mu\nu}$ satisfying $j^\mu=\partial_\nu\kappa^{\mu\nu}$, whose existence is ensured by Noether's second theorem; for this purpose we can set $J^{\mu\nu}=0$ without loss of generality.
Integrating by parts each term in \eqref{jmu_spin2}, employing the equations of motion $\partial_\alpha \partial_\beta H^{\mu\alpha\nu\beta} = 0$, and renaming the indices appropriately we get
\be
j^\mu =  \frac{1}{2} \left\{ \partial_\alpha \left[ H^{\mu\alpha\nu\beta}\partial_\nu\xi_\beta - \partial_\nu \left( H^{\mu\alpha\nu\beta}+ H^{\mu\nu\alpha\beta} + H^{\mu\beta\nu\alpha} \right)\xi_\beta \right]\right\},
\ee
so that thanks to the cyclic identity 
\be
j^\mu = \partial_\alpha \kappa^{\mu\alpha},\qquad 
\kappa^{\mu\alpha} = \frac{1}{2}H^{\mu\alpha\nu\beta} \partial_\nu \xi_\beta + \xi_\nu \partial_\beta H^{\mu\alpha\nu\beta}.
\ee
Now, we may think to have obtained these expressions in a given locally inertial frame: to covariantize them we simply replace ordinary derivatives with covariant derivatives and note that no ambiguity arises in their ordering, since the connection defining them is given by the flat background metric and hence such derivatives commute; thus 
\begin{equation}\label{js=2}{
		j^c =
		\frac{1}{2}\left( H^{cadb} \nabla_d \nabla_a \xi_b - \nabla_d H^{cadb} (\nabla_a \xi_b + \nabla_b \xi_a) \right)+ J^{cd}\xi_{d},}
\end{equation}
and
\begin{equation}\label{ks=2}{
		\kappa^{ca} = \frac{1}{2}H^{cadb} \nabla_d \xi_b + \xi_d \nabla_b H^{cadb}.}
\end{equation}

From the perspective of linearized gravity (\emph{i.e.} of a generic spin-2 massless field) the Bondi gauge is fixed by the following choice of boundary conditions, which as we saw stems from considerations in the nonlinear theory:
\be\label{definizione_Bondi_gauge}
h_{ab}dx^a dx^b = \frac{2m_B}{r}du^2 - 2U_z du dz - 2 U_{\bar z} du d\bar z + r C_{zz} dz^2 + r C_{\bar z \bar z}d\bar z^2\,,
\ee
where $u,r,z,\bar z$ are the usual retarded coordinates. 

Considering now the asymptotic symmetries we have evaluated in  Chapter \ref{chap:basics} and restricting to supertranslations, we have the infinitesimal symmetry generators
\be\begin{aligned}
	\xi_a dx^a &= - (T+ D^z D_z T )du - T(z, \bar z)dr - r D_z T dz - r D_{\bar z}T d\bar z,\\
	\xi^a \partial_a  &= T(z, \bar z) \partial_u  + D^z D_z T \partial_r - \frac{1}{r} (D^zT \partial_z + D^{\bar z}T \partial_{\bar z} )
\end{aligned}\ee
which indeed leave the ``Bondi gauge'' defined by \eqref{definizione_Bondi_gauge} invariant.

We may now compute the charge associated with this residual supertranslation gauge symmetry, starting either with the Noether tensor $k^{ab}$ or from the current $j^a$ itself. In any case, the explicit computation of the nonvanishing components of the tensor $H^{abcd}$ is quite useful: to leading order,
\be\begin{aligned}
\label{useful_table}
		H^{ur zr}= \frac{U_{\bar z}}{\gamma_{z \bar z}r^2}\,,&\qquad 
		H^{uzrz} = -\frac{C_{\bar z\bar z}}{\gamma^2_{z\bar z}r^3}\,,\qquad 
		H^{rzrz}= \frac{C_{\bar z \bar z}}{\gamma^2_{z\bar z}r^3}\,,\\
		&H^{rz z\bar z}= \frac{U_{\bar z}}{(\gamma_{z \bar z}r^2)^2}\,,\qquad 
		H^{rzr\bar z} = \frac{2m_B}{\gamma_{z\bar z}r^3}\,,
\end{aligned}\ee
where the components with $\bar z$ and $z$ interchanged are obtained by formal conjugation of all indices.
It is also convenient to compute the ``commutators'' $\xi_{[a;b]}=\xi_{[a,b]}$:
\be\label{tabella_dei_commutatori}
\xi_{[u,r]}=0,\qquad \xi_{[u,z]} = D_z(T + D^zD_zT),\qquad \xi_{[r,z]}=0,\qquad \xi_{[z,\bar z]}=0.
\ee

We start computing $\kappa^{ur}$ from \eqref{ks=2}, since this component is selected by the measure element of $\scrip_-$. Observe that
$
\frac{1}{2}H^{urdb}\nabla_d\xi_b = \frac{1}{4}H^{urdb} \xi_{[b, d]}
$
by the antisymmetry of $H^{urdb}$ in $db$; by \eqref{tabella_dei_commutatori} the only potentially surviving term would be $\frac{1}{4}H^{uruz}\xi_{[u,z]}$, which vanishes anyway since $H^{uruz}$ is itself zero. The other contribution to the $\kappa^{\mu\nu}$ form from \eqref{ks=2} is
\be
\xi_d \nabla_b H^{urdb} = \xi_d \partial_b H^{urdb} + 
\xi_d \Gamma^{u}_{eb}H^{e r db} + \xi_d \Gamma^r_{eb} H^{uedb} + \xi_d 
\Gamma^d_{eb}H^{ur eb} + \xi_d \Gamma^b_{be}H^{urde};
\ee
the fourth term on the right-hand side vanishes by symmetry/antisymmetry in the summed indices while $\Gamma^\beta_{\beta\rho}=\partial_\rho \log \sqrt{g}$ in the last term. Taking into account the nonvanishing Christoffel symbols and $H^{abcd}$ components, we get
\be\begin{aligned}
	\kappa^{ur}=&
	\left[{2}\partial_z T \frac{U_{\bar z}}{\gamma_{z\bar z}r^2} + T \partial_z \left(\frac{U_{\bar z}}{\gamma_{z\bar z}r^2} \right)+z\leftrightarrow \bar z\right]\\
	&+\left[ \left( \partial_z T \frac{U_{\bar z}}{\gamma_{z\bar z}r^2} + z\leftrightarrow\bar z \right) + 2 T \frac{2m_B}{r^2}\right]\\
	&+\left[-2 \partial_zT \frac{U_{\bar z}}{\gamma_{z\bar z}r^2} + T \frac{U_{\bar z}}{\gamma^2_{z\bar z}r^2} \partial_z \gamma_{z\bar z}+z\leftrightarrow\bar z\right],
\end{aligned}\ee
were ``$z\leftrightarrow\bar z$'' refers to formal complex conjugation in the $z$ and $\bar z$ indices.
Hence, after expanding the derivative in the second term,
\be
\kappa^{ur} = 2 T\, \frac{2m_B}{r^2} + \frac{1}{\gamma_{z\bar z}r^2}\left[\partial_z(T U_{\bar z}) + z\leftrightarrow\bar z\right];
\ee
integrating this expression as
\be
Q^+ = \int_{\scrip_-} \kappa^{ur} \gamma_{z\bar z} r^2d^2\!z,
\ee
and recalling that the sphere has no boundary, we obtain
\be
{
	Q^+ = 4 \int_{\scrip_-} T(z, \bar z) m_B(u, z, \bar z) \gamma_{z\bar z}d^2\!z.
}
\ee
Again, the factor $r^2$ from the measure element gets canceled and the charge is meaningfully expressed as an integral over the boundary of null infinity. 

The computation of $j^r$ from \eqref{js=2}, instead, goes as follows. Note that
\be H^{radb}\nabla_d\nabla_a\xi_b = H^{radb}\partial_a\partial_d\xi_b = \frac{1}{2}H^{radb}\partial_a \xi_{[b,d]}\ee by the vanishing of the Riemann tensor and by antisymmetry in $db$. Therefore, due to \eqref{useful_table}, the only relevant component is $H^{rzuz}\sim 1/r^3$: this term gives a sub-leading contribution. Altogether, always taking \eqref{useful_table} and \eqref{tabella_dei_commutatori} into account, one finds that the only leading contribution to $j^a$ comes from the following term
\be
\partial_u H^{rzuz} \nabla_{(z}\xi_{z)} + z\leftrightarrow\bar z = \frac{2}{\gamma^2_{z\bar z}r^2}\left[ \partial_u C_{\bar z \bar z} D^2_zT(z,\bar z) + z\leftrightarrow\bar z\right].
\ee
Thus,
\be
j^r = - \frac{1}{\gamma^2_{z\bar z}r^2}\left[ \partial_u C_{ z  z} D^2_{\bar z}T(z,\bar z)+\partial_u C_{\bar z \bar z} D^2_zT(z,\bar z)\right] - J^{rr}(u,r, z, \bar z)T(z, \bar z)
\ee
and
\be
{
	Q^+ =  \int_{\scrip} T(z, \bar z) \left[-\partial_u \left( D^z D^zC_{zz} + D^{\bar z} D^{\bar z} C_{\bar z \bar z} \right) - J(u,z,\bar z) \right]\gamma_{z\bar z} d^2\!z du,
}
\ee
where
\be
J(u, z, \bar z) \equiv \lim_{r\to\infty}r^{2}J^{rr}(u, r, z, \bar z).
\ee
Since supertranslations act on matter fields by $i T(z, \bar z) \partial_u$ at $\scrip$, we get by LSZ reduction
\be\label{Q+SSQ-}
\langle\text{out}| (Q^+ S - S Q^-) |\text{in}\rangle = \sum_{n=1}^N \eta_{n} f_n E_n T (z_n, \bar z_n) \langle\text{out}| S |\text{in}\rangle,
\ee 
where $f_n$ is the gravitational coupling of each field. Using the auxiliary boundary condition 
\be
D^zD^z C_{zz} =D^{\bar z} D^{\bar z} C_{\bar z \bar z}\quad \text{ at }\scrip_\pm,
\ee
we have
\be\label{T_s=2}
Q^+ = -2 \int_{\scrip} T(z, \bar z) \partial_u D^z D^zC_{zz} \gamma_{z\bar z} d^2\!z du.
\ee
Now, in order to make contact with Weinberg's soft theorem, we choose as $T(z,\bar z)$ an angular function of the following type:
\be\label{TzzMIO}
T(z, \bar z) = \frac{1}{w-z}\frac{1+w\bar z}{1+z\bar z}.
\ee
Then the left-hand side of \eqref{Q+SSQ-}, after an integration by parts in $\partial_{\bar z}$, involves computing
\be\begin{aligned}
	\partial_{\bar z} \left( \frac{1}{w-z}\frac{1+w\bar z}{1+z\bar z} \right) =& -2\pi \delta^2(z-w)\frac{1+w\bar z}{1+z\bar z} + \frac{1}{w-z}\frac{w(1+z\bar z) - (1+w\bar z)z}{(1+z\bar z)^2}\\
	=& -2\pi \delta^2(z-w) + \frac{1}{(1+z\bar z)^2}\\
	=& -2\pi \delta^2(z-w) + \frac{1}{2}\gamma_{z\bar z}.
\end{aligned}\ee
Therefore
\be
Q^+ = -4\pi \int du D^w C_{ww} + \int D^z C_{zz} \gamma_{z\bar z} d^2\!z du, 
\ee
where the second term is a boundary contribution on the sphere and hence gives zero.
To sum up,
\be\label{Ward_s=2}\begin{aligned}
&-4\pi D^z\langle\text{out}|\left[\left(\int du \partial_u C_{zz} \right) S - S \left( \int dv \partial_v C_{zz} \right)\right]|\text{in}\rangle \\
&=
\sum_{n=1}^N \eta_{n}  \frac{f_nE_n}{z-z_n}\frac{1+z\bar z_n}{1+z_n\bar z_n} \langle\text{out}| S |\text{in}\rangle.
\end{aligned}\ee
As in the spin-one case, we perform a stationary phase approximation as $r\to\infty$ to express $C_{zz}$ in terms of soft graviton creation and annihilation operators, which yields
\be
C_{zz} = - \frac{i}{8\pi^2} \frac{2}{(1+z\bar z)^2}\int_0^{+\infty}
d\omega_{\mathbf q} \left[ a^{\text{out}}_+(\omega_{q}\hat x) e^{-i\omega_{\mathbf q}u}
- a^{\text{out}\dagger}_-(\omega_{q}\hat x) e^{i\omega_{\mathbf q}u}\right],
\ee
and
\be
\int du \partial_u C_{zz} = - \frac{1}{8\pi} \frac{ 2}{(1+z\bar z)^2} \lim_{\omega\to 0^+}\left[ \omega a^{\text{out}}_+(\omega \hat x) + \omega a^{\text{out}\dagger}_-(\omega \hat x)\right].
\ee
Using crossing symmetry and the matching condition, we also have
\be
-4\pi \langle\text{out}|\left[\left(\int du \partial_u C_{zz} \right) S - S \left( \int dv \partial_v C_{zz} \right)\right]|\text{in}\rangle = \frac{2}{(1+z\bar z)^2} \lim_{\omega\to 0}\langle \text{out}|\omega a^{\text{out}}_+(\omega \hat x) |\text{in}\rangle,
\ee
and this implies, by comparison with \eqref{Ward_s=2},
\be{
	\lim_{\omega\to 0}\langle \text{out}|\omega a^{\text{out}}_+(\omega \hat x) |\text{in}\rangle= \lim_{\omega\to0} (1+z\bar z)\sum_n \eta_n f_n \frac{E_n(\bar z - \bar z_n)}{(z-z_n)(1+z_n \bar z_n)},}
\ee
since
\be
\gamma_{z\bar z}\partial_{\bar z} \frac{2}{1+z\bar z}\sum_n \eta_n f_n \frac{E_n(\bar z - \bar z_n)}{(z-z_n)(1+z_n \bar z_n)} = \sum_n \eta_n f_n \frac{E_n(1+z\bar z_n)}{(z-z_n) (1+z_n \bar z_n)};
\ee
note that we omitted the $\partial_{\bar z}\frac{1}{z-z_n}$ term, since here the delta multiplies a function which vanishes when $\bar z = \bar z_n$. 

This shows the supertranslation Ward identity to be equivalent to Weinberg's factorization formula \eqref{WeinbergPSPACE}, without assuming from the beginning $f_n=$ constant.
Notice also that our choice \eqref{T_s=2} of $T$ is not restrictive, since we may always write
\be
f(z, \bar z) = \int \frac{d^2w}{2\pi} f(w, \bar w) \partial_{\bar w} \frac{1}{w-z}\frac{1+w\bar z}{1+z \bar z}
\ee
and then use the linearity of the Ward identity to recover the full supertranslation invariance from Weinberg's theorem. 

\section{Memory Effects}
By memory effects, we mean a class of observable phenomena that characterize the passage of radiation impinging on a test charge and persist after said radiation has died out. 
For instance, a pair of test masses may undergo a nonzero relative displacement after the passage of gravitational radiation      \cite{Zeldovich,Frauendiener,memory,Bieri:2015yia} or a small electric charge, initially at rest, may display a nonzero velocity after it is invested by electromagnetic radiation \cite{Bieri:2013hqa, Mao_em, Mao_EvenD}.
Analogous Yang-Mills memory effects have also been proposed \cite{Jokela:2019apz} and a similar phenomenon has been identified in the context of the interaction of a two-form with a test string \cite{Afshar:2018sbq}. Memory effects can be also stored in the quantum states of superconducting condensates \cite{Susskind} and quarks \cite{StromingerColor}. In all these cases, such effects have the property of leaving on the test systems a permanent imprint of the radiation itself. 

Memory effects can be induced on a particle sitting near null infinity both by the radiation emitted by the movement of charged sources in the interior of the spacetime, or by the outflow of charged massless matter, which travels along null rays. The former case is usually referred to as \emph{linear} or \emph{ordinary} memory and can be regarded as a picture of the movement of the bulk charges that is stored in properties of the test particles, while the latter has been termed \emph{nonlinear} or \emph{null} memory, because it signals the passage of charged radiation, as occurs both in nonlinear field theories containing self-interacting massless particles and in linearized theories with charged massless sources.

We will now illustrate some examples of memory effects and then turn to explaining the connection between these phenomena and asymptotic symmetries, the former being interpreted as observable consequences of the latter.

\subsection{Scalar and electromagnetic memory}
Although the main interest has been on memory effects in gauge theories, a simple example thereof is already provided by scalar memory. To illustrate it, in its simplest realization, it is sufficient to consider a particle, charged under a scalar field $\varphi$, that is created at rest in the origin at $t=0$. This situation is described by the equation
\be
-\Box \varphi(t,\mathbf x)=q\theta(t)\delta(\mathbf x)\,,
\ee
(we stick to the mostly-plus convention $\Box = \eta^{\mu\nu}\partial_\mu \partial_\nu=-\partial_t^2+\nabla^2$)
which is solved by convolution of the right-hand side with the retarded propagator
\be
D_{\text{ret}}(x)=\frac{\delta(t-|\mathbf x|)}{4\pi|\mathbf x|}
\ee
and, switching to retarded coordinates, yields
\be
\varphi(u,r)=q\,\frac{\theta(u)}{4\pi r}\,.
\ee
This process will induce the following change in the energy of a test particle with charge $Q$ which is held at a distance $r$ from the source
\be
\Delta P_u(u) = Q\int_{-\infty}^{u} \partial_u \varphi\,du' = qQ\,\frac{\theta(u)}{4\pi r}\,.
\ee
This is just the expected variation in the Coulombic interaction energy due to the creation of the new particle in the origin. Naturally, the effect only takes place after $u=0$, namely after the world-line of the test particle crosses the wave of radiation induced by the particle's creation in the origin, which is a spherical delta-like impulse traveling on the light-cone. 

Another example in the scalar case is that of a particle, initially at rest in the origin, that suddenly starts moving with nonzero velocity $\mathbf v$ at $t=0$. The corresponding solution to the field equation can be obtained by combining a time-reversal and a Lorentz boost of the previous solution, namely, by superposing the field generated by a particle that is destroyed in the origin and that of a particle that is created with velocity $\mathbf v$. 
The needed boost in this case is  
\be
t\mapsto \gamma(\mathbf v)(t-\mathbf v \cdot \mathbf x)\,,\qquad
\mathbf x \mapsto \mathbf x + \mathbf v (\gamma(\mathbf v)-1)\frac{\mathbf v \cdot \mathbf x}{\mathbf v^2}-\gamma(\mathbf v)\mathbf v t
\ee   
or, in terms of retarded coordinates,
\be
u\mapsto \frac{u}{\gamma(\mathbf v)(1-\mathbf n\cdot \mathbf v)}+\mathcal O(r^{-1})\,,\qquad
t\mapsto r\, \gamma(\mathbf v)(1-\mathbf n\cdot \mathbf v)+\mathcal O(1)\,,
\ee
where $\gamma(\mathbf v)=(1-\mathbf v^2)^{-1/2}$ and $\mathbf n = \mathbf x/r $. The solution is then given, to leading order, by 
\be
\varphi(u,r,\mathbf n) = q\,\frac{\theta(-u)}{4\pi r}+\frac{q\,\theta(u)}{4\pi r\,\gamma(\mathbf v)(1-\mathbf n\cdot \mathbf v)}+\mathcal O(r^{-2})\,.
\ee
Performing the integral of $\partial_u\varphi$ with respect to retarded time, we arrive at the memory effect
\be
\Delta P_u =qQ\, \frac{\theta(u)}{4\pi r}\left(-1+ \frac{1}{\gamma(\mathbf v)(1-\mathbf n\cdot \mathbf v)} \right)+\mathcal O(r^{-2})\,,
\ee
which is to be interpreted as the fact that, after the particle starts moving, the interaction energy becomes modified as a consequence of the relativistic length contraction. Solutions to the field equation, and hence memory effects, associated to a generic, idealized scattering process involving a number of incoming and outgoing particles and taking place near the origin can be constructed by superposition of solutions in a similar manner \cite{Garfinkle:2017fre}, but do not introduce qualitatively new elements.

While until now we have been concerned only with ordinary memory, it is also possible to provide an explicit example of null memory effect, by considering a point-like source moving at the speed of light in the $\mathbf x_0$ direction ($|\mathbf x_0|=1$)
\be
-\Box \varphi = q\,\delta(\mathbf x - \mathbf x_0 t)\,.
\ee 
Rewriting this equation in retarded coordinates near future null infinity, we have
\be
2\left(\partial_r + \frac{1}{r} \right)\partial_u \varphi - \left(\partial_r^2+\frac{2}{r}\,\partial_r + \frac{\Delta}{r^2} \right)\varphi= \frac{q}{r^2}\,\delta(u)\delta(\mathbf n, \mathbf x_0)\,,
\ee
where we recall that $\Delta$ is the Laplace-Beltrami operator on the Euclidean unit sphere and $\delta(\mathbf n, \mathbf x_0)$ is the normalized delta function thereon. We see that a solution is furnished by writing $\varphi$ in terms of the formal expansion
\be\label{formalsolutionmemory}
\varphi(u,r,\mathbf n) = \sum_{k=0}^\infty r^k\delta^{(k+1)}(u) C^{(k)}(\mathbf n)
\ee 
provided that\footnote{
	We added a uniform surface charge density $-\frac{q}{4\pi}$ on the sphere, which has no influence on $\partial_i\varphi$.
}  
\be\label{C0memory}
-\Delta C^{(0)}(\mathbf n) = q\,
\left(
\delta(\mathbf n, \mathbf x_0)-\frac{1}{4\pi}
\right)\,,
\ee
and the other coefficients satisfy the recursion relation
\be\label{recursionnull4}
[\Delta+(k+1)(k+2)]C^{(k+1)}(\mathbf n)=2(k+1) C^{(k)}(\mathbf n)\,.
\ee
Equation \eqref{C0memory} is solved by means of the Green's function of the Laplace-Beltrami operator
\be
C^{(0)}(\mathbf n) = -\frac{q}{4\pi}\,\log(1-\mathbf n\cdot \mathbf x_0)\,,
\ee
while equations \eqref{recursionnull4} can be solved recursively for $C^{(k+1)}(\mathbf n)$, up to terms proportional to the spherical harmonics, which satisfy $-\Delta Y_{k+1}^m(\mathbf n)=(k+1)(k+2)Y_{k+1}^m(\mathbf n)$ and annihilate the left-hand side. The field \eqref{formalsolutionmemory} will then give rise to the following leading-order memory effect  on a test charge $Q$:
\be
\Delta P_i(u) = Q \int_{-\infty}^u \partial_i \varphi\, du' = qQ\,\frac{(x_0)_i}{4\pi(1-\mathbf n \cdot \mathbf x_0)}\,,
\ee
provided that $u>0$, while $\Delta P_i=0$ for $u<0$, where we have made use of the fact that the all terms in the expansion of $\varphi$, except $k=0$, are multiplied by higher-order derivatives of $\delta(u)$ and hence give a (singular) contribution with support in $u=0$.

Similar effects are present in the case of the electromagnetic field and can be conveniently calculated in the Lorenz gauge $\partial^\mu \mathcal A_\mu  =0$, where the equations of motion reduce to a set of scalar wave equations
\be
\Box\, \mathcal A_\mu = j_\mu\,.
\ee
Indeed, first considering the case of a static particle created in the origin,\footnote{Strictly speaking, this equation is not well-posed, since the right-hand side has a nonzero divergence. However, the final process in which will be interested does not share this problem.}
\be
\Box\, \mathcal A^\mu = u^\mu \,q\theta(t)\delta(\mathbf x)\,,
\ee
where $u^\mu=(1,0,0,0)$, hence
$
\mathcal A^\mu = - u^\mu \varphi
$,
where $\varphi$ denotes the corresponding solution for the scalar field.
Boosting this solution yields $\mathcal A^\mu=(\mathcal A^0, \mathbf A)=-\gamma(\mathbf v)(1,\mathbf v)\varphi$ and going to retarded components yields
\be
\mathcal A_u = \gamma(\mathbf v)\varphi\,,\qquad
\mathcal A_r = \gamma(\mathbf v)(1-\mathbf n \cdot \mathbf v)\varphi\,,\qquad
\mathcal A_i = -r\,\gamma(\mathbf v) v_i \varphi\,.
\ee
Conjoining as before the solution corresponding to a particle destroyed in the origin and to a particle created with velocity $\mathbf v$, we obtain 
\be \begin{aligned}
\mathcal A_u &= 
q\,\frac{\theta(-u)}{4\pi r}+\frac{q\,\theta(u)}{4\pi r\,(1-\mathbf n\cdot \mathbf v)}+\mathcal {O}(r^{-2})\,,\\
\mathcal A_r &= \frac{q}{4\pi r}+\mathcal {O}(r^{-2})\\
\mathcal A_i &= -\frac{q\,v_i\theta(u)}{4\pi(1-\mathbf n\cdot \mathbf v)}+\mathcal {O}(r^{-1})\,.
\end{aligned} \ee
The change four-momentum of a test charge $Q$ will be subject to 
\be\label{memoryD=4}
\Delta  P_i(u) = Q \int_{-\infty}^{u} \mathcal F_{iu} du' = \frac{qQ\,v_i\theta(u)}{4\pi(1-\mathbf n\cdot \mathbf v)}+\mathcal {O}(r^{-1})\,.
\ee 
Let us note that the leading memory effect is proportional to the variation of the field component $\mathcal A_i$ between $u>0$ and $u<0$ (where it vanishes), which can be rewritten in the following way
\be\label{totalgradsphere}
\mathcal A_i\big|_{u>0}-\mathcal A_i\big|_{u<0} = -\frac{q\,v_i}{4\pi(1-\mathbf n\cdot \mathbf v)}=q\,\partial_i \log(1-\mathbf n\cdot \mathbf v)\,.
\ee  
Note that this difference takes the form of a derivative on the sphere, \emph{i.e.} of a gauge transformation.
More details on the interpretation of this equation in connection  with asymptotic symmetries in the Lorenz gauge will be provided in the next section.

Null memory can be instead displayed by considering 
\be
\Box \mathcal A^\mu = q\, v^\mu \delta(\mathbf x- \mathbf x_0 t)\,,
\ee
with $v=(1,\mathbf x_0)$. Taking into account the corresponding solution \eqref{formalsolutionmemory} for the scalar field $\varphi$, we then have $A^\mu = - v^\mu \varphi $ and, moving to retarded components,
\be
\mathcal A_u = \varphi\,,\qquad
\mathcal A_r = (1-\mathbf n \cdot \mathbf x_0)\varphi\,,\qquad
\mathcal A_i = -r\, (x_0)_i \varphi\,.
\ee
Consequently
\be\begin{aligned}\label{Anull4}
\mathcal A_u &= -\frac{q\delta(u)}{4\pi}\,\log(1-\mathbf n \cdot \mathbf x_0)+\cdots\,,\\
\mathcal A_r &= \frac{q\delta(u)}{4\pi}(1-\mathbf n \cdot \mathbf x_0)\log(1-\mathbf n \cdot \mathbf x_0)+\cdots\,,\\
\mathcal A_i &= - r\,\frac{q (x_0)_i \delta(u)}{4\pi}\,\log(1-\mathbf n \cdot \mathbf x_0)+\cdots\,,
\end{aligned}\ee
where we have omitted terms proportional to higher-order derivatives of the delta function $\delta(u)$.
The null memory formula then reads, for $u>0$,
\be
\Delta P_i (u)= Q\int_{-\infty}^u \mathcal F_{iu}\, du' = \frac{qQ(\mathbf x_0)_i}{4\pi(1-\mathbf n \cdot \mathbf x_0)}\,,
\ee
while $\Delta P_i=0$ for $u<0$;
indeed, the terms omitted from \eqref{Anull4} have support in $u=0$ and hence do not contribute to the memory effect. 

Note that this result is formally identical to the analogous formula for ordinary memory, upon substituting $\mathbf v$ with $\mathbf x_0$. A key difference is that, in the former case, the formula is smooth over the whole celestial sphere, while in the latter it displays a puncture at the point where the sphere is pierced by the outgoing massless charge.

\subsection{Memory and asymptotic symmetries}\label{ssec:memoryas}
Aside from being physically interesting in their own right, memory effects possess an additional piece of interest in that they can be interpreted as observable effects associated to asymptotic symmetries. More precisely, one observes that the configurations of the system before and after the passage of radiation are mapped into one another by the action of a large gauge symmetry transformation; it is this underlying transition between inequivalent radiative vacua that can be held responsible for a nontrivial memory effect.

This feature is clearly exhibited by electromagnetism in radial gauge in four dimensions, where one considers solutions to the Maxwell field equations subject to $\mathcal A_r = 0$ and
\be
\mathcal A_u=\mathcal O(r^{-1})\,,\qquad
\mathcal A_i=\mathcal O(1)\,,
\ee
as $r\to\infty$, where $i=1$, $2$ denote two angular coordinates. The asymptotic symmetries of this system are given by gauge parameters $\epsilon  =  T(x^1,x^2)$ that only depend on the angular coordinates and hence generate the transformations
\be
\delta \mathcal A_i = \partial_i T\,,
\ee
while $\mathcal A_u$ is gauge-invariant. 

For a generic solution of the field equations, a test particle with unit electric charge, which is initially at rest at a large distance $r$ from the origin, will feel a leading-order electric field given by 
\be
\mathcal F_{ui}=\partial_u \mathcal A_i+\cdots\,,\qquad
\mathcal F_{ur}=\mathcal O(r^{-2})\,.
\ee
Hence, assuming it is subject to a radiation train with support between retarded times $u_0$ and $u_1$, it develop a momentum kick
\be
\Delta P_i = \int_{u_0}^{u_1} \mathcal F_{ui} \,du = \mathcal A_i\big|_{u_1}-\mathcal A_i\big|_{u_0}+\mathcal O(r^{-1})
\ee
in the direction tangent to the celestial sphere.
In this step, it has been assumed that $u_1-u_0$ is sufficiently small and allows us to neglect the contribution due to the magnetic field, which will be further suppressed by the particle's velocity.

On the other hand, before and after the passage of radiation, the particle must be in a radiative vacuum configuration and hence there must exist a function $T(x^1, x^2)$ such that
\be
\mathcal A_i\big|_{u_1}-\mathcal A_i\big|_{u_0}=\partial_i T\,.
\ee
We thus finally see that the momentum kick memory effect, signaled by the test particle, can be interpreted as the action of a large gauge transformation on the underlying gauge field, which connects two different ``vacua'' of the theory.

A similar line of reasoning can be applied to the Lorenz gauge. Indeed, the memory formula \eqref{memoryD=4} strongly suggests, by \eqref{totalgradsphere}, that also in this case a momentum kick is actually proportional to a gauge transformation relating the leading term of $\mathcal A_i$ before and after the passage of radiation. In order to see that this is indeed the case, one must find solutions of 
\be
\Box \epsilon=0\,,
\ee
which identifies the residual gauge freedom after imposing $\partial^\mu \mathcal A_\mu=0$, that approach asymptotically a given function $T(\mathbf n)$ on the celestial sphere. For instance, in the case of \eqref{memoryD=4} and \eqref{totalgradsphere},
\be
T(\mathbf n) = q\,\log(1-\mathbf n \cdot \mathbf v)\,.
\ee
This can be achieved by taking the following the integral on the Euclidean unit sphere \cite{Hirai:2018ijc}
\be\label{epsilonLorenz4}
\epsilon(x) = \oint G(x,\mathbf q) T(\mathbf q)\,d\Omega(\mathbf q)\,,
\ee
where the propagator $G$ is defined by
\be
G(x,\mathbf q) = -\frac{1}{4\pi}\,\text{Re}\, \frac{x^2}{(x\cdot q+i\varepsilon)^2}\,,
\ee
with $q=(1,\mathbf q)$, the limit $\varepsilon\to0^+$ being understood. The introduction of this small imaginary part in the denominator is needed in order to avoid the poles at $\mathbf n \cdot \mathbf q=t/r$, which occur when $x$ lies outside of the light-cone, $|t|<r$. In order to verify that \eqref{epsilonLorenz4} indeed furnishes the desired result, first, we can immediately check that $\Box G=0$. Second, aligning the direction $\mathbf n=\mathbf x/r$ along the $z$ axis, which is not restrictive due to the rotational invariance of the measure $d\Omega$, we have, for any $u\neq0$,
\be\begin{aligned}
\oint G(x,\mathbf q) \lambda(\mathbf q)\,d\Omega(\mathbf q)
&=
\frac{1}{4\pi}\, \text{Re}
\int_0^{2\pi} d\phi \int_0^{\pi} \sin\theta d\theta\, \frac{u(u+2r)T(\mathbf q(\theta, \phi))}{[(u+r(1-\cos\theta)+i\varepsilon]^2}\\
&=
\left(1+\frac{u}{2r}\right)
\text{Re}
\int_0^{2\pi}\frac{d\phi}{2\pi}
\int_0^{2r/u}d\tau\, \frac{T(\mathbf q(\arccos(1-\frac{u\tau}{r}), \phi))}{(1+\tau+i\varepsilon)^2}\\
&
\xrightarrow[r\to\infty]{}
\text{Re}
\int_0^{u \cdot \infty} \frac{d\tau}{(1+\tau+i\varepsilon)^2}\,T(\mathbf n)=\text{Re}\,\frac{T(\mathbf n)}{1+i\epsilon}=T(\mathbf n)\,.
\end{aligned}\ee
This integral gives instead zero when evaluated on the light-cone $u=0$, so that the pointwise limit $r\to\infty$ yields $T(\mathbf n)$ except for $u=0$, where it vanishes.
Since this is simply a removable discontinuity on $\mathscr I^+$, we may as well ignore it for all practical purposes, and simply state
\be
\lim_{r\to\infty}\epsilon(u+r,r \mathbf n)=T(\mathbf n)\,,
\ee
which is the property we needed, together with $\Box\epsilon=0$, in order to interpret the memory  effect as a large gauge transformation in Lorenz gauge. 

Since the explicit expression for $\epsilon(x)$ is valid everywhere in Minkowski space, it also allows us to evaluate its limit to \emph{past} null infinity, and to address a point we raised earlier while discussing the antipodal matching condition. Indeed, performing the large $r$ limit for fixed advanced time $v=t+r=u+2r$, we find
\be
\lim_{r\to\infty}\epsilon(v-r,r\mathbf n) = T(-\mathbf n)\,,
\ee
which is precisely the antipodal matching.

\section{Gravitational Memory}
We conclude this chapter on observable effects linked to asymptotic symmetries with a brief discussion of memory effects induced by the passage of a gravitational wave. Originally introduced by Zel'dovich and Polnarev \cite{Zeldovich} (see also \cite{Frauendiener} for a concise treatment within the Newman-Penrose framework \cite{Newman-Penrose}), gravitational memory has been the first type of memory effect to be revisited in connection with asymptotic symmetries, in this case $BMS$ supertranslation symmetry \cite{memory}.

As in the case of electromagnetic memory, one considers a situation in which radiation crosses the sphere at large radius $r$ only between two given retarded times $u_0$ and $u_1$. In other words, the gravitational field is, asymptotically, in a radiative vacuum before $u_0$ and after $u_1$. Such radiative vacua are characterized, in terms of the $BMS$ data \eqref{BMSFalloffs}, by
\be\label{vacuumconditions}
\partial_u C_{ij}=0\,,\qquad
D_{[i}D\cdot C_{j]}=0\,.
\ee
The first condition is the vanishing of the Bondi news $N_{ij}=\partial_u C_{ij}$, which ensures that the energy flux is zero, while the second condition guarantees in addition that no angular momentum flux be present \cite{Barnich_BMS/CFT,Ashtekar_Lectures, Ashtekar_New_Lectures}. Equations \eqref{vacuumconditions} imply that $C_{ij}$ is $u$-independent and  that there exists an angular function $C(x^i)$ such that \cite{memory}
\be\label{pureST}
C_{ij}=(-2D_i D_j+\gamma_{ij}\Delta)C\,.
\ee 
Recalling \eqref{Bondieq},
\be\label{gravwave}
\partial_u m_B = \frac{1}{4}\partial_u D\cdot D\cdot C-\frac{1}{8} N_{ij}N^{ij}\,,
\ee
we see also that
\be\label{mBvacuum}
\partial_u m_B=0
\ee 
when equations \eqref{vacuumconditions} hold. To summarize, a radiative vacuum is specified by assigning $m_B(x^i)$ and $C(x^k)$, where  $C_{ij}=(-2D_i D_j+\gamma_{ij}\Delta)C$.

However, the specific values attained by $C$ and $m_B$ before $u_0$ and after $u_1$ will in general differ due to the evolution induced by the passage of radiation at intermediate times and are obtained by integrating equation \eqref{gravwave} for a specific radiation profile:
\be
m_B\big|_{u_1}-m_B \big|_{u_0} = -\frac{1}{4} \Delta (\Delta-2)\left( C\big|_{u_1}-C\big|_{u_0}\right)-\frac{1}{8}\int_{u_0}^{u_1}  N_{ij} N^{ij}du
\ee
On the other hand, we note that supertranslations map the space of $BMS$ vacua to itself, since an infinitesimal supertranslation parametrized by a function $T(x^i)$ (recalling \eqref{BMSFalloffs} and \eqref{BMSor}) induces the transformation
\be
\delta_T m_B=T \partial_u m_B\,,\qquad \delta_T C_{ij}=T \partial_uC_{ij}+(-2D_iD_j+\gamma_{ij}\Delta )T\,,
\ee
which leaves \eqref{vacuumconditions} and \eqref{mBvacuum} invariant. Moreover, supertranslations generically map a given vacuum to an inequivalent one, since they shift the function $C$ by $T$; naturally, the corresponding shift of $C_{ij}$ vanishes precisely when $T$ parametrizes an ordinary translation (\emph{i.e.} when $T$ is a spherical harmonic with $l=0$ or $l=1$). On the other hand, supertranslations do not alter the value of the Bondi mass aspect $m_B$, consistently with the fact that they commute with translations and hence cannot alter the Bondi four-momentum.

The relevance of the above structure of radiative vacua in relation to observable effects can be seen by considering the proper distance $\delta L$ that is measured between two detectors held fixed in the positions $(r,x^i)$ and $(r,x^i+\delta x^i)$ for large $r$ and $\delta x^i=\mathcal O(r^{-1})$. Assuming for simplicity that $C_{ij}$ vanishes before $u_0$ and attains a nonzero value after $u_1$, where it takes the form of a ``pure supertranslation'' \eqref{pureST}, we will have
\be
\delta L\big|_{u_0} = r \sqrt{\gamma_{ij}\delta x^i \delta x^j} = \mathcal O(r^{-1}) 
\ee
and 
\be
\delta L \big|_{u_1} = \delta L\big|_{u_0} \left[1+\frac{1}{2r}\left(\frac{C_{kl}\delta x^k \delta x^l}{\gamma_{ij}\delta x^i \delta x^j}\right)+\cdots\right]\,.
\ee
In stereographic coordinates \eqref{stereo}, recalling that $C_{z\bar z}=0$ by the trace constraint, this result takes the simple explicit form
\be
\delta L\big|_{u_0} = \frac{2r\,|\delta z|}{1+z\bar z}\,,\qquad 
\delta L \big|_{u_1} = \delta L\big|_{u_0} \left[1+\frac{(1+z\bar z)^2}{8r}\left( C_{zz}\frac{\delta z}{\delta \bar z}+ C_{\bar z \bar z}\frac{\delta \bar z}{\delta z}\right) \right]+ \cdots \,.
\ee
The above formulas highlight a first kind of gravitational memory effect: two static test detectors will experience a change to order $\mathcal O(r^{-1})$ in their relative proper distance after being invested by a gravitational wave burst.

A related type of gravitational memory effect can be revealed by considering the geodesic deviation of two test detectors, \emph{initially} placed in the positions $(r,x^i)$ and $(r,x^i+\delta x^i)$, which are then left free to move and exposed to a radiation train \cite{StromingerGravmemevD}. In this case the general equation for geodesic deviation,
\be
\frac{D^2 \xi^a}{d\tau^2}+R\indices{^a_{bcd}} u^b u^c \xi^d=0\,,
\ee
where $\xi^a$ denotes the coordinate separation between the two particles and $u^a$ the corresponding four-velocity, reduces to
\be
\nabla^2_u \delta x^i + R\indices{^i_{uuj}}\delta x^j=0
\ee
upon restricting to the $i$th angular component, up to higher-order corrections in $1/r$, since $u^a = \delta^a_u+\cdots$ . Substituting the falloffs \eqref{BMSFalloffs} into the expression for the Riemann tensor on the right-hand side and in the Christoffel symbols arising on the left-hand side then affords
\be
\partial_u^2 \delta x_i = \frac{1}{r}\, \partial_u^2 C_{ij}\delta x^j
\ee
to leading order, where $\delta x_i = \gamma_{ij}\delta x^j$. This equation shows that there can be no gravitational memory \emph{kick} to leading order, contrary to the scalar and spin-one case, and can be integrated, up to further subleading corrections, to obtain the finite displacement
\be
\delta x_i\big|_{u_1}-\delta x_i\big|_{u_0} = \frac{1}{r}\left(
C_{ij}\big|_{u_1}-C_{ij}\big|_{u_0}
\right)
\delta x^j\big|_{u_0} + \cdots\,.
\ee
We note that this displacement effect again appears to order $\mathcal O(r^{-1})$.

Similar arguments allow to prove that static or geodesic detectors also undergo a relative permanent time delay due to the passage of a gravitational radiation train \cite{memory}, thus providing yet another example of gravitational memory effect.

%%%%%%%%%%%%%%%%%%%%%%%%%%%%%%%%%%%%%%
\chapter{Maxwell and Yang-Mills Theory in Higher Dimensions}\label{chap:spin1higherD}

In this chapter we will explore the structure of asymptotic symmetries and charges in connection with memory effects and soft theorems in higher spacetime dimensions. We shall do so by providing an original analysis of linear and self-interacting spin-one theories, Maxwell and Yang-Mills, in $D\ge4$ \cite{Cariche-io,Memory-io}. As already remarked in the four-dimensional case, the asymptotic structure of such theories shares many interesting features with the gravitational case: an infinite-dimensional asymptotic symmetry group, associated soft surface charges and soft theorems, ordinary and null memory effects, and, in the Yang-Mills case, the phenomenon of color leak. Therefore, besides being interesting in its own right, the analysis of such theories provides a conceptual laboratory for ideas and phenomena associated with the gravitational case. It also furnishes a convenient working ground in view of the higher-spin extension which we will discuss in Chapter \ref{chap:HSP}.

In Section \ref{sec:YMGlobal} we shall address the main novel feature of the asymptotic expansion in dimensions greater than four, namely the separation into two distinct branches of Coulombic and radiation terms, in the context of a conservative analysis performed in the radial gauge. This will allow us to explicitly calculate the expressions for the power radiated by a Yang-Mills field, for the total color charge at a given retarded time, and to derive a formula expressing the time-dependence of the latter due to nonlinearities \cite{Cariche-io}. In particular, let us anticipate that Coulombic terms, appearing to order $\mathcal O(r^{3-D})$ in the asymptotic expansion of the gauge fields, will be responsible for finite and nonvanishing color charges. Conversely, radiation terms, which scale as $\mathcal O(r^{\frac{2-D}{2}})$ and are therefore leading compared to Coulombic order, will give no contribution to color charges, while giving rise to nonzero charge leak and energy flux.

While providing satisfactory answers as far as global charges and fluxes are concerned, no infinite-dimensional asymptotic symmetry enhancement will stem from the above analysis, in striking contrast with the situation in four dimensions. 

This puzzling feature would seemingly leave open the question regarding the ultimate origin of the Weinberg photon theorem. The validity of this theorem is independent of the dimension of spacetime and, as we have seen in Chapter \ref{chap:obs}, it can be understood as a manifestation of the invariance of the $S$ matrix under asymptotic symmetries in four dimension. 

Another class of phenomena related to the invariance of four-dimensional electrodynamics under an infinite-dimensional family of asymptotic symmetries is provided by memory effects, which can be interpreted as signaling the transition between two inequivalent radiative vacua. Therefore, it appears natural to investigate the relation between memory effects, if any, and the classical vacuum structure of the theory in higher dimensions as well.

For these reasons, Section \ref{sec:memoryexamples} is then devoted to the exploration of  memory effects in higher-dimensional spacetimes by means of explicit examples in the context of scalar and electromagnetic theories. Indeed, both ordinary \cite{Mao_EvenD} and null \cite{Memory-io} kick memory effects, which are precluded to radiation order, contrary to what was preliminarily conceived  \cite{Garfinkle:2017fre,WaldOddD}, can be shown to appear to Coulombic order. We shall take the occurrence of memory effects as an additional piece of evidence indicating that the classical vacuum structure of gauge theories in higher dimension is far from the trivial picture that the analysis performed in Section \ref{sec:YMGlobal} would apparently suggest.

This fact, together with the very existence of Weinberg's theorem, provides the basic motivation for performing further investigations of the asymptotic structure of electromagnetic and Yang-Mills theory in Section \ref{sec:em}, where we will be able to uncover a link between a suitable subclass of residual symmetries of the Lorenz gauge and memory effects. More specifically we shall see how ordinary and null kick memories can be indeed interpreted as vacuum transitions from the gauge theory perspective. This also holds for a type of memory which we did not discuss in Chapter \ref{chap:obs}: phase memory in Maxwell theory and color memory in Yang-Mills, which can be interpreted as phase/color rotations induced in the Hilbert space of two asymptotic test charges \cite{Memory-io}.

The key-points of this refined analysis are a different treatment of falloff conditions and a more flexible gauge-fixing choice, identified in the Lorenz gauge condition, which will allow us to establish a connection, on the one hand, between a suitable family of residual symmetries and memory effects, and on the other hand between \emph{bona fide} asymptotic symmetries and nontrivial asymptotic charges responsible for the Weinberg theorem \cite{Memory-io}. 
The analysis presented in this section is partly inspired by \cite{StromingerGravmemevD}, for the treatment of memory, and by \cite{StromingerQEDevenD}, for the proposal of alternative falloff conditions.

\section{Asymptotic Charges for Yang-Mills}\label{sec:YMGlobal}

In this section we analyze the equations of motion for Yang-Mills theory in Minkowski spacetime, for any dimension $D$, by means of an expansion of the fields in powers of $1/r$, thereby identifying the data that contribute to color charge and to color or energy flux at null infinity \cite{Cariche-io}. In particular, we will complement the related discussions in \cite{Strominger_YM, Strominger_QED, StromingerKac, StromingerColor, Maxwelld=3Barnich, Campiglia, Adamo:2015fwa, Mao_note} by providing a unified treatment of all spacetime dimensions, and those in \cite{Einstein-YM.Barnich, Mao_EvenD} by explicitly checking the finiteness of asymptotic charges in any dimension while also including radiation for $D=3$.

We adopt the usual retarded Bondi coordinates  $u$, $r$, $x^i$, where $x^i$, for $i=1,2,\ldots,D-2$, denotes the $D-2$ angular coordinates on the sphere at null infinity.
The Yang-Mills connection is denoted by
$
\mathcal A_a = \mathcal A_a^A T^A\, ,
$ 
where the $T^A$ are the anti-Hermitian generators of the $su(N)$ algebra, and its gauge transformation is  $ \delta_\epsilon\mathcal A_a = \nabla_{a} \epsilon + \left[ \mathcal A_a, \epsilon \right]$. The corresponding field strength reads 
\be
\mathcal F_{ab} = \partial_{a} \mathcal A_b - \partial_{b} \mathcal A_a + \left[ \mathcal A_a, \mathcal A_b \right]\, ,
\ee
while the field equations are $\mathcal G^a=0$ with
\be\label{EEOM}
\mathcal G^b 
= \nabla_a \mathcal F^{ab} + \left[ \mathcal A_a, \mathcal F^{ab}\right]\,.
\ee
In terms of retarded Bondi coordinates, we have\footnote{It can be convenient to employ $\sqrt{-g}\,\nabla_a \mathcal F^{ab} =\partial_a(\sqrt{-g}\,\mathcal F^{ab})$.}
\be\begin{aligned}\label{eomYMD}
\mathcal G^u &=\Big(\partial_r + \frac{D-2}{r}\Big)\mathcal F_{ur} + \frac{1}{r^2}D^i \mathcal F_{ri} + [\mathcal A_r, \mathcal F_{ur}]+\frac{1}{r^2}\gamma^{ij}[\mathcal A_i, \mathcal F_{rj}]\,,\\
\mathcal G^r &= \partial_u \mathcal F_{ru}+ \frac{1}{r^2}D^i(\mathcal F_{ir}-\mathcal F_{iu})+[\mathcal A_u,\mathcal F_{ru}]+\frac{1}{r^2}\gamma^{ij}[\mathcal A_i, \mathcal F_{jr}-\mathcal F_{ju}]\,,\\
\mathcal G_i &= \partial_u \mathcal F_{ir} + \Big(\partial_r + \frac{D-4}{2}\Big) (\mathcal F_{ri}-\mathcal F_{ui})+\frac{1}{r^2}D^j \mathcal F_{ji}\\
&+ [\mathcal A_u, \mathcal F_{ir}]+[\mathcal A_r, \mathcal F_{ri}-\mathcal F_{ui}]+\frac{1}{r^2}\gamma^{jl}[\mathcal A_j, \mathcal F_{li}]\,.
\end{aligned}\ee
Furthermore, we enforce the radial gauge
\be \label{bondi1}
\mathcal A_r=0 \,,
\ee
which completely fixes the gauge in the bulk. 

%%%%%%%%%%%%%%%%%%%%%%%%%%%%%%%%%%%%%%%%%%
\subsection{Boundary conditions}
%%%%%%%%%%%%%%%%%%%%%%%%%%%%%%%%%%%%%%%%%%

For $D > 3$ we consider field configurations  $\mathcal A_\mu$ whose asymptotic null behavior  is captured by an asymptotic expansion in powers of $1/r$, for $r\to\infty$, while, as we shall briefly discuss in Section \ref{sec:3-4dim_1}, in the three-dimensional case, we shall also consider a logarithmic dependence on $r$. 

In order to generalize the falloffs adopted in four dimensions (see Section \ref{sec:YMD=4}) and provide a suitable guess for the form of the asymptotic expansions, we shall adhere to the following guiding principle: namely that our field configurations should furnish finite and nonzero energy flux and color energy at null infinity. 

The energy per unit time flowing across a section $S_u$ of $\mathscr I^+$ at constant retarded time $u$ can be calculated as follows. Starting from the Yang-Mills Lagrangian  $\cL = \frac{1}{4}\mathrm{tr}(\mathcal F_{ab} \mathcal F^{ab})$,  the stress-energy tensor takes the form $T_{ab}=-\mathrm{tr}\left(\mathcal F_{ac}\mathcal F\indices{_b^c}\right)+\frac{1}{4}g_{ab}\mathrm{tr}\left(\mathcal F_{cd}\mathcal F^{cd}\right)$. The energy flux across $S_u$ is then given by $-\int_{S_u} T\indices{^r_u} r^{D-2} d\Omega=\int_{S_u} (T_{uu}-T_{ur}) r^{D-2} d\Omega\, $ as $r\to\infty$. Therefore,
\be\label{power_u}
\mathcal P(u)=\lim_{r\to\infty}\int_{S_u} \gamma^{ij}\, \mathrm{tr}\left(\mathcal F_{ui}(\mathcal F_{rj}-\mathcal F_{uj})\right) r^{D-4} d\Omega\,.
\ee
The request that \eqref{power_u} be finite imposes that the integrand must go to zero at infinity in order to compensate for the factor of $r^{D-4}$. Furthermore, to saturate this bound and furnish a nonzero energy flux, the slowest decaying terms in the asymptotic expansion of the field strength, which we name \emph{radiation} terms, should scale as $\mathcal O(r^{\frac{4-D}{2}})$, so as to cancel the measure factor $r^{D-4}$ exactly.

On the other hand, a quick calculation of the color charge at null infinity, on which we shall provide further details in sect.~\ref{subsec: Charges}, allows us to express it as the following integral over $S_u$,
\be\label{color_u}
\mathcal Q^A(u)=\lim_{r\to\infty}\int_{S_u} \mathcal F_{ur}^A r^{D-2} d\Omega\,,
\ee
where we have employed the formula $\kappa_\epsilon^{ab}=\sqrt{-g}\,\mathrm{tr}(\mathcal F^{ab}\epsilon)$ derived in \ref{sec:YMD=4}. We conclude, that, in order for the above expression to be nonvanishing, the asymptotic expansion of $\mathcal F_{ur}$ must contain terms scaling $\mathcal O(r^{2-D})$ (equivalently, $\mathcal O(r^{3-D})$ in the field component $\mathcal A_u$ once we adopt radial gauge), which we will refer to as \emph{Coulombic} contributions.

The above discussion of energy flux and color charge immediately highlights an interesting feature, namely that radiation terms give contributions that diverge at face value when inserted in the expression for the color charge. As we shall see below, the solution to this apparent paradox is obtained by first evaluating the charge \emph{on-shell} on a sphere of finite $r$, and only then sending $r$ to infinity. This is in fact trivially the case as far as the linear theory is concerned, since there the equations of motion reduce to $\nabla_a \mathcal  F^{ab}=0$, which ensures the independence of the  flux $\int_{S_{u,r}}\mathcal F^{ab}\,d\sigma_{ab}$ both on $r$ and on $u$ by the Stokes formula. The situation is more interesting in the case of the nonlinear theory, since in this case $\nabla_a \mathcal F^{ab}=[F^{ab},\mathcal A_a]$ and the behavior of the surface charge as we vary $r$ and $u$ crucially depends on the falloffs of the field, precisely due to the nonlinear nature of the theory.

This cancellation of potentially divergent terms due to the equations of motion will turn out to be possible on account of the fact that, since in higher dimensions the field strength components must decay as $r$ grows, the nonlinear terms cannot appear in the leading-order equations, provided that strictly decaying falloff conditions are also assigned to the gauge potential. This mechanism can be regarded as an \emph{asymptotic linearization} of the equations of motion. 

Another interesting result of the above comparison between radiation and Coulombic terms is that the asymptotic expansion must include both integers and half-odd powers of $r$ in odd-dimensional spacetimes in order to account for both radiation and the presence of charges. This choice is forced by the fact that, as remarked above, the leading-order radiation components of the field strength scale as $\mathcal O(r^{-\frac{D}{2}})$ while Coulombic ones behave as $\mathcal O(r^{2-D})$. A bonus feature that will emerge from the resulting analysis is the description of the color flux for large $r$, namely the interplay occurring between radiation and Coulombic terms.

Before moving to the explicit calculations, let us stress once again that the presence of such two distinct classes of terms in the solutions, radiation and Coulombic, is a genuinely new feature of $D>4$, while in the four-dimensional case they effectively coincide.  

%%%%%%%%%%%%%%%%%%%%%%%%%%%%%%%%%%%%%%%%%%
\subsubsection{Equations of motion}
%%%%%%%%%%%%%%%%%%%%%%%%%%%%%%%%%%%%%%%%%%

When $D > 4$ is even, in order to saturate the bound set by \eqref{power_u}, it is natural to consider an expansion of the following type in the radial gauge $\mathcal A_r=0$:
\be\label{expansion_D_even}
\mathcal A_u = \sum_{k=\frac{D-2}{2}}^\infty \frac{A_u^{(k)}}{r^{k}}\,,\qquad
\mathcal A_i = \sum_{K=\frac{D-4}{2}}^\infty \frac{A_i^{(k)}}{r^k}\,,
\ee
where $A_u^{(k)}$ and $A_i^{(k)}$ are functions of retarded time $u$ and of the angles $x^i$ and the index $k$ takes integer values. On the basis of the previous discussion we expect $A_u^{(\frac{D-2}{2})}$ and $\mathcal A_i^{(\frac{D-4}{2})}$ to play the role of leading radiation terms, and indeed
\be
\mathcal P(u)=-\lim_{r\to\infty}\int_{S_u} \gamma^{ij}\, \mathrm{tr}\Big(\partial_u A_i^{(\frac{D-4}{2})} \partial_u A_j^{(\frac{D-4}{2})}\Big) d\Omega\,.
\ee
This is the expression for energy flux, which depends quadratically on the time derivative of the radiation terms. Furthermore, it is also positive definite, according to our choice of anti-Hermitian generators.

The seeming discrepancy of a factor of $r$ in the $u$ and $i$ components, which is shared by the four-dimensional case, is actually due to the fact that we are employing a non-normalized, coordinate basis for the spherical components. A normalized basis is obtained by defining the noncoordinate vectors $\mathbf e_i=\frac{1}{r}\,\partial_i$ and similarly $\tilde{\mathbf e}^i=r\,dx^i$ for their duals. Correspondingly 
\be
\mathcal A = \mathcal A_u du + A_i dx^i = \mathcal A_u du + \frac{1}{r}\,A_i \tilde{\mathbf e}^i\,,
\ee
which explains the extra factor of $r$.

We shall now substitute the above asymptotic expansion into the equations of motion \eqref{eomYMD}, starting from $\mathcal G^u=0$, which gives the recursion relations
\be\label{Furr}
k(D-3-k)A_u^{(k)}-(k-1)D^i A_i^{(k-1)}-\sum_l\gamma^{ij}[A_i^{(k-l)}, (l-1)A_j^{(l-1)}]=0\,.
\ee
This equation will be crucial for the discussion of the finiteness of charges. It also allows us to illustrate explicitly the phenomenon of asymptotic linearization. Indeed, the nonlinear term in the above equation will appear when both $l-1\ge\frac{D-4}{2}$ and $k-l\ge \frac{D-4}{2}$, which implies $l\ge \frac{D-2}{2}$ and $k\ge D-3$. The latter condition in particular tells us that the nonlinear terms do not enter the first few instances of the recursion relations, namely those with $\frac{D-2}{2}\le k \le D-4$. In other words, the nonlinear contributions are absent at radiation order and start appearing at Coulombic order.

Furthermore, we extract the term proportional to $r^{2-D}$ in $\mathcal G^r=0$, which will be instrumental to calculating the time-dependence of charges, namely
\be\label{Furu}
(D-3)\partial_u A_u^{(D-3)}=\gamma^{ij}[A_i^{(\frac{D-4}{2})}, \partial_u A_j^{(\frac{D-4}{2})}]+\cdots\,,
\ee
which, remarkably, links the time dependence of the Coulombic term $A_u^{(D-3)}$ precisely to a commutator of radiation terms.
The terms omitted in \eqref{Furu} are of the form $D^i f_i$ and will not give contributions to the results in the next section.

In the case of odd dimensions $D>4$, as anticipated, we have to include two distinct expansions in $1/r$ in order to capture both radiation and Coulombic terms:
\be\begin{aligned}
	\mathcal A_u = \sum_{k=\frac{D-2}{2}}^\infty \frac{A_u^{(k)}}{r^{k}}
	+\sum_{k=D-3}^\infty \frac{A_u^{(k)}}{r^k}\,,\qquad
 	\mathcal A_i = \sum_{k=\frac{D-2}{2}}^\infty \frac{A_i^{(k)}}{r^{k}}
 	+\sum_{k=D-4}^\infty \frac{A_i^{(k)}}{r^k}\,.
\end{aligned}
\ee
Aside from this observation, the discussion proceeds essentially in the same way as in the even-dimensional case, and in particular \eqref{Furr} and \eqref{Furu} still hold.

%%%%%%%%%%%%%%%%%%%%%%%%%%%%%%%%%%%%%%%%%%
\subsubsection{Three and four spacetime dimensions}\label{sec:3-4dim_1}
%%%%%%%%%%%%%%%%%%%%%%%%%%%%%%%%%%%%%%%%%%

In $D = 4$, which we discussed in  Section \ref{sec:YMD=4}, the leading radiation term and the Coulombic term coincide and the main information regarding the dynamics is contained in 
\be\label{flux_four}
\partial_u A_u = \partial_u D^i A_i + \gamma^{ij}\left[A_i, \partial_u A_j\right]\,.
\ee

The situation for  $D = 3$ is rather different with respect to the previous cases, mainly because of two features. First, the factor of $r^{-1}$ in \eqref{power_u} tells us that, in order to produce a finite energy flux across $S_u$, the field components need not necessarily decay at infinity; consequently, one expects no clear distinction between radiation and Coulombic terms in the solution, because no asymptotic linearization  occurs in the equations of motion.
Second, the expression \eqref{color_u}, and more specifically the factor of $r$ due to the line element on the circle, suggests that $\mathcal A_u$  should behave as 
$\log\frac{1}{r}$
in order to give a nonvanishing color charge. 

These considerations motivate the following leading-order ansatz in three dimensions:
\be
\mathcal A_u (u,r,\varphi) \sim  q \log \frac{1}{r}+p\,,\qquad
\mathcal A_\varphi (u,r,\varphi) \sim \frac{\sqrt r}{\log r}\, C\,,
\ee
where $q$, $p$ and $C$ are $r$-independent functions. Indeed, with this choice, the color    flux and the energy flux read
\be \label{QP3}
Q^A (u) = \int_{S_u} q^A d\varphi\,,\qquad
\mathcal P (u)  = -\int_{S_u} \mathrm{tr}([q,C][q,C]) d\varphi\, .
\ee
Using this ansatz we find 
\be\label{flux_three}
\partial_u q =- \left[q,p\right] ,
\ee
as the only relevant constraint arising from the equations of motion. This equation describes the $u$-evolution of $q$ at null infinity and hence, together with the first of \eqref{QP3}, will lead to a formula for the color    flux.
%

%%%%%%%%%%%%%%%%%%%%%%%%%%%%%%%%%%%%%%%%%%
\subsection{Global symmetries and charges}\label{subsec: Charges}
%%%%%%%%%%%%%%%%%%%%%%%%%%%%%%%%%%%%%%%%%%

In this section we would like to discuss the form \eqref{color_u} of the color    charge at null infinity in the various dimensions. For related analyses see  \cite{abbott-deser, Barnich-Brandt, Avery_Schwab, Mao_note}. 

To begin with, let us discuss which large gauge symmetries are admissible at null infinity. The residual gauge symmetry within the radial gauge is parameterized by an $r$-independent gauge parameter, since
\be
0=\delta_\epsilon A_r = \partial_{r}\, \epsilon+ \left[\mathcal A_r,\epsilon\right] 
\ee
but $\mathcal A_r=0$, hence $\partial_{r}\epsilon=0$. Then, we look for those parameters $\epsilon$ which preserve the leading falloff conditions imposed on the field $\mathcal A_a$, again distinguishing the case of $D>4$ from $D=4$ and $D=3$.

When $D>4$, where radiation gives the dominant behavior    at infinity, we find, to leading order
\be
\begin{aligned}
r^{\frac{2-D}{2}} \, \delta_\epsilon A_u =  \partial_u  \epsilon + r^{\frac{2-D}{2}}[A_u, \epsilon]\, ,
\end{aligned}\ee
which requires $\partial_u\epsilon=0$ (because $D>2$). Furthermore
\be
r^{\frac{4-D}{2}} \,\delta_\epsilon A_i = \partial_i  \epsilon+ r^{\frac{4-D}{2}}[A_i, \epsilon]\, ,
\ee
but, since $D>4$, this also implies $\partial_i\epsilon=0$, which reduces $\epsilon$ to a constant. Hence, in $D>4$, asymptotic symmetries coincide with the global part of the gauge group and the asymptotic charge is the ordinary color charge computed via \eqref{color_u}.
Now, using \eqref{Furr} and recalling by the above discussion that the nonlinear terms do not contribute for $k<D-3$, we have
\be\begin{aligned}\label{ColorChargeglobale}
Q^A(u) 
&= \lim_{r\to\infty }\int_{S_u} \mathrm{tr}(\mathcal F_{ur} T^A)\, r^{D-2} d\Omega\\
&= (D-3)\int_{S_u}\left(A_u^{(D-3)}\right)^A\,d\Omega\,,
\end{aligned}\ee
where we have used the fact that the terms of the type $D^i A_i^{(k-1)}$ give no contribution to the integral since integrate to zero on the sphere. This formula is valid for any spacetime with dimension $D>4$ of either parity.

The time dependence of $\mathcal Q^A(u)$ is furnished by taking the derivative with respect to $u$ and recalling \eqref{Furu}
\be
\frac{d}{du} Q^A(u) = \int_{S_u} \gamma^{ij} \Big[A_i^{(\frac{D-4}{2})}, \partial_u A_j^{(\frac{D-4}{2})}\Big]^A d\Omega\,.
\ee
This is again consistent with the interpretation of $A_i^{(\frac{D-4}{2})}$ as the leading radiation term: this formula describes how Yang-Mills radiation, made of ``classical gluons'', flowing to null infinity induces a change in the total color at successive retarded times $u$.

In $D=4$, the gauge parameter must satisfy:
\be\begin{aligned}
r^{\, -1} \, \delta_\epsilon A_u & = \partial_u  \epsilon + r^{-1}[A_u, \epsilon] \, ,\\
\delta_\epsilon A_i & = \partial_i  \epsilon + [A_i, \epsilon] \, .
\end{aligned}\ee
The first equation again enforces $\partial_u \epsilon=0$, whereas the second allows for an $\epsilon(x^1, x^2)$ with arbitrary dependence on the angles on the celestial sphere. The corresponding asymptotic charge is therefore
\be
Q_\epsilon(u) = \lim_{r\to\infty}\int_{S_u} \mathrm{tr}(\mathcal F_{ur}\epsilon)\, r^2 d\Omega = \int_{S_u} \mathrm{tr}(A_u \epsilon)\, d\Omega\,.
\ee
Taking into account \eqref{flux_four},
\be
\frac{d}{du} Q_\epsilon(u) = \int_{S_u} \mathrm{tr}\left[\big(\partial_u D^i A_i+\gamma^{ij}[A_i, \partial_u A_j]\big) \epsilon\right] d\Omega\,.
\ee

To complete the picture, let us now turn to the situation in $D=3$. There, neither $\mathcal A_u$ nor $\mathcal A_\varphi$ fall off at infinity, and hence any $\epsilon(u,\varphi)$ generates an allowed gauge transformation (the same result, in a slightly different setting, was already obtained in \cite{Maxwelld=3Barnich}). Thus, using the notation of the previous section,
\be\begin{aligned}
Q_\epsilon(u) &= \int_{S_u} \mathrm{tr}(q \epsilon) d\varphi\,,\\
\frac{d}{du}Q_\epsilon(u) &= \int_{S_u} \mathrm{tr}(q \partial_u\epsilon) d\varphi - \int_{S_u} \mathrm{tr}([q, p] \epsilon )d\varphi\,.
\end{aligned}\ee

Let us observe that these charges indeed form a representation of the underlying algebra: for $D\ge4$, since $\delta_\epsilon A_u= [A_u,\epsilon]$,
\be\label{commutation_rel}
[Q_{\epsilon_1},Q_{\epsilon_2}]=\delta_{\epsilon_1} Q_{\epsilon_2}= \int_{S_u} \mathrm{tr}([A_u, \epsilon_1]\epsilon_2) d\Omega =  \int_{S_u} \mathrm{tr}(A_u [\epsilon_1,\epsilon_2]) d\Omega= Q_{[\epsilon_1,\epsilon_2]}\,.
\ee
The same result holds for $D=3$, noting that $\delta_\epsilon q= [q,\epsilon]$ and $\delta_\epsilon p = \partial_u \epsilon$, but $p$ does not enter the charge formula.  While \eqref{commutation_rel} holds in any dimension, it should be stressed that, when $D>4$, the corresponding charge algebra coincides with $su(N)$, whereas in $D=4$ and $D=3$ it is in fact an infinite-dimensional Kac-Moody algebra, owing to the arbitrary gauge parameters $\epsilon(x^1,x^2)$ and $\epsilon(u,\varphi)$. In particular, we note the absence of a central charge, which could however emerge by performing the analysis for the linearized theory around a nontrivial background, as pointed out in \cite{Avery_Schwab}.

Let us conclude this section by comparing the calculation of the Noether charge with the one obtained by means of covariant phase space techniques. In particular, we will check that, for Yang-Mills theory, the quantity
\be
Q_\epsilon= \int_{\partial \Sigma} dx_{\mu\nu}\, \mathrm{tr}\left(F^{\mu\nu} \epsilon\right)\,,
\ee
where $\Sigma$ is a Cauchy surface, provides not only the conserved charge as obtained by the Noether algorithm, but also the Hamiltonian generator of the gauge symmetry parameterized by $\epsilon$ on the space tangent to the surface of solutions, as calculated via covariant phase space methods. 

Indeed, a generic variation of the Yang-Mills Lagrangian, after integrating by parts, reads
\be\label{delta L}
\delta \mathcal L=-\,\mathrm{tr}\left(\mathcal G^{\mu} \delta A_\mu\right) + \partial_\mu \mathrm{tr}\left(F^{\mu\nu}\delta A_\nu\right) = -\mathrm{tr}\left(\mathcal G^{\mu} \delta A_\mu\right) +\partial_\mu \theta^\mu(\delta A) \,,
\ee
where we defined the symplectic potential $\theta^\mu(\delta A)=\mathrm{tr}\left(F^{\mu\nu} \delta A_\nu\right)$, while $\mathcal G^\mu$ denotes the Euler-Lagrange derivatives of $\mathcal L$, given in   \eqref{EEOM}. The presymplectic form is then given by 
\be
\omega^{\mu}(\delta_1 A, \delta_2 A)=\delta_{[1}\theta^\mu(\delta_{2]}A)\, ,
\ee
with square brackets denoting antisymmetrization, and correspondingly the formal variation of the Hamiltonian generator of the gauge symmetry  
$H_\epsilon$ is
\be
\slashed{\delta} H_\epsilon = \int_\Sigma dx_\mu \omega^{\mu}(\delta A, \delta_\epsilon A)=\delta \int_{\partial\Sigma} dx_{\mu\nu} \mathrm{tr}(F^{\mu\nu}\epsilon)- \int_{\Sigma} dx_\mu \mathrm{tr}(\delta\mathcal G^\mu \epsilon).
\ee
Noting that the last term is proportional to the linearized equations of motion, i.e.\ that it vanishes on the space tangent to the surface of solutions, we can write
$$
\slashed{\delta} H_\epsilon \approx \delta Q_\epsilon\,,
$$
which explicitly shows that $\slashed{\delta} H_\epsilon$ is integrable and that we may choose to set $H_\epsilon = Q_\epsilon$ by requiring a flat connection to have zero charge. 
Furthermore, the Noether charge is simply
\be
\int_{\Sigma}dx_\mu\, \theta^{\mu}(\delta_\epsilon A) = Q_\epsilon - \int_\Sigma dx_\mu\, \mathrm{tr}(\mathcal G^\mu \epsilon) \approx Q_\epsilon\,,
\ee
so that the two approaches agree in this case. 
The definition of $Q_\epsilon$ is in principle subject to ambiguities stemming from $\theta^\mu \mapsto \theta^\mu + \partial_\nu \lambda^{\mu\nu}$, where $\lambda^{\mu\nu}=-\lambda^{\nu\mu}$, which anyway do not alter \eqref{delta L}. In the spirit of \cite{Wald-Zoupas}, we may choose to set to zero the corresponding additional terms, precisely because this choice defines an integrable Hamiltonian, as shown above. Further motivation for the absence of these terms is provided by the agreement with the general analysis of \cite{Barnich-Brandt} and by the fact that they play no role in the generation of Ward identities for residual gauge freedom \cite{Avery_Schwab}. A possible way to eliminate this ambiguity could be to study its role in connection with the  the addition of boundary terms to the Yang--Mills action and the well--definiteness of the corresponding variational principle  \cite{CompereBoundaryAmb}.  

\section{Classical Solutions and Memory Effects in Higher $D$}\label{sec:memoryexamples}

As already stressed, the absence of infinite-dimensional asymptotic symmetries in $D>4$ would be at odds with the persistence of universality relations in soft amplitudes. Additional arguments in this respect actually come from a proper investigation inquiring on the presence or absence of higher-dimensional memory effects. Indeed, this latter aspect was analyzed in the literature from various angles, with emphasis on the gravitational case, leading sometimes to opposite conclusions on the status of memory effects in higher $D$ \cite{Garfinkle:2017fre, WaldOddD,Mao_EvenD,SatishchandranWald}.

Here, we shall consider the issue from the perspective of scalar and vector fields and provide explicit calculations of full-fledged memory effects in any even dimension, while also commenting on the difficulties associated with the nature of odd-dimensional memory effects. Our conclusions further motivate reconsidering the problem of asymptotic symmetries from a broader perspective.

\subsection{Scalar fields in even $D$}\label{sec:spin0}
Let us first consider a particle with scalar charge $q$ that is created at the origin at $t=0$. The scalar field $\varphi$ generated by this process is obtained by solving the wave equation 
\be\label{scalar-theta}
-\Box \varphi(t,\mathbf x) = q\, \theta(t)\delta(\mathbf x)
\ee 
(recalling $-\Box=-\eta^{\mu\nu}\partial_\mu\partial_\nu=\partial_t^2-\nabla^2$).
The solution is given by the convolution of the source on the right-hand side with the ($D$--dimensional) retarded wave propagator $D_\text{ret}(x)$, \emph{i.e.} in this case
\be
\varphi(t,\mathbf x) 
= q\, \int_{-\infty}^{t} D_\text{ret}(\tau, \mathbf x)\,d\tau\,.
\ee
The field generated by a particle that is destroyed at the origin can then be obtained by time reversal of the above solution, while the field for a moving particle can be calculated by applying a Lorentz boost. 

Let us recall that, for even $D\ge2$, the retarded propagator is 
\be\label{prop-even}
D_\text{ret}(x)= 
\dfrac{1}{2\pi^{D/2-1}}\delta^{(\frac{D-4}{2})}(x^2)\theta(x^0)
\ee
(see \emph{e.g.} \cite{Friedlander}),
where $\theta$ is the Heaviside distribution.
Restricting to even $D\ge4$, and using the chain rule for the distribution $\delta^{(\frac{D-4}{2})}(x^2)$, the propagator \eqref{prop-even} can be recast as 
\be
D_\mathrm{ret}(t, \mathbf x)= 
\sum_{k=0}^{D/2-2} c_{D,k}\,
\frac{\delta^{(D/2-2-k)}(u)}{r^{D/2-1+k}}\,,
\ee
where $u=t-r$ and $r=|\mathbf x|$, while with $c_{D,k}$ we denote the coefficients
\be
c_{D,k} = \frac{1}{2(2\pi)^{\frac{D}{2}-1}}\,\frac{\left(\frac{D}{2}-2+k\right)!}{2^k\left(\frac{D}{2}-2-k\right)!\,k!}\,.
\ee 
In particular note that
\be
c_{D,0}=\frac{1}{2(2\pi)^{\frac{D}{2}-1}}\,,\qquad
c_{D,\frac{D}{2}-3} = c_{D,\frac{D}{2}-2}= \frac{1}{(D-3)\Omega_{D-2}}\,,
\ee
where $\Omega_{D-2}$ is the area of the $(D-2)$-dimensional Euclidean unit sphere.
The resulting scalar field is thus
\be\label{scalar-basic}
\varphi(u,r)=q
\sum_{k=0}^{D/2-2} c_{D,k}\,
\frac{\theta^{(D/2-2-k)}(u)}{r^{D/2-1+k}}\,.
\ee
Notice that only the term associated to $k=D/2-2$ gives rise to a persistent field for fixed $r$, while the other terms have support localized at $u=0$, namely on the future-directed light-cone emanating from the particle's creation event. This is a general consequence of the recursion relation \eqref{residual_r}, namely
\be
(D-2k-2)\partial_u \varphi^{(k)}
=
[\Delta+(k-1)(k-D+2)]\varphi^{(k-1)}\,,
\ee 
obeyed by $\varphi$ near future null infinity, which requires $[\Delta +k(k-D+3)]\varphi^{(k)}=0$ for $0<k<D-3$, and hence $\varphi^{(k)}=0$, in that range, for any stationary solution.

Now, a test particle with charge $Q$, held in place at a distance $r$ from the origin, will be subject to a force $f_a =Q\,\partial_a \varphi(u,r)$ at a given retarded time $u$ due to the presence of the scalar field. Hence, its $D$--momentum $P_a$ will in general be subject to the leading-order variation
\be
P_a\big|_{u}-P_\mu\big|_{u=-\infty}=Q\int_{-\infty}^u \partial_a \varphi(u',r) du'\,.
\ee
For this very simple example, this quantity can be calculated explicitly for any even $D$. The variations of $P_u$ and $ P_r$ in particular yield
\be\label{scalarmem0_u}
P_u\big|_{u>0}-P_u\big|_{u<0}= 
Q\int_{-\infty}^{+\infty} \partial_u\varphi(u', r)\,du' = 
\frac{Qq}{(D-3)\Omega_{D-2}r^{D-3}}\,,
\ee
and
\be\label{scalarmem0_r}
P_r\big|_{u>0}-P_r\big|_{u<0}=
Q\int_{-\infty}^{+\infty} \partial_r\varphi(u', r)\,du'=
-\frac{(D-4)Qq}{(D-3)\Omega_{D-2}r^{D-3}}\,.
\ee
Equation \eqref{scalarmem0_u} simply expresses the fact that the test particle will start feeling the Coulombic interaction energy with the newly created particle at the origin as soon as it crosses the light-cone subtended by the origin of spacetime. On the other hand, \eqref{scalarmem0_r} tells us that the test particle will feel an instantaneous, radial momentum kick, for even dimensions greater than four. Consistently with the spherical symmetry of this process, the variations of the angular components $P_i$ vanish identically.

The field emitted by a particle destroyed in the origin at $t=0$ is obtained by sending $u\mapsto -u$ in \eqref{scalar-basic}. The case of a particle moving with velocity $\mathbf v$ can be instead obtained by boosting \eqref{scalar-basic}: 
\be\label{standardboost}
t \mapsto \gamma(\mathbf v)(t-\mathbf v \cdot \mathbf x)\,,\qquad
\mathbf x \mapsto \mathbf x + \mathbf v (\gamma(\mathbf v)-1)\frac{\mathbf v \cdot \mathbf x}{\mathbf v^2}- \gamma(\mathbf v)\mathbf v t\,,
\ee
which gives, for large $r$, denoting $\mathbf n = \mathbf x/r$, 
\be\label{transf-ur}
u \mapsto u\, \gamma(\mathbf v)^{-1}(1-\mathbf n \cdot \mathbf v)^{-1}+\mathcal O(r^{-1})\,,\qquad
r \mapsto r\, \gamma(\mathbf v)(1-\mathbf n \cdot \mathbf v)+\mathcal O(1)\,.
\ee
We can then cast the boosted solution in the following form:
\be\label{boosted-scalar}
\varphi(u,r, \mathbf n) = \,\frac{q\,\theta(u)}{(D-3)\Omega_{D-2}[\gamma(\mathbf v)(1-\mathbf n \cdot \mathbf v)r]^{D-3}} 
+\bar \varphi(u, r, \mathbf n)+\mathcal O(r^{2-D})\,,
\ee
where $\bar\varphi$ is a sum of terms of the type
\be
f_\alpha(r, \mathbf n)\delta^{(\alpha)}(u)\,,\qquad \text{ with }\alpha\ge0\,,
\ee
that is to say, whose support is localized on the light-cone.
Let us stress that the terms in $\bar \varphi$ formally dominate the asymptotic expansion of $\varphi$ as $r\to\infty$. However, these terms will not contribute to the leading $u$-component of the momentum kick due to the presence of $\delta(u)$ and its derivatives. We can therefore conclude that
\be
P_u\big|_{u>0}-P_u\big|_{u<0} =Q\int_{-\infty}^{+\infty} \partial_u \varphi\,du'= \frac{qQ}{(D-3)\Omega_{D-2}[\gamma(\mathbf v)(1-\mathbf n \cdot \mathbf v)r]^{D-3}}\,,
\ee
in any even $D$. This is, not surprisingly, just the analog of equation \eqref{scalarmem0_u} for the Coulombic energy in which one needs to account for the relativistic length contraction.

For a more general scattering process involving a number of ``in''  and ``out'' particles destroyed or created in the origin, the result is obtained by linearly superposing solutions and therefore reads ($\eta_a=-1$ for an incoming particle and $\eta_a=+1$ for an outgoing one)
\be
P_u\big|_{u>0}-P_u\big|_{u<0} =
\sum_{a\in\text{in/out}}\frac{\eta_a\,q_a Q}{(D-3)\Omega_{D-2}[\gamma(\mathbf v_a)(1-\mathbf n \cdot \mathbf v_a)r]^{D-3}}\,.
\ee

Calculating radial and angular components of $P_\mu$ requires more effort, since they arise instead from the terms proportional to  $\delta(u)$ in the expansion of $\bar \varphi$, whose number increases with the spacetime dimension. They have been given for any even dimension in \cite{Mao_EvenD} in terms of derivatives of a generating function. For our present, illustrative, purposes, it suffices to consider the first relevant case $D=6$, where the exact solution in the case of the particle created in the origin with velocity $\mathbf v$ is given by 
\be\label{exactscalarD=6}
8\pi^2 \varphi = \frac{\delta(u)}{\gamma(\mathbf v)(1-\mathbf n\cdot\mathbf v)r^2}+\frac{\theta(u)}{\gamma(\mathbf v)^3(1-\mathbf n\cdot\mathbf v)^3r^3}\,\Delta(u,r)^{-3/2} 
\ee
with
\be
\Delta(u,r) = 1
+\frac{2u(\mathbf v^2-\mathbf n \cdot \mathbf v)}{r(1-\mathbf n\cdot\mathbf v)^2}
+\frac{u^2\mathbf v^2}{r^2(1-\mathbf n\cdot\mathbf v)^2}\,.
\ee
The corresponding radial and angular memory effects in $D=6$ are then, to leading order,
\be\begin{split}
P_r\big|_{u>0}-P_r\big|_{u<0} &= \frac{-2Qq}{8\pi^2\gamma(\mathbf v)(1-\mathbf n \cdot \mathbf v)r^3} \,,\\
P_i\big|_{u>0}-P_i\big|_{u<0} &= \frac{v_i Qq}{8\pi^2 \gamma(\mathbf v)(1-\mathbf n\cdot \mathbf v)r^2}\,,
\end{split}
\ee
where $v_i = \partial_i \mathbf n \cdot \mathbf v$ is the component of the particle's velocity in the $i$-th angular direction.

While the above examples illustrate the phenomenon of ordinary memory, associated with the field emitted by massive charges that move in the bulk of the spacetime, we can also consider the wave equation with a source term characterizing the presence of a massless charged particle \cite{Memory-io}, moving along a given direction $\mathbf x_0$:
\be
-\Box \varphi = q\, \delta(\mathbf x - \mathbf x_0 t)\,,
\ee 
with $|\mathbf x_0|=1$. This equation can be conveniently solved for any even $D\ge6$ by going to retarded coordinates, where it reads
\be
\left(2\partial_r + \frac{D-2}{r}\right)\partial_u \varphi = \left(\partial_r^2 + \frac{D-2}{r}\partial_r + \frac{1}{r^2}\Delta\right)\varphi+\frac{q}{r^{D-2}}\,\delta(u)\,\delta(\mathbf n, \mathbf x_0)
\ee
and performing the usual asymptotic expansion $\varphi(u,r,\mathbf n)= \sum \varphi^{(k)}(u, \mathbf n)r^{-k}$, which gives
\be\label{recursionnullD}
(D-2k-2)\partial_u \varphi^{(k)} = [\Delta + (k-1)(k-D+2)]\varphi^{(k-1)}+\delta_{k,D-3}\,\delta(u)\,\delta(\mathbf n, \mathbf x_0)\,.
\ee
The latter equation is solved by setting $\varphi^{(k)}=0$ for $k\le \frac{D}{2}-2$ and for $k\ge D-3$, while, for $\frac{D}{2}-1\le k \le D-4$,
\be\label{nullmemk}
\varphi^{(k)}(u, \mathbf n)=\delta^{(D-4-k)}(u)C_k(\mathbf n)\,,
\ee
where the functions $C_k(\mathbf n)$ are determined recursively by
\be\begin{aligned}
	(D-2k-2) C_k(\mathbf n) &= [\Delta + (k-1)(k-D+2)]C_{k-1}(\mathbf n)\,,\\
	C_{D-4}(\mathbf n) &= -(\Delta-D+4)^{-1}(\mathbf n,\mathbf x_0)\,.
\end{aligned}\ee
Here, $(\Delta-D+4)^{-1}$ is the Green's function for the operator $\Delta-D+4$, which is unique for $D>4$. As a consequence, the field gives rise to the null memory effect
\be
P_i\big|_{u>0}-P_i\big|_{u<0} = \int_{-\infty}^{+\infty} \partial_i \varphi\, du = -\frac{1}{r^{D-4}}\partial_i(\Delta-D+4)^{-1}(\mathbf n,\mathbf x_0)
\ee
(note that only the term with $k=D-4$ contributes),
consisting in a kick along a direction tangent to the celestial sphere.

\subsection{Electromagnetic fields in even $D$} \label{sec:memories}

Since we shall work in the Lorenz gauge, much of the calculations are essentially the same as those of the scalar case, which we have discussed in the previous section.
In Cartesian coordinates, where the gauge condition reads $\partial^\mu \mathcal A_\mu  =0$, the equations of motion reduce to a set of scalar wave equations
\be
\Box\, \mathcal A_\mu = j_\mu\,.
\ee
First, we consider the case of a static point-like source created at the origin,\footnote{Strictly speaking, equation \eqref{createdcha} is not well-posed, since the right-hand side has a nonzero divergence. We shall take this aspect in due account when calculating the solution \eqref{dim6vectorpot}, where in particular the right-hand side of \eqref{createdcha} occurs just as part of a full source that respects the continuity equation.}
\be\label{createdcha}
\Box\, \mathcal A^\mu = u^\mu q\,\theta(t)\delta(\mathbf x)\,,
\ee
where $u^\mu=(1,0,\ldots,0)$, hence the retarded solution is
\be
\mathcal A^\mu = - u^\mu \varphi\,,
\ee
where $\varphi$ denotes the corresponding solution \eqref{scalar-basic} for the scalar field.
Boosting this solution according to \eqref{standardboost} yields $\mathcal A^\mu=(\mathcal A^0, \mathbf A)=-\gamma(\mathbf v)(1,\mathbf v)\varphi$ and going to retarded coordinates gives
\be
\mathcal A_u = \gamma(\mathbf v)\varphi\,,\qquad
\mathcal A_r = \gamma(\mathbf v)(1-\mathbf n \cdot \mathbf v)\varphi\,,\qquad
\mathcal A_i = -r\,\gamma(\mathbf v) v_i \varphi\,.
\ee
Focusing for simplicity on the case of $D=6$, let us consider the radiation field generated by a massive particle with charge $q$ sitting at rest in the origin for $t<0$ that starts moving with velocity $\mathbf v$ at $t=0$. Such field is obtained by matching the solution for a charge destroyed at the origin to that of a charge  there created with velocity $\mathbf v$. Proceeding as before, one obtains, for large values of $r$,
\be \begin{aligned}\label{dim6vectorpot}
	8\pi^2\mathcal A_u &= 
	\frac{q\,\delta(u)}{r^2}\left(\frac{1}{1-\mathbf n \cdot \mathbf v}-1 \right)
	+\mathcal O(r^{-3})
	\,,\\
	8\pi^2\mathcal A_r &=
	\mathcal {O}(r^{-3})\,\\
	8\pi^2\mathcal A_i &= -\frac{q\,v_i\delta(u)}{(1-\mathbf n\cdot \mathbf v)r}-\frac{q\,v_i\theta(u)}{\gamma(\mathbf v)^2(1-\mathbf n\cdot \mathbf v)^3r^2}+\mathcal {O}(r^{-3})\,,
\end{aligned} \ee
where we have kept the orders relevant to the calculation of the memory effect.

The change in the angular components of the momentum of a test charge $Q$, initially at rest, gives rise to a {linear} ({ordinary}) memory effect that, to leading order, reads
\be\label{memoryformulaD=6}
P_i\big|_{u>0}-P_i\big|_{u<0} = Q \int_{-\infty}^{+\infty} \mathcal F_{iu}\, du = \frac{qQ\,v_i(2-\mathbf v^2-\mathbf n \cdot \mathbf v)}{8\pi^2(1-\mathbf n\cdot \mathbf v)^3r^2}+\mathcal {O}(r^{-3})\,,
\ee 
where the magnetic force does not contribute to subleading orders, because it will be further suppressed by the powers of $1/r$ appearing in the velocity.

With hindsight, having in mind in particular the results of \cite{Mao_EvenD}, in order to interpret this leading memory effect in terms of a symmetry, it is useful to perform a further gauge fixing procedure in order to get rid of the residual symmetries affecting the radiation order. Choosing a gauge parameter of the form
\be
8\pi^2\epsilon = -\frac{q\,\theta(u)}{r^2} \left(\frac{1}{1-\mathbf n \cdot \mathbf v}-1 \right)+\mathcal O(r^{-3})
\ee 
allows us to cancel the leading term of $\mathcal A_u$, compatibly with the condition $\Box \epsilon=0$.
The resulting field after performing this gauge transformation satisfies
\be
\mathcal A_u = \mathcal O({r^{-3}})\,,\qquad
8\pi^2\mathcal A_i= 
-\frac{q\,v_i\delta(u)}{(1-\mathbf n \cdot \mathbf v)r}
-\frac{q\,v_i\theta(u)(2-\mathbf v^2-\mathbf n \cdot \mathbf v)}{(1-\mathbf n \cdot \mathbf v)^3r^2}
+\mathcal O(r^{-3})\,.
\ee 
In particular, the $\mathcal O(r^{-2})$ component of $\mathcal A_u$ is now zero\footnote{More explicitly, $\mathcal A_u = \tfrac{q}{8\pi^2r^3}+\mathcal O(r^{-4})$.}, so that $\mathcal A_i$ is fully responsible for the memory formula \eqref{memoryformulaD=6}. The effect 
is proportional to the variation of the latter field component between $u>0$ and $u<0$ and takes the form of a total derivative  on the celestial sphere (\emph{i.e.} a gauge transformation): 
\be\begin{aligned}\label{specificexamplememory}
	\mathcal A_i\big|_{u>0}-\mathcal A_i\big|_{u<0} 
	&= -\frac{q\,v_i(2-\mathbf v^2-\mathbf n \cdot \mathbf v)}{(1-\mathbf n \cdot \mathbf v)^3r^2}\\
	&=-q\,\partial_i \left(
	\frac{1-\mathbf v^2}{2(1-\mathbf n \cdot \mathbf v)^2}+\frac{1}{1-\mathbf n \cdot \mathbf v}
	\right)\,.
\end{aligned}\ee  
This result provides an explicit connection between the memory effect and a residual symmetry acting, for large values of $r$, at Coulombic order.

Let us turn our attention to the case of \emph{null} memory. Let us consider, in any even dimension $D\ge6$, the field generated by a charge moving in the $\mathbf x_0$ direction at the speed of light: in Cartesian coordinates,
\be
\Box \mathcal A^\mu = q\, v^\mu \delta(\mathbf x- \mathbf x_0 t)\,,
\ee
with $v^\mu=(1,\mathbf x_0)$ and $|\mathbf x_0|=1$. Taking into account the corresponding retarded solution \eqref{nullmemk} for the scalar field $\varphi$, we then have $A^\mu = - v^\mu \varphi $ and, moving to retarded coordinates,
\be
\mathcal A_u = \varphi\,,\qquad
\mathcal A_r = (1-\mathbf n \cdot \mathbf x_0)\varphi\,,\qquad
\mathcal A_i = -r\, (x_0)_i \varphi\,.
\ee
Consequently
\be\begin{aligned}\label{AnullD}
	\mathcal A_u &\sim -\frac{q\delta(u)}{r^{D-4}}\,(\Delta-D+4)^{-1}(\mathbf n, \mathbf x_0)\,,\\
	\mathcal A_r &\sim -\frac{q\delta(u)}{r^{D-4}}(1-\mathbf n \cdot \mathbf x_0)(\Delta-D+4)^{-1}(\mathbf n, \mathbf x_0)\,,\\
	\mathcal A_i &\sim r\,\frac{q (x_0)_i \delta(u)}{r^{D-4}}\,(\Delta-D+4)^{-1}(\mathbf n, \mathbf x_0)\,,
\end{aligned}\ee
where $(\Delta-D+4)^{-1}$ denotes the Green function for the operator $\Delta-D+4$ and we have omitted terms proportional to higher-order derivatives of the delta function $\delta(u)$ (see \eqref{nullmemk}), which are \emph{leading} with respect to those displayed in \eqref{AnullD}, but which do not contribute to the memory effect.
The null memory formula then reads
\be\label{nullmemDeven}
P_i\big|_{u>0}-P_i\big|_{u<0}= Q\int_{-\infty}^{+\infty} \mathcal F_{iu}\, du' = -\frac{qQ}{r^{D-4}}\partial_i(\Delta-D+4)^{-1}(\mathbf n, \mathbf x_0)\,.
\ee

\subsection{Overview of the odd-dimensional case}
In odd dimensions $D\ge3$ the retarded propagator is given by \cite{Friedlander}
\be
D_\mathrm{ret}(x)=
c\,
(-x^2)_+^{1-\frac{D}{2}}\theta(x^0)\,,
\ee
where
$
c^{-1}=2\pi^{\frac{D}{2}-1}\Gamma(2-\frac{D}{2})
$,
while $(\kappa)_+^\alpha$ is the distribution defined as
\be
\langle 
(\kappa)_+^\alpha, \chi(\kappa)
\rangle =
\int_0^\infty \kappa^\alpha \chi(\kappa)\,d\kappa\,\qquad \text{ for }\alpha>-1\,,
\ee
$\chi(\kappa)$ denoting a generic test function, and analytically continued to any $\alpha\neq -1$, $-2$, $-3$, $\ldots$ by 
\be
\langle 
(\kappa)_+^\alpha, \chi(\kappa)
\rangle =
\frac{(-1)^n}{(\alpha+1)(\alpha+2)\cdots (\alpha+n)}\,
\langle
(\kappa)_+^{\alpha+n}, \chi^{(n)}(\kappa)
\rangle\,
\qquad
\text{for }n>-1-\alpha\,.
\ee
A relevant feature of the wave propagator in odd dimensional spacetimes is that its support is not localized on the  light-cone $|t|=r$, in contrast with the case of even dimensions, as it is nonzero also for $|t|>r$. This is to be interpreted as the fact that even an ideally sharp perturbation, $\delta(t,\mathbf x)$, will not give rise to an ideally sharp wave-front, but rather the induced radiation will display a dispersion phenomenon and nontrivial disturbances will linger on even after the first wave-front has passed.

The solution to equation \eqref{scalar-theta} is then furnished by 
\be\begin{aligned}
	\varphi(t,\mathbf x) = 
	\frac{cq}{2}
	\left\langle 
	\kappa^{1-D/2}_+,
	\theta(t-\sqrt{\kappa+|\mathbf x|^2})/\sqrt{\kappa+|\mathbf x|^2}
	\right\rangle\,.
\end{aligned}\ee
Integrating by parts, and assuming $t>r=|\mathbf x|$ (otherwise the field vanishes by causality), one obtains the following expansion
\be\begin{aligned}\label{phioddD}
	\frac{2}{cq}\,\varphi &= 
	\frac{(t^2-r^2)^{2-\frac{D}{2}}}{(2-\frac{D}{2})t}
	+
	\frac{(t^2-r^2)^{3-\frac{D}{2}}}{(2-\frac{D}{2})(3-\frac{D}{2})t^3}\,\tfrac{1}{2}
	+
	\frac{(t^2-r^2)^{4-\frac{D}{2}}}{(2-\frac{D}{2})(3-\frac{D}{2})(4-\frac{D}{2})t^5}\,\tfrac{1}{2}\cdot\tfrac{3}{2}
	+\cdots\\
	&+\frac{(t^2-r^2)^{-\frac{1}{2}}}{(2-\frac{D}{2})(3-\frac{D}{2})\cdots (-\frac{3}{2})(-\frac{1}{2})t^{D-4}}\,\tfrac{1}{2}\cdot\tfrac{3}{2}\cdots \left(\tfrac{D}{2}-3\right)\\
	&+(-1)^{\frac{D-3}{2}}\int_0^{t^2-r^2}\frac{d\kappa}{\sqrt\kappa (\kappa+r^2)^{\frac{D}{2}-1}}\,.
\end{aligned}\ee 
Moving to retarded coordinates, this result can be recast as
\be\label{scalar-odd}
\varphi(u,r) = 
\bar\varphi(u,r)
+\frac{cq}{2}
(-1)^{\frac{D-3}{2}}\theta(u)
\int_0^{u(u+2r)}
\frac{d\kappa}{\sqrt{\kappa}(\kappa+r^2)^{\frac{D}{2}-1}} 
\,,
\ee
where $\bar \varphi(u,r)$ is given by a sum of terms proportional to
\be
\frac{\theta(u)}{(u(u+2r))^\alpha(u+r)^\beta}\,,
\ee
with $\alpha$, $\beta$ positive and $\alpha+\beta$ half odd. In particular, 
it is then clear that the limit of this field as $r\to\infty$ for any fixed $u$ does not display any term with the Coulombic behavior   $r^{3-D}$ and hence that there is no memory effect on $\mathscr I^+$ to that order, since 
\be
\int_0^{u(u+2r)}
\frac{d\kappa}{\sqrt{\kappa}(\kappa+r^2)^{\frac{D}{2}-1}} \sim \frac{1}{r^{D-3}}\int_0^{\frac{2u}{r}}\frac{dx}{\sqrt x(1+x)^{\frac{D}{2}-1}}\sim \frac{2\sqrt{2u}}{r^{D-\frac{5}{2}}}\,.
\ee 
Considering instead the limit of $\varphi$ as $t\to+\infty$ for fixed $r$, one sees that only the last term in \eqref{phioddD} survives and yields
\be\begin{split}
\frac{cq}{2}(-1)^{\frac{D-3}{2}}\int_0^{\infty}\frac{d\kappa}{\sqrt\kappa (\kappa+r^2)^{\frac{D}{2}-1}} &= \frac{cq}{2}\frac{(-1)^{\frac{D-3}{2}}}{r^{D-3}}B\big(\tfrac{1}{2},\tfrac{D-3}{2}\big) \\
&= \frac{q}{(D-3)\Omega_{D-2}r^{D-3}}\,.
\end{split}
\ee
This means that the Coulombic energy due to the newly created particle is felt by the test charge only after one has waited (for an infinite time) at a fixed distance $r$ that the perturbations due to the dispersion occurring in odd spacetime dimensions have died out. To some extent, this is to be regarded as a smeared-out memory effect, as opposed to memory effects occurring sharply at $\mathscr I^+$ near $u=0$ in even dimensions \cite{SatishchandranWald}.

The situation does not improve if one considers a particle that is created with a nonzero velocity $\mathbf v$. Indeed, boosting the exact solution \eqref{phioddD} by means of \eqref{standardboost}, one sees that $\varphi$ goes to zero for fixed $r$ as $t\to+\infty$. The reason is that, while one waits for the dispersion to die out, the source, moving at a constant velocity, has traveled infinitely far from the test charge. 

Shifting our attention to the case of null memory, we see that it is possible to provide the following formal solution to the recursion relations \eqref{recursionnullD}, which hold in any dimension. We consider $\varphi= \sum \varphi^{(k)}r^{-k}$, setting $\varphi^{(k)}=0$ for $k\ge D-3$, while, for $k \le D-4$,
\be\label{nullmemkodd}
\varphi^{(k)}(u, \mathbf n)=\delta^{(D-4-k)}(u)C_k(\mathbf n)\,,
\ee
with the functions $C_k(\mathbf n)$ determined recursively by
\be\begin{aligned}
	(D-2k-2) C_k(\mathbf n) &= [\Delta + (k-1)(k-D+2)]C_{k-1}(\mathbf n)\,,\\
	C_{D-4}(\mathbf n) &= -(\Delta-D+4)^{-1}(\mathbf n,\mathbf x_0)\,.
\end{aligned}\ee
Thus, although the field is highly singular at $u=0$, the resulting null memory effect will be formally identical to the one occurring in even dimensions.

%%%%%
\section{Electromagnetic Memory and Residual Symmetries}\label{sec:em}
%%%%%

In the previous section, we saw how to establish a connection between local symmetries acting at large $r$ and memory effects for specific matter configurations. 
Now we would like to investigate this connection beyond those examples by studying the general structure of the solution space in Lorenz gauge. 

As remarked above, the nontrivial nature of higher-dimensional memory effects suggests the existence of infinite-dimensional asymptotic symmetries, in spite of the analysis performed in Section \eqref{sec:YMGlobal}.
We will therefore analyze once more the asymptotic behavior of the fields in higher dimensions, adopting now the Lorenz gauge. First, we shall choose to adopt radiation falloff, which is sufficient to the description of the memory effect. A more general possibility will be discussed in Section \ref{sec:poly}, in order to explore the full structure of the asymptotic symmetry group in Lorenz gauge. 

The choice of Lorenz gauge is mainly motivated by the fact that it allows us to make a direct comparison with the above examples. However, as we will see, another important feature of this gauge is its greater flexibility with respect to the $r$-dependence of fields and gauge parameters, as opposed to the rigidity of radial gauge where no $r$-dependence of the gauge transformation is allowed. This will play a key role in the interpretation of memory effects as transitions between field configurations that differ by a residual symmetry acting at Coulombic order.

%%%%%
\subsection{Electromagnetism in the Lorenz gauge}\label{ssec:lorenz}
%%%%%

The Lorenz gauge condition reads 
\be\label{Lorenz}
\nabla^a \mathcal A_a
=
-\partial_u \mathcal A_r + \left(\partial_r + \frac{D-2}{r}\right)(\mathcal A_r - \mathcal A_u)+\frac{1}{r^2}D\cdot \mathcal A=0\,.
\ee
The residual symmetry parameters then satisfy $\Box \epsilon=0$, namely
\be\label{residual}
\left(2\partial_r + \frac{D-2}{r}\right)\partial_u \epsilon = \left(\partial_r^2 + \frac{D-2}{r}\partial_r + \frac{1}{r^2}\Delta\right)\epsilon\,.
\ee
The equations of motion reduce to $\Box \mathcal A_a=0$, which, component by component, reads
\be\begin{aligned}\label{eom}
&\left[\partial_r^2-2\partial_u\partial_r - 
\frac{D-2}{r} (\partial_u-\partial_r)+\frac{1}{r^2}\Delta\right]\mathcal A_u=0\, ,\\
&\left[\partial_r^2-2\partial_u\partial_r \, - \, \frac{D-2}{r}(\partial_u-\partial_r)\, + \, \frac{1}{r^2}\Delta \right]\mathcal A_r+\frac{D-2}{r^2}(\mathcal A_u - \mathcal A_r)-\frac{2}{r^3} D\cdot \mathcal A=0\, ,\\
&\left[\left(\partial_r^2-2\partial_u\partial_r + \frac{1}{r^2}\Delta\right)-\frac{D-4}{r}(\partial_u-\partial_r)-\frac{D-3}{r^2}\right]\mathcal A_i-\frac{2}{r}D_i(\mathcal A_u -\mathcal A_r)=0\,.
\end{aligned}\ee

We assume now the expansions 
\be
\mathcal A_a = \sum_k A_a^{(k)} r^{-k}\,,
\qquad
\epsilon = \sum_k \epsilon^{(k)}r^{-k}\,,
\ee
where the summation ranges are, for the moment, unspecified.
Equations \eqref{Lorenz} and \eqref{residual}  then give
\begin{align}
	\partial_u A^{(k+1)}_r&=(k-D+2)(A_u^{(k)}-A_r^{(k)})+D\cdot A^{(k-1)}
	\label{Lorenz_r}\, ,\\
	(D-2k-2)\partial_u \epsilon^{(k)}&=[\Delta + (k-1)(k-D+2)]\epsilon^{(k-1)}
	\label{residual_r}\, , 
\end{align}
while from \eqref{eom} one obtains
\begin{align}
	&(D-2k-2)\partial_u A_u^{(k)}=[\Delta + (k-1)(k-D+2)]A_u^{(k-1)}
	\label{equ_r}\, ,\\
	&(D-2k-2)\partial_u A^{(k)}_r = [\Delta+k(k-D+1)]A_r^{(k-1)}\, +\, (D-2)\, A_u^{(k-1)}-2 D\cdot A^{(k-2)}
	\label{eqr_r}\, ,\\
	&(D-2k-4)\partial_u A_i^{(k)}=[\Delta+k(k-D+3)-1]A_i^{(k-1)}-2D_i(A_u^{(k)}-A_r^{(k)})
	\label{eqi_r}\,.
\end{align}
Equations \eqref{Lorenz_r}---\eqref{eqi_r} appear in particular to order $r^{-k-1}$ in the asymptotic expansions of the original equations.

To the purposes of analyzing the electromagnetic memory effect, we may keep the leading falloffs to match the corresponding radiation falloffs, 
\be\label{radiationfall}
\mathcal A_u = \mathcal O(r^{-(D-2)/2})\,,\qquad
\mathcal A_r = \mathcal O(r^{-(D-2)/2})\,,\qquad
\mathcal A_i = \mathcal O(r^{-(D-4)/2})\,.
\ee 
More general options are possible and influence the structure of the asymptotic symmetry group. We will be concerned with these more general aspects of the discussion in Section \ref{sec:poly}.

The significance of the choice \eqref{radiationfall} lies in the fact that the derivatives with respect to $u$ of the field components are unconstrained to leading order: these conditions for the asymptotic expansion are well-suited to identifying the boundary data for a radiation solution with an arbitrary wave-form. Such components also provide the energy flux at a given retarded time, as happened in the radial gauge, according to
\be
\mathcal P(u) = \int_{S_u}  \gamma^{ij} \partial_u A^{(\frac{D-4}{2})}_i \partial_u A^{(\frac{D-4}{2})}_j d\Omega\,,
\ee
where $S_u$ is the section of $\mathscr I^+$ at fixed $u$ and $d\Omega$ is the measure element on the unit $(D-2)$-sphere.\footnote{
	The falloff conditions \eqref{radiationfall} can also be  heuristically justified as follows. A solution $\mathcal A_u(u,r,x^i)$ of eq. \eqref{equ_r} with nontrivial $u$ dependence, as required for radiation, must behave as $r^{-(D-2)/2}$ to leading order. If one then assumes the bounds $\mathcal A_r \lesssim A_r r^{-(D-6)/2}$ and $\mathcal A_i \lesssim A_i r^{-(D-4)/2}$, which provide a sufficient conditions for the finiteness of the energy flux, then equations \eqref{eqr_r} and \eqref{eqi_r} actually impose $\mathcal A_r \sim A_r r^{-(D-2)/2}$ and $\mathcal A_i \sim A_i r^{-(D-4)/2}$.
}

The asymptotic behavior of radiation differs in higher dimensions with respect to the characteristic Coulombic falloff $r^{3-D}$ that, in its turn, can be identified as the leading falloff for $u$-independent solutions. (See also the discussion in Section \ref{ssec:em_memory} on this point.) As we shall see,  Coulomb fields give nonvanishing contributions to the surface integral associated with the electric charge as well as to the memory effects.

%%%%%
\subsubsection{Recursive gauge fixing}\label{ssec:emgaugefix}
%%%%%

The gauge variations
\be
\delta A_u^{(k)}=\partial_u \epsilon^{(k)}\,,\qquad
\delta A_r^{(k)}=-(k-1)\epsilon^{(k-1)}\,,\qquad
\delta A_i^{(k)}=\partial_i \epsilon^{(k)}\,
\ee 
imply  a number of restrictions on the allowed gauge parameters, in order to keep the falloffs of such field configurations. From the $r$-variation, we read off $\epsilon^{(k)}=0$ for $k<(D-4)/2$ and $k\neq 0$. The $u$-variation additionally requires that $\epsilon^{(\frac{D-4}{2})}$ be independent of $u$, whereas the angular variation does not give rise to further constraints at this stage. This leads to a gauge parameter of the following, provisional, form
\be \label{epsilon}
\epsilon(u,r,x^i)=1+\epsilon^{(\frac{D-4}{2})}(x^i)/r^{\frac{D-4}{2}}+\cdots
\,.
\ee
This is not the actual form of the residual gauge parameter, however, since it does not satisfy \eqref{residual_r}\footnote{
	Interestingly, the corresponding putative charge
	\be
	\mathcal {\tilde Q}_\epsilon = \lim_{r\to\infty}\int_{S_u} \epsilon\, \mathcal F_{ur} r^{D-2} d\Omega
	= \int D\cdot A^{(\frac{D-4}{2})} \epsilon^{(\frac{D-4}{2})} d\Omega
	\ee
	is finite and nonvanishing as $r\to\infty$.
	Unfortunately, \eqref{noepsilonrad} projects $\epsilon^{(\frac{D-4}{2})}$ to zero. 
}:
\be\label{noepsilonrad}
\Box \epsilon = \frac{1}{r^{\frac{D}{2}}}\left[\Delta - \frac{(D-2)(D-4)}{4} \right]\epsilon^{(\frac{D-4}{2})}+\cdots\,.
\ee

Thus, we need to search further down in the asymptotic expansion of $\epsilon$. We do so by employing a recursive on-shell gauge-fixing procedure \cite{StromingerGravmemevD}. Setting to zero \eqref{noepsilonrad},  implies $\epsilon^{(\frac{D-4}{2})}=0$ since the Laplacian on the $(D-2)$-sphere is negative semidefinite.
Therefore, the residual symmetry is parametrized as follows
\be
\epsilon(u,r,x^i)=1+\frac{\epsilon^{(\frac{D-2}{2})}(u,x^i)}{r^{(D-2)/2}}+\cdots\,,
\ee
where we conventionally set the global part of $\epsilon$ to $1$. 
Equation \eqref{residual_r} leaves the $u$-dependence of $\epsilon^{(\frac{D-2}{2})}$ unconstrained, and therefore we may use it to set 
\be
A_u^{(\frac{D-2}{2})}=0\,,
\ee
leaving 
$\epsilon^{(\frac{D-2}{2})}(x^i)$ arbitrary.

We may now proceed by noting that, setting $k=D/2$, \eqref{equ_r} and $\eqref{residual_r}$ reduce to
\be
\partial_u A_u^{(\frac{D}{2})}=0\,,\qquad
\partial_u \epsilon^{(\frac{D}{2})}+\left[\Delta-\frac{(D-2)(D-4)}{4}\right]\epsilon^{(\frac{D-2}{2})}=0\,,
\ee
respectively. Thus, $A_u^{(\frac{D}{2})}$ is a function of the angles $x^i$ only, while $\delta A_u^{(\frac{D}{2})}=\partial_u \epsilon^{(\frac{D}{2})}$ can be expressed in terms of $\epsilon^{(\frac{D-2}{2})}$, which
can be used to set $A_u^{(\frac{D}{2})}=0$, while still leaving
$\epsilon^{(\frac{D}{2})}(x^i)$ arbitrary.

This procedure may now be repeated recursively. Assuming
\be
A_u^{(\frac{D-2}{2})}=A_u^{(\frac{D}{2})}=\cdots=A_u^{(q-1)}=0\,,\qquad \epsilon^{(q-1)}(x^i)\quad\text{arbitrary}\,,
\ee
for some $q>D/2$.
Then, for $k=q-1$, \eqref{equ_r} and $\eqref{residual_r}$ give
\be\label{induction-equations}
(D-2q-2)\partial_u A_u^{(q)}=0\,,\qquad
(D-2q-2)\partial_u\epsilon^{(q)}=\left[\Delta-(q-1)\left(D-q-2\right)\right]\epsilon^{(q-1)}\,.
\ee
Therefore, we may employ $\epsilon^{(q-1)}(x^i)$ to set $A_u^{(q)}$ to zero provided that the differential operator on the right-hand side is invertible, which is true for any $q<D-2$.

We shall now consider two options. We may first choose to truncate the recursive gauge fixing right after the step labeled by $q=D-4$, which leaves us with the asymptotic expansions
\be\label{falloffmemory1}
\mathcal A_u = \sum_{k=D-3}^\infty A_u^{(k)}r^{-k}\,,\qquad
\mathcal A_r = \sum_{k=\frac{D}{2}}^\infty A_r^{(k)}r^{-k}\,,\qquad
\mathcal A_i = \sum_{k=\frac{D-4}{2}}^\infty A_i^{(k)}r^{-k}\,,
\ee
where $A_r^{(\frac{D-2}{2})}=0$ on shell.
The residual symmetry is given by
\be\label{res_even}
\epsilon(u,r,x^i) \, =\, 1+ \epsilon^{(D-4)}(x^i)r^{4-D}\, + \cdots\,,
\ee
whose corresponding charge, evaluated in the absence of radiation close to the past boundary $\mathscr I^+_-$ of $\mathscr I^+$, reads\footnote{A similar procedure is adopted in Section \ref{sec:poly}.}
\be\label{ChargeStrom2}
\mathcal Q_\epsilon = \frac{1}{r^{D-4}}\int_{\mathscr I^+_-}
\left(
\partial_u A^{(D-2)}_r+(D-3)A_u^{(D-3)}
\right)
\epsilon^{(D-4)}d\Omega\,.
\ee

Alternatively, we may also perform the recursive gauge-fixing until the very last allowed step, $q=D-3$. In which case,
\be\label{falloffmemory2}
\mathcal A_u = \sum_{k=D-2}^\infty A_u^{(k)}r^{-k}\,,\qquad
\mathcal A_r = \sum_{k=\frac{D}{2}}^\infty A_r^{(k)}r^{-k}\,,\qquad
\mathcal A_i = \sum_{k=\frac{D-4}{2}}^\infty A_i^{(k)}r^{-k}
\ee
and
\be
\epsilon(u,r,x^i) \, =\, 1+ \epsilon^{(D-3)}(x^i)r^{3-D}\, + \cdots\,.
\ee
The latter choice highlights the possibility of making the components
\be\label{gaugeinvtower}
A_i^{(\frac{D-4}{2})},\cdots, A_{i}^{(D-5)}, A_i^{(D-4)}
\ee
gauge-invariant, and hence in principle responsible for any observable effect due to radiation impinging on a test charge placed at a large distance $r$ from a source.
Indeed, as we shall verify explicitly in Section \ref{sec:memories}, electromagnetic memory effects appear at Coulombic order $A_i^{(D-4)}$.

%%%%%
\subsection{Memory and residual symmetries}\label{ssec:em_memory}
%%%%%

A test particle with charge $Q$, initially at rest at a large distance $r$ from the origin, will experience a leading-order momentum kick due to the presence of an electric field according to  
\be\label{velocity}
P_{i}\big|_{u_1}-P_i\big|_{u_0} = Q\int_{u_0}^{u_1} \mathcal F_{iu}\, du\,.
\ee 
We are neglecting the contribution from the magnetic field by the assumption that they will always give rise to subleading contributions in $1/r$, due to their proportionality to the velocity of the test particle, as we explicitly checked in Section \ref{sec:memories} in the case of an ideally sharp wavefront.

Now we shall consider solutions that are stationary before $u=u_0$ and after $u=u_1$. For such solutions, the Maxwell tensor before $u_0$ and $u_1$ is not zero, in general, because static forces are present. However, it does not contain radiation, and thus all the components of the gauge potential associated to radiation are to vanish, or, more generally, are to be pure gauge. In particular, the radiation field components before $u_0$ and $u_1$ are to be identical or are to differ by a gauge transformation.

Now, let us combine the information of the previous recursive gauge-fixing with the requirement that the solution be stationary before $u_0$ and after $u_1$.
The asymptotic expansion of the electric field $\mathcal F_{ui}=\partial_i \mathcal A_u-\partial_u \mathcal A_i$ satisfies
\be
F_{iu}^{(k)} = -\partial_u A^{(k)}_i(u,x^k)\,,  
\ee
for $k=\frac{D-4}{2},\ldots,D-4$
since, then, $A_u^{(k)}$ is zero, thanks to the gauge-fixing procedure. Thus, 
\be\label{DeltaPDeltaA}
P_i\big|_{u_1}-P_i\big|_{u_0} =-Q \sum_{k=\frac{D-4}{2}}^{D-4}\frac{1}{r^k}\left( A^{(k)}_i\big|_{u_1}-A_i^{(k)}\big|_{u_0} \right)+ \mathcal O(r^{3-D})\,.
\ee
With respect to our previous observation, let us notice that the components of $\mathcal A_i$ that enter the subleading terms $\mathcal O(r^{3-D})$ in \eqref{DeltaPDeltaA} are those connected with stationary properties of the field (we assume the absence of permanent long-range magnetic fields, which would induce $\mathcal O(r^{4-D})$ contributions to $\mathcal F_{ij}$), while all the leading components explicitly written enter the radiation behavior and thus their difference after $u_1$ and before $u_0$ can be at most the angular gradient of given functions. Actually, we shall immediately see that combining this information with the equations of motion will allow us to conclude that they all vanish with the exception of the last one $A_i^{(D-4)}$.

Indeed, let us note that for a stationary solution, in our gauge,
equations \eqref{Lorenz_r}, \eqref{equ_r}, \eqref{eqr_r} and \eqref{eqi_r} read, for $k<D-2$,
\begin{align}
	(D-k-1)(A_r^{(k-1)}-A_u^{(k-1)})+D\cdot A^{(k-2)}&=0
	\label{Lorenz_ru=0}\,,\\
	[\Delta + (k-1)(k-D+2)]A_u^{(k-1)}&=0
	\label{equ_ru=0}\, ,\\
	[\Delta+k(k-D+1)]A_r^{(k-1)}\, +\, (D-2)\, A_u^{(k-1)}-2 D\cdot A^{(k-2)}&=0
	\label{eqr_ru=0}\,,\\
	[\Delta+(k-1)(k-D+2)-1]A_i^{(k-2)}-2D_i(A_u^{(k-1)}-A_r^{(k-1)})&=0
	\label{eqi_ru=0}\,.
\end{align}
For $1<k<D-2$, equation \eqref{equ_ru=0} implies $A_u^{(k-1)}=0$, compatibly with the outcome of recursive gauge-fixing. Equations \eqref{Lorenz_ru=0} and \eqref{eqr_ru=0} then give, for $1<k<D-2$,
\be
[\Delta+(k-2)(k-D+1)]A_r^{(k-1)}=0
\ee
so that $A_r^{(k-1)}=0$ for $2<k<D-2$. Considering finally equation \eqref{eqi_ru=0}, for $2<k<D-2$, we have
\be
[\Delta+(k-1)(k-D+2)-1]A_i^{(k-2)}=0
\ee
and hence $A_i^{(k-2)}=0$ provided provided that $k$ also satisfies $k_-(D)<k<k_+(D)$ with 
\be
k_\pm(D) = \frac{1}{2}\left[D-1\pm \sqrt{(D-3)^2+4} \right]\,;
\ee
actually, $k_-(D)<1$ and $k_+(D)>D-2$ for any $D>3$, so we conclude that stationary solutions obey $A_i^{(k-2)}=0$ for $2<k<D-2$.

To summarize, $A_u^{(k)}=0$ for $0<k<D-3$, while $A_r^{(k)}=0$ for $1<k<D-3$ and 
\be
A_i^{(k)}=0\text{ for }0<k<D-4\,.
\ee 
By equation \eqref{DeltaPDeltaA}, the condition on $\mathcal A_i$ implies that the memory effect appears to leading order $r^{4-D}$,
\be\label{memoryform_even}
P_i\big|_{u_1}-P_i\big|_{u_0} = -\frac{Q}{r^{D-4}}\left( A^{(D-4)}_i\big|_{u_1}-A_i^{(D-4)}\big|_{u_0} \right)+\mathcal O(r^{3-D})\,.
\ee
In view of the discussion below \eqref{DeltaPDeltaA}, we conclude that the momentum shift must take the form 
\be\label{memoryformulaf}
P_i\big|_{u_1}-P_i\big|_{u_0} = \frac{Q}{r^{D-4}}\partial_ig(x^k)+\mathcal O(r^{3-D})\,,
\ee
with 
$g(x^i)$ a $u$-independent function, which will depend on the shape of the radiation train and in particular on $u_0$ and $u_1$ (see for instance  \eqref{specificexamplememory}). 

Let us note that, as it must be, this difference is not affected by the action of the residual gauge transformation \eqref{res_even}, as can be understood by the fact that the latter is $u$-independent and thus does not alter the difference $A_i^{(D-4)}\big|_{u_1}-A_i^{(D-4)}\big|_{u_0}$.
In this sense, whether or not one performs the last step of the recursive gauge fixing is irrelevant to the extent of calculating the electromagnetic memory.

To conclude, we have established a formula that exhibits a momentum kick characterizing the transition between the initial and final vacuum configurations, parametrized by the gauge transformation $g(x^i)$, induced by the exposure to electromagnetic radiation crossing null infinity. In particular, the norm of this effect scales as $r^{3-D}$.

Up to this point we have only been dealing with ordinary/linear memory effect. In order to encompass a null/nonlinear memory effect, we must modify the equations of motion \eqref{eom} by adding suitable source terms on the right-hand sides, namely a current density $J^\mu$ allowing for the outflow to future null infinity of charged massless particles. The falloff conditions on such a current can be taken as follows
\be
J_u=\mathcal O(r^{2-D})\,,\qquad
J_r=\mathcal O(r^{2-D})\,,\qquad
J_i=\mathcal O(r^{3-D})\,.
\ee
This is clearly displayed by the example of a single massless charge $q$ moving in the $\mathbf x_0$ direction, whose current reads in Minkowski components
\be
J^0 = q\, \delta(\mathbf x - \mathbf x_0 t)\,,\qquad
\mathbf J =q\, \mathbf x_0 \delta(\mathbf x - \mathbf x_0 t)
\ee 
and in retarded components (for $t=u+r>0$)
\be\begin{aligned}
	J_u &= - \frac{q}{r^{D-2}}\,\delta(u)\delta(\mathbf n, \mathbf x_0)\,,\\
	J_r &= -\frac{q(1-\mathbf n \cdot \mathbf x_0)}{r^{D-2}}\,\delta(u)\delta(\mathbf n, \mathbf x_0)\,,\\
	J_i &= \frac{q (\mathbf x_0)_i}{r^{D-3}}\,\delta(u)\delta(\mathbf n, \mathbf x_0)\,.
\end{aligned}\ee
As far as the discussion of the previous section is concerned, namely, for the purposes of the recursive gauge fixing, the only modification is thus the introduction of a source term $J^{(D-2)}_u$ in the right-hand side of \eqref{equ_r} when $k=D-3$, which now actually forces us to stop the gauge fixing after the use of said equation for $k=D-4$ (the step labeled by $q=D-4$ in the previous section) and leaves us with the falloff \eqref{falloffmemory1}. On the contrary, reaching \eqref{falloffmemory2} is not allowed, and thus \eqref{falloffmemory1} comprises a complete gauge fixing. However, also in view of the above considerations, the discussion of the memory effect and its relation to the symmetry acting at Coulombic order thus remain unaltered.

\subsection{Phase memory}\label{ssec:phasemem}
Let us consider a pair of electric charges $q$ and $-q$ that are pinned in the positions $(r,\mathbf n_1)$ and $(r,\mathbf n_2)$, for large $r$. We will now derive the expression for an imprint that the passage of a radiation train leaves on the properties of these particles that is encoded in the phase of their states. See \cite{Susskind} for the discussion a very similar phenomenon in a four-dimensional setup. A quantum treatment of electromagnetic kick memory is instead given in \cite{Afshar:2018sbq}.

To this purpose, let us assume that, as in the previous section, radiation impinges on the charges only during the interval between two given retarded times $u_0$ and $u_1$. As we have seen, this means that
\be\label{gaugevacua}
\mathcal A_i\big|_{u_1}-\mathcal A_i\big|_{u_0}=\frac{1}{r^{D-4}}\,\partial_i g + \mathcal O(r^{3-D})
\ee
for a suitable angular function $g(\mathbf n)$. We will assume for simplicity that the gauge field before the onset of radiation is the trivial one.

Let $|\psi_1\rangle=|q\rangle$ and $|\psi_2\rangle=|-q\rangle$ be the initial states in which the charged particles are prepared, which are uniquely labeled by their charges since translational degrees of freedom have been suppressed. Before $u_0$, the state $|\psi_2,\mathbf n_1\rangle$ obtained by the parallel transport of the second state $|\psi_2\rangle$ to the position $\mathbf n_1$ of the first the charge is 
\be
|\psi_2,\mathbf n_1\rangle=|\psi_2\rangle\,,
\ee 
because $\mathcal A_\mu=0$, so that the corresponding tensor state evaluated in $\mathbf n_1$ is given by 
\be
|\psi_1,\mathbf n_1\rangle \otimes |\psi_2,\mathbf n_1\rangle=|\psi_1\rangle\otimes |\psi_2\rangle\,.
\ee 

At $u_1$, instead, the same operation must be performed by calculating
\be
|\psi_2,\mathbf n_1\rangle = 
\exp\left[iq\int_{\mathbf n_2}^{\mathbf n_1}\mathcal A_i \,dx^i\right]|\psi_2\rangle = 
\exp\left[iq\,\frac{g(\mathbf n_1)-g(\mathbf n_2)}{r^{D-4}}\right]|\psi_2\rangle+\mathcal O(r^{3-D})\,,
\ee
where we have employed \eqref{gaugevacua} to establish the second equality.
Therefore, after the passage of radiation,
\be
|\psi_1, \mathbf n_1\rangle \otimes |\psi_2,\mathbf n_1\rangle = \exp\left[iq\,\frac{g(\mathbf n_1)-g(\mathbf n_2)}{r^{D-4}}\right]|\psi_1\rangle\otimes |\psi_2\rangle+\mathcal O(r^{3-D})\,,
\ee
which displays how the transition between two different radiative vacua, already experimentally detectable by the occurrence of a nontrivial velocity kick for a test charge, is also signaled by the variation of the relative phases in the states obtained by parallel transport of charged particles. Such a phase can be nontrivial provided that the function $g$ is not a constant, namely when there is a nontrivial memory kick \eqref{memoryformulaf}. 

A point that should be underlined is that, in this setup, the states $|\psi_1\rangle$ and $|\psi_2\rangle$ do not evolve, since each particle is kept fixed in its position (its translational quantum numbers are \emph{frozen}) and electromagnetic radiation cannot change its charge. The relative phase difference occurs entirely as an effect of the evolution of $\mathcal A_\mu$, which undergoes a transition between two underlying radiative vacua. We shall see in the following how this aspect is qualitatively different in a non-Abelian theory, where radiation can alter the color charge.

%%%%%
\section{Yang-Mills Memory} \label{sec:color}
%%%%%

We now turn to the extension of the above analysis of memory effects to the non-Abelian case.

%%%%%
\subsection{Yang-Mills theory in Lorenz gauge}\label{ssec:lorenzYM}
%%%%%
We consider pure Yang-Mills theory of an anti-Hermitian gauge field, adopting the same conventions as in Section \ref{sec:YMGlobal} (in particular, we again adopt anti-Hermitian $su(N)$ generators).
We will however impose the Lorenz gauge condition $\nabla^a \mathcal A_a=0$, instead of radial gauge,
which leaves as residual gauge parameters those that satisfy
$
\Box \epsilon +[\mathcal A_a, \partial^a \epsilon]=0
$.
Furthermore, the equations of motion reduce to $\Box \mathcal A_a + [\mathcal A^a, \nabla_a A_b+\mathcal F_{ab}]=0$.

Adopting retarded Bondi coordinates, the Lorenz gauge condition reads
\be\label{Lorenz_YM}
\partial_u \mathcal A_r = \left(\partial_r + \frac{D-2}{r}\right)(\mathcal A_r - \mathcal A_u)+\frac{1}{r^2}D\cdot \mathcal A\,,
\ee
while the constraint on residual transformations is
\be\begin{aligned} \label{residual_YM}
&\left(\partial_u^2 - 2 \partial_u \partial_r + \frac{1}{r^2} \Delta \right) \epsilon + \frac{D-2}{r}(\partial_r - \partial_u)\epsilon \\
&=
[\partial_r \epsilon, \mathcal A_r - \mathcal A_u] - [\partial_u \epsilon, \mathcal A_r] + \frac{1}{r^2}\gamma^{ij}[D_i \epsilon, \mathcal A_j]\,.
\end{aligned}\ee
The equations of motion give instead
\be\begin{aligned}[]\label{eom_YM}
&\left[\partial_r^2-2\partial_u\partial_r - 
\frac{D-2}{r} (\partial_u-\partial_r)+\frac{1}{r^2}\Delta\right]\mathcal A_u\\
&=
[\mathcal A_u-\mathcal A_r, \partial_r \mathcal A_u + \mathcal F_{ru}]+[\mathcal A_r, \partial_u \mathcal A_u]-\frac{\gamma^{ij}}{r^2}[\mathcal A_i, D_j\mathcal A_u + \mathcal F_{ju}]\, ,
\\
&\left[\partial_r^2-2\partial_u\partial_r \, - \, \frac{D-2}{r}(\partial_u-\partial_r)\, + \, \frac{1}{r^2}\Delta \right]\mathcal A_r+\frac{D-2}{r^2}(\mathcal A_u - \mathcal A_r)-\frac{2}{r^3} D\cdot \mathcal A\\
&=
[\mathcal A_u- \mathcal A_r , \partial_r \mathcal A_r] + [\mathcal A_r, \partial_u \mathcal A_r + \mathcal F_{ur}] - \frac{\gamma^{ij}}{r^2}[\mathcal A_i, D_j \mathcal A_r+ \mathcal F_{jr}]
\, ,
\\
&\left[\left(\partial_r^2-2\partial_u\partial_r + \frac{1}{r^2}\Delta\right)-\frac{D-4}{r}(\partial_u-\partial_r)-\frac{D-3}{r^2}\right]\mathcal A_i-\frac{2}{r}D_i(\mathcal A_u -\mathcal A_r)\\
&=
\left[\mathcal A_u- \mathcal A_r, \left(\partial_r - \frac{2}{r}\right)\mathcal A_i + \mathcal F_{ri}\right] + [\mathcal A_r, \partial_u \mathcal A_i + \mathcal F_{ui}]-\frac{\gamma^{jk}}{r^2}[\mathcal A_j, D_k\mathcal A_i+\mathcal F_{ki}]\,.
\end{aligned}\ee

Performing the usual asymptotic expansion in inverse powers of the radial coordinate $r$, one obtains the following set of equations.
Equations \eqref{Lorenz_YM} and \eqref{residual_YM} give
\be
\partial_u A^{(k+1)}_r=(D-k-2)(A_r^{(k)}-A_u^{(k)})+D\cdot A^{(k-1)}
\label{Lorenz_r_YM}\, 
\ee
and
\be\begin{aligned}
&(D-2k-2)\partial_u \epsilon^{(k)}-[\Delta - (k-1)(D-k-2)]\epsilon^{(k-1)}\\ 
&= \sum_{l+m=k}
\left(
-l[\epsilon^{(l)}, A_u^{(m)}-A_r^{(m)}]+[\partial_u \epsilon^{(l+1)}, A_r^{(m)}]-\gamma^{ij}[D_i \epsilon^{(l-1)}, A_j^{(m)}]
\right)
\label{residual_r_YM}\, ,
\end{aligned}\ee 
respectively, while from \eqref{eom_YM} one obtains
\begin{align}
	\label{equ_r_YM}
	&(D-2k-2)\partial_u A_u^{(k)}-[\Delta - (D-k-2)(k-1)]A_u^{(k-1)}
	\\ \nonumber
	= & \sum_{l+m=k}
	\Big(
	[A_r^{(m)}- A_u^{(m)}, -l A_u^{(l)}+ F_{ru}^{(l+1)}]-[A_r^{(m)}, \partial_u A_u^{(l+1)}]
	\\ \nonumber
	&+\gamma^{ij}[A_i^{(m)}, D_j A_u^{(l-1)} + F_{ju}^{(l-1)}]
	\Big)
	\, ,\\
	\label{eqr_r_YM}
	&(D-2k-2)\partial_u A^{(k)}_r \!-\! [\Delta-k(D-k-1)]A_r^{(k-1)}\! -\! (D-2) A_u^{(k-1)}+2 D\cdot A^{(k-2)}
	\\ \nonumber
	= & \sum_{l+m=k}
	\Big([A_r^{(m)}-A_u^{(m)}, -l A_r^{(l)}]-[A_r^{(m)}, \partial_u A_r^{(l+1)}+F_{ur}^{(l+1)}]
	\\ \nonumber
	&+\gamma^{ij}[A_i^{(m)}, D_j A_r^{(l-1)}+F_{jr}^{(l-1)}]
	\Big)
	\, ,\\
	\label{eqi_r_YM}
	&(D-2k-4)\partial_u A_i^{(k)}-[\Delta-k(D-k-3)-1]A_i^{(k-1)}+2D_i(A_u^{(k)}-A_r^{(k)})
	\\ \nonumber
	= & \sum_{l+m=k}
	\Big([A_r^{(m)}-A_u^{(m)}, -(l+2) A_i^{(l)}+F_{ri}^{(l+1)}]-[A_r^{(m)}, \partial_u A_i^{(l+1)}+F_{ui}^{(l+1)}]
	\\ \nonumber
	&+\gamma^{jk}[A_j^{(m)}, D_k A_i^{(l-1)}+F_{ki}^{(l-1)}]
	\Big)
	\, ,
\end{align}
for the corresponding components of the equations of motion. In particular, \eqref{Lorenz_r_YM}---\eqref{eqi_r_YM} appear to order $\mathcal O(r^{-k-1})$ in the asymptotic expansions.

We choose to adopt the same radiation falloff conditions \eqref{radiationfall} that we imposed in the linearized theory: the expansions of $\mathcal A_u$ and $\mathcal A_r$ start at order $\mathcal O(r^{-(D-2)/2})$ and that of $\mathcal A_i$ starts at order $\mathcal O(r^{-(D-4)/2})$.

For completeness, in order to verify the consistency of this choice, we now calculate the color charge $\mathcal Q(u) = \mathcal Q^A(u) T^A$ at a given retarded time $u$, as we did in retarded radial gauge in Section \ref{sec:YMGlobal}. This quantity is given by the surface integral
\be\begin{aligned}
\mathcal Q^A(u) 
=& \int_{S_u} F_{ur}^A r^{D-2}d\Omega\\
=& \sum_k r^{D-2-k} \int_{S_u} \Big(\partial_u A_r^{(k)} + (k-1) A_u^{(k-1)} + \sum_{l+m=k}[A_u^{(m)}, A_r^{(l)}] \Big)^A d\Omega\,,
\end{aligned}\ee 
in the limit $r\to\infty$.
Combining the Lorenz condition \eqref{Lorenz_r_YM} and the $r$-equation of motion \eqref{eqr_r_YM}, one obtains
\be\begin{aligned} \label{finite-charge-instr_YM}
&(D-2-k)\left(
\partial_u A_r^{(k)}+(k-1)A_u^{(k)}
\right)
-
D^i \left(
D_i A_r^{(k-1)}+(k-2)A_i^{(k-2)}
\right)\\
&=
\sum_{l+m=k}\Big(
[A_r^{(m)}-A_u^{(m)}, -l A_r^{(l)}]-[A_r^{(m)}, \partial_u A_r^{(l+1)}+F_{ur}^{(l+1)}]\\
&
+\gamma^{ij}[A_i^{(m)}, D_j A_r^{(l-1)}+F_{jr}^{(l-1)}]
\Big)
\,.
\end{aligned}\ee
We see that \eqref{Lorenz_r_YM} implies $\partial_u A^{(\frac{D-2}{2})}_r=0$ and that \eqref{finite-charge-instr_YM} reduces to 
\be
(D-2-k)\left(
\partial_u A_r^{(k)}+(k-1)A_u^{(k-1)}
\right)
=
D^i \left(
D_i A_r^{(k-1)}+(k-2)A_i^{(k-2)}
\right)
\ee
for $k<D-2$. This allows us to conclude that the color charge is finite in the limit $r\to\infty$ and that it equals
\be\label{ColorChargeLor}
\mathcal Q^A(u) = \int_{S_u} \Big( 
\partial_u A_r^{(D-2)}
+(D-3)A_u^{(D-3)}
+[A_u^{(\frac{D-2}{2})}, A_r^{(\frac{D-2}{2})}]
\Big)^A d\Omega\,,
\ee
since all other terms either integrate to zero on the $(D-2)$-sphere or vanish in the limit $r\to\infty$. 
The final expression for the color charge may be equivalently rewritten in the form 
\be
\mathcal Q^A(u) = \int_{S_u} \Big( 
A_r^{(D-3)}
+(D-4)A_u^{(D-3)}
+[A_u^{(\frac{D-2}{2})}, A_r^{(\frac{D-2}{2})}]
\Big)^A d\Omega\,,
\ee
by means of the Lorenz condition \eqref{Lorenz_r_YM}. It is amusing to note that \eqref{ColorChargeLor} reduces to the expression we obtained in Section \ref{sec:YMGlobal}, namely \eqref{ColorChargeglobale}, for the color charge upon formally setting $\mathcal A_r=0$.

Let us now investigate the dependence of $\mathcal Q^A$ on retarded time.
Recalling that $\partial_u A_r^{(\frac{D-2}{2})}=0$ by the Lorenz condition, 
\be\label{uder-Q}
\frac{d}{du}\mathcal Q^A(u) = \int_{S_u} \Big( 
\partial_u A_r^{(D-3)}
+(D-4)\partial_u A_u^{(D-3)}
+[\partial_u A_u^{(\frac{D-2}{2})}, A_r^{(\frac{D-2}{2})}]
\Big)^A d\Omega\,.
\ee 
By \eqref{finite-charge-instr_YM} evaluated for $k=D-3$, we have
\be
\partial_u A_r^{(D-3)}+(D-4)A_u^{(D-4)}
=
D^i \left(
D_i A_r^{(D-4)}+(D-5)A_i^{(D-5)}
\right)\,,
\ee 
while equation \eqref{equ_r_YM} with $k=D-3$ gives
\be\label{ddotAu}
(D-4)(\partial_u A_u^{(D-3)}-A_u^{(D-4)}) + [\partial_u A_u^{(\frac{D-2}{2})}, A_r^{(\frac{D-2}{2})}] = 
\Delta A_u^{(D-4)} + \gamma^{ij} [A_i^{(\frac{D-4}{2})}, \partial_u A_j^{(\frac{D-4}{2})}]\,.
\ee
Hence, substituting into \eqref{uder-Q} (and disregarding total divergences), we get
\be\label{GeneralColorEvolution}
\frac{d}{du}\mathcal Q^A(u) = \int_{S_u} \gamma^{ij} [A_i^{(\frac{D-4}{2})}, \partial_u A_j^{(\frac{D-4}{2})}]^A d\Omega\,.
\ee 
This formula provides the flux of the total color charge due to nonlinearities of the theory, \emph{i.e.} self-interaction effects. Note that in particular the right-hand side involves the radiation components, representing the flux of classical gluons across null infinity. We thus retrieve all the physically consistent global features of the analysis performed in \ref{sec:YMGlobal}.

%%%%%
\subsection{Yang-Mills kick memory}\label{ssec:YM_memory}
%%%%%

Starting from the radiation falloffs, we may employ the residual gauge symmetry of the theory to perform a further, recursive gauge fixing in the same spirit of Section \ref{ssec:emgaugefix}. In fact, since the nonlinear corrections to equations \eqref{induction-equations} appear to order $q=D-3$ or higher, the discussion of this gauge fixing is completely identical to that of Section \ref{ssec:emgaugefix}. This is another manifestation of the mechanism of asymptotic linearization we highlighted in Section \ref{sec:YMGlobal}.

 Moreover, in accordance with the fact that Yang-Mills theory must encompass both ordinary/linear and null/nonlinear memory, the gauge fixing stops at $q=D-4$ and cannot be performed up until $q=D-3$, as happened in the case of null electromagnetic memory. 

The resulting falloffs after this procedure are thus
\be
\mathcal A_u = \sum_{k=D-3}^\infty A_u^{(k)}r^{-k}\,,\qquad
\mathcal A_r = \sum_{k=\frac{D-2}{2}}^\infty A_r^{(k)}r^{-k}\,,\qquad
\mathcal A_i = \sum_{k=\frac{D-4}{2}}^\infty A_i^{(k)}r^{-k}
\ee
with residual symmetry parameter
\be\label{res_even_YM}
\epsilon(u,r,x^i) \, =\, c^A T^A+ \epsilon^{(D-4)}(x^i)r^{4-D}\, + \cdots\,,
\ee
where $c^A$ are constant coefficients.

A colored test particle with charge $Q = Q^A T^A$ interacts with the background Yang-Mills field by the Wong equations \cite{Wong}
\be\label{Wong}
\dot P^a = \mathrm{tr} (Q \mathcal F^{ab}) \dot x_b\,,\qquad
\dot Q +\dot x^a [\mathcal A_a, Q] =0\,.
\ee 
Focusing on the region near null infinity, a test ``quark'', initially at rest, subject to radiation between the retarded times $u_0$ and $u_1$ will therefore experience a leading-order momentum kick according to  
\be\label{velocity_YM}
P'_{i}-P_i = 
\int_{u_0}^{u_1} \mathrm{tr}(Q \mathcal F_{iu}) du 
=
-\frac{1}{r^{D-4}}\mathrm{tr}(Q {A'_i}^{(D-4)})
=
-\frac{1}{r^{D-4}}\mathrm{tr} [Q\, e^{-\epsilon^{(D-4)}}\partial_i e^{\epsilon^{(D-4)}}]\,,
\ee
where we have chosen the vacuum configuration at $u=u_0$ to be $\mathcal A_\mu =0$.
On the other hand, the color of the test quark will change, to leading order, according to 
\be
Q'-Q = - \int_{u_0}^{u_1} [\mathcal A_u, Q]\,du = - \frac{1}{r^{D-3}} \int_{u_0}^{u_1}[A_u^{(D-3)},Q]\,du\,,
\ee
where the $u$-dependence of $\mathcal A_u$ is governed, according to equation \eqref{ddotAu}, by
\be\label{ColorDelta}
\partial_u A_u^{(D-3)} = \frac{1}{D-4}\, \gamma^{ij}[A_i^{(\frac{D-4}{2})}, \partial_u A_j^{(\frac{D-4}{2})}]
\ee
after recursive gauge fixing. Equation \eqref{ColorDelta} characterizes the evolution of the charge of the quark in terms of the leading outgoing radiation terms. 

In order to better understand the dependence of this momentum kick and color rotation on the incoming radiation near null infinity, let us further analyze the equations of motion. Combining equation \eqref{Lorenz_r_YM} and \eqref{eqr_r_YM} we see that
\be\label{recrad_1}
\partial_u A_r^{(k+1)} = \mathscr D_k A_r^{(k)}
\ee
for $k<D-3$, where we have introduced the following self-adjoint differential operators on the $(D-2)$-sphere 
\be\label{operatoresfera}
\mathscr D_k = [\Delta-(D-2-k)(k-1)]/(D-2-2k)\,,
\ee  
and 
\be\label{recrad_2}
\partial_u A_r^{(D-2)} = \mathscr D_{D-3} A_r^{(D-3)} - A_u^{(D-3)} - \frac{1}{D-4}\mathcal J\,,
\ee
for $k=D-3$ in dimensions $D>4$, where we have defined
\be
\mathcal J = 
2 \gamma^{ij} [A_i^{(\frac{D-4}{2})}, D_j A_r^{(\frac{D-2}{2})}]
-2 [A_r^{(\frac{D-2}{2})}, \partial_u A_r^{(\frac{D}{2})}] \,.
\ee 
Equations \eqref{Lorenz_r_YM} and \eqref{eqr_r_YM}, evaluated for $k=D-3$ and $k=D-2$ respectively, actually imply the following constraint on $\mathcal J$,
\be \label{constr_J}
\mathcal J = -D^i [D_i A_r^{(D-3)}+(D-4)A_i^{(D-4)}]\,,
\ee
and, taking the derivative of the previous equation with respect to $u$,
\be\label{Coul-rad_1}
(D-4) \partial_u D\cdot A^{(D-4)}=-\Delta \partial_u A_r^{(D-3)}
+2 [A_r^{(\frac{D-2}{2})}, \partial_u^2 A_r^{(\frac{D}{2})}] - 2\gamma^{ij}[\partial_u A_i^{(\frac{D-4}{2})}, D_j A_r^{(\frac{D-2}{2})}]\,.
\ee
Starting from eq. \eqref{recrad_2} and employing \eqref{recrad_1} recursively, we find that
\be
\partial_u^{(\frac{D}{2})}A_r^{(D-2)}
=
\prod_{l=\frac{D-2}{2}}^{D-3}\mathscr D_l \partial_u A_r^{(\frac{D-2}{2})}
-\partial_u^{\frac{D-2}{2}} A_u^{(D-3)} - \frac{1}{D-4} \partial_u^{\frac{D-2}{2}} \mathcal J\,,
\ee
and, by equation \eqref{constr_J},
\be\begin{aligned}\label{Coul-rad_2}
\partial_u^{\frac{D-2}{2}}D\cdot A^{(D-4)} &= \partial_u^{\frac{D-2}{2}} (\partial_u A_r^{(D-2)}+A_u^{(D-3)})
\\
&-\frac{1}{D-4} \partial_u^{\frac{D-2}{2}} \Delta A_r^{(D-3)} - \prod_{l=\frac{D-2}{2}}^{(D-3)} \mathscr D_l \partial_u A_r^{(\frac{D-2}{2})}\,.
\end{aligned}\ee
Equations \eqref{Coul-rad_1} and \eqref{Coul-rad_2} encode the dependence of $A_i^{(D-4)}$, which as we have seen is responsible for the memory kick both in the linear theory and in the nonlinear theory, on the outflowing radiation, here encoded in particular in $A_i^{(\frac{D-4}{2})}$ and $A_r^{(\frac{D-2}{2})}$. In dimension $D=4$, the above equations can be cast in the form
\be
D^i\partial_u[A_i^{(0)}, A_r^{(1)}]+\gamma^{ij}[A_i^{(0)},\partial_u[A_j^{(0)},A_r^{(1)}]]=0\,,
\qquad
\partial_u D\cdot A^{(0)} = \partial_u (\partial_u A_r^{(2)}+A_u^{(1)})\,.
\ee

\subsection{Color memory}

Let us now consider a pair of color charges that are pinned in the positions $(r,\mathbf n_1)$ and $(r,\mathbf n_2)$, for large $r$. 
Analyzing the effect of the passage of Yang-Mills radiation on their states will provide the non-Abelian analog of the memory effect highlighted Section \eqref{ssec:phasemem}, namely a relative rotation between their states in color space. This provides a generalization to even dimensions of \cite{StromingerColor}. We consider colored particles in the fundamental representation and, to simplify the presentation, we  first focus on the case of the gauge group $SU(2)$.

Assuming that radiation is nontrivial only between two given retarded times $u_0$ and $u_1$, as we have seen, the gauge field satisfies
\be\label{gaugevacua1}
\mathcal A_i\big|_{u_1}-\mathcal A_i\big|_{u_0}=\frac{1}{r^{D-4}}\,e^{-\alpha}\partial_i e^{\alpha}+ \mathcal O(r^{3-D})
\ee
for a suitable $\alpha(\mathbf n)=\alpha^A(\mathbf n)T^A$, where $T^A$ are the $su(2)$ generators. We have assumed for simplicity that the pure gauge configuration before $u_0$ is the trivial one.

Let $|\psi_1\rangle$ and $|\psi_2\rangle$ be the initial states in which the colored particles are prepared. Before $u_0$, the state $|\psi_2,\mathbf n_1\rangle$ that results from the parallel transport of the second state $|\psi_2\rangle$ to the position $\mathbf n_1$ of the first the charge is
\be
|\psi_2,\mathbf n_1\rangle=|\psi_2\rangle\,,
\ee 
because $\mathcal A_\mu=0$. In particular, one can consider as initial state $|+\rangle$ or $|-\rangle$, namely the standard eigenstates of $T^3$. In order to build a color singlet state in $\mathbf n_1$ it then suffices to prepare the superposition 
\be\label{prepare-singlet}
\frac{|+,\mathbf n_1\rangle \otimes |-,\mathbf n_1\rangle-|-,\mathbf n_1\rangle \otimes |+,\mathbf n_1\rangle}{\sqrt2} =\frac{|+\rangle\otimes|-\rangle-|-\rangle\otimes|+\rangle}{\sqrt2}\,.
\ee 
We remark that, in order for the concept of singlet to be well-defined when $SU(2)$ transformations can depend on the position, it is crucial to build a singlet out of states that have been parallel-transported to the same point.

A given state $|\psi\rangle$ evolves according to the covariant conservation equation \cite{ClassicalFTRusso}\footnote{The covariant conservation equation for the color states, which in general reads $\left(\tfrac{d}{d\tau}+\dot x^a \mathcal A_a\right)|\psi\rangle=0$, implies the Wong equation \eqref{Wong} for the color charge $\tfrac{d}{d\tau}Q+[\dot x^a \mathcal A_a,Q]=0$, where $Q=Q^AT^A$ and $Q^A=\langle \psi|T^A|\psi\rangle$.}
\be
\partial_u |\psi\rangle = - \mathcal A_u |\psi\rangle
\ee
which means that, recalling $\mathcal A_u = \mathcal O(r^{3-D})$ for large $r$, the state after the passage of radiation differs with respect to its initial value by
\be\label{colour-evolution}
|\psi\rangle\big|_{u_1}-|\psi\rangle \big|_{u_0}=\mathcal O(r^{3-D})\,.
\ee
In particular, a state prepared in $|\pm\rangle$ before $u_0$ will evolve into 
$
|\pm\rangle+\mathcal O(r^{3-D})
$ 
at $u_1$, namely when radiation has died out.

Now, at $u_1$, one must also take into account that parallel transport from $\mathbf n_2$ to $\mathbf n_1$ on the sphere is defined by
\be\begin{aligned}
	|\psi_2,\mathbf n_1\rangle &= 
	\mathcal P\exp\left[-\int_{\mathbf n_2}^{\mathbf n_1}\mathcal A_i \,dx^i\right]|\psi_2\rangle \\
	&= 
	|\psi_2\rangle
	-
	\frac{1}{r^{D-4}}\int_{\mathbf n_2}^{\mathbf n_1}e^{-\alpha}\partial_i e^{\alpha}dx^i\,|\psi_2\rangle+\mathcal O(r^{3-D})\,,
\end{aligned}\ee
where $\mathcal P$ denotes path ordering and we have employed \eqref{gaugevacua1} to establish the second equality.
We conclude that the state obtained by parallel transport after radiation has passed is no longer the singlet \eqref{prepare-singlet}, but rather:
\be\begin{aligned}\label{falsesinglet}
	&\frac{|+,\mathbf n_1\rangle \otimes |-,\mathbf n_1\rangle-|-,\mathbf n_1\rangle \otimes |+,\mathbf n_1\rangle}{\sqrt2} 
	= 
	\frac{|+\rangle\otimes|-\rangle-|-\rangle\otimes|+\rangle}{\sqrt2}\\
	&-
	\frac{1}{r^{D-4}\sqrt2}
	\left[
	|+\rangle\otimes 
	\int_{\mathbf n_2}^{\mathbf n_1}e^{-\alpha}\partial_i e^{\alpha}dx^i\,|-\rangle
	-
	|-\rangle\otimes
	\int_{\mathbf n_2}^{\mathbf n_1}e^{-\alpha}\partial_i e^{\alpha}dx^i\,|+\rangle
	\right]+\mathcal O(r^{3-D})
	\,.
\end{aligned}\ee

Comparison between the expressions \eqref{prepare-singlet} and \eqref{falsesinglet} shows that the interaction of the color charges with the external Yang-Mills radiation induces a rotation of the initial state that manifests itself  to order $\mathcal O(r^{4-D})$ in the final state. In other words, a pair of particles initially prepared in a singlet will no longer be in a singlet after the passage of radiation. 
This color memory can be nontrivial provided that the function $\alpha$ is not constant, namely whenever there is also a nontrivial memory kick \eqref{memoryformulaf}. 

From a technical point of view, it should be noted that the effect of radiation on each color state is to induce a time evolution to order $\mathcal O(r^{3-D})$, according to \eqref{colour-evolution}, which allows one to disentangle it from the leading effect due to the vacuum transition \eqref{gaugevacua1} undergone by $\mathcal A_i$, which enters \eqref{falsesinglet} to order $\mathcal O(r^{4-D})$. 

A very similar derivation allows one to extend the result to more general states and gauge groups. Adopting an orthonormal basis $|n\rangle$ for the fundamental representation of $SU(N)$, we can consider the superposition
\be\label{beforenm}
\sum_{n,m} c_{n,m}|n\rangle\otimes|m\rangle
\ee
with $\sum_{n,m}|c_{n,m}|^2=1$,
prepared before $u_0$. Time evolution will induce modifications of $|n\rangle$ that appear to order $\mathcal O(r^{3-D})$ by $\mathcal A_u=\mathcal O(r^{3-D})$. On the other hand, the effect of parallel transport from $\mathbf n_2$ to $\mathbf n_1$ performed after $u_1$ gives rise to the leading-order correction 
\be
-\frac{1}{r^{D-4}}\sum_{n,m} c_{n,m}|n\rangle\otimes \int_{\mathbf n_2}^{\mathbf n_1}e^{-\alpha}\partial_i e^{\alpha} dx^i\, | m\rangle
\ee
to \eqref{beforenm},
due to the nontrivial configuration \eqref{gaugevacua1} attained by $\mathcal A_i$.

\section{More on Asymptotic Symmetries for the Maxwell Theory}
\label{sec:poly}

So far, we have focused on those symmetries directly related to the memory effects. As we saw, they act at Coulombic order and the corresponding charge \eqref{ChargeStrom2} displays a falloff $1/r^{D-4}$; therefore they do not comprise, strictly speaking, asymptotic symmetries at $\mathscr I^+$.
In this section we shall perform a more complete analysis of the asymptotic symmetry group for the Maxwell theory in higher dimensions.

As we shall see,
in even dimensions, the standard power-like ansatz is not sufficient in order to highlight the presence of infinite-dimensional asymptotic symmetries with nonvanishing charge in the Lorenz gauge, which requires instead the introduction of logarithmic terms in the asymptotic expansion.
Finiteness of the soft surface charges is also not manifest at face value and requires further specifications of the definition of the charges themselves.

%%%%%
\subsection{Polyhomogeneous expansion for $D\ge4$}
%%%%%

In order to investigate the possible existence of large gauge symmetries acting at $\mathscr I^+$, we try to solve the wave equation \eqref{residual} with the boundary condition
\be\label{boundaryfuncI}
\lim_{r\to\infty}\epsilon(u,r,\mathbf n) = \epsilon^{(0)}(u,\mathbf n)\,,
\ee 
for some nonconstant function $\epsilon^{(0)}(u,\mathbf n)$.
As it turns out, a power-law ansatz
\begin{equation}
	\epsilon = \sum_{k=0}^\infty \frac{\epsilon^{(k)}(u,\mathbf n)}{r^k}\,,
\end{equation}
effectively selects the global symmetry, namely the constant parameter, whenever $D$ is even. Indeed, considering \eqref{residual_r} with $k=0,1,\ldots,\frac{D-2}{2}$, we have
\be\begin{aligned}\label{no-symmetry-even}
(D-2)\partial_u \epsilon^{(0)}&=0\,,\\
(D-4)\partial_u \epsilon^{(1)}&=\Delta\epsilon^{(0)}\,,\\
&\,\,\,\vdots\\
2\partial_u\epsilon^{(\frac{D-4}{2})}&=\left[\Delta-\frac{D(D-6)}{4}\right]\epsilon^{(\frac{D-6}{4})}\,,\\
0&=\left[\Delta-\frac{(D-4)(D-2)}{4}\right]\epsilon^{(\frac{D-4}{2})}\,.
\end{aligned}\ee
As already observed (see Appendix \ref{app:Laplaciano}), the last equation sets $\epsilon^{(\frac{D-4}{2})}$ to zero, by the invertibility of the differential operator occurring on the right-hand side, for any even $D\ge6$ and to a constant for $D=4$. Then, the above equations recursively set to zero the other components $\epsilon^{(k)}$ all the way up to $\epsilon^{(0)}$ which need be a constant by $\Delta \epsilon^{(0)}=0$ and $\partial_u \epsilon^{(0)}=0$.

One concludes that, in even dimensions, the power-law ansatz does not allow for an enhanced asymptotic symmetry sitting at the order $1/r^0$, as it occurs instead for instance in the radial gauge \cite{StromingerQEDevenD}. In order to retrieve them, one needs a more general ansatz. We find that it is sufficient, to this purpose, to consider the following type of asymptotic expansion involving also a logarithmic dependence on $r$:
\be\label{expepsilonlog}
\epsilon(u,r,\mathbf n)=
\sum_{k=0}^\infty \frac{\epsilon^{(k)}(u,\mathbf n)}{r^k}
+
\sum_{k=\frac{D-2}{2}}^\infty\frac{\lambda^{(k)}(u,\mathbf n)\log r}{r^k}\,.
\ee
In this fashion, the last equation \eqref{no-symmetry-even} becomes modified by the presence of the logarithmic branch and yields
\be\label{matching-logpot}
2\partial_u \lambda^{(\frac{D-2}{2})}=\left[\Delta-\frac{(D-4)(D-2)}{4}\right]\epsilon^{(\frac{D-4}{2})}\,,
\ee 
an equation that determines the $u$-dependence of $\lambda^{(\frac{D-2}{2})}$. Therefore, an arbitrary $\epsilon^{(0)}(\mathbf n)$ is allowed. The need of introducing logarithms in even dimensions was also recognized in \cite{He-Mitra-photond+2,HenneauxQEDanyD}.

Indeed, the recursion relations expressing the equation $\Box\epsilon=0$ read
\be\begin{aligned}\label{expansioneq-with-logs}
(D-2-2k)\partial_u \epsilon^{(k)}
+
2\partial_u \lambda^{(k)}
&=
[\Delta +(k-1)(k-D+2)]\epsilon^{(k-1)}
+
(D-1-2k)\lambda^{(k-1)}\\
(D-2-2k)\partial_u \lambda^{(k)}
&=
[\Delta +(k-1)(k-D+2)]\lambda^{(k-1)}\,,
\end{aligned}\ee
and hence can be solved by direct integration with respect to $u$. Explicitly, for $k<\frac{D-2}{2}$, these equations reduce to the familiar expression
\be
(D-2-2k)\partial_u \epsilon^{(k)}=[\Delta+(k-1)(k-D+2)]\epsilon^{(k-1)},
\ee  
so that the solution is given by a polynomial in $u$ with angle-dependent coefficients $C_{j,k}(\mathbf n)$, with $0\le j\le k$, and is uniquely determined by specifying the integration functions $\hat \epsilon^{(k)}(\mathbf n)$:
\be
\epsilon^{(k)}(u,\mathbf n)=\sum_{l=0}^k C_{j,k}(\mathbf n) u^j\,,
\ee
with  
\be
C_{j,k}(\mathbf n) = 
\begin{cases}
	\hat \epsilon^{(k)}(\mathbf n) &\text{if }j=0\\
	\frac{1}{j!}\prod_{l=k-j+1}^k \mathscr D_l \hat \epsilon^{k-l}(\mathbf n)&\text{otherwise}\,.
\end{cases}
\ee
If $k=\frac{D-2}{2}$, equation \eqref{expansioneq-with-logs} reduces to \eqref{matching-logpot}, so that
\be
\lambda^{(\frac{D-2}{2})}(u,\mathbf n)=\frac{1}{2}\int_0^u\left[\Delta-\frac{(D-4)(D-2)}{4}\right]\epsilon^{(\frac{D-4}{2})}(u',\mathbf n)\, du'+\hat \lambda^{(\frac{D-2}{2})}(\mathbf n) \,,
\ee
with a suitable integration function, while $\epsilon^{(\frac{D-2}{2})}(u,\mathbf n)$ is unconstrained. For $k>\frac{D-2}{2}$, we can recast \eqref{expansioneq-with-logs} in the more suggestive form
\be\begin{aligned}
(D-2-2k)\partial_u \lambda^{(k)}
&=
[\Delta +(k-1)(k-D+2)]\lambda^{(k-1)}\\
(D-2-2k)\partial_u \epsilon^{(k)}
&=
[\Delta +(k-1)(k-D+2)]\epsilon^{(k-1)}
\\
&+\left[
\frac{2}{2k-D+2}(\Delta+(k-1)(k-D+2))+(D-1-2k)
\right]\lambda^{(k-1)}\,.
\end{aligned}\ee
In this way, it becomes clear that these two equations can always be solved by first finding an integral of the first equation and then substituting it in the second one where it acts as ``source'' term on the right-hand side. 

To summarize, a solution of \eqref{expansioneq-with-logs} in even $D$ is specified by assigning a set of integration functions $\hat \epsilon^{(k)}(\mathbf n)$, for $k\ge0$, and $\hat \lambda^{(k)}(\mathbf n)$, for $k\ge\frac{D-2}{2}$, together with an arbitrary function $\epsilon^{(\frac{D-2}{2})}(u,\mathbf n)$. In particular, this achieves the boundary condition \eqref{boundaryfuncI}
as $r\to\infty$, where $\epsilon^{(0)}(\mathbf n)$ is an arbitrary function on the celestial sphere.

The discussion simplifies in odd dimensions, where the left-hand side of \eqref{residual_r} never vanishes for integer $k$, and hence the recursion relations can be integrated compatibly with \eqref{boundaryfuncI} with no need of introducing a logarithmic branch.

By means of ``gauge'' transformations
\be\label{exact...}
\mathcal A_a \mapsto \mathcal A_a + \partial_a \epsilon
\ee 
identified by the gauge parameters thus obtained, we can then act on solutions to Maxwell's equations in Lorenz gauge characterized by the radiation falloffs \eqref{radiationfall} and generate a wider solution space. In particular, this type of solutions will generically exhibit the following asymptotic behavior   as $r\to\infty$:
\be\label{dimindep-fall}
\mathcal A_u = \mathcal O(r^{-1})\,,\qquad
\mathcal A_r = \mathcal O(r^{-2})\,,\qquad
\mathcal A_i = \mathcal O(r^{0})\,,
\ee 
independently of the dimension $D$ of the spacetime, in view of \eqref{expepsilonlog}, provided $D\ge5$, while
\be\label{diminde4}
\mathcal A_u = \mathcal O(r^{-1}\log r)\,,\qquad
\mathcal A_r = \mathcal O(r^{-1})\,,\qquad
\mathcal A_i = \mathcal O(r^{0})\,,
\ee 
in dimension $D=4$.

Nonetheless, such solutions will still retain all the desirable physical properties of those characterized by \eqref{radiationfall} (also recently discussed in \cite{SatishchandranWald}), such as the finiteness of the energy flux at any given retarded time. This is due to the fact that they differ from this more restrictive solution space only by a transformation of the type \eqref{exact...}, which does not alter the electromagnetic field, nor, by definition, any other physical observable quantity. Now, from this point of view only, it would be possible to perform \eqref{exact...} with parameters $\epsilon$ that obey arbitrary asymptotics near $\mathscr I^+$; however, as we shall see explicitly below, the ones satisfying \eqref{boundaryfuncI} are precisely those that will give rise to finite and nonvanishing asymptotic surface charges and are therefore the true asymptotic symmetries of the theory.

Going back to equation \eqref{residual} with the boundary condition \eqref{boundaryfuncI}, let us note that it is actually possible to provide a general solution thereof in the following more compact form in any even dimension $D$:
\be
\epsilon(x) = \frac{\Gamma(D-2)}{\pi^{\frac{D-2}{2}}\Gamma(\frac{D-2}{2})} \text{Re} \oint \frac{(-x^2)^{\frac{D-2}{2}}}{(-2x\cdot q+i\varepsilon)^{D-2}}\,\epsilon^{(0)}(\mathbf q)\, d\Omega(\mathbf q)\,,
\ee
where $q= (1,\mathbf q)$ and the limit $\varepsilon\to0^+$ is understood. The introduction of this small imaginary part is needed in order to avoid the singularities occurring in the angular integration for $|t|<|\mathbf x|$, namely outside the light-cone. Indeed, it is straightforward to verify that
\be
\Box \frac{(-x^2)^{\frac{D-2}{2}}}{(-2x\cdot q)^{D-2}} = 0\,,
\ee
while, aligning $\mathbf n$ along the $(D-1)$th direction, we have
\be\begin{aligned}
	&\mathrm{Re}
	\oint \frac{(-x^2)^{\frac{D-2}{2}}}{(-2x\cdot q+i\varepsilon)^{D-2}}\,\epsilon^{(0)}(\mathbf q)\, d\Omega(\mathbf q)
	\\
	&=\mathrm{Re}\int_{0}^\pi d\theta (\sin\theta)^{D-3}\oint d\Omega'(\mathbf q') \frac{2^{2-D}[u(u+2r)]^{\frac{D-2}{2}}}{[u+r(1-\cos\theta)+i\varepsilon]^{D-2}}\epsilon^{(0)}(\mathbf q' \sin\theta, \cos\theta)\,,
\end{aligned}
\ee
where $d\Omega'(\mathbf q')$ denotes the integral measure on the $(D-3)$-sphere, and letting $\tau=r(1-\cos\theta)/u$, for $u\neq0$, the previous expression becomes
\be\label{arccos_asymptD}
\text{Re}
\int_0^{2r/u}\frac{u \,d\tau}{r}\,\frac{(1+\frac{2r}{u})^{\frac{D-2}{2}}(
	\tfrac{2u\tau}{r}-\tfrac{u^2\tau^2}{r^2})^{\frac{D-4}{2}}}{2^{D-2}(1+\tau+i\varepsilon)^{D-2}}
\oint d\Omega'(\mathbf q')\, 
\epsilon^{(0)}\left(
\mathbf q' \sqrt{
	\tfrac{2u\tau}{r}-\tfrac{u^2\tau^2}{r^2}}
,
1-\tfrac{u \tau}{r}
\right)\,,
\ee
which, as $r\to\infty$, tends to 
\be
\frac{1}{2}\,
\mathrm{Re} \int_0^{u\cdot\infty} \frac{\tau^{\frac{D-4}{2}}d\tau}{(1+\tau + i\varepsilon)^{D-2}}
\oint d\Omega'(\mathbf q') \epsilon^{(0)}(\mathbf n)=\frac{\pi^{\frac{D-2}{2}}\Gamma(\frac{D-2}{2})}{\Gamma({D-2})}\,\epsilon^{(0)}(\mathbf n)\,.
\ee
This solution is indeed compatible with the asymptotic expansion \eqref{expepsilonlog} and generalizes the expression given in \cite{Hirai:2018ijc}.

\subsection{Finite and nonvanishing charges on $\mathscr I$}\label{ssec:Finitecharges}
As is usually the case in the presence of radiation, the surface charge associated to the symmetry \eqref{expepsilonlog}, namely
\be
\mathcal Q_\epsilon(u) = \lim_{r\to\infty} \oint_{S_u}\mathcal F_{ur}(u,r,\mathbf n)\epsilon(u,r,\mathbf n)r^{D-2}\,d\Omega(\mathbf n)\,,
\ee
is formally ill-defined, because the right-hand side contains terms of the type \emph{e.g.}
\be
r^{\frac{D-4}{2}}\oint_{S_u}F_{ur}^{(\frac{D}{2})}\epsilon^{(0)}d\Omega\,,
\ee
which do not vanish, even after imposing the equations of motion, precisely due to the presence of an \emph{arbitrary} parameter $\epsilon^{(0)}(\mathbf n)$.

Such difficulties are absent in the case of the \emph{global} charge $\epsilon=1$, namely the electric charge, because the equations of motion always ensure that the surface charge is actually independent of the specific surface on which it is calculated, as long as no sources are crossed.
In fact, this feature is shared by all global symmetries associated to linearized gauge theories \cite{Barnich-Brandt}.

It should be noted, however, that the above difficulties only arise if one attempts to calculate the surface charge, in the case of the general transformation \eqref{expepsilonlog}, by integrating over a sphere at a given retarded time and radius $r$ and then lets $r$ tend to infinity. Instead, the calculation of the charge on a Cauchy surface still gives a well-defined result \cite{HenneauxQEDanyD}. In particular, under the simplifying assumption that the electromagnetic field due to radiation vanish for $u<u_0$ in a neighborhood  of future null infinity so that $S_{u}$ is the boundary of a Cauchy surface for $u<u_0$, the calculation of the surface charge then yields
\be
\mathcal Q_\epsilon(u) 
= \oint_{S_u} F_{ur}^{(D-2)}\epsilon^{(0)}\,d\Omega
=\oint_{S_u}
\big[A_r^{(D-3)}+(D-4)A_u^{(D-3)}\big]\epsilon^{(0)}\,d\Omega\,,
\ee
for \emph{fixed} $u<u_0$, since indeed all radiation components $F_{ur}^{(k)}$, for $k<D-2$, vanish for a stationary solution (see the discussion in Section \ref{ssec:em_memory}). Letting then $u$ approach $-\infty$, one has 
\be
\mathcal Q_\epsilon (-\infty)
= \oint_{\mathscr I^+_-} F_{ur}^{(D-2)}\epsilon^{(0)}\,d\Omega
=\oint_{\mathscr I^{+}_-}
\big[ A_r^{(D-3)}+(D-4)A_u^{(D-3)}\big]\epsilon^{(0)}d\Omega\,.
\ee
For $u<u_0$, the quantity $\mathcal Q_\epsilon(u)$ must match the analogous surface integral calculated at spatial infinity because in both cases the Noether two form is integrated over the boundary of a Cauchy surface, in view of the requirement that no radiation be present in a neighborhood of $\mathscr I^+$ for $u<u_0$.

The evolution of $\mathcal Q_\epsilon(u)$ \emph{along} $\mathscr I^+$ can be then defined, even in the presence of radiation, by the equations of motion. Indeed, the Maxwell equation $\nabla\cdot \mathcal F_r=0$ gives
\be
\Big(\partial_r + \frac{D-2}{r}\Big) \mathcal F_{ur} = \frac{1}{r^2}D\cdot \mathcal F_r\,,
\ee
while $\nabla\cdot \mathcal F_u=0$ reads
\be
\partial_u \mathcal F_{ur}=\Big(\partial_r + \frac{D-2}{r}\Big) \mathcal F_{ur}+\frac{1}{r^2}D\cdot \mathcal F_u 
\implies
\partial_u \mathcal F_{ur} = \frac{1}{r^2}D\cdot(\mathcal F_u+ \mathcal F_r)\,.
\ee 
and hence
\be
\frac{d}{du}\,\mathcal Q_\epsilon (u) = \oint_{S_u} D\cdot(F_u^{(D-4)}-F_r^{(D-4)})\epsilon^{(0)}\,d\Omega\,.
\ee
Analogous considerations allow one to introduce well-defined surface charges evaluated on $\mathscr I^-$.

From the perspective of the analysis performed in Section \ref{sec:memories}, it is possible to explicitly perform the calculation of soft charges according to the above strategy. Restricting to the case of a massive charge in dimension $D=6$ that starts moving at $t=0$, the integral of $\mathcal F_{ur} \epsilon^{(0)}$ on a sphere at fixed retarded time $u$ and radius $r$ yields 
\be\begin{aligned}
	\mathcal Q_\epsilon(r,u) &= 
	r\, \frac{\delta(u)}{8\pi^2}\oint \frac{\mathbf n \cdot \mathbf v(3 \mathbf n \cdot \mathbf v - 4)-\mathbf v^2}{(1-\mathbf n \cdot \mathbf v)^2}\,\epsilon^{(0)}(\mathbf n) \,d\Omega\\
	&+\frac{3}{8\pi^2}\oint \left[\theta(-u)+\theta(u)\frac{(1-\mathbf v^2)^2}{(1-\mathbf n \cdot \mathbf v)^4}\right]\,\epsilon^{(0)}(\mathbf n) \,d\Omega\,.
\end{aligned}\ee
Except for the case of the electric charge $\epsilon^{(0)}=1$, where the integral in the first line vanishes identically, the limit of this surface charge as $r\to\infty$ is ill-defined in the presence of radiation, namely on the forward light-cone $u=0$, due to the appearance of a linear divergence. However, the charge is well-defined on $\mathscr I^+$ before and after the passage of radiation, $|u|>\varepsilon$, and reads
\be
\mathcal Q_\epsilon(u)=\frac{3}{8\pi^2}\oint \left[\theta(-u)+\theta(u)\frac{(1-\mathbf v^2)^2}{(1-\mathbf n \cdot \mathbf v)^4}\right]\,\epsilon^{(0)}(\mathbf n) \,d\Omega\,.
\ee
For $\epsilon^{(0)}=1$, this quantity reduces to the (constant) electric charge $Q=1$, while for more general parameters $\epsilon^{(0)}$, the soft charge exhibits a jump discontinuity at $u=0$, which measures the fact that the particle is no longer static for $u>0$ in a manner akin to the memory effect itself.

Performing instead the limit $r\to\infty$ at fixed time $t$, it is also possible to verify the matching between the surface charge evaluated at null infinity before the onset of radiation, for $u<0$ (or, equivalently, at $\mathscr I^+_-$), and the Hamiltonian charge $\mathcal H_\epsilon(t)$, obtained by integrating on a slice at fixed time $t$. Indeed, taking \eqref{exactscalarD=6} into account and writing the result in terms of polar coordinates $t$, $r$ and $\mathbf n$, we have, for the scalar field,
\be\begin{aligned}\label{exactscalarD=6polar}
	8\pi^2 \varphi &= \frac{\delta(t-r)}{\gamma(\mathbf v)(1-\mathbf n\cdot\mathbf v)r^2}+\frac{\theta(t-r)}{\gamma(\mathbf v)^3(1-\mathbf n\cdot\mathbf v)^3r^3}\,\Delta(t-r,r)^{-3/2}\\
	&-\frac{\delta(t-r)}{r^2}+\frac{\theta(r-t)}{r^3}\,.
\end{aligned}\ee 
The corresponding electromagnetic potential is given by $\mathcal A^\mu=(\mathcal A^0,\mathbf A)=-\gamma(\mathbf v)(1,\mathbf v)\varphi$, for $t>r$, and $A^\mu=(\varphi,0)$, for $t<r$. The radial component of the electric field then yields $\mathcal F_{tr}=3 r^{-4}$ as $r\to\infty$ for fixed $t$
and hence
\be
\mathcal H_\epsilon (t)= \frac{3}{8\pi^2}\oint \epsilon^{(0)}(\mathbf n) \,d\Omega = \mathcal Q_\epsilon(u<0)\,.
\ee

Similar arguments showing the finiteness of the Hamiltonian charge in higher-dimensions have been given, in the case of linearized spin-two in retarded Bondi gauge, in \cite{Aggarwal2018}, while a renormalization procedure has been recently proposed, for Maxwell theory in the radial gauge, in \cite{FreidelHopfmuellerRiello}.

\subsection{Soft photon theorem}\label{ssec:Weinberg}

The surface charge associated to \eqref{expepsilonlog}, evaluated at the past boundary $\mathscr I^+_-$ of $\mathscr I^+$ reads
\be\label{Strominger1Charge}
\mathcal Q_\epsilon = \int_{\mathscr I^+_-}
\left(
\partial_u A^{(D-2)}_r+(D-3)A_u^{(D-3)}
\right)
\epsilon^{(0)}d\Omega\,,
\ee
where we have taken into account the absence of radiation terms for $u\to-\infty$. Recasting this as an integral over the whole $\mathscr I^+$, and assuming no contribution comes from $\mathscr I^+_+$, which is the case if we assume that there are no stable massive charges in the theory,
\be\label{Strominger-charge}
\mathcal Q_\epsilon= -\frac{1}{r^{D-4}}\int_{\mathscr I^+} \partial_u^2 A_r^{(D-2)} \epsilon^{(0)} du\, d\Omega\,,
\ee 
where we have used the fact that $A_u^{(D-3)}$ is independent of $u$ on shell (thanks to the recursive gauge-fixing).
We would like to express \eqref{Strominger-charge} in terms of the leading radiation field, which, as we shall see below, indeed contains the creation and annihilation operator of asymptotic photons.
To this end, we first combine \eqref{eqr_r} and \eqref{Lorenz_r} and obtain
\be
D\cdot A^{(k-1)} = \frac{\Delta-(D-2-k)(D-3-k)}{D-2-2k}A_r^{(k)}+(D-3-k)A_u^{(k)}\,;
\ee
employing \eqref{equ_r} as well,
\be
\partial_u A_r^{(k+1)} = \mathscr D_k A_r^{(k)} - A_u^{(k)}\,,
\ee
where the operator $\mathscr D_k$ was introduced in \eqref{operatoresfera}.
 
Employing this relation recursively, we find
\be\label{recursive-u}
\partial_u^{D/2} A_r^{(D-2)} = \prod_{l=D/2}^{D-3} \mathscr D_l\,\, \partial_u D\cdot A^{(\frac{D-4}{2})}\,,
\ee
where we have used \eqref{eqr_r} to deduce $\partial_u A_r^{(\frac{D}{2})} = D\cdot A^{(\frac{D-4}{2})}$. In the above writing, we adopt the convention that for $D=4$ the product $\prod_l \mathscr D_l$ (which in this case has a formally ill-defined range) reduces to the identity. We can then use \eqref{recursive-u} to recast \eqref{Strominger-charge} as
\be
\mathcal Q_\epsilon = -\frac{1}{r^{D-4}}\int_{-\infty}^{+\infty}\left(\int_{-\infty}^u du\right)^{D/2-2} \partial_u D\cdot A^{(\frac{D-4}{2})} \prod_{l=D/2}^{D-3}\mathscr D_l \epsilon^{(0)}du\,d\Omega\,. 
\ee
On the other hand, the asymptotic expansion of the free electromagnetic field operator, expressed in terms of creation and annihilation operators, yields, to leading order,
\be
\mathcal A_i(u,r,x^k) = \frac{i^{1-D/2}}{8\pi^{2}r^{(D-4)/2}} \int_0^{+\infty} \left(\frac{\omega}{2\pi}\right)^{(D-4)/2}e^{-i\omega u} \epsilon^\sigma_i (\hat x) a_\sigma(\omega\hat x)\,d\omega+h.c.\,,
\ee
where
$\hat x^\mu = x^\mu/r$, while $\epsilon^\sigma$ are polarization tensors for the $D-2$ propagating helicities. This formula provides thus an explicit expression for $A_i^{(\frac{D-4}{2})}$ and hence allows us to make explicit the relation between the charge $\mathcal Q_\epsilon$ and the soft photon creation and annihilation operators as follows (we employ the prescription $\int_{-\infty}^{+\infty}du \int_{0}^{+\infty}d\omega\, e^{i\omega u} f(\omega)=f(0)/2$) 
\be
\mathcal Q_\epsilon = \frac{1}{8(2\pi)^{\frac{D-2}{2}} r^{D-4}} \lim_{\omega\to0^+} \int_{S^{D-2}}
D^i[\epsilon_i^\sigma(\hat x) \omega a_\sigma(\omega\hat x) + h.c.] 
\prod_{l=D/2}^{D-3} \mathscr D_l\,\, \epsilon^{(0)}(\hat x) d\Omega(\hat x)\,.
\ee
Assuming that the charge $\mathcal Q$, together with its counterpart at $\mathscr I^-$, generates the residual symmetry $\delta \psi(u,r,\hat x) = i\epsilon^{(0)}(\hat x)+\cdots$ in a canonical way, and using suitable matching and crossing symmetry conditions, we have the Ward identity
\be\begin{aligned}\label{Ward-identity-per-Weinberg}
&\frac{1}{2(2\pi)^{(D-2)/2}} \int_{S^{D-2}}  \epsilon_i^\sigma(\hat x) \lim_{\omega\to0^+} D^i\langle\text{out}|\omega a_\sigma(\omega\hat x) S |\text{in}\rangle  \prod_{l=D/2}^{D-3} \mathscr D_l\,\, \epsilon^{(0)}(\hat x)\, d\Omega(\hat x) \\ 
&= 
\sum_{n} e_n \epsilon^{(0)}(\hat x_n) \langle\text{out}| S |\text{in}\rangle,
\end{aligned}\ee
where the sum on the right-hand side extends to all charged external particles in the amplitude and $e_n$ is the electric charge of the $n$th particle (taking into account with a suitable sign the fact that the particle is outgoing, resp. incoming).
Notably, the left-hand side contains exactly the combination $\mathcal P[\,\cdot\,] = \lim_{\omega\to 0^+}[\omega\,\cdot\,]$ that selects the pole in the amplitude with the soft insertion.

On the other hand, the Weinberg theorem for an amplitude involving external massless particles with momenta $p_n=E_n(1,\hat x_n)$ and a soft photon emitted with helicity $\sigma$ pointing along the $\hat n$ direction on the celestial sphere (see Chapter \ref{chap:obs}) reads
\be\label{Weinberg-massless}
\lim_{\omega\to0^+} \langle \text{out}| \omega a_{\sigma}(\omega \hat n)  S |\text{in}\rangle = \sum_n \frac{e_n \epsilon^\sigma(\hat n)^\ast\cdot p_n}{p_n\cdot(1,\hat n)}
\langle \text{out}|  S |\text{in} \rangle\,.
\ee
Multiplying this relation by $\epsilon^\sigma_i(\hat n)$ and summing over $\sigma$, we see that this is equivalent to
\be\label{identity_i}
\epsilon^{\sigma}_i(\hat n) 
\lim_{\omega\to0^+}\langle \text{out}|\omega a_\sigma(\omega \hat n) S |\text{in}\rangle
=\sum_{n}e_n D_i \alpha(\hat x_n, \hat n) \langle \text{out}| S |\text{in}\rangle\,,
\ee
where we have used the completeness relation for polarization vectors and defined a function
\be
\alpha(\hat x, \hat n) = \log(1-\hat x \cdot \hat n)\,.
\ee
This function $\alpha$ satisfies
the following identity (see \cite{StromingerQEDevenD})
\be\label{delta-identity}
\frac{1}{2(2\pi)^{(D-2)/2}}\prod_{l=D/2}^{D-3}\mathscr D_l\,\, \Delta \alpha(\hat x, \hat n) = \delta(\hat x, \hat n)\,,
\ee
where $\hat x$ is here a treated as a constant vector on the unit sphere and
$\delta(\hat x, \hat n)$ is the invariant delta function on the $(D-2)$-sphere. Now, acting with the differential operator 
\be
\frac{1}{2(2\pi)^{(D-2)/2}}\prod_{l=D/2}^{D-3}\mathscr D_l\,\cdot D^i
\ee
on equation \eqref{identity_i}, multiplying by an arbitrary $\epsilon^{(D-4)}(\hat x)$ and integrating over the unit sphere then allows one to retrieve the Ward identity \eqref{Ward-identity-per-Weinberg}, thanks to the relation \eqref{delta-identity}. This proves that the Weinberg factorization implies the existence of the asymptotic symmetry Ward identities. 

Remarkably, the charge \eqref{ChargeStrom2} associated to the symmetry \eqref{res_even}, which is responsible for the memory effect, formally differs from \eqref{Strominger1Charge} only by a factor of $1/r^{D-4}$ (other than by the substitution $\epsilon^{(0)}\leftrightarrow\epsilon^{(D-4)}$), which makes it vanish on $\mathscr I^+$. However, the corresponding symmetry transformation of the matter fields would be $\delta \psi(u,r,\hat x) = i\epsilon^{(D-4)}(\hat x)/r^{D-4}+\cdots$, and hence would give rise to Ward identities completely equivalent to \eqref{Ward-identity-per-Weinberg}, with the factors of $1/r^{D-4}$ canceling each other on the two sides. This indicates that both the large gauge symmetry \eqref{expansioneq-with-logs} and residual symmetry \eqref{res_even}, acting at Coulombic order, can be seen equally as consequences of the validity of Weinberg's soft theorem. This is also reflected in the observation that the Fourier transform of the soft factor occurring in Weinberg's theorem is strictly related to the memory formulas \cite{Mao_EvenD}.

%%%%%%%%%%%%%%%%%%%%%%%%%%%%%%%%%%%%%%

\part{Higher Spins}

%%%%%%%%%%%%%%%%%%%%%%%%%%%%%%%%%%%%%%
\chapter{Scalar Soft Theorems and Two-Form Asymptotic Symmetries}\label{chap:scalar2form}

We begin the part on higher spins with a chapter devoted to the scalar case. While this may seem contradictory, one can be justified to do so on account of the structural differences exhibited by this situation with respect to spin-one and spin-two theories, which we have dealt with in the previous chapters. Such differences indeed allow one to include the scalar case in the framework of ``exotic'' instances of asymptotic behavior.

A puzzling feature of theories containing massless scalar particles, as we will briefly review in Section \ref{sec:scalrtheorems}, is that they exhibit a nontrivial asymptotic structure at null infinity, and in particular allow for the definition of soft charges that account for factorization theorems, while on the other hand lacking any underlying gauge symmetry of which such charges could be a manifestation. In Section \ref{sec:twoformss} we propose a way in which this puzzle can be addressed by appealing to the duality that, in four spacetime dimensions, links the scalar field to a two-form gauge theory \cite{twoform-io}. While consistent as far as the matching between the scalar and two-form solution spaces is concerned, our choice of working in the radial gauge does not allow for a fully satisfactory matching between the scalar soft charges and nonvanishing two-form charges on $\mathscr I$, as opposed to a similar approach based on the Lorenz gauge \cite{Campigliascoop} which we discuss in Section \ref{sec:scoop}.

The reason for this shortcoming of the radial gauge can be traced back to the fact that, in order to reach such a gauge, one may need perform a large gauge transformation. As a consequence, this gauge freezes the asymptotic symmetries and effectively hides their presence. In this respect, such a failure to highlight the correct asymptotic structure of the two-form theory appears to be structurally different compared to the difficulties arising in calculation of asymptotic symmetries in higher dimensions for spin-one theories, which we discussed in the previous chapter. Indeed the latter can actually be ascribed to the adoption of too-stringent falloff conditions, namely radiation falloffs for $D>4$, while the former seems an intrinsic problem of the gauge-fixing condition, \emph{i.e.} the radial gauge. 

This phenomenon begs for a deeper understanding of the gauge dependence of the nature of asymptotic symmetries or at least for a systematic criterion enabling one to know \emph{a priori} whether a given gauge-fixing is acceptable from the point of view of the asymptotic analysis (see \cite{Riello} for a recent proposal in this direction).

\section{A Soft Theorem for Scalar Quanta}\label{sec:scalrtheorems}

The derivation of soft theorems carried out in Chapter \ref{chap:obs} admits a rather direct extension to the case of a massless scalar mediator \cite{CampigliaCoitoMizera, Campiglia_scalars,Hamada:2017atr}. Indeed, a scattering amplitude involving, say, the emission of a very soft scalar will receive its leading contributions from the diagrams in which the soft particle is attached to an external line, schematically:

\begin{fmffile}{MyOtherDiagram}
	\begin{align*}
	\begin{gathered}
	\begin{fmfgraph*}(90, 65)
	\fmfleft{i1,d1,i2,i3}
	\fmfv{label={\color{blue}\vdots}, label.dist=-4mm}{d1}
	\fmfright{o1,d2,o2,o4,o3}
	\fmfv{label={\color{blue}\vdots}, label.dist=-4mm}{d2}
	\fmf{fermion, fore=blue}{i1,v1}
	\fmf{fermion, fore=blue}{i2,v1}
	\fmf{fermion, fore=blue}{i3,v1}
	\fmf{fermion, fore=blue}{v1,o1}
	\fmf{fermion, fore=blue}{v1,o2}
	\fmf{fermion, fore=blue}{v1,o3}
	\fmf{dashes, fore=red, tension=0}{v1,o4}
	%\fmfblob{.35w}{v1}
	\fmfv{d.sh=circle,d.f=empty,d.si=.25w,b=(.5,,0,,1)}{v1}
	\end{fmfgraph*}
	\end{gathered}\ 
	=
	\
	\begin{gathered}
	\begin{fmfgraph*}(90, 65)
	\fmfleft{i1,d1,i2,i3}
	\fmfv{label={\color{blue}\vdots}, label.dist=-4mm}{d1}
	\fmfright{o1,d2,o2,u3,o3}
	\fmfv{label={\color{blue}\vdots}, label.dist=-4mm}{d2}
	\fmf{fermion, fore=blue}{i1,v1}
	\fmf{fermion, fore=blue}{i2,v1}
	\fmf{fermion, fore=blue}{i3,v1}
	\fmf{fermion, fore=blue}{v1,o1}
	\fmf{fermion, fore=blue}{v1,u2}
	\fmf{dashes, fore=red, tension=0}{u2,u3}
	\fmf{fermion, fore=blue}{u2,o2}
	\fmf{fermion, fore=blue}{v1,o3}
	%\fmfblob{.35w}{v1}
	\fmfv{d.sh=circle,d.f=empty,d.si=.25w,b=(.5,,0,,1)}{v1}
	\end{fmfgraph*}
	\end{gathered}
	\ +\ \cdots
	\end{align*}
\end{fmffile} 
Such contributions take a factorized form analogous to Weinberg's leading soft theorem. Explicitly, adopting the same notation employed in \eqref{WeinbergsWeinberg},
\be
S_{\beta \alpha}^{0}(\mathbf q) \approx \frac{1}{(2\pi)^{3/2}\sqrt{2|\mathbf q|}}\left[ \sum_n \eta_n g_n \frac{1}{p_n\cdot q}\right]S_{\beta\alpha}\,,
\ee
where $g_n$ denotes the coupling of the $n$th particle with the massless spin-zero mediator.
To  leading order and at tree level, it is also possible to recast this identity for the emission of a scalar particle in the form of a  Ward identity for the corresponding $S$ matrix
\be\label{Id_CCM}
Q^+ S - S Q^-=0\,,
\ee
with $Q^\pm$ suitable operators expressed in terms of creation and annihilation operators of external physical quanta. As usual, $Q^\pm$ can be split into their hard parts $Q^\pm_h$ and soft parts $Q^\pm_s$. 

Adopting the usual retarded Bondi coordinates in four-dimensional Minkowski space, one can perform the standard asymptotic expansion for the free scalar field equation $\Box\varphi=0$, 
\be\label{wavescalar4}
2\Big(\partial_r+\frac{1}{r}\Big)\partial_u\varphi= \Big(\partial_r+\frac{2}{r}\Big)\partial_r\varphi + \frac{1}{r^2}\Delta \varphi\,,
\ee
and verify that the falloff condition 
\be
\varphi(u, r, z, \bar z)=b(u, z, \bar z)/r + o(\tfrac{1}{r})
\ee
satisfies \eqref{wavescalar4} to leading order, thereby correctly highlighting the free propagating mode $b(u,z,\bar z)$.
The soft ``scalar charges'' can then be expressed  in terms of the massless scalar radiative mode as follows
\be
Q^+_s= \int_{S^2} b(u, z, \bar z) \Lambda(z, \bar z)\gamma_{z\bar z}dz d\bar z\,, 
\ee
where $\Lambda(z,\bar z)$ is an arbitrary function of the two angular coordinates $z$ and $\bar z$ on the unit sphere while $\gamma_{z\bar z}$ is the corresponding metric.

\section{Soft Scalar and Dual Two-Form Charges}\label{sec:twoformss}
 
The interpretation of the above Ward identity in terms of an underlying symmetry, however, remained elusive for some time. Indeed, differently from the analogous results holding for the case of soft particles with spin $s \geq 1$, which are the subject of chapters \ref{chap:basics}, \ref{chap:obs}, \ref{chap:spin1higherD} and \ref{chap:HSP}, for the case of soft scalars it is not clear a priori what symmetry, if any, could be underlying the conservation of the corresponding charges. 

This puzzle was investigated in \cite{Campigliascoop, twoform-io}, where it was conjectured that the scalar soft charges $Q^\pm_s$, in four spacetime dimensions, could be identified with the Noether charges associated to the large gauge symmetries of a two-form gauge field, to be interpreted as propagating the same massless scalar degree of freedom, in a dual picture. According to this scenario \eqref{Id_CCM} would be naturally interpreted as the Ward identity arising from the large gauge symmetry of a two-form field. 

While this idea turned out to be essentially correct, the different gauges employed in \cite{Campigliascoop} and \cite{twoform-io} determined relevant differences in the asymptotic behavior of the corresponding charges, on which we shall comment at the end of section \ref{sec:scoop}. Before detailing the procedure followed and the results obtained, let us add a few remarks on the general perspective and the possible lessons that may be learned from this exploration.

The possibility to analyze the relation between asymptotic symmetries and soft theorems from the perspective of dual theories may be worth exploring in a number of additional contexts. While already approached to some extent for the case of asymptotic $U(1)$ symmetries \cite{Strominger-dual, Shahin-dual} and supertranslations \cite{GodazgarDual1, GodazgarDual2, PorratiDual}, in $D=4$, it would be interesting to reconsider from this vantage point the issue of higher-dimensional asymptotic symmetries for gravity and for higher spins. Some symmetries may be easier to identify in a given dual description rather than in other, on-shell equivalent, pictures, a possibility that is conceivable on account of the typically nonlocal relation that connects two dual covariant descriptions of the same degrees of freedom. On the other hand, one ought to keep in mind that dualities typically only hold at the free level, and hence may only hold between asymptotic fields.

Coming back to soft scalars, let us also observe that, while the main focus here is on the four-dimensional case, the very existence of analogous duality relations between free massless scalars and $(D-2)$-forms in $D$ dimensions provides natural candidate explanations for the corresponding soft scalar charges identified in any even $D$ \cite{Campiglia_scalars}, while also possibly indicating the existence of analogous results in odd dimensions as well.

Let us also mention that factorization theorems involving soft scalars can actually be derived, according to the strategy described at the beginning of Chapter \ref{chap:obs}, in a specific class of models characterized by spontaneously broken scale invariance \cite{DiVecchia:2015jaq,DiVecchia:2017uqn}, where the soft scalar is interpreted as the corresponding Goldstone boson.

%%%%%
\subsection{Asymptotic symmetries for two-form gauge fields} \label{ssec:2formasympt}
%%%%%

We consider the gauge field described by a two-form
\be
B=\frac{1}{2}B_{ab}dx^a \wedge dx^b\,,
\ee 
\emph{i.e.} an antisymmetric rank-two tensor $B_{a b}= - B_{ba}$, subject to the reducible gauge transformation $\delta B = d\epsilon$, in components $\epsilon=\epsilon_a dx^a$ and
\be
\d B_{ab}  = \pr_{a}  \e_{b}  -  \pr_{b}  \e_{a}\,.
\ee
In its turn, the one-form parameter $\e$ is subject to the gauge-for-gauge symmetry $\delta \epsilon = df$, \emph{i.e.}  $\d\e_{a} \, = \, \pr_{ a} f$, where $f$ is a scalar parameter. The gauge-invariant field strength is $H=dB$, equivalently 
\be
H_{a b c}  =  \pr_{a}  B_{b c}  +  \pr_{ c}  B_{a b} +  \pr_{ b}  B_{c a} \,,
\ee
while Lagrangian and equations of motion are given by
\be
\cL = - \fr{1}{6}  H_{ a b c}  H^{ a b c} \,,  \qquad  \nabla_{ a}  H^{ a b c}  =  0 \,,
\ee
where the latter, in components of $B_{ab}$, take the form
\be \label{eomB}
\Box B_{ab} +  \nabla_{a} \nabla^c  B_{ bc} - \nabla_b  \nabla^c  B_{ ac} = 0\,.
\ee

Our goal in this section is to investigate the asymptotic symmetries of this theory, much in the spirit of what we have done for the Maxwell theory and for (linearized) gravity. The asymptotic analysis for $p$-form fields was actually already explored in \cite{Shahinpforms}, but in the ``critical dimension'' in which radiation order $\mathcal O(r^{\frac{2-D}{2}})$ and Coulombic order $\mathcal O(r^{2-D+p})$ coincide, namely $D=2(p+1)$ , while here we need to investigate a different situation where $D=4$ and $p=2$.

Adopting the standard retarded coordinates, we start by exploiting the gauge-for-gauge symmetry to set $\e_r=0$, thus fixing the the scalar parameter $f$, up to an $r$-independent but otherwise arbitrary function $f (u, z, \bar{z})$. Then, we employ the gauge transformations 
\be
\delta B_{ru} = \partial_r \e_u \,,\qquad \delta B_{ri} = \partial_r \e_i\,,
\ee
to reach the ``radial gauge'' 
\be
B_{ru}=0=B_{ri}\, ,
\ee
where $x^i$, with $i=1,2$, stand for $z$, $\bar z$. This leaves a residual gauge freedom with parameters $\e_u(u,z,\bar z)$ and $\e_i(u,z,\bar z)$, and the gauge-for-gauge redundancy $f  (u,z,\bar z)$. We may then further exploit the $u$-dependence of $\e (u,z,\bar z)$ to set $\e_u (u,z,\bar z)=0$. The result of this gauge-fixing strategy is the following: one is left with the gauge-field components
\be
B_{ui}(u,r,z,\bar z)\,,\qquad
B_{z \bar z}(u,r,z,\bar z)\,,
\ee
while still keeping the residual gauge parameters
\be\label{transform-gauge}
\e_i  (z, \bar z)\,,
\ee
together with the residual gauge-for-gauge symmetry encoded in
\be
f  (z,\bar z)\,.
\ee
Expanding the equations \eqref{eomB} in the above gauge yields
\begin{align}
& \partial_r D^j B_{ju} \, =\, 0\,,\\
& \partial_r^2 B_{ui}+\frac{1}{r^2}\partial_r D^{j} B_{ij} \, =\, 0\,,\\
& \partial_u \partial_r B_{ui} \, - \,  \frac{1}{r^2}\partial_u D^j B_{ij} \, - \, \partial_r^2 B_{ui}\, - \, \frac{\Delta-1}{r^2} B_{ui} \, - \, \frac{1}{r^2}D_iD^j B_{ju} \, =\, 0\,,\\ \label{r-eq}
& 2\left(\partial_r - \frac{1}{r}\right)\partial_u B_{ij}  \, - \,  \frac{\Delta}{r^2}  B_{ij}
\, + \, \left( \partial_r - \frac{2}{r} \right) \left(D_{[i}B_{j]u} \, - \, \partial_rB_{ij}\right) \, - \, \frac{1}{r^2}D_{[i}D^l B_{j]l} \, =\, 0\,,
\end{align}
where we recall that $D_i$ denotes the covariant derivative on the unit sphere and $\Delta = D^i D_i$ is the Laplacian on the unit sphere.

In order to impose consistent falloff conditions, we adopt two guiding criteria: we consider field configurations that radiate a finite energy per unit time across any spherical section $S_u$ of null infinity and we check compatibility with the free equations of motion to leading order as $r\to\infty$.

Finiteness of the energy flux at infinity imposes that the limit
\be\label{energyflux}
\cP\, (u) =
\lim_{r\to\infty}\int_{S_u}\gamma^{ij}\gamma^{jk} H_{uil}(H_{ujk}-H_{rjk}) r^{-2} d\Omega
\ee
be finite, hence indicating\footnote{Following the discussion of the previous chapter, it is worth keeping in mind that one may weaken these radiation falloffs, if needed, provided one does so while still ensuring the finiteness of the gauge-invariant quantity \eqref{energy-flux}.} that both $B_{ij}$ and $B_{uj}$ should scale at most like $r$, as $r\to\infty$. Equation \eqref{r-eq} further suggests that $B_{ij}$ should scale precisely like $r$, thus saturating the energy bound, so that the leading component of $\partial_u B_{ij}$ be unconstrained on-shell.
Indeed, we find that the free equations of motion are solved to leading order as $r\to\infty$ by
\be\label{falloff-radiation}
B_{ui} = D^j C_{ij} \log r+ \cdots\,,\qquad
B_{ij} = r C_{ij}+ \cdots\,,
\ee
where $C_{ij}(u,z,\bar z)$ is an antisymmetric tensor on the sphere. In particular, this class of asymptotic solutions highlights $C_{z\bar z}$ as the single on-shell propagating degree of freedom carried by the two-form field being the only independent function of the leading solution space. Moreover,  it carries a finite amount of energy to null infinity,
\be
\cP\, (u) \, = \, \int_{S_u} \gamma^{ij}\gamma^{jk} \partial_u C_{il}\partial_uC_{jk}\, d \O\,,
\ee
as required.

The falloff conditions \eqref{falloff-radiation} are  invariant under any gauge transformation \eqref{transform-gauge}, which we thus identify as providing the set of residual symmetries of the theory.  We can compute the corresponding surface charge \cite{Barnich-Brandt, HS-charges-cov, Avery_Schwab}
\be\label{generic-charge}
\tilde Q^+ = \oint_{S_u} \kappa^{ur}  dz\,d\bar z\,,
\ee
where the integration is performed on a sphere $S_u$ at fixed retarded time $u$ and for a large value of the radial coordinate $r$, while the Noether two-form \cite{Avery_Schwab} $\kappa^{ab}$ is given by 
\be\label{twoformtwoform}
\kappa^{ab}=\sqrt{-g}\, H^{bac}\epsilon_{c}
\ee
and hence satisfies 
\be
\kappa^{ur} = r^{2}\gamma_{z\bar z} \e_\mu H^{\mu ru}  = \gamma_{z\bar z}\, \e_i \gamma^{ij} H_{jur}  = -\gamma_{z\bar z}\,\e_i \gamma^{ij}\partial_r B_{uj}\,.
\ee
Making use of the equations of motion we can further rewrite the charge as follows
\be\label{charge_two_form}
\begin{split}
	\tilde Q^+ & =  \frac{1}{r}\oint_{S_u} \gamma^{ij}\gamma^{lk} D_{i}\e_l\, C_{kj}\,d\Omega \\
	& = \frac{1}{r} \int \gamma^{z\bar z}(\partial_z\e_{\bar z}-\partial_{\bar z}\e_z)C_{z\bar z}\,dzd\bar z\, .
\end{split}
\ee

%%%%%
\subsection{Duality and scalar charges} \label{ssec:duality}
%%%%%

A two-form gauge field  $B_{\mu\nu}$ in $D=4$ is dual, on shell, to a scalar field $\vf$ via  the relation $\ast dB = d\vf$, or equivalently $dB=\ast d\varphi$, where $d$ is the exterior derivative and $\ast$ the Hodge dual\footnote{We recall that, given the $p$-form
	$$
	\alpha= \frac{1}{p!}\,\alpha_{a_1\cdots a_p}dx^{a_1}\wedge\cdots\wedge dx^{a_p}\,,
	$$ 
its Hodge dual is defined in terms of the metric tensor $g_{ab}$ as
$$
\ast\alpha = \frac{\sqrt{-g}}{p!(D-p)!}\,\alpha^{a_1\cdots a_p} \epsilon_{a_1\cdots a_p b_1 \cdots b_{D-p}}dx^{b_1}\wedge\cdots\wedge dx^{b_{D-p}}
$$ with indices raised by $g^{ab}$.} in $D=4$; in components,
\be
\frac{1}{2}\,r^2\, \gamma_{z\bar z}\, \epsilon_{abcd}\,\partial^a B^{bc} = \partial_d \vf
\ee
or, equivalently,
\be\label{gaugeinvduality}
H_{abc}=r^2 \gamma_{z\bar z} \epsilon_{abcd}\partial^d\varphi\,.
\ee
In fact, this duality corresponds to the identification between the singlet representation $\bullet$ and the antisymmetric rank-$2$ form  {\tiny$\young(\hfill,\hfill)$} of SO(2), {\it i.e.} the little group for massless particles in $D=4$. A remarkable, albeit elementary, feature of the above duality relation is that it trades the equations of motion of $\varphi$ for the Bianchi identities of $B$ and vice-versa
\be
d\ast d\varphi=dH=0\,,\qquad d\ast H=d^2\varphi=0\,.
\ee
This means that, if we find a two-form $B$ that satisfies the duality, it must automatically satisfy the equations of motion $d\ast H=0$, which we can use to cross check the solution obtained above.

In components, the duality relation yields
\be\begin{aligned}
	&\partial_r B_{z\bar z} = r^2\gamma_{z\bar z} \partial_r \varphi\,,\\
	&\partial_r B_{uz} = -\partial_z \varphi\,,\\
	&\partial_r B_{u\bar z} = \partial_{\bar z} \varphi\,,\\
	&\partial_u B_{z\bar z}+D_{[z}B_{\bar z]u} - \partial_r B_{z\bar z} = -r^2 \gamma_{z\bar z} \partial_u \varphi\,.
\end{aligned}\ee
Comparing with the falloffs for the two-form \eqref{falloff-radiation}, we see that these equations are compatible to leading order with the standard falloff condition for the massless scalar
\be\label{scalarasympt4}
\varphi(u,r,z,\bar z) = \frac{b(u,z,\bar z)}{r}+\cdots\,,
\ee 
which in particular satisfies the equation of motion $\Box\varphi=d\ast d\varphi=0$ identically to leading order,
provided one identifies
\be\label{link}
b\, \gamma_{z\bar z} = - C_{z\bar z}\,.
\ee
This relation furnishes the desired connection between the on-shell degree of freedom $C_{ij}$ of the two-form field and the propagating component $b$ of the massless scalar. Equation \eqref{link} can be rewritten covariantly as $C_{ij} =i b\, \Omega_{ij}$, where $\Omega$ is the standard symplectic form on the Euclidean sphere.

Let us now compare ``charge'' operators arising from scalar soft theorems \cite{CampigliaCoitoMizera} with the surface charges given by two-form asymptotic symmetries \eqref{charge_two_form}, in order to connect the former to the latter by means of the duality transformation. 
We recall that the soft part of the scalar charges can be expressed as
\be\label{scalar-charge4}
Q^+_s= \int_{S^2} b(u, z, \bar z) \Lambda(z, \bar z)\gamma_{z\bar z}dz d\bar z\,, 
\ee
where $\Lambda(z,\bar z)$ is an arbitrary function of the two angular coordinates.
In view of \eqref{link}, we may identify
\be\label{charge-link}
Q^+_s = r \tilde Q^+\,,
\ee
and correspondingly for the residual symmetry parameters
\be
\Lambda\gamma_{z\bar z} = \partial_z \epsilon_{\bar z} \, - \, \pr_{\bar{z}} \, \epsilon_{z}\,.
\ee
To summarize, this picture leads to an identification between the propagating scalar mode $b (u,z,\bar z)$ and the two-form physical degree of freedom $C_{z\bar z}(u,z,\bar z)$ as 
$C_{z\bar z} = - \gamma_{z\bar z}b$. On the other hand, the asymptotic charge for the two-form residual symmetry takes the form
\be
\tilde Q^+ = \frac{1}{r}\int_{S^2} C_{z \bar z} \gamma^{z\bar z}(\partial_z \e_{\bar z}- \partial_{\bar z}\e_z) dz d \bar z\,,
\ee
and are conjectured to be dual to the scalar soft charges \eqref{scalar-charge4} by means of \eqref{charge-link}.  

A puzzling, although not completely unfamiliar\ft{See {\it e.g.} \cite{StromingerGravmemevD, Conde:2016csj, Conde:2016rom}, where charges depending on some inverse power of the radial coordinate were considered and interpreted as connecting asymptotic symmetries to soft theorems. Asymptotically vanishing surface charges were also encountered in the previous chapter, in connection with residual symmetries associated with memory effects in even dimensions.} feature of the identification is the fact that, while  $Q^+_s$ is nontrivial on $\mathscr I^+$, \emph{i.e.} even after performing the limit $r\to\infty$, our two-form charge $\tilde Q^+$ vanishes in the large-$r$ limit. This seems to be a consequence of the fact that, in radial gauge, symmetry parameters are not allowed to grow with $r$, and hence are unable to compensate for the falloff $\partial_r B_{ui}\sim 1/r$. 

Indeed, in order to avoid this shortcoming, one could try relaxing some of the restriction that have been imposed on the gauge parameter $\epsilon_r$ by means of gauge-for-gauge transformations. That is, one could still impose the radial gauge $B_{ar}=0$, allowing however for a generic $\epsilon_a$ satisfying $\partial_{[r}\epsilon_{a]}=0$. In fact, the matching condition \eqref{gaugeinvduality} between the field strength of the two-form theory and the on-shell scalar is gauge invariant and can be substituted into \eqref{twoformtwoform} to obtain the charge
\be\label{chargequasi}
\tilde Q^+=\frac{1}{r}\int(\partial_{\bar z}\epsilon_z-\partial_z\epsilon_{\bar z})b\, dzd\bar z\,,
\ee 
whose value on $\mathscr I^+$ is nonzero provided that $\epsilon_i$ scales like $\mathcal O(r)$. However, upon performing the asymptotic expansion for $\epsilon_a$, the condition that radial gauge be preserved imposes
\be
\partial_{[r}\epsilon_{i]}=0 \implies \epsilon^{(1)}_i = \partial_i \epsilon_r^{(0)}\,,  
\ee
which spoils the order $\mathcal O(r^{0})$ of above charge.

Another possible way out of this inconvenience could be to add terms of the type 
\be\label{Coulombs}
\partial_zf(z)\,r \,,\qquad
\partial_{\bar z} g(\bar z)\,r \, ,
\ee 
to the two-form components $B_{uz}$, $B_{u\bar z}$ respectively. These terms are indeed allowed by the leading equations of motion and give no contribution to the energy flux at infinity. These new terms would give rise to the modified charge
\be \label{modified charge}
\int \big[\epsilon_z \partial_{\bar z}g(\bar z)+\epsilon_{\bar z} \partial_zf(z)\big]
dzd\bar z
-\frac{1}{r} \int \gamma^{z\bar z}(\partial_z\epsilon_{\bar z}-\partial_{\bar z}\epsilon_z)C_{z\bar z}\,dzd\bar z\,,
\ee
which no longer goes to zero as $r\to\infty$. On the other hand, in this limit, it appears to become independent of the physical degree of freedom $C_{z\bar z}$ and, in the dual interpretation, of the radiative mode $b$ of the massless scalar. Indeed, the $f(z)$ and $g(\bar z)$, appearing in the first term of \eqref{modified charge}, are related by duality to a scalar field $\varphi(u,r,z, \bar z) = -f(z)+g(\bar z)+\cdots$, which is static to leading order as $r\to\infty$.

The terms \eqref{Coulombs} admit a natural interpretation if one phrases the problem of studying the two-form falloffs in a spacetime of generic dimension $D$. In this setup, the asymptotic analysis of the equations of motion highlights two classes or ``branches'' of solutions: denoting by $x^i$ coordinates on the celestial $(D-2)$-sphere, one has a radiation branch
\be\label{radiation-branch}
B_{ui} =  \frac{2}{4-D} U_{i}(u,x^k)\, r^{(4-D)/2}+\cdots\,,\qquad
B_{ij} = C_{ij}(u,x^k)\, r^{(6-D)/2}+\cdots\,,
\ee	
subject to $U_i=D^j C_{ij}$ (unless $D=6$, in which case only $\partial_u U_i = \partial_u D^j C_{ij}$ need be imposed)
and a Coulomb-like branch
\be\label{Coulomb-branch}
B_{ui} = \tilde U_{i}(x^k) r^{5-D} +\cdots\,,\qquad
B_{ij} = \frac{1}{D-4}\tilde C_{ij}(u,x^k)r^{5-D}+\cdots\,,
\ee
where $\partial_u \tilde C_{ij} = D_{[i}\tilde U_{j]}$ and $(D-5)D^j \tilde U_{j}=0$. Solutions of the first type give rise to nonzero energy flux across sections of null infinity, $\mathcal P(u)\neq 0$, and only give vanishing contributions to the (global) charges as $r\to\infty$. The second class, on the other hand, does not contribute to the energy flux, while giving nonzero contributions to charge integrals. In $D=4$,  the above expressions exhibit singularities and \eqref{radiation-branch} reduces to \eqref{falloff-radiation}, while \eqref{Coulomb-branch} gives rise to \eqref{Coulombs}. 

The scalar radiative mode $b$, in four dimensions, appears thus to be dual to the a radiation solution for its two-form counterpart. From this observation, it appears natural that its soft charge may be dual to an asymptotically vanishing two-form charge. 

\section{Lorenz Gauge Approach}\label{sec:scoop}
A similar logic can be employed, with minor modifications, to discuss the problem at stake in the Lorenz gauge \cite{Campigliascoop}. The main advantage of this choice, as opposed to radial gauge employed above, is that it allows for the presence of a finite nonzero asymptotic charge.

One starts from the duality condition $H=\ast d\varphi$, namely \eqref{gaugeinvduality}, which affords
\be\begin{aligned}\label{falloffH}
	H_{urz}&=-\frac{\partial_zb}{r}+\cdots
	\,,\\ 
	H_{ur\bar z}&=\frac{\partial_{\bar z} b}{r}+\cdots\,,\\ 
	H_{uz\bar z}&=r\gamma_{z\bar z}\partial_ub+\cdots
	\,,\\ 
	H_{rz\bar z}&=\gamma_{z\bar z}b+\cdots\,,
\end{aligned}\ee
upon substituting the asymptotics \eqref{scalarasympt4} for the scalar field. One then looks for a two-form $B_{ab}$ satisfying $\nabla^a B_{ab}=0$ whose field strength respects \eqref{falloffH}. This is achieved to leading order by 
\be\begin{aligned}\label{falloffB}
	B_{ij}&=rB_{ij}^{(-1)}(u,z,\bar z)+\cdots
	\,,\\ 
	B_{ui}&=B_{ui}^{(0)}(u,z,\bar z)+\cdots\,,\\ 
	B_{ri}&=B_{ri}^{(0)}(z,\bar z)+\frac{1}{r}\,B_{ri}^{(1)}(u,z,\bar z)+\cdots
	\,,\\ 
	B_{ur}&=\frac{1}{r}\,D\cdot B_r^{(0)}(z,\bar z)+\cdots\,,
\end{aligned}\ee
provided
\be
B_{z\bar z}^{(-1)}+\partial_{[z}B_{\bar z]r}^{(0)}=\gamma_{z\bar z}b
\ee
and
\be
\partial_u B_{rz}^{(1)}+\partial_z D\cdot B_r^{(0)}=\gamma_{z\bar z}\partial_z b\,,\qquad
\partial_u B_{r\bar z}^{(1)}+\partial_{\bar z} D\cdot B_r^{(0)}=-\gamma_{z\bar z}\partial_{\bar z} b\,.
\ee
Residual symmetries can be parametrized by closed two-forms $\alpha$ such that $d\alpha=0$ and $\nabla^a\alpha_{ab}=0$, which also respect the falloffs \eqref{falloffB}. These conditions are solved to leading order by 
\be\begin{aligned}\label{falloffalpha}
	\alpha_{ij}&=r\partial_{[i}\alpha_{j]r}^{(0)}(z,\bar z)+\cdots
	\,,\\ 
	\alpha_{ui}&=\alpha_{ui}^{(0)}(u,z,\bar z)+\cdots\,,\\ 
	\alpha_{ri}&=\alpha_{ri}^{(0)}(z,\bar z)+\frac{1}{r}\,\alpha_{ri}^{(1)}(u,z,\bar z)
+\cdots
	\,,\\ 
	\alpha_{ur}&=\frac{1}{r}\,D\cdot \alpha_r^{(0)}(z,\bar z)+\cdots\,,
\end{aligned}\ee
subject to
\be
\partial_u \alpha_{ri}^{(1)}+\partial_i D\cdot \alpha_r^{(0)}=0\,.
\ee
Then, the expression for the charge obtained in \eqref{chargequasi} immediately yields
\be
\tilde Q^+ = \int \partial_{[i}\alpha_{j]r}^{(0)}(z,\bar z)b(u,z,\bar z) dzd\bar z\,.
\ee
While this result indeed provides a more satisfactory answer to our initial question compared to the one we obtained in the radial gauge, since it allows for a matching between the soft scalar charges and \emph{nonvanishing} two-form asymptotic charges, it shows that the calculation of asymptotic symmetries may be strongly influenced not only by the choice of falloffs, as already highlighted by the higher-dimensional spin-one and spin-two theories, but also by the more basic choice of gauge fixing. In this setup, the radial gauge effectively trivializes the asymptotic symmetries, by forbidding nontrivial $\alpha_{ir}\sim\mathcal O(1)$, and hence hides the presence of finite soft charges. One may interpret this fact by observing that reaching radial gauge from a generic field configuration (before gauge fixing) may involve a \emph{large} gauge transformation and, therefore, that such a gauge choice is not feasible if one wants to have control on the full structure of asymptotic symmetries.

%%%%%%%%%%%%%%%%%%%%%%%%%%%%%%%%%%%%%%
\chapter{Asymptotic Symmetries of Higher Spins}\label{chap:HSP}

The discussion of the preceding chapters allowed us to exhibit a link between asymptotic symmetries and soft theorems in the context of electrodynamics, Yang-Mills theory, gravity and, maybe less directly, in the case of scalar field theories and their duals. In these cases, the spin of the boson mediator of the interaction was thus respectively 1, 2 and 0. The goal of this chapter is to extend this type of analysis to  theories containing a spin-$s$ gauge boson, namely a massless particle carrying a kind of generalized electromagnetic/gravitational force, with arbitrary integer $s$ \cite{carlo_tesi,super-io,Cariche-io}.

The interest in this type of theories derives primarily from the analysis of the string theory spectrum of elementary excitations, which contains particles with arbitrary integer spin and with mass proportional to the string tension. In order to better understand the properties of these excitations in the massless limit, which is conjectured to be linked to a suitably-defined tensionless limit of the string, it is natural and worthwhile to try to set up a purely field-theoretical description for them, \emph{i.e.} a higher-spin gauge theory.

Such theories admit a consistent Lagrangian formulation on maximally symmetric backgrounds up to the level of cubic vertices. The possibility to construct a Lagrangian to all order in the vertices is subject to notorious obstructions of both technical and conceptual nature, in particular to a conflict with locality. On the other hand, the fully nonlinear construction of Vasiliev \cite{Vasi} (see \cite{BCIValgebras,Did} for reviews) does not admit Minkowski spacetime among its vacuum solutions. Still, as we stressed in our discussion of the lower-spin cases, one may expect that the linear theory \cite{Fronsdal, ML} already contains salient features related to the asymptotic symmetry transformations.

In our analysis, we will thus propose a definition for the notion of asymptotic symmetries of a free spin-$s$ field theory on Minkowski background coupled to a suitable external source, representing unspecified nondynamical matter. As we shall see, in four spacetime dimensions, these asymptotic symmetries on the one hand give rise to Ward identities and allow us to make contact with the corresponding Weinberg theorem, while on the other hand they display a structure which can be regarded as interesting \emph{per se} and in relation with the conjectured existence of an underlying non-Abelian higher-spin algebra. 

\section{A Bondi-like Gauge}
As repeatedly stressed, the first point we need to address, while attempting to define asymptotic symmetries in a given gauge theory, is the assignment of a suitable set of gauge-fixing conditions and falloff conditions on the gauge fields. Before tackling this issue, we will recall the basic definitions for a free spin-$s$ gauge theory in the Fronsdal formulation \cite{Fronsdal} (see also \cite{Sagnotti_e_co} for a review).

\subsection{Fronsdal formulation}
A spin-three gauge field is described kinematically by a tensor $\varphi_{\mu\nu\rho}$ that is completely symmetric in its indices and is subject to the gauge transformation (we adopt here Cartesian coordinates $x^\mu$ for simplicity)
\be\label{spin3-transf}
\delta_\epsilon\varphi_{\mu\nu\rho}
=
\partial_\mu \epsilon_{\nu\rho}
+\partial_\nu \epsilon_{\rho\mu}
+\partial_\rho \epsilon_{\mu\nu}\,,
\ee
where the gauge parameter $\epsilon_{\mu\nu}$ is symmetric in its two indices and has vanishing trace,
\be\label{spin3-trace}
\eta^{\mu\nu}\epsilon_{\mu\nu}=0\,.
\ee
Notice that, while eq. \eqref{spin3-transf} is just the natural generalization of the gauge transformation $h_{\mu\nu}\mapsto h_{\mu\nu}+\partial_\mu \xi_\nu + \partial_\nu \xi_\mu$ for a linearized metric perturbation, the trace constraint constitutes a novel feature.  

In the following, we will often use a shorthand notation to denote symmetrization and contraction of indices. A round bracket enclosing a set of indices denotes their (\emph{unweighted}) symmetrization obtained by the least possible number of terms: for instance, 
\be
A_{(\mu\nu)}=A_{\mu\nu} + A_{\nu\mu}
\ee
if $A$ is generic rank-two tensor, while
\be
P_{(\mu}P_{\nu)}=P_\mu P_\nu\,,
\ee 
if $[P_\mu,P_\nu]=0$.
A ``prime'' symbol denotes a trace, \emph{e.g.} $\varphi'_{\rho}=\eta^{\mu\nu}\varphi_{\mu\nu\rho}$. With these conventions, the spin-three gauge symmetry can be expressed as
\be
\delta_\epsilon\varphi_{\mu\nu\rho}=\partial_{(\mu}\epsilon_{\nu\rho)}\,,\qquad
\epsilon'=0\,.
\ee
Finally, a divergence, namely a contraction between a derivative and an index of the tensor on which the derivative acts, is denoted by a ``dot'', for example
$
\partial\cdot\varphi_{\mu\nu}=\eta^{\sigma\rho}\partial_\sigma\varphi_{\mu\nu\rho}
$.

The dynamics for a spin-three field freely propagating in $D$ spacetime dimensions, coupled to an external source, can be defined by means of the gauge-invariant action $S=\int \mathcal L\, d^D\!x$,
\be\label{Fronsdalacs3}
\mathcal L=
\frac{1}{2} \mathcal E^{\mu\nu\rho}\varphi_{\mu\nu\rho}
-  J^{\mu\nu\rho}\varphi_{\mu\nu\rho}\,,
\ee
where the $\mathcal E^{\mu\nu\rho}$ is given by
\be
\mathcal E^{\mu\nu\rho} = \mathcal F^{\mu\nu\rho} - \frac{1}{2}\eta^{(\mu\nu}\mathcal F'^{\rho)}
\ee
and $\mathcal F^{\mu\nu\rho}$, also called the Fronsdal tensor, is defined as
\be
\mathcal F^{\mu\nu\rho} = \Box \varphi^{\mu\nu\rho} - \partial^{(\mu}\partial\cdot\varphi^{\nu\rho)} + \partial^{(\mu}\partial^\nu\varphi'^{\rho)}\,.
\ee
The symmetric tensor $J^{\mu\nu\rho}$ defines the external source and it is subject to the local conservation condition
\be\label{tracelesscons3}
\partial\cdot J^{\mu\nu}-\frac{1}{D}\,\eta^{\mu\nu}\partial\cdot J'=0\,,
\ee
where on the left-hand side we recognize the traceless projection of $\partial\cdot J^{\mu\nu}$.

In order to handle symmetric tensors with an arbitrary number of indices, it is sometimes useful to introduce a more abstract notation in which all spacetime indices are implicit \cite{FSagn}. In which case, the last three equations take the compact form
\be\label{FronsdalanyS}
\mathcal E = \mathcal F - \frac{1}{2} \eta \mathcal F'\,,
\qquad \mathcal F = \Box \varphi - \partial \partial\cdot\varphi + \partial^2 \varphi'\,,
\qquad
\partial\cdot J - \frac{1}{D}\,\eta\, \partial\cdot J'=0\,.
\ee
We will not, however, employ this notation extensively in the following.

The equations of motion obtained from the variation of the action \eqref{Fronsdalacs3} are 
\be
\mathcal E^{\mu\nu\rho} = J^{\mu\nu\rho}\,.
\ee 
Under a gauge transformation, the Fronsdal tensor transforms into
\be
\delta_\epsilon \mathcal F^{\mu\nu\rho} = 3 \partial_\mu \partial_\nu \partial_\rho \epsilon'\,,
\ee
and the last term vanishes by the constraint $\epsilon'=0$. This is is sufficient, for spin three, in order to prove the gauge invariance of the equations of motion.

Since this is needed for the evaluation of the Noether charge, let us explicitly verify, as an exercise, the gauge invariance of the Fronsdal action. The trace of $\mathcal F$ is given by
$
\mathcal F'_\rho = 2 \Box \varphi'_\rho - 2 \partial\cdot\partial\cdot\varphi_\rho + \partial_\rho \partial\cdot\varphi'
$,
while the $\mathcal E$ tensor satisfies the following identities, 
\be\label{anomalousBianchi}
\partial_\rho \mathcal E^{\mu\nu\rho} = - \frac{1}{2}\eta^{\mu\nu}\partial\cdot \mathcal F'\,,
\ee
which may be regarded as an ``anomalous'' analog of the (linearized) Bianchi identities, due to the nonzero term on the right-hand side.
Therefore, employing the identity \eqref{anomalousBianchi}, 
\be
\frac{1}{2}\delta_\epsilon ( \mathcal E^{\mu\nu\rho}\varphi_{\mu\nu\rho})  = 
\frac{3}{2} \mathcal E^{\mu\nu\rho} \partial_\mu \epsilon_{\nu\rho} =\frac{3}{2} \partial_\mu \left(\mathcal E^{\mu\nu\rho} \epsilon_{\nu\rho}\right) + \frac{3}{4}  \partial\cdot \mathcal F' \epsilon' \,,
\ee
which is a boundary term plus a contribution that vanishes by the constraint $\epsilon'=0$. For the source term we have
\be
-\delta_\epsilon ( J^{\mu\nu\rho}\varphi_{\mu\nu\rho} )=-3  \partial_{\mu}\left( J^{\mu\nu\rho} \epsilon_{\nu\rho}\right)\,,
\ee
by the traceless conservation condition \eqref{tracelesscons3}. Thus,
\be
\delta_\epsilon \mathcal L = -\frac{3}{2} {\partial_\mu} (\mathcal E^{\mu\nu\rho}\epsilon_{\nu\rho})\,,
\ee 
so that the action is indeed invariant, up to a boundary term.
The standard Noether current is then given by
\be\label{jspin}
j^\mu_\epsilon
=
\frac{\partial \mathcal L}{\partial \varphi_{\rho\alpha\beta,\mu\nu}}\delta\varphi_{\rho\alpha\beta,\nu}
-
\partial_\nu \frac{\partial \mathcal L}{\partial \varphi_{\rho\alpha\beta,\mu\nu}}\,\delta\varphi_{\rho\alpha\beta}+\frac{3}{2}{\partial_\mu} (\mathcal E^{\mu\nu\rho}\epsilon_{\nu\rho})\,,
\ee
where, in compact notation,\footnote{I am grateful to Dario Francia for sharing his unpublished notes on higher-spin conserved currents.}
\be\begin{aligned}\label{secondderj}
	\frac{\delta \mathcal L}{\delta \varphi_{,\mu\nu}} =&
	\frac{1}{2}\left\{ \eta^{\mu\nu} \varphi - \frac{1}{2}\left( \eta^\mu \varphi^\nu + \eta^\nu  \varphi^\mu\right) + \eta \varphi^{\mu\nu} - \eta^{\mu\nu} \varphi' \eta \right.\\
	&\left. +\frac{1}{2}\left( \eta^{\mu}\eta^\nu \varphi' + \eta^\nu \eta^\mu \varphi'\right) - \frac{1}{4}\left( \varphi'^\mu \eta^\nu \eta + \varphi'^\nu \eta^\mu \eta\right) \right\}\,.
\end{aligned}\ee

The formulation outlined in this section for a spin-three field can be extended to any integer spin $s$. In particular, equations \eqref{FronsdalanyS} hold for arbitrary $s$, the action is given by $S=\int \mathcal L\, d^D\!x$,
\be
\mathcal L = 
\frac{1}{2} \mathcal E^{\mu_1\cdots\mu_s}\varphi_{\mu_1\cdots \mu_s}
+
J^{\mu_1\cdots \mu_s}\varphi_{\mu_1\cdots\mu_s}\,
\ee 
and the gauge transformation is given by $\delta_\epsilon \varphi_{\mu_1\cdots \mu_s}=\partial_{(\mu_1}\epsilon_{\mu_2\cdots\mu_s)}$ (or $\delta_\epsilon\varphi=\partial\epsilon$ in compact notation). An important technical point is that the consistency of the theory requires the field to be doubly traceless for spin four or higher, namely $\varphi''=0$, together with $J''=0$. Once this point is taken into account, the equations of the theory take the form \eqref{FronsdalanyS} for any $s$. As far as the corresponding Noether current is concerned, \eqref{secondderj} and the obvious generalization of \eqref{jspin} remain valid for any spin.

Alternatively, one may employ the Maxwell-like formulation of free higher spins \cite{ML,DarioGabrieleKarapet}. In this approach, the trace constraint for the gauge parameter $\epsilon$ is traded for the transversality condition $\partial\cdot \epsilon=0$, while trace constraints may or may not be enforced, depending on the type of spectrum that one wishes to describe. It should be noted that the conditions we are going to impose on the gauge fields in the next section effectively make the two frameworks not just equivalent but actually identical as far as the asymptotic analysis is concerned.

\subsection{Boundary conditions}
In analogy with the lower-spin cases, we propose the following set of gauge-fixing and falloff conditions for a spin-three field, focusing for the moment on the case of four spacetime dimensions. Employing the usual retarded Bondi coordinates, with stereographic coordinates on the sphere \eqref{stereo}, we impose
\be\label{radtr3}
\varphi_{rab}=0=\varphi_{a z\bar z}\,,
\ee
for any $a$, $b$,
together with the conditions that the normalized field components, namely
\be
\varphi_{uuu}\,,\qquad 
\frac{\varphi_{uuz}}{r}\,,\qquad
\frac{\varphi_{uzz}}{r^2}\,,\qquad
\frac{\varphi_{zzz}}{r^3}
\ee
(and similarly for $z\leftrightarrow \bar z$), scale asymptotically as 
$
\mathcal O(r^{-1})
$
as $r\to\infty$ for fixed $u$. These conditions provide a natural generalization of the radial gauge $\mathcal A_r=0$ of Maxwell's theory and the (sharp) Bondi gauge adopted in the discussion of asymptotic symmetries of asymptotically flat spaces, as we have seen in Chapter \ref{chap:basics}. Notice that, in the chosen coordinates,
\be
\varphi'_a = -2 \varphi_{ura}+\varphi_{rra}+\frac{2}{r^2\gamma_{z\bar z}} \varphi_{z\bar z a}\,,
\ee
so the trace of the field is identically zero in view of the conditions \eqref{radtr3}.

In the case of a spin-$s$ field, we impose the analogous conditions
\be\label{Bondilikegau}
\varphi_{ra_2\cdots a_s}=0=\varphi_{z\bar z a_3\cdots a_s}\,,\qquad
\frac{\varphi_{uu\cdots uz z \cdots z}}{r^d}=\mathcal O(r^{-1})\,,
\ee
where $d$ denotes the number of $z$ indices appearing in the last formula, and analogously for $z\leftrightarrow\bar z$. In the following, we shall refer to this set of conditions as the ``Bondi-like gauge''.

As a first consistency check, it is worthwhile to calculate the flux of energy carried by a spin-$s$ field as measured on a sphere $S_u$ at large radius $r$ and fixed retarded time $u$. The expression for this quantity in Bondi-like gauge takes the following form:
\be
\mathcal P(u)=\oint_{S_u} (\gamma_{z\bar z})^{-s} \partial_u \varphi_{zz\cdots z} \partial_u \varphi_{\bar z \bar z \cdots \bar z}\,\gamma_{z\bar z}dzd\bar z\,.
\ee
This is generically nonvanishing and, most importantly, finite, even in the large-$r$ limit, thus providing support to the choice of the conditions \eqref{Bondilikegau}.

Another important consistency check that needs to be performed on the Bondi-like gauge conditions is their compatibility with the (free) equations of motion as $r\to\infty$. This ensures that the leading field components indeed specify the free boundary data associated to radiation and charge measurements performed near null infinity.

Let us elaborate on this point, focusing for the moment on the spin-three case. Imposing $\varphi_{ab r}=0$ and $\varphi_{z\bar zr}=0$, the equations of motion take the form
\be\label{Fronsdaleq}
\mathcal F_{abc} \, = \, \Box \varphi_{abc}-\nabla_{(a}\nabla\cdot \varphi_{bc)}=0 \,.
\ee
Now, $\mathcal F_{rrr}=0$ is identically solved, and similarly for $\mathcal F_{a z \bar z}=0$. 
From $\mathcal F_{uur}=0$ we have 
\be
\begin{split}
	& \frac{2}{r^2}\, \varphi_{uuu} + 
	2\, \partial_r \left( \frac{\varphi_{uuu}}{r} \right) +
	\partial^2_r \varphi_{uuu}-
	\frac{2}{r^2}\left(D^z \varphi_{zuu} + D^{\bar z}\varphi_{\bar z uu} \right)\\
	& - 
	\partial_r\! \left[ \frac{1}{r^2}(D^z \varphi_{zuu}+ D^{\bar z}\varphi_{\bar zuu})\right]=0 \,,
\end{split}
\ee
and expanding $\varphi_{uuu}=B\, r^{\alpha}$, $\varphi_{zuu}=U_z\, r^{\beta}$, with $\beta=\alpha+1$, to leading order we have
\be
(\alpha+1)\left[\alpha B - (D^z U_z + D^{\bar z}U_{\bar z}) \right]=0\,.
\ee
By comparison with $\mathcal F_{uuu}=0$, which reads
\be
\frac{4}{r}\, \partial_u \varphi_{uuu}+ \partial_u \partial_r \varphi_{uuu} - \frac{3}{r^2} \left(D^z \varphi_{zuu}+D^{\bar z}\varphi_{\bar z uu} \right)=0 \,,
\ee
and yields, upon expansion,
\be
(\alpha+4)\,\partial_u B - 3\, \partial_u (D^z U_z + D^{\bar z}U_{\bar z})=0 \,,
\ee
we have two possible behaviors: a ``growing mode'' $\alpha=2$, $\beta =3$ and a ``decaying mode'' $\alpha=-1$, $\beta=0$. We choose the latter, obtaining $\varphi_{uuu}=B/r$ and $\varphi_{zuu}=U_z$, together with the additional condition
\be\label{interr_1}
\partial_u B=  \partial_u \left(D^z U_z + D^{\bar z}U_{\bar z} \right) .
\ee
From $\mathcal F_{zur}=0$, we have
\be
-\frac{2}{r^3}\,D^z\varphi_{zzu}+\frac{4}{r^2}\,\varphi_{zuu} + \left(\partial_r - \frac{2}{r}\right)\!\left( 
\partial_r \varphi_{zuu}+\frac{2}{r}\,\varphi_{zuu}-\frac{1}{r^2}\,D^z\varphi_{zzu}\right)=0 \,.
\ee 
Which is solved by setting $\varphi_{zzu}= r\, C_{zz}$, where
\be\label{interr_2}
U_z = \frac{1}{2}\,D^z C_{zz}\,.
\ee
Finally, from $\mathcal F_{zzr}=0$, we have
\be
-\frac{2}{r^3}\,D^z\varphi_{zzz}+\frac{6}{r^2}\,\varphi_{zzu}-\left(\partial_r - \frac{4}{r}\right)\!\left(\frac{1}{r^2}\,D^z \varphi_{zzz}-\frac{2}{r}\,\varphi_{zzu}-\partial_r\varphi_{zzu}\right)=0 \,,
\ee
which is solved by setting $\varphi_{zzz}=B_{zzz}\,r^{2}$, with
\be\label{interr_3}
C_{zz} = \frac{1}{3}\,D^zB_{zzz}\,.
\ee
This completes the consistency check for the spin-three Bondi-like gauge against the equations of motion, since the number of independent equations in four spacetime dimensions is 10, owing to the Bianchi identities. To summarize, in the Bondi-like gauge, the nonvanishing field components read, to leading order in $r$,
\be
\varphi_{uuu}=\frac{B}{r}\,,\qquad
\varphi_{uuz}=U_z\,,\qquad
\varphi_{uzz}=r C_{zz}\,,\qquad
\varphi_{zzz}=r^2B_{zzz}\,,\ee 
and are subject to the constraints \eqref{interr_1}, \eqref{interr_2} and \eqref{interr_3}.

The consistency of the Bondi-like gauge for the generic spin-$s$ case can be checked in a similar manner, which we now sketch. The equations of the form 
\be
\mathcal F_{rr a_{1}\cdots a_{s-2}} = 0 
\ee
are identically solved once we choose $\varphi_{ra_{1}\cdots a_{s-1}}=0$ and $\varphi_{z\bar z a_{1}\cdots a_{s-2}}=0$. The equations $\mathcal F_{ru\cdots u}=0$, $\mathcal F_{uu\cdots u}=0$ and $\mathcal F_{ru\cdots uz}=0$ have the same form as the analogous equations of the spin-three case obtained by removing $s-3$ indices $u$ from them, the technical reason being that the connection coefficients $\Gamma^a_{b u}$ vanish identically. The equation $\mathcal F_{ru\cdots u z\cdots z}=0$ with a given number $1<d<s$ of $z$ indices  reads explicitly 
\be\begin{aligned}
&\frac{2}{r^3}\,D^z\varphi_{z_{d+1}u_{s-d-1}}- \frac{2(d+1)}{r^2}\,\varphi_{z_d u_{s-d}}\\
&-\left(\partial_r- \frac{d+2}{r}\right)\!\left(\partial_r\varphi_{ z_d u_{s-d}}-\frac{1}{r^2}D^z\varphi_{z_{d+1}u_{s-d-z}}+\frac{2}{r}\varphi_{z_d u_{s-d}}\right)=0\,,
\end{aligned}\ee
where for brevity $\varphi_{z_du_{s-d}}$ denotes $\varphi_{u\cdots u z \cdots z}$ with $d$ indices $z$ and $s-d$ indices $u$.
Altogether these equations impose
\be
\varphi_{z_du_{s-d}}=B_{z_d}\, r^{d-1}\,,
\ee
where the tensors $B_{z_d}$ have to satisfy
\be
B_d = \frac{1}{d+1}\, D^zB_{z_{d+1}}\,,
\ee
whereas the other equations are identically satisfied to leading order.

\section{Higher-Spin Supertranslations and Superrotations}
We now turn to the illustration of the asymptotic symmetries of higher spins in four dimensions. These are determined as the large gauge symmetries that leave the Bondi-like gauge \eqref{Bondilikegau} invariant, namely, restricting for the moment to spin three, as the solutions of the asymptotic Killing tensor equations
$
\delta_{\epsilon}\varphi_{r ab}=0=\delta_\epsilon \varphi_{z\bar z a}
$,
\be
\delta_\epsilon \varphi_{uuu}=\mathcal O(r^{-1})\,,\qquad
\delta_\epsilon \varphi_{uuz}=\mathcal O(r^{0})\,,\qquad
\delta_\epsilon \varphi_{uzz}=\mathcal O(r^{1})\,,\qquad
\delta_\epsilon \varphi_{zzz}=\mathcal O(r^{2})\,.
\ee 

\subsection{A higher-spin analog of supertranslations}

To begin with, we shall restrict to the case of gauge parameters that do not depend on retarded time. 
Then, the answer is provided by the following family of tensors, parameterized by the arbitrary function $T(z, \bar z)$:
\be\label{supertransl_spin3}
\begin{aligned}
	&\epsilon_{ab}dx^a dx^b = 
	- \left( \frac{3}{4}\,T + D^z D_z T + \frac{1}{4}\,(D^zD_z)^2T \right)du^2 
	- 2 \left( \frac{3}{4}\, T + \frac{1}{4}\,D^zD_z T \right) du dr\\ 
	&- 2r \left( \frac{3}{4}\,D_zT + \frac{1}{4}\,D_z^2 D^z T \right) du dz - 2r \left( \frac{3}{4}\,D_{\bar z}T + \frac{1}{4}\,D_{\bar z}^2 D^{\bar z} T \right) du d\bar z - T dr^2 \\
	& - r \left( D_zT dz + D_{\bar z}T d\bar z \right) dr - \frac{r^2}{2} \left( D^2_z T dz^2 
	+ D^2_{\bar z}T d\bar z^2 \right) - \frac{r^2}{2}\,\gamma_{z\bar z} \left( T + D^zD_z T \right) dz d\bar z \,.
\end{aligned}\ee
We refer to Appendix \ref{app:spin3symm} for the details of this derivation.

Note that the corresponding contravariant tensor on $\scrip$ is given by
\be
\epsilon^{ab}\partial_a \partial_b = -\, T(z, \bar z) \partial_u^2 \,,
\ee
so that this residual symmetry indeed furnishes a spin-three generalization of the ordinary gravitational supertranslation symmetry $\xi^a\partial_a= T(z,\bar z)\partial_u$. 
The nonvanishing gauge variations generated by \eqref{supertransl_spin3} are:
\begin{subequations}
	\begin{align}
	\delta \varphi_{uuz} &= - D_z \left( \frac{3}{4}\,T + D^z D_z T + \frac{1}{4}\,(D^zD_z)^2T \right) , \\
	\delta \varphi_{uzz} &= - \frac{r}{2}\, D^2_z\left( 3\, T + D^z D_z T \right) , \\
	\delta \varphi_{zzz} &= - \frac{3}{2}\, r^2 D^3_z T \,,
	\end{align}
\end{subequations}
together with their conjugates. 

Moving on to arbitrary spins, we must look for spin-$s$ gauge transformations that leave invariant our Bondi-like gauge, as summarized by
\be \label{constr_spin-s}
\varphi_{r a_{s-1}}=0=\varphi_{z \bar z a_{s-2}}
\ee
and
\be \label{boundary_spin-s}
\varphi_{u_{s-d} z_d} = r^{d-1} B_{z_d}(u,z,\bar z)\,,
\ee
where we have adopted the following multi-index notation, which will be understood from now on for symmetrized indices \cite{massive}: a subscript attached to a spacetime index indicates the number of times that type of index appears. For instance, for a spin-five field,
\be
\varphi_{u_2z_3}=\varphi_{uuzzz}\,,\qquad \varphi_{ua_4}=\varphi_{uabcd}\,.
\ee 
As already remarked, the Bondi-like gauge conditions actually imply that the field is traceless.

The solutions to the asymptotic Killing tensor condition may be conveniently labeled by the following numbers:
\begin{itemize}
	\item the number $p$ of ``$u$'' indices appearing,
	\item the number $d$ of ``$z$'' indices appearing without $\bar z$ counterpart,
	\item the number $c$ of pairs ``$z \bar z$'', counted ignoring their order.
\end{itemize}
Assuming a power-law dependence on $r$, the residual gauge symmetries that do not depend on $u$ admit then the following parametrization: 
\begin{align}\label{pd0}
\epsilon_{u_pz_d}=&- \frac{r^d D^d_zT_p(z, \bar z)}{\prod_{k=1}^d (s-p-k)}\,,\\
\label{c+1c}
\epsilon_{u_pz_{d+c+1}\bar z_{c+1}} =& - \frac{r^2}{2}\,\gamma_{z\bar z} 
\left( 
\epsilon_{u_pz_{d+c}\bar z_{c}} 
- 
2\, \epsilon_{u_{p+1}z_{d+c}\bar z_{c}}
\right),
\end{align}
where $T_p(z, \bar z)$ for $p=0,\ldots,s-1$ is a set of angular functions satisfying
\be
\label{p+1p}
T_{p+1} =\frac{s-p}{s[s-(p+1)]}\,T_p + \frac{1}{[s-(p+1)]^2}\,D^zD_zT_p\,.
\ee
The details of this derivation are available in \cite{carlo_tesi}.

We thus note that the $T_p$ are actually determined recursively in terms of only one angular function $T_0(z, \bar z) \equiv T(z,\bar z)$. The corresponding, nonvanishing, gauge variations are, for $s = p + d$,
\be
\delta\varphi_{u_p z_d} = d\, D_{z}\epsilon_{u_pz_{d-z}} = -  \frac{d\,r^{d-1} D^{d}_zT_p}{\prod_{k=1}^{d-1} (s-p-k)}\,.
\ee
This concludes the presentations of the set of ``higher-spin supertranslations'', namely those asymptotic symmetries whose parameters do not depend on retarded time. To summarize, we have seen that they can be parametrized by a single angular function $T(z,\bar  z)$, in analogy with gravitational supertranslations and spin-one asymptotic symmetries. As we shall see in the next section, these symmetries will play a distinguished role in connection with the leading soft theorem. Furthermore, their identification as the proper higher-spin analogs of gravitational supertranslations will receive additional support in the analysis of the general solution to the asymptotic Killing equation, which we now turn to describe.

\subsection{Higher-spin superrotations and more}

In the remainder of this section, we shall explore the full set of asymptotic symmetries in the case of spin three. From now on, we will adopt the notation $x^i$ for generic angular coordinates on the celestial sphere, in terms of which the Bondi-like gauge conditions take the form:
\be\label{r+traces'}
\varphi_{rab}=0\,,\qquad \gamma^{ij}\varphi_{ija}=0\,,
\ee
together with the requirement that the normalized field components
\be\label{eqWeakest'}
\varphi_{uuu}\,,\qquad \frac{\varphi_{uui}}{r}\,,\qquad \frac{\varphi_{uij}}{r^2}\,,\qquad \frac{\varphi_{ijk}}{r^3}
\ee
scale as $\mathcal O(r^{-1})$ near future null infinity. As we have already seen in Chapter \ref{chap:spin1higherD}, this ``more covariant'' notation is better suited to the generalization of the present discussion to a higher-dimensional context.  
In fact, with this notation, it is easy and worthwhile to solve the asymptotic Killing equation dictated by requirement that \eqref{r+traces'} and \eqref{eqWeakest'} be preserved for a generic spacetime dimension 
\be
D=n+2\,,
\ee  
\emph{i.e.} when the coordinates $x^i$ take $n$ values, $i=1$, $2$,$\ldots$, $n$. However, one should keep in mind that \eqref{r+traces'} and \eqref{eqWeakest'} provide finite energy flux and specify consistent boundary data for the dynamical problem only when $D=4$, namely $n=2$.

Keeping in mind these considerations, let us quote here the solution to the asymptotic Killing equation and to the trace constraint $\eta^{ab}\epsilon_{ab}=0$, whose derivation is detailed in Appendix \ref{app:spin3symm}: 
\be\begin{aligned}\label{gaugepars3} 
\epsilon_{rr}&=
\frac{u^2D\cdot D\cdot K}{(n+1)(n+2)} 
-\frac{2u}{n+1} D\cdot \rho
+T\,,\\
\epsilon_{ur}&=
-\frac{ur\, D\cdot D\cdot K}{(n+1)(n+2)} 
+ \frac{r}{n+1}D\cdot \rho
-\frac{u[\Delta+2(n+1)]}{(n+1)(n+2)} D\cdot \rho 
+ \frac{\Delta+2(n+1)}{2(n+2)}T\,,\\
\epsilon_{uu}&=
\frac{r^2D\cdot D\cdot K}{(n+1)(n+2)}
+\frac{2r\,[\Delta+2(n+1)]}{(n+1)(n+2)} D\cdot \rho
+\frac{(\Delta+n)[\Delta+2(n+1)]}{2n(n+2)} T\,,\\
\epsilon_{ri}&=
-\frac{ur^2D\cdot K_i}{n+2}
+r^2\rho_i
+\frac{u^2r\, D_iD\cdot D\cdot K}{2(n+1)(n+2)}
-\frac{ur}{n+1}D_iD\cdot\rho
+\frac{r}{2}D_iT\,,\\
\epsilon_{ui}&=
\frac{r^3}{n+2}D\cdot K_i
-\frac{ur^2 D_iD\cdot D\cdot K}{2(n+1)(n+2)}
+\frac{r^2}{n+2}\left[\frac{n}{n+1}D_iD\cdot\rho+(\Delta+2n+1)\rho_i\right]\\
&-\frac{ur}{(n+1)(n+2)}D_i[\Delta+2(n+1)]D\cdot\rho
+\frac{r}{2(n+2)}[\Delta+2(n+1)]T\,,\\
\epsilon_{ij}&=
r^4 K_{ij}
-\frac{ur^3}{n+2}\left[
D_{(i}D\cdot K_{j)}-\frac{2\gamma_{ij}}{n+1}D\cdot D\cdot K
\right]
+r^3 \left[
D_{(i}\rho_{j)}-\frac{2\gamma_{ij}}{n+1}D\cdot \rho
\right]\\
&+\frac{u^2r^2}{2(n+1)(n+2)}(D_iD_j+2\gamma_{ij})D\cdot D\cdot K
\\
&-\frac{ur^2}{n+1}\left[
D_iD_j - \frac{\gamma_{ij}}{n+2}(\Delta-2)
\right]D\cdot \rho
+\frac{r^2}{2}\left[
D_iD_j - \frac{\gamma_{ij}}{n+2}(\Delta-2)
\right]T\,,
\end{aligned}\ee
while the corresponding nonvanishing symmetry variations read
\be\begin{aligned}
\delta_\epsilon\varphi_{uui}&=
\frac{1}{2n(n+2)}D_i(\Delta+n)[\Delta+2(n+1)]T\,,\\
\delta_\epsilon\varphi_{uij}&=
-\frac{2ur}{(n+2)(n+2)}(D_iD_j+\gamma_{ij})[\Delta+2(n+1)]D\cdot\rho
\\
&+\frac{r}{n+2}\left(
D_iD_j-\tfrac{1}{n}\gamma_{ij}\Delta
\right)
[\Delta+2(n+1)]T\,,\\
\delta_\epsilon\varphi_{ijk}&=
\frac{u^2r^2}{2(n+1)(n+2)}\left[D_{(i}D_j D_{k)}+4\gamma_{(ij}D_{k)}\right]D\cdot D\cdot K
\\
&-\frac{ur^2}{d+1}\left[
D_{(i}D_jD_{k)}-\frac{1}{n+2}\gamma_{(ij}D_{k)}[3\Delta+2(n-1)]
\right]D\cdot\rho\\
&+\frac{r^2}{2}\left[
D_{(i}D_jD_{k)}-\frac{1}{n+2}\gamma_{(ij}D_{k)}[3\Delta+2(n-1)]
\right]T\,.
\end{aligned}\ee
The key-point is that the full residual symmetry is parameterized by the tensors 
\be\label{spin3tensors5}
T(x^k)\,,\qquad
\r^i(x^k)\,,\qquad
K^{ij}(x^k)\,,
\ee
defined on the celestial sphere, bound to satisfy a set of constraints that we now turn to illustrate and interpret.

The function $T(x^i)$ is completely arbitrary and specifies the family of asymptotic symmetries we have already identified as spin-three supertranslations above. Furthermore, the subset of these transformations that do not change the field, \emph{i.e.} spin-three translations, is characterized by the equation (see Appendix \ref{app:spin3exact})
\be\label{l=012}
D_{(i}D_j D_{k)}T-\frac{1}{n+2}\gamma_{(ij}\left[ D_{k)}\Delta T + 2\Delta D_{k)}T \right]=0\,,
\ee 
whose solutions are given by the spherical harmonics (see Appendix \ref{app:Laplaciano}) with $l=0,1,2$. The solution space thus contains $1+(n+1)+\left[\tfrac{(n+2)(n+1)}{2}-1\right]=\tfrac{(n+1)(n+4)}{2}$ independent solutions. 

If we want to provide a tentative interpretation of this kind of symmetry in terms of generators of an underlying algebra, if any, we may draw the following analogy. In the case of the $BMS$ algebra, which can be viewed as an infinite-dimensional enhancement of the Poincar\'e algebra, supertranslations arise as an enlarged version of translation symmetry, associated to the Poincar\'e generator $P^\mu$. In the spin-three case, it is then natural to interpret these generalized supertranslations as the asymptotic symmetry analog of the exact Killing symmetry corresponding to the traceless projection of $P^{(\mu}P^{\nu)}$. Indeed, the number of these generators, $\tfrac{(n+2)(n+3)}{2}-1=\tfrac{(n+1)(n+4)}{2}$, matches the number of independent solutions of \eqref{l=012}: in particular, $9$ generators in the relevant case of four dimensions. 

Turning our attention to the other tensors identified in \eqref{spin3tensors5}, let us observe that
the tensor $K^{ij}(x^k)$ must have vanishing trace
\be
\gamma^{ij}K_{ij}=0
\ee 
and must satisfy the conformal Killing tensor equation
\be\label{confKillt2}
D^{(i}K^{jk)}-\frac{2}{n+2}\, \gamma^{(ij}D\cdot K^{k)}=0\,.
\ee
These equations take the following particularly simple form in stereographic coordinates $z$, $\bar z$ in four spacetime dimensions
\be
K^{z\bar z}=0\,,\qquad
\partial_{\bar z}K^{zz}=0\,,
\qquad
\partial_z K^{\bar z \bar z}=0\,,
\ee
and hence locally admit two infinite-dimensional family of solutions expressed as holomorphic and anti-holomorphic functions
\be
K^{z\bar z}=0\,,\qquad
K^{zz}=K(z)\,,\qquad
K^{\bar z \bar z}=\tilde K(\bar z)\,.
\ee
Thus, these tensors are the natural spin-three analogs of the conformal Killing vectors parametrizing ordinary gravitational superrotations, which correspond to the enhancement of rotations and boosts generated by $M_{\mu\nu}$ at the level of finite transformations. This identification as spin-three superrotations can be further supported by counting the number of independent \emph{global} solutions to equation \eqref{confKillt2}. Consider for instance the Laurent expansion of $K=K^{zz}\partial_z^2$ near $z=0$
\be
K^{zz}\partial_z^2=K(z)\partial_z^2 = \sum_{n\in\mathbb Z} K_n z^n \partial_z^2\,.
\ee 
For this tensor to be regular, we need to restrict the sum to  nonnegative values of $n$, namely $K_n=0$ for $n\le-1$. Performing now the change of variable $z=1/w$, we have
\be
K^{zz}\partial_z^2=\sum_{n\in\mathbb Z}K_n w^{4-n}\partial_w^2\,,
\ee
from which $K_n=0$ for $n\ge5$ by regularity near $w=0$. This leaves us with five independent coefficients $K_0$, $K_1$, $K_2$, $K_3$, $K_4$, plus the additional five arising from the anti-holomorphic sector $\tilde K_0$, $\tilde K_1$, $\tilde K_2$, $\tilde K_3$, $\tilde K_4$ and hence a total of ten independent global spin-three superrotations. This is precisely the number of components of the traceless projection of $M^{\mu(\nu}M^{\rho)\sigma}$, the generators of the exact spin-three Killing symmetries corresponding to the (traceless projections of symmetrized) products of two rotations/boosts.

Finally, the vector $\rho^i$ must satisfy 
\be\label{confrho2}
D^{(i}D^j \rho^{k)}-\frac{2}{n+2}\,\left[\gamma^{(ij}\Delta \rho^{k)}+\gamma^{(ij}\{D^{k)},D^l\}\rho_l\right]=0\,,
\ee
where the curly brackets denote the anti-commutator. Again, going to stereographic coordinates $z$, $\bar z$ in four spacetime dimensions allows us to recast this equation in the form
\be
\partial_{\bar z}(\gamma^{z\bar z}\partial_{\bar z}\rho^{z})=0\,,\qquad
\partial_{z}(\gamma^{z\bar z}\partial_{z}\rho^{\bar z})=0\,,
\ee 
which admits solutions expressed in terms of holomorphic and anti-holomorphic functions
\be
\rho^{z}=a(z)\partial_z \mathcal K(z,\bar z)+b(z)\,,\qquad
\rho^{\bar z}=\tilde a(\bar z)\partial_{\bar z}\mathcal K(z,\bar z)+\tilde b(\bar z)\,,
\ee
where $\mathcal K(z,\bar z)$ is a K\"ahler potential for the metric on the unit sphere, \emph{e.g.} $\mathcal K(z,\bar z)=2\log(1+z\bar z)$. We may identify this infinite-dimensional family as the one corresponding to ``mixed'' supertranslations/superrotations, namely the infinite-dimensional enhancement of the symmetries generated by the traceless part of $P^{(\mu}M^{\nu)\rho}$. Indeed, the number of independent generators of this form is 16, while on the other hand
\be \begin{aligned}
\rho^z\partial_z = 
\left[
a(z)\frac{2z}{1+z\bar z}+b(z)\right]\partial_z&=
\sum_{n\in\mathbb Z}\left[\frac{2a_n z^{n+1}}{1+z\bar z}+b_n z^n\right]\partial_z\\
&=
-\sum_{n\in\mathbb Z}\left[\frac{2a_n w^{2-n}\bar w}{1+w\bar w}+b_n w^{2-n}\right]\partial_w
\end{aligned}
\ee
so that the global regularity of $\rho$ allows for $a_{-1}$, $a_0$, $a_1$, $a_2$, $a_3$, $b_0$, $b_1$ and $b_2$, together with the corresponding anti-holomorphic coefficients: precisely 16 independent global transformations.

To summarize, we have identified three families of spin-three asymptotic symmetries, and put them in one-to-one correspondence with elements of the universal enveloping algebra of the Poincar\'e algebra given by the traceless projections of the combinations $P^{(\mu}P^{\nu)}$, $P^{(\mu}M^{\nu)\rho}$ and $M^{\mu(\nu}M^{\rho)\sigma}$. 
These products are expected to be identified with the spin-three generators of a would-be higher-spin algebra, if any, with Poincar\'e subalgebra (see \emph{e.g.}\ \cite{flat-algebras_1,flat-algebras_2} for  discussions on higher-spin algebras possibly related to four-dimensional Minkowski space). The asymptotic symmetries generated by $T$, $\r^{i}$ and $K^{ij}$ can thus be interpreted as the infinite-dimensional enhancement of the corresponding global Killing symmetries.

\section{Weinberg's Soft Theorem as a Ward Identity for Any Spin}

We shall now address the relation between the asymptotic symmetries of higher spins that we have described so far and Weinberg's soft theorem, which we now briefly recall.
As we discussed in Chapter \ref{chap:obs}, Weinberg considered the $S$-matrix element $S_{\beta\alpha} (\mathbf q)$, for arbitrary asymptotic particle states  $\alpha \rightarrow \beta$, also involving an extra soft massless particle with momentum $q^{\, \mu}\equiv (\omega, \mathbf q) \to 0$ and helicity $s$. The leading contribution to this process takes a particularly simple factorized form that, in the notation of \cite{Weinberg_64, Weinberg_65}, can be written as
\be \label{wfact}
\lim_{\omega\to0^+}\omega\, S_{\beta \alpha}^{\pm s}(\mathbf q) = 
-\lim_{\omega\to0^+}\left[\, \omega\sum_i \eta_i^{\phantom{(s)}}\!\!\! g_i^{(s)} \frac{(p_i \cdot \varepsilon^{\pm}(\mathbf q))^s}{p_i\cdot q}\,\right]S_{\beta\alpha} \,,
\ee
with $\eta_i$ being $+1$ or $-1$ according to whether the particle $i$ is incoming or outgoing ($g_n^{(s)}$ is the spin-$s$ coupling of the $n$th particle while $\varepsilon^\pm_a$ is the polarization tensor of the soft particle).

In the case of higher spins, the constraints imposed by Weinberg's theorem are in contradiction with a nontrivial interaction in the soft regime, thereby implying the absence of a long-range force associated to higher-spin quanta, if any.
Nonetheless, it can be insightful to see whether or not this theorem can be associated to the presence of an underlying large gauge symmetry, as was the case for the less exotic spin $1$ and $2$ theories we considered in the previous chapters.

Our aim is now to derive a Ward identity associated to spin-three supertranslation symmetry and show that it is equivalent to the corresponding Weinberg theorem, confining our attention to the case of four spacetime dimensions. In order to do this, let us first note that the only nonvanishing contribution to the Noether (surface) charge comes from $\delta\varphi_{zzz}$, and reads
\be \label{Q+spin3}
Q^+ = \frac{3}{4}\int_{\scrip} \gamma_{z\bar z}\,\partial_u \left[(D^z)^3 B_{zzz} + \text{c.c.}\right]  T(z,\bar z) d^2z du - \frac{3}{2}\int_{\scrip} \gamma_{z\bar z}\, J(u, z, \bar z) d^2z du \,,
\ee
where
\be
J(u, z, \bar z) \equiv \lim_{r\to\infty}r^2 J^{rrr}(u, z, \bar z) \, .
\ee
The surface charge thus computed is in agreement, in particular, with that obtainable from the results of \cite{HS-charges-cov}.

Under the assumption that the residual symmetry generators act on  scalar matter fields as follows, 
\be
[Q^+, \Phi] = \frac{3}{2}\, g^{\, (3)}_i T(i\partial)^2_u \Phi \,,
\ee
where $g^{\, (3)}_i$ is the coupling of the corresponding matter, in the frequency domain we get
\be
\langle\text{out}| (Q^+ S - S Q^-) |\text{in}\rangle = \frac{3}{2}\sum_{i} \eta_{i}^{\phantom{(3)}}\!\!\! g^{(3)}_i E^2_i T (z_i, \bar z_i) \langle\text{out}| S |\text{in}\rangle \, .
\ee
In addition, we shall impose the following auxiliary boundary condition at $\scri^\pm_\mp$ 
\be
(D^z)^3B_{zzz} = (D^{\bar z})^3 B_{\bar z\bar z \bar z} \,.
\ee
This condition can be interpreted as the higher-spin analog of the condition that no long-range magnetic fields be present on $\mathscr I^+$.
We also leave aside the $J$ term, which acts trivially on the vacuum, thus obtaining
\be
Q^+ = \frac{3}{2} \int_{\scrip} T(z, \bar z) \partial_u (D^z)^3 B_{zzz} \gamma_{z\bar z}\, d^2z du\, .
\ee
An analogous result holds for $Q^-$. For the function $T(z, \bar z)$ we choose 
\be
T(z, \bar z) = \frac{1}{w-z}\left(\frac{1+w\bar z}{1+z\bar z}\right)^{\!2},
\ee
so that, after an integration by parts in $\partial_{\bar z}$, the computation of the charge involves 
\be
\partial_{\bar z} \left( \frac{1}{w-z}\left(\frac{1+w\bar z}{1+z\bar z}\right)^{\!2}\right) =
-2\pi \delta^2(z-w) + \frac{1}{2}\,\gamma_{z\bar z}\, \frac{1+w\bar z}{1+z\bar z}\, .
\ee
Therefore
\be
Q^+ = 3\pi \int du D^w D^w B_{www} - \frac{3}{4} \int D^zD^z B_{zzz} \gamma_{z\bar z}\, \frac{1+w\bar z}{1+z\bar z}\, d^2z du\, ,
\ee
where in particular the last term is a vanishing boundary contribution.
To sum up:
\be\label{Ward_s=3}\begin{aligned}
	&2\pi (D^z)^2\langle\text{out}|\!\left[\left(\int\! du \partial_u B_{zzz} \right) S - S \left( \int\! dv \partial_v B_{zzz} \right)\right]\!|\text{in}\rangle \\
	&=
	\sum_{i} \eta_{i}\,  \frac{g^{(3)}_i E^2_i}{z-z_i}\left(\frac{1+z\bar z_i}{1+z_i\bar z_i}\right)^{\!2} \langle\text{out}| S |\text{in}\rangle \,.
\end{aligned}\ee
The large-$r$ limit performed for a free spin-three field in Fourier space gives, to leading order, 
\be
B_{zzz} = - \frac{i}{8\pi^2} \frac{2^{3/2}}{(1+z\bar z)^3}\int_0^{+\infty}
d\omega_{\mathbf q} \left[ a^{\text{out}}_+(\omega_{q}\hat x) e^{-i\omega_{\mathbf q}u}
- a^{\text{out}\dagger}_-(\omega_{q}\hat x) e^{i\omega_{\mathbf q}u}\right]
\ee
so that
\be\begin{aligned}
\int du\, \partial_u B_{zzz} 
= - \frac{1}{8\pi} \frac{ 2^{3/2}}{(1+z\bar z)^3} \lim_{\omega\to 0^+}\left[ \omega a^{\text{out}}_+(\omega \hat x) + \omega a^{\text{out}\dagger}_-(\omega \hat x)\right] .
\end{aligned}\ee
Thus, using crossing symmetry, we also have 
\be\begin{aligned}
&-4\pi \langle\text{out}|\!\left[\left(\int\! du \partial_u B_{zzz} \right) S - S \left( \int\! dv \partial_v B_{zzz} \right)\right]\!|\text{in}\rangle \\
&= \frac{2^{3/2}}{(1+z\bar z)^3} \lim_{\omega\to 0^+}\langle \text{out}|\,\omega\, a^{\text{out}}_+(\omega \hat x) |\text{in}\rangle \,,
\end{aligned}\ee
and this implies, by comparing with \eqref{Ward_s=3},
\be
\lim_{\omega\to 0^+}\langle \text{out}|\,\omega\, a^{\text{out}}_+(\omega \hat x) |\text{in}\rangle= -\lim_{\omega\to0^+} \sqrt{2} (1+z\bar z)\sum_i \eta_i g_i\, \frac{E^2_i\, (\bar z - \bar z_i)^2}{(z-z_i)(1+z_i \bar z_i)^2} \,,
\ee
since
\be
(D^z)^2 \frac{4}{(1+z\bar z)^2}\sum_i \eta_i g_i\, \frac{E^2_i(\bar z - \bar z_i)^2}{(z-z_i)(1+z_i \bar z_i)^2} = 2 \sum_i \eta_i g_i\, \frac{E^2_i(1+z\bar z_i)^2}{(z-z_i) (1+z_i \bar z_i)^2} \,.
\ee 
This shows that the Ward identity of the residual spin-three gauge symmetry implies Weinberg's formula \eqref{WeinbergPSPACE}.

We shall now sketch the generalization of these steps to the case of an arbitrary integer spin $s$.
The relevant contribution to the Noether current in this case is given by
\be
\delta \varphi_{z\ldots zz} = - \frac{s\,r^{s-1}}{(s-1)!}\,D^s_z T \, .
\ee
Using the auxiliary boundary condition $(D^z)^sB_{z\ldots zz} = (D^{\bar z})^sB_{\bar z\ldots\bar z \bar z}$ and integrating by parts, the charge corresponding to our family of large gauge transformation is therefore
\be
Q^+ = (-1)^s\frac{s}{2(s-1)!}\int_{\scrip} \partial_{\bar z} T (D^z)^{s-1}\partial_u B_{z\ldots zz} d^2z du - \frac{s}{2} \int_{\scrip} \gamma_{z\bar z} J(u, z, \bar z)d^2z du \,.
\ee
Choosing
\be
T(z, \bar z) = \frac{1}{w-z}\left(\frac{1+w\bar z}{1+z\bar z}\right)^{\!s-1}
\ee
yields
\be\begin{split}
	&-4\pi\, \frac{(-1)^s}{(s-1)!}(D^z)^{s-1}\langle\text{out}|\!\left[\left(\int\! du \partial_u B_{z\ldots zz}\right) S - S \left( \int\! dv \partial_v B_{z\ldots zz}\right) \right]\!|\text{in}\rangle
	\\
	& = \sum_{i} \eta_i\, \frac{g_i^{(s)}E_i^{s-1}}{z-z_i}\left( \frac{1+z\bar z_i}{1+z_i\bar z_i} \right)^{\!s-1} \langle \text{out} | S | \text{in}\rangle\,,
\end{split}\ee
where we have used the action 
\be
[Q^+, \Phi] = \frac{s}{2}\,g_i^{(s)}T(i\partial_u)^{s-1}\Phi
\ee
on matter fields.
The $r\to\infty$ limit gives to leading order
\be\begin{split}
	&-4\pi\, \langle\text{out}|\!\left[\left(\int\! du \partial_u B_{z\ldots zz}\right) S - S \left( \int\! dv \partial_v B_{z\ldots zz}\right) \right]\!|\text{in}\rangle \\
	&= \frac{2^{s/2}}{(1+z\bar z)^s}\lim_{\omega\to0^+} \left[\,\omega\langle \text{out} | a^{\text{out}}_+S | \text{in}\rangle \right]
\end{split}\ee
and hence
\be
\lim_{\omega\to0^+} \left[\,\omega\langle \text{out} | a^{\text{out}}_+S | \text{in}\rangle \right] = 
(-1)^s 2^{s/2-1} (1+z\bar z) \sum_i \eta_i\, \frac{g_i^{(s)}E_n^{s-1}}{z-z_i} \left(\frac{\bar z - \bar z_i}{1+z_i \bar z_i}\right)^{\!s-1} ,
\ee
where we have employed the following identity
\be\begin{aligned}
	&\frac{1}{(s-1)!}\,(D^z)^{s-1} \frac{2^{s-1}}{(1+z\bar z)^{s-1}} \sum_i \eta_i\, \frac{g_i^{(s)}E_i^{s-1}}{z-z_i} \left(\frac{\bar z - \bar z_i}{1+z_i \bar z_i}\right)^{\!s-1} \\
	&=  \sum_{i} \eta_i\, \frac{g_i^{(s)}E_i^{s-1}}{z-z_i}\left( \frac{1+z\bar z_i}{1+z_i\bar z_i} \right)^{\!s-1} \langle \text{out} | S | \text{in}\rangle \,. 
\end{aligned}\ee
Thus, Weinberg's factorization can be understood as a manifestation of an underlying spin-$s$ large gauge symmetry acting on the null boundary of Minkowski spacetime, namely the set of higher-spin supertranslations we discussed in the previous section.

\section{Higher Spins in Arbitrary Dimensions}

So far, the discussion of higher-spin asymptotic symmetries and of their link to the soft theorem has essentially focused on the case of four spacetime dimensions. This is due to the fact that, although the conditions \eqref{r+traces'} and \eqref{eqWeakest'} can be in principle adopted for any $D=n+2$, where they indeed give rise to infinite-dimensional asymptotic symmetries, they only ensure the finiteness of the energy flux and the compatibility with the equations of motion in $D=4$.

In this section, we shall consider the dynamics of a higher-spin field on Minkowski background of arbitrary dimension and identify a set of gauge and boundary conditions at null infinity that are compatible with these two requirements. According to the terminology already discussed in Chapter \ref{chap:spin1higherD}, we will thus adopt the so-called ``radiation falloffs'' as the leading falloffs of our solution space, $\varphi\sim r^{-n/2}$. 

As we shall see, in analogy with the situation discussed in the context of Yang-Mills theory in Chapter \ref{chap:spin1higherD}, adopting radiation falloffs in higher dimensions effectively trivializes asymptotic symmetries. In other words, this choice selects the global Killing symmetries as solutions of the asymptotic Killing tensor equation.
Therefore, the analysis will only allow us to calculate and discuss the surface charges associated to these global higher-spin Killing symmetries, and not an infinite-dimensional family of surface charges, in contrast with the four-dimensional case. 

However, the lower-spin examples suggest a possible way out of the seeming lack of infinite-dimensional asymptotic symmetries in higher dimensions, which would leave open the question regarding the existence of a symmetry underlying Weinberg's theorem. According to the strategy discussed in Section \ref{sec:poly}, one may first establish the physical properties of the solution space characterized by radiation falloffs, by ensuring that they give rise to finite fluxes and global charges, and then act on this space by means of transformations \eqref{gaugepars3}, which preserve the equations of motion. By construction, the wider solution space thus obtained shares all observables with the former one, but it also possesses infinite-dimensional symmetries given precisely by \eqref{gaugepars3} \cite{highspinsrel-io}.

%%%%%%%%%%%%%%%%%%%%%%%%%%%%%%%%%%%%%%%%%%
\subsection{Bondi-like gauge, reloaded}
%%%%%%%%%%%%%%%%%%%%%%%%%%%%%%%%%%%%%%%%%%

Restricting for the moment to the spin-three case, we shall impose the conditions 
\be \label{bondi3}
\vf_{rab} = 0 \,, \qquad \g^{ij} \vf_{ija} = 0\,.
\ee
Following the same logic as in the four-dimensional case, one can substitute a power-law ansatz in the equations of motion and realize that they are satisfied to leading order provided that 
\be \label{boundary3}
\vf_{uuu}  \, , \qquad
\frac{\vf_{uui}}{r}  \, , \qquad
\frac{\vf_{uij}}{r^2}  \, , \qquad
\frac{\vf_{ijk}}{r^3}  \, 
\ee
display the asymptotic behavior of radiation, namely $\mathcal O(r^{-\frac{n}{2}})$. 
  
We shall analyze separately even and odd spacetime dimensions, due to the different nature of radiation and Coulombic terms in the asymptotic expansion in these two cases: the former are associated to half-integer powers of the radial coordinate, as already clearly displayed by the previous equation, while the latter occur at the, integer, order $r^{1-n}$. Due to their peculiarities, also the cases of $D=3$ and $D=4$ will be discussed in a separate section.

Although technically and conceptually very simple, a step-by-step  solution of the equations of motion would be quite lengthy. For this reason, we will only quote here the main features and outcomes of their discussion, referring to \cite{super-io, Cariche-io} for a more detailed derivation.

%%%%%%%%%%%%%%%%%%%%%%%%%%%%%%%%%%%%%%%%%%
\subsubsection{Even spacetime dimension}
%%%%%%%%%%%%%%%%%%%%%%%%%%%%%%%%%%%%%%%%%%

When $n$ is even we consider the following ansatz for the fields in the Bondi gauge \eqref{bondi3}:
\begin{subequations} \label{ansatz3}
	\begin{alignat}{5}
	\vf_{uuu} & = \sum_{l\,=\,0}^\infty r^{-\frac{n}{2}-l} B^{(l)}(u,x^m) \, , \qquad & 
	\vf_{uui} & = \sum_{l\,=\,0}^\infty r^{-\frac{n}{2}-l+1} U^{(l)}_i(u,x^m) \, , \\
	\vf_{uij} & = \sum_{l\,=\,0}^\infty r^{-\frac{n}{2}-l+2} V^{(l)}_{ij}(u,x^m) \, , \qquad &
	\vf_{ijk} & = \sum_{l\,=\,0}^\infty r^{-\frac{n}{2}-l+3} C^{(l)}_{ijk}(u,x^m)  \, ,
	\end{alignat}
\end{subequations}
with $\g^{ij}V_{ij}^{(l)} = \g^{ij} C_{ijk}^{(l)} = 0$.
This choice is mainly motivated by the observation that, as already anticipated, a solution of this type gives rise to a finite and generically nonzero energy flux through null infinity:
\be \label{energy-flux3}
\cP(u) = \lim_{r\to \infty} \int_{S_u} \left(T_{uu}-T_{ur}\right) r^n d\Omega =
\int_{S_u} \gamma^{i_1j_1}\gamma^{i_2 j_2}\gamma^{i_3 j_3}\,\partial_u{C}^{(0)}_{i_1i_2i_3} \partial_u{C}^{(0)}_{j_1 j_2j_3} d\Omega \, ,
\ee
where $T_{ab}$ denotes the energy-momentum tensor of the solution, which can be therefore interpreted as a spin-three wave reaching null infinity.

Substituting the ansatz \eqref{ansatz3} in \eqref{Fronsdaleq}, one obtains
\begin{subequations} \label{sol3}
	\begin{alignat}{5}
	B^{(k)} & = \frac{2(n+2k-2)}{(n+2k)(n-2k-2)}\, \Ddot U^{(k)} & \quad \textrm{for}\ k & \neq \frac{n-2}{2} \, , \label{solB3}\\[10pt]
	U^{(k)}_i & = \frac{2(n+2k)}{(n+2k+2)(n-2k)}\, \Ddot V^{(k)}_i \quad & \textrm{for}\ k & \neq \frac{n}{2} \, , \label{solU3}\\[10pt]
	V^{(k)}_{ij} & = \frac{2(n+2k+2)}{(n+2k+4)(n-2k+2)}\, \Ddot C^{(k)}_{ij} \quad & \textrm{for}\ k & \neq \frac{n+2}{2} \, . \label{solV3}
	\end{alignat}
\end{subequations}
In the cases excluded from the previous formulas the coefficients  of, respectively, $B^{(\frac{n-2}{2})}$, $U^{(\frac{n}{2})}$ and $V^{(\frac{n+2}{2})}$ vanish and the equations of motion imply instead
\be \label{constrC3}
(n-2)\, \Ddot\Ddot\Ddot C^{(\frac{n-2}{2})}_{\phantom{i}} = 0 \, , \qquad
\Ddot\Ddot C^{(\frac{n}{2})}_i = 0 \, , \qquad
\Ddot C^{(\frac{n+2}{2})}_{ij} = 0 \, .
\ee
Note in particular the factor $(n-2)$ in the first constraint, which shows that $C^{(0)}$ actually remains arbitrary even when $n = 2$, namely in four spacetime dimensions. Cancellations of this type are ubiquitous in this type of analysis, exhibiting the peculiarity of the $D=4$ case.

Substituting the same ansatz in the equation $\cF_{ijk} = 0$ one obtains, for $l \neq \frac{n+2}{2}$,
\be \label{solC3}
\begin{split}
	\partial_u{C}^{(l+1)}_{ijk} & = -\, \frac{1}{2(l+1)}\, \bigg\{ \left[\, \Delta - \frac{n(n-2)}{4} + l(l+1) - 3 \,\right] C^{(l)}_{ijk} \\
	& - \frac{4}{(n+2l+4)(n-2l+2)} \left[\, (n+2)\, D_{(i}\Ddot C_{jk)}{}^{\!\!(l)} - 2\, \g_{(ij} \Ddot\Ddot C_{k)}{}^{\!\!(l)} \,\right] \bigg\} \, .
\end{split}
\ee
The value of $l$ excluded from \eqref{solC3} shows that, from there on, the tensors $C^{(l)}$ also depend on $V^{(\frac{n+2}{2})}$,
\be \label{solC3-special}
\partial_u{C}^{(\frac{n+4}{2})}_{ijk} = \frac{1}{n+4}\, \bigg\{ D^{\phantom{(\frac{n}{2})}}_{(i} \!\!\!\! V^{(\frac{n+2}{2})}_{jk)} - \frac{2}{n+2}\, \g_{(ij}^{\phantom{(\frac{n}{2})}}\!\! \Ddot V_{k)}^{(\frac{n+2}{2})} - \left( \Delta + 2n - 1 \right) C^{(\frac{n+2}{2})}_{ijk} \bigg\} \,.
\ee

The $u$-evolution of $B^{(\frac{n-2}{2})}$, $U^{(\frac{n}{2})}$ and $V^{(\frac{n+2}{2})}$ is fixed instead by the equations $\cF_{uab} = 0$ (with $a, b \neq r$) as
\begin{subequations} \label{u3}
	\begin{align}
	B^{(\frac{n-2}{2})} & = \cM - \frac{n-3}{6(n+1)n} \int_{-\infty}^u \!\!du'\!\left( \Delta - n + 2 \right) \Ddot\Ddot\Ddot C^{(\frac{n-4}{2})} ,\label{uB3} \\[10pt]
	U_i^{(\frac{n}{2})} & = \cN_i + \frac{u}{n+2}\, \pr_i \cM - \frac{n-1}{2(n+1)(n+2)} \int_{-\infty}^u \!\!du'\! \left( \Delta - 1 \right) \Ddot\Ddot C_i^{(\frac{n-2}{2})} \nn \\
	& - \frac{n-3}{6n(n+1)(n+2)} \int_{-\infty}^u \!\!du'\! \int_{-\infty}^{u'} \!\!du''\, D_i \left( \Delta - n + 2 \right) \Ddot\Ddot\Ddot C^{(\frac{n-4}{2})} \, , \label{uU3} \\[10pt]
	\nonumber
	V_{ij}^{(\frac{n+2}{2})} & = \cP_{ij} + \frac{u}{n+3} \left( D_{(i\,} \cN_{j)} - \frac{2}{n}\, \g_{ij} \Ddot \cN \right) \\ &+\frac{u^2}{2(n+2)(n+3)} \left( D_{(i} D_{j)} \cM - \frac{2}{n}\, \g_{ij}\, \Delta \cM \right) \nn \\
	& - \frac{n+1}{(n+2)(n+3)} \int_{-\infty}^u \!\!du'\! \left( \Delta + n - 2 \right) \Ddot C_{ij}^{(\frac{n}{2})} + \cdots \, ,\label{uV3}
	\end{align}
\end{subequations}
where $\cM$, $\cN_i$ and $\cP_{ij}$ are arbitrary functions and tensors on the celestial sphere.

The omitted terms in the last equation correspond to the multiple integrals in the retarded time that one obtains by integrating the differential equation
\be \label{diff-eq-V}
\partial_u{V}_{ij}^{(\frac{n+2}{2})} = \frac{1}{n+3} \left\{ D_{(i}^{\phantom{(\frac{n}{2})}}\!\!\!\! U_{j)}^{(\frac{n}{2})} - \frac{2}{n}\, \g_{ij}^{\phantom{(\frac{n}{2})}}\!\!\!\! \Ddot U^{(\frac{n}{2})} - \frac{n+1}{n+2} \left( \Delta + n - 2 \right) \Ddot C_{ij}^{(\frac{n}{2})} \right\} 
\ee
given by the equation $\cF_{uij} = 0$. At any rate, in Section \ref{sec:symm3} we shall show that they do not contribute to the linearized charges.

We anticipate that the calculation of the surface charges will essentially boil down to determining their precise dependence on the ``integration constants'' or, more precisely, ``integration functions''
$\cM$, $\cN_i$ and $\cP_{ij}$, which admit an arbitrary dependence on the coordinates on the sphere at null infinity and are indeed needed to specify a solution to the recursion relations dictated by the equations of motion. They all appear at order $r^{n-1}$ in the expansions \eqref{ansatz3} and enter \eqref{u3} in combinations with a fixed polynomial dependence on the retarded time $u$.

For all other powers of $1/r$, the equations $\cF_{uab} = 0$ (with $a, b \neq r$) reduce to divergences of \eqref{solC3} and are therefore identically satisfied. The divergences of \eqref{solC3} also imply the constraints \eqref{constrC3}. As a result, the latter do not impose any further condition on the $C^{(l)}$ with lower values of $l$. 

Let us stress once again that some of the above conclusions are only valid for $n > 2$ due to a number of cancellations that we have highlighted in the four-dimensional case. Below we will see how these results need to be emended in the four- and three-dimensional cases.

%%%%%%%%%%%%%%%%%%%%%%%%%%%%%%%%%%%%%%%%%%
\subsubsection{Odd spacetime dimension}
%%%%%%%%%%%%%%%%%%%%%%%%%%%%%%%%%%%%%%%%%%

When $n$ is odd one has to add further terms to the ansatz \eqref{ansatz3} in order to obtain nontrivial asymptotic charges at null infinity. The necessity of this completion is clearly displayed by the fact that such charges are obtained by integrating a quantity linear in the fields over a sphere with measure element scaling as $r^{n}$, while the radiation ansatz only involves half-odd powers of $1/r$. In other words, one must add Coulombic contributions to the asymptotic series. We therefore consider the ansatz
	\begin{alignat}{5}\label{ansatz3-odd}
	\vf_{uuu} & = \vf_{uuu}[B] + \sum_{l\,=\,0}^\infty r^{1-n-l} \tilde{B}^{(l)} \, , \qquad & 
	\vf_{uui} & = \vf_{uui}[U] + \sum_{l\,=\,0}^\infty r^{1-n-l} \tilde{U}^{(l)}_i \, , \\
	\vf_{uij} & = \vf_{uij}[V] + \sum_{l\,=\,0}^\infty r^{1-n-l} \tilde{V}^{(l)}_{ij} \, , \qquad &
	\vf_{ijk} & = \vf_{ijk}[C] + \sum_{l\,=\,0}^\infty r^{1-n-l} \tilde{C}^{(l)}_{ijk}  \, ,
	\end{alignat}
where $\vf_{uuu}[B]$ and so on denote the terms introduced in \eqref{ansatz3}, which are still necessary if one desires to describe radiation, that is if one wishes to have a nonvanishing energy flux through null infinity (which is still given by \eqref{energy-flux3}). The new contributions to the expansion of the field components satisfy $\g^{ij} \tilde{V}_{ij}^{(l)} = \g^{ij} \tilde{C}_{ijk}^{(l)} = 0$.

Since $n$ is odd, all factors entering the expansion of the equations of motion in powers of $\sqrt{r}$ are different from zero. As a result, the tensors $B^{(l)}$, $U^{(l)}$ and $V^{(l)}$ satisfy the same conditions as in \eqref{sol3}, but without any constraint on the allowed values of $l$. Similarly, the tensors $C^{(l)}$ satisfy \eqref{solC3} for any $l$. The tensors appearing at the leading order of the new, Coulombic branches of our ansatz must satisfy
\begin{subequations} \label{u3-odd}
	\begin{align}
	\tilde{B}^{(0)} & = \cM \, , \qquad 
	\tilde{U}_i^{(0)} = \cN_i + \frac{u}{n+2}\, \pr_i \cM \, ,  \\[10pt]\nonumber
	\tilde{V}_{ij}^{(0)} & = \cP_{ij} + \frac{u}{n+3} \left( D_{(i\,} \cN_{j)} - \frac{2}{n}\, \g_{ij} \Ddot \cN \right) \\
	&+ \frac{u^2}{2(n+2)(n+3)} \left( D_{(i} D_{j)} \cM - \frac{2}{n}\, \g_{ij}\, \Delta \cM \right) , 
	\end{align}
\end{subequations}
on account of the equations of motion $\cF_{uab} = 0$ (with $a, b \neq r$). Notice the similarity with \eqref{u3}: the only difference is that, when $n$ is odd, there is no contribution from the data of the solution that encode radiation (here stored in the $\sqrt{r}$ branch). As we shall see below, the latter terms anyway do not contribute to the linearized charges. The subleading $\cO(r^{-n})$ terms in our ansatz do not contribute as well to the linearized charges. 

%%%%
\subsubsection{Three and four spacetime dimensions}\label{sec:3-4dim_3}
%%%%%%%%%%%%%%%%%%%%%%%%%%%%%%%%%%%%%%%%%%

When $n = 1$ or $n = 2$, as we have already anticipated, the previous analysis must be modified. We begin by revisiting the four-dimensional case, where, in the component $\vf_{uuu}$ the leading order of the radiation and of the Coulomb-like solutions now coincide. Moreover, the equation $\cF_{ruu} = 0$ does not impose any constraint on the triple divergence of $C^{(0)}$. As a result, the equations fixing the dependence on $u$ of $B^{(0)}$, $U^{(1)}$ and $V^{(2)}$ are slightly modified as follows:
\begin{subequations} \label{int4D-spin3}
	\begin{align}
	B^{(0)} & = \cM + \frac{1}{6}\, \Ddot\Ddot\Ddot C^{(0)} \, , \\[10pt]
	U_i{}^{(1)} & = \cN_i + \frac{u}{4}\, \pr_i \cM - \frac{1}{24} \int_{-\infty}^u \!\!du'\! \left[ \left( \Delta - 1 \right) \Ddot\Ddot C_i{}^{(0)} - 2\, D_i \Ddot\Ddot\Ddot C^{(0)} \right] , \\[10pt]
	V_{ij}{}^{(2)} & = \cP_{ij} + \frac{u}{5} \left( D_{(i\,} \cN_{j)} - \g_{ij} \Ddot \cN \right) + \frac{u^2}{40} \left( D_{(i} D_{j)} \cM -  \g_{ij}\, \Delta \cM \right) + \cdots \, . \label{V4D}
	\end{align}
\end{subequations}
$\cM$, $\cN_i$ and $\cP_{ij}$ are arbitrary functions of $x^i$ as in \eqref{u3} while in \eqref{V4D} we omitted the integrals that are obtained by the substitution of the previous formulas in the differential equation \eqref{diff-eq-V}, which is not modified even when $n =2$. In particular, these additional terms in \mbox{$\Ddot\Ddot\Ddot C^{(0)}$} will be instrumental in building the charges associated to the spin-three  supertranslations and superrotations identified in the previous section.  

When $n = 1$, namely in a three-dimensional spacetime, the radiation branch becomes subleading with respect to the Coulombic one in $\vf_{uuu}$. Moreover, fields of  spin greater than one do not propagate local degrees of freedom. It is therefore natural to ignore the radiation branch and work with boundary conditions that only encompass Coulomb-type solutions of the equations of motion. The only nonvanishing components of the field in the Bondi gauge \eqref{bondi3} are
\be 
\vf_{uuu} = \cM(\theta) + \cO(r^{-1}) \, , \qquad
\vf_{uu\theta} = \cN(\theta) + \frac{u}{3}\, \pr_\theta \cM(\theta) +\cO(r^{-1}) \, ,
\ee
where $\theta$ denotes again the angular coordinate on the circle at null infinity and we already included the constraints on the leading terms imposed by the equations of motion.

%%%%%%%%%%%%%%%%%%%%%%%%%%%%%%%%%%%%%%%%%%
\subsection{Asymptotic symmetries and charges for spin three}\label{sec:symm3}
%%%%%%%%%%%%%%%%%%%%%%%%%%%%%%%%%%%%%%%%%%

We now identify the key features of the gauge transformations preserving the Bondi-like gauge conditions \eqref{boundary3} and calculate the associated charges. The general expression for these charges, expressed in terms of the fields and the parameters of gauge symmetries is given by
\be \label{spin3_cov}
\begin{split}
	Q(u) & = -\lim_{r \to \infty} \frac{r^{n-1}}{2} \int_{S_u} \sqrt{\g}\, 
	\bigg\{\,  r\,\vf_{uuu} \,\pr_r\, \e_{rr} + \e_{rr} \left( r\pr_r + 2n \right) \vf_{uuu} - \frac{2}{r}\, \e_{rr} \Ddot \vf_{uu} \\
	& - \frac{2}{r^2}\, \g^{ij} \left[ \vf_{uui} \left( r\pr_r + n \right) \e_{rj} - \frac{1}{r}\, \e_{ri} \Ddot \vf_{uj} \right] \\
	& + \frac{1}{r^3}\, \g^{ik}\g^{jl} \left[\, \vf_{uij} \,\pr_r\, \e_{kl} - \e_{ij} \,\pr_r\,  \vf_{ukl} \,\right] \bigg\} \,,
\end{split}
\ee
where we recall the expression \eqref{gaugepars3} for the allowed gauge parameters.
However, while in four-dimensions these quantities are allowed to take an infinite number of independent values, as we shall see, the further constraints arising in higher dimensions will reduce them to a finite-dimensional family to be identified with global Killing symmetries.

The tensors $K_{ij}$, $\r_i$ and $T$ that characterize the asymptotic symmetries are not arbitrary: for $n > 2$ they are bound to satisfy the  differential equations (see Appendix \ref{app:spin3exact})
\begin{align}
\cK_{ijk} & \equiv D_{(i} K_{jk)} - \frac{2}{n+2}\, \g_{(ij} \Ddot K_{k)} = 0 \, , \qquad \g^{ij} K_{ij} = 0 \, , \label{K} \\[5pt]
\cR_{ijk} & \equiv D_{(i} D_j \r_{k)} - \frac{2}{n+2} \left(\, \g_{(ij} \Delta \r_{k)} + \g_{(ij} \big\{ D_{k)} , D_l \big\} \r^l \,\right) = 0 \, , \label{R} \\[5pt]
\cT_{ijk} & \equiv D_{(i} D_j D_{k)} T - \frac{1}{n+2} \left(\, \g_{(ij} \Delta D_{k)} T + \g_{(ij} \big\{ D_{k)} , D_l \big\} D^l T \,\right) = 0 \, . \label{T}
\end{align}
When the dimension of spacetime is equal to four, \emph{i.e.} when $n=2$, the last condition simply does not apply and the function $T(x^m)$ is arbitrary. In this case, the corresponding symmetry is the analogue of gravitational supertranslations, from which we derived Weinberg's theorem in the previous section. Eq.~\eqref{K} is the conformal Killing tensor equation of rank two on the $n$-dimensional celestial sphere. For $n > 2$ this equation admits a finite number of solutions, while when $n = 2$ it admits an infinite-dimensional family of local solutions, which we discussed, generalizing gravitational superrotations. The same is true for the less familiar eq.~\eqref{R} satisfied by $\r_i$: when $n>2$ it admits a finite number of solutions, while for $n = 2$ locally one can build infinitely many independent solutions.  All combinations appearing in the above equations are traceless, and hence they vanish identically when the dimension of spacetime is equal to three, \emph{i.e.} $n = 1$. This implies that in three dimension $T(\theta)$ and $\r(\theta)$ are arbitrary functions, while the symmetry generated by the traceless $K_{ij}$ is absent. 

Substituting \eqref{gaugepars3} into the expression for the charges \eqref{spin3_cov} one obtains
\be \label{charge3-mid}
\begin{split}
	Q(u) = \lim_{r \to \infty} \frac{r^{n-1}}{2} \int_{S_u} \!\sqrt{\g} \, \bigg\{\, \chi & \left( T - \frac{2u}{n+1}\, \Ddot \r + \frac{u^2}{(n+1)(n+2)}\, \Ddot\Ddot K \right) \\
	+\, \chi^i & \left( \r_i - \frac{u}{n+2}\, \Ddot K_i \right) + \chi^{ij} K_{ij} \,\bigg\} \, ,
\end{split}
\ee
with
\begin{subequations}
	\begin{align}
	\chi & = - \left( r\pr_r + 2n \right) \vf_{uuu} - \frac{n-1}{r}\, \Ddot \vf_{uu} + \frac{1}{2r}\, \pr_r \Ddot\Ddot \vf_u \, , \\
	\chi_i & = 2(n+2)\,\vf_{uui} - \frac{2}{r} \left( r\pr_r - 2 \right) \Ddot \vf_{ui} \, , \\[3pt]
	\chi_{ij} & = \left( r\pr_r -4 \right) \vf_{uij} \, .
	\end{align}
\end{subequations}
The next task is to evaluate \eqref{charge3-mid} on the solutions of the equations of motion discussed above. For $n$ even and greater than two \eqref{sol3}, \eqref{constrC3} and \eqref{u3} lead to
\begin{subequations} \label{all-chi}
	\begin{align}
	\chi & = - \sum_{k\,=\,0}^{\frac{n-4}{2}} r^{-\frac{n}{2}-k} \frac{n+2k-2}{2(n-2k-2)}\, \Ddot\Ddot\Ddot C^{(k)} - r^{1-n} (n+1) \cM \nn \\
	& + r^{1-n}\, \frac{n-3}{6n} \int_{-\infty}^u \!\!du'\!\left( \Delta - n + 2 \right) \Ddot\Ddot\Ddot C^{(\frac{n-4}{2})} + \mathcal O(r^{-n}) \, , \label{chi} \\[5pt]
	\chi_i & = \sum_{k\,=\,0}^{\frac{n-2}{2}} r^{-\frac{n}{2}-k+1}\, \frac{2(n+2k)}{n-2k}\, \Ddot\Ddot C_i{}^{\!(k)} + 2\, r^{1-n} \Big\{ (n+2)\, \cN_i + u\, \pr_i \cM \Big\}  \nn \\
	& - \frac{r^{1-n}}{n+1} \int_{-\infty}^u \!\!\!\!du'\! \Big\{ (n-1) \left( \Delta - 1 \right) \Ddot\Ddot C_i{}^{\!(\frac{n-2}{2})} \nn \\
	&+ \frac{n-3}{3n} \int_{-\infty}^{u'} \!\!\!\!du'' D_i\! \left( \Delta - n + 2 \right)(\Ddot)^3 C^{(\frac{n-4}{2})} \Big\} , \label{chi1}+ \mathcal O(r^{-n}) \\[5pt]
	\chi_{ij} & = - \sum_{k\,=\,0}^{\frac{n}{2}} r^{-\frac{n}{2}-k+2} \frac{n+2k+2}{n-2k+2}\, \Ddot C_{ij}{}^{\!(k)} \label{chi2} \nn\\
	& - r^{1-n} \Big\{ (n+3)\, \cP_{ij} + u \left( D_{(i\,} \cN_{j)} - \frac{2}{n}\, \g_{ij} \Ddot \cN \right) \nn\\
	&+ \frac{u^2}{2(n+2)} \left( D_{(i} D_{j)} \cM - \frac{2}{n}\, \g_{ij}\,\Delta \cM \right) \Big\} + \cdots \, , 
	\end{align}
\end{subequations}
where in \eqref{chi2}, besides the terms $\mathcal O(r^{-n})$, we also omitted the integrals in the retarded time that one obtains by substituting \eqref{uV3}.  For \mbox{$n=2$} the first two expressions are modified as follows,
\begin{subequations} \label{chi4D}
	\begin{align}
	\chi & = - \frac{1}{r} \left\{ 3\, \cM + \frac{1}{2}\, \Ddot\Ddot\Ddot C^{(0)} \right\} + \cO(r^{-2}) \, , \\[5pt]
	\chi_i & = 2\,\Ddot\Ddot C_i{}^{\!(0)} \nn\\
	&+ \frac{2}{r} \left\{ 4\, \cN_i + u\, \pr_i \cM + \frac{1}{6} \int_{-\infty}^u \!\!\!\!du'\! \left[ \left( \Delta - 1 \right) \Ddot\Ddot C_i{}^{\!(0)} - 2\,D_i \Ddot\Ddot\Ddot C^{(0)} \right] \right\} \nn \\
	& + \cO(r^{-2}) \,,
	\end{align}
\end{subequations}
while $\chi_{ij}$ is instead obtained by setting $n = 2$ in \eqref{chi2} and by correcting the integral terms according to \eqref{V4D}. For $n$ odd the extrema of the sums become, respectively, $\frac{n-3}{2}$, $\frac{n-1}{2}$ and $\frac{n+1}{2}$, while the terms in the second and third lines of eqs.~\eqref{all-chi} are absent.

Looking only at the $r$-dependence, the sums in the previous formulas would naively give rise to divergent contributions to the charges. This is a standard feature due to the presence of radiation fields, that roughly speaking always induce terms of the type
\be
\int \frac{\kappa(u,x^i)}{r^\alpha}\,r^{n}d\Omega\,,\qquad \frac{n}{2}\le \alpha \le n
\ee
in the calculation of the surface charges. 
These terms vanish, however, thanks to the equations of motion and to  the differential constraints on the parameters in \eqref{K}, \eqref{R} and, when $n > 2$, \eqref{T}. Let us begin by exhibiting this mechanism in the simplest case: the term $\chi^{ij} K_{ij}$ in \eqref{charge3-mid} contains divergent contributions that, integrating by parts, can be cast in the form
\be
C_{ijk}^{(l)}\, D^{(i} K^{jk)} = C_{ijk}^{(l)} \left\{ \frac{2}{n+2}\,\g^{(ij} \Ddot K^{k)} - \cK^{ijk} \right\} = 0 \, ,
\ee
where we recall that $\cK_{ijk}$ is the shorthand introduced in \eqref{K} to denote the differential equation satisfied by $K_{ij}$.
The next cancellation is slightly more involved: integrating by parts one obtains
\be \begin{aligned}
&\int_{S_u} \!\!\!\sqrt{\g}\ \chi^i \left( \r_i - \frac{u}{n+2}\, \Ddot K_i \right) \\
&\sim \sum_{l\,=\,0}^{[\frac{n-1}{2}]} \!r^{-\frac{n}{2}-l}\! \int_{S_u} \!\!\!\sqrt{\g}\  C_{ijk}^{(l)} \left( D^{(i} D^{j} \r^{k)} - \frac{u}{n+2}\, D^{(i} D^{j} \Ddot K^{k)} \right) + \cdots \, .
\end{aligned}\ee
To cancel the contribution in $\r_i$ one can use eq.~\eqref{R}, which again allows one to substitute the symmetrized gradient with a term in $\g_{ij}$. To cancel the contribution in $K_{ij}$ one can instead use the following consequence of the conformal Killing tensor equation \eqref{K}: 
\be
\begin{split}
	D_{(i} D_j \Ddot K_{k)} & = -\, 2\, \g_{(ij} \Ddot K_{k)} + \frac{3}{n+1}\, \g_{(ij} D_{k)} \Ddot\Ddot K \\
	& - \frac{n+2}{n} \left\{ \Delta \cK_{ijk} - D_{(i} \Ddot \cK_{jk)} + \frac{1}{n+1}\, \g_{(ij} \Ddot\Ddot \cK_{k)} + (n-3)\, \cK_{ijk} \right\} .
\end{split}
\ee
Similar considerations apply to the integral terms in the second line of \eqref{chi1}.
The remaining contribution in the charge formula \eqref{charge3-mid} contains three addenda whose divergent parts can be cast in the following form by integrating by parts:
\be
C_{ijk}^{(l)} D^{(i} D^{j} D^{k)} T \, , \qquad
C_{ijk}^{(l)} D^{(i} D^{j} D^{k)} \Ddot \r \, , \qquad
C_{ijk}^{(l)} D^{(i} D^{j} D^{k)} \Ddot\Ddot K \, .
\ee
These terms are actually absent when $n=2$. For $n>2$ the first contribution vanishes thanks to \eqref{T}. The other two types of terms vanish thanks to consequences of the equations satisfied by $\r_i$ and $K_{ij}$, obtained by taking symmetrized gradients thereof.

We have therefore proven that, in the Bondi-like gauge \eqref{bondi3}, a spin-three field with the falloffs \eqref{boundary3} at null infinity and satisfying the field equations up to the contributions of order $r^{n-1}$ to its components admits finite asymptotic linearized charges. For any $n > 2$, these depend on the ``integration functions'' specifying the solution as
\be \label{finalcharge3}
Q = - \frac{1}{2} \int_{S^n} \!\sqrt{\g}\, \Big\{ (n+1)\, T \cM - 2(n+2)\, \r^i \cN_i + (n+3)\, K^{ij} \cP_{ij} \Big\} \, ,
\ee 
where $\cM$, $\cN_i$ and $\cP_{ij}$ are the tensors on the sphere at null infinity introduced in \eqref{u3} (see \eqref{u3-odd} for $n$ odd). As anticipated, the charges are constant along null infinity when the dimension of spacetime is greater than four. The same is true also in three dimensions: in this case both $K^{ij}$ and $\cP_{ij}$ are not present and the asymptotic charges take the form
\be
Q = - \int d\theta \Big\{ T(\theta) \cM(\theta) - 3\, \r(\theta) \cN(\theta) \Big\}\, ,
\ee
in agreement with the result derived in the Chern-Simons formulation \cite{spin3-3D_1,3D-flat_2}. 

When $n = 2$, the modifications in the dependence on $u$ of the leading terms in the Coulomb-type solution recalled in \eqref{int4D-spin3} (and \eqref{chi4D}) lead to the following expression for the asymptotic charges:
\be \label{charge3-4D}
Q(u) = - \frac{1}{2} \int_{S_u} \!\!\sqrt{\g} \left\{ 3\, T \left( \cM + \frac{1}{6}\, \Ddot\Ddot\Ddot C^{(0)} \right) - 8\, \r^i \cN_i + 5\, K^{ij} \cP_{ij} + \cdots \right\} \, .
\ee
In this formula we omitted other $u$--dependent terms in $C^{(0)}$, whose form is not particularly illuminating and can be readily obtained by substituting \eqref{chi4D} in \eqref{charge3-mid}. The main information is anyway that in four dimensions a dependence on the retarded time appears already in the linearized theory, thanks to the contribution to the charges of the radiation solution. As shown in the previous section, where the terms in $T$ in \eqref{charge3-4D} have been already presented, the dependence on radiation data is instrumental in deriving Weinberg's theorem for spin-three soft quanta from the Ward identities of the supertranslation symmetry generated by the arbitrary function $T(x^i)$.

%%%%%%%%%%%%%%%%%%%%%%%%%%%%%%%%%%%%%%%%%%%%%%%%%%%%%%%%%%%%%%%%%%%%%
\section{Arbitrary Spin}\label{sec:spin-s}
%%%%%%%%%%%%%%%%%%%%%%%%%%%%%%%%%%%%%%%%%%%%%%%%%%%%%%%%%%%%%%%%%%%%%

Without delving too much into the details, and restricting for simplicity to the higher dimensions $n>2$, we shall now illustrate the generalization of the results of the previous section to the case of an arbitrary integer spin $s$.
The Bondi-like gauge in this case reads
\be \label{Bondi-s}
\vf_{ra_{s-1}} = 0 \, , \qquad
\g^{ij} \vf_{ij  a_{s-2}} = 0 \, .
\ee
together with the ansatz
\be \label{s-ansatz_even}
\vf_{u_{s-k}i_k} = \sum_{l\,=\,0}^\infty r^{-\frac{n}{2}+k-l} C_{i_k}{}^{\!\!\!(k,l)}(u,x^m) \,,
\ee
for even-dimensional spacetimes. We recall that we are conventionally employing a multi-index notation for symmetrized indices: for instance $\varphi_{ij a_{s-2}}$ is a shorthand notation for any component of the type $\varphi_{ij a_3 a_4 \cdots a_s}$, with two angular indices and $s-2$ arbitrary indices.

As in the previous examples, the leading behavior   of our ansatz is designed to give a finite flux of energy per unit time across the sphere $S_u$ at fixed $u$, a feature that we interpret as radiation flowing through null infinity. Indeed, the canonical energy-momentum tensor of the higher-spin Lagrangian in Bondi-like gauge,
\be
\mathcal L = \frac{1}{2}\,  \varphi^{a_s} \left(\, \Box \varphi_{a_s} - \nabla_{a} \nabla\cdot \varphi_{a_{s-1}}\right) ,
\ee
reads
\be
T_{ab}= \nabla_{a} \varphi_{c_s} \nabla_{b} \varphi^{c_s} - s\, \nabla\cdot \varphi^{c_{s-1}} \nabla_{b} \varphi_{a c_{s-1}}+ \eta_{ab} \mathcal L \, .
\ee
The corresponding power flowing through null infinity is then ($\gamma^{i_sj_2}=\gamma^{ij}\cdots\gamma^{kl}$ with $s$ factors)
\be \label{energy-flux}
\cP(u) =  \lim_{r\to \infty} \int_{S_u} \left(T_{uu}-T_{ur}\right) r^n d\Omega
= \int_{S_u} \gamma^{i_s j_s} \partial_u{C}^{(s,0)}_{i_s} \partial_u {C}^{(s,0)}_{j_s} d\Omega\,.
\ee

The asymptotic analysis of the field equations shows that, for all even values of $n$, the tensors entering the ansatz \eqref{s-ansatz_even} are fixed in terms of the $C^{(s,l)}$, with the exception of 
\be \label{def-charges}
C_{i_k}{}^{\!\!\!(k,\frac{n}{2}+k-1)} \equiv Q_{i_k}{}^{\!\!\!(k)} \qquad \textrm{for}\ k < s \, .
\ee
The tensors $C^{(s,l)}$ are then determined (up to integrations functions) in terms of an arbitrary tensor $C^{(s,0)}(u,x^m)$, in particular via the equation $\cF_{i_s} = 0$.
The remaining components of the equations of motion fix the $u$-evolution of the tensors $\mathcal Q^{(k)}$ defined in \eqref{def-charges}. 
Furthermore, on-shell, the $\mathcal Q^{(k)}$ are $n$-divergences as the other $C^{(k,l)}$, up to a set of integration functions $\cM^{(k)}(x^m)$. More specifically, $\mathcal Q^{(k)}$ depends on the integrations functions of all $\mathcal Q^{(l)}$ with $l < k$ with a precise polynomial dependence on $u$. As we have seen in the spin-three context, this is instrumental in making the independence on the retarded time of the asymptotic charges explicit. Concretely, the $Q^{(k)}$ depend on the $\cM^{(k)}$ in the following way 
\be \label{finalQ}
\mathcal Q_{i_k}{}^{\!\!(k)} = \sum_{l\,=\,0}^k \frac{(n+s+k-l-2)!}{l!(n+s+k-2)!}\, u^l \underbrace{D_{i} \cdots D_i}_{\textrm{$l$ terms}} \cM_{i_{k-l}}{}^{\!\!\!\!\!\!\!\!(k-l)} + \cdots\,,
\ee
where we omitted terms that will not contribute to the surface charges.

When $n$ is odd, one has to consider an ansatz containing both integer and half-integer powers of $r$. Restricting to $n > 1$, we thus set
\be \label{s-ansatz_odd}
\vf_{u_{s-k}i_k} = \sum_{l\,=\,0}^\infty r^{-\frac{n}{2}-l+k} C^{(k,l)}(u,x^m) + \sum_{l\,=\,0}^\infty r^{1-n-l} \tilde{C}^{(k,l)}(u,x^m) \, ,
\ee
so that the leading order has the same form as in \eqref{s-ansatz_odd}. Eq.~\eqref{energy-flux} thus guarantees that we have a finite flux of energy per unit time across $S_u$ in this case too. 
The leading order of the Coulomb branch is in this case not constrained by the equations of motion, so that we can define
\be
\tilde{C}^{(k,0)} \equiv \mathcal Q^{(k)} \qquad \textrm{for}\ k < s \, .
\ee
The remaining $\tilde{C}^{(k,l)}$ are again fixed in terms of the $\tilde{C}^{(s,l)}$, provided that one identifies $\tilde{C}^{(k,l)} = C^{(k,\frac{n}{2}+k+l-1)}$, that is
\be \label{gen-Ctilde}
\tilde{C}^{(k,l+1)} = - \frac{n+2k+l-1}{(l+1)(n+2k+l)}\, \Ddot \tilde{C}^{(k+1,l)} \, .
\ee
The field equations then determine $\tilde{C}^{(s,l)}$ in terms of an arbitrary $\tilde{C}^{(s,0)}(u,x^m)$ up to integration functions. 
The $\mathcal Q^{(k)}$ satisfy a relation analogous to eq.~\eqref{finalQ} where the omitted terms, which anyway do not contribute to the charges, are actually absent.

To summarize, the solution of the equations of motion up to Coulombic order can be cast into the following form
\be \label{boundary-cond-s}
\begin{split}
	\vf_{u_{s-k}i_k} = & \sum_{l\,=\,0}^{\left[\frac{n+1}{2}\right]+k-2} r^{-\frac{n}{2}+k-l}\, \frac{2^{s-k}(n+2(k-l-2))!!(n+2(k+l-1))}{(n+2(s-l-2))!!(n+2(s+l-1))}\, (D \cdot)^{s-k} C_{i_k}{}^{\!\!\!(s,l)} \\
	& + r^{1-n} \mathcal Q_{i_k}{}^{\!\!\!(k)} + \cO(r^{-\frac{n}{2}-\left[\frac{n}{2}\right]}) \,.
\end{split}
\ee
Below, we will show that \eqref{boundary-cond-s} indeed leads to the cancellation of some of the potentially divergent contributions to the linearized charges, while also proving that the $\mathcal Q^{(k)}$'s give a finite contribution to them. 

%%%%%%%%%%%%%%%%%%%%%%%%%%%%%%%%%%%%%%%%%%
\subsection{Asymptotic symmetries and charges for any spin}
%%%%%%%%%%%%%%%%%%%%%%%%%%%%%%%%%%%%%%%%%%
In the Bondi-like gauge \eqref{Bondi-s} the surface charges be expressed in terms of the nonvanishing field components as
\be \label{charges-s} 
\begin{split}
	Q(u) & = -\,\lim_{r\to\infty} \int_{S_u} \frac{r^{n-1} \sqrt{\g}}{(s-1)!}\, \sum_{p\,=\,0}^{s-1} \binom{s-1}{p} \bigg\{ \vf_{u_{s-p}i_{p}}\left( r\pr_r + n + 2p \right) \e^{u_{s-p-1}i_p} \\
	& +  \e^{u_{s-p-1}i_p} \left[ (s-p-2)  \left( r\pr_r + n \right) \vf_{u_{s-p}i_{p}} - \frac{s-p-1}{r}\, D\cdot \vf_{u_{s-p-1}i_p} \right] \bigg\} \, .
\end{split}
\ee
Therefore, they only depend on the components $\e_{r_{s-k-1}i_k} = (-1)^{s-k-1} \e^{u_{s-k-1}}{}_{i_k}$ of the gauge parameters of asymptotic symmetries. Analyzing the asymptotic Killing equations obtained by enforcing that these parameters preserve the Bondi-like gauge, one obtains that these components are parametrized as
\be\begin{aligned} \label{K-u}
\e_{r_{s-k-1}i_k} &= r^{2k} \left( K_{i_k}{}^{\!\!\!(k)} + \sum_{m\,=\,1}^{s-k-1}  \frac{(-1)^m u^m}{(n+s+k-2)_m} \binom{s-k-1}{m} (D \cdot)^m K_{i_k}{}^{\!\!\!(k+m)} \right)
\\
& + \cO(r^{2k-1}) \, ,
\end{aligned}\ee
where the tensors $K^{(k)}(x^m)$ only depend on the coordinates on the $n$-dimensional sphere at null infinity, while $(a)_n \equiv a (a+1) \cdots (a+n-1)$ is the Pochhammer symbol. Moreover, the $K^{(k)}_{i_k}$ are traceless
and must satisfy suitable differential constraints, which generalize those displayed in \eqref{K} and \eqref{R} for the spin-three case.

While substituting the explicit form of the residual symmetry parameters into the formula for the surface charge, one first has to keep track of the cancellation of all potentially divergent terms in the limit $r\to\infty$. As we have seen in the previous sections, these usually occur after integrations by parts; for instance, the coefficient of the leading power in $r$ is of the form
\be
\int_{S_u} \sqrt{\g}\: K^{(s-1)}_{i_{s-1}} \Ddot C^{(s,0)\,i_{s-1}} = \int_{S_u} \sqrt{\g}\:  
D_i^{\phantom{(s)}}\!\!\!\! 
K^{(s-1)}_{i_{s-1}} C^{(s,0)\,{i_s}} \, ,
\ee 
and hence vanishes because $K^{(s-1)}$ satisfies the conformal Killing tensor equation of rank $s-1$.
Since this type of cancellation must occur in general, as we have verified in detail in the spin-three case, the finite contribution to the charges is determined by the Coulomb-like terms $\mathcal Q^{(k)}$ in the boundary conditions \eqref{boundary-cond-s}. Substituting them into \eqref{charges-s}, while taking into account their dependence on the integration constants in \eqref{finalQ}, one obtains after integration by parts
\begin{align}
& Q(u) = \int_{S_u} \!\frac{\sqrt{\g}}{(s-1)!} \sum_{p=0}^{s-1} \sum_{m=0}^{s-p-1} \sum_{l=0}^{p} \nn\\ 
&\binom{s-1}{p} \binom{s-p-1}{m} \binom{p}{l} \frac{(s+p+n-2)(s+n+p-l-2)!}{(s+n+p-2)_m (s+n+p-2)!} \nn \\
& \phantom{Q(u) = \int_{S_u}} \times (-1)^{s+m+p+l}\, u^{l+m} \cM^{(p-l)}_{i_{p-l}} (\Ddot)^{l+m} K^{(m+p)\,i_{p-l}} \label{charges-int-s} \\[5pt]
& = \int_{S_u} \sum_{k,q} \frac{\sqrt{\g}\,(-1)^{s+k}(s+n+q-2)!}{(s-1)!(s+n+k+q-3)!}\nn\\
& \left[ \sum_{p=0}^{s-1} (-1)^p \binom{s-1}{p} \binom{s-p-1}{k+q-p} \binom{p}{q} \right] u^{k} \cM^{(q)}_{i_{q}} (D\cdot)^{k} K^{(k+q)\,i_{q}} \, . \nn
\end{align}
The final expression has been obtained by introducing the new labels $k = l+m$ and $q = p-l$ and by swapping the sums, whose ranges precisely correspond to the values of the labels for which the integrand does not vanish (with the convention $\binom{N}{n} = 0$ for $n < 0$ and $n > N$). One can eventually verify that the sum within square brackets in the second line of \eqref{charges-int-s} vanishes for any $k > 0$, thus confirming that the charges do not depend on $u$ for any value of the spin. The $u$-independent contribution then reads
\be \label{charges-final}
Q = \int_{S^n} \frac{\sqrt{\g}}{(s-1)!} \sum_{q\,=\,0}^{s-1} (-1)^{s+q} (s+n+q-2) \binom{s-1}{q}\, K^{(q)}_{i_q} \cM^{(q)\,i_q} \, ,
\ee
in full analogy with the result that we presented for $s=3$ in \eqref{finalcharge3}.

\chapter*{Conclusions}
\addcontentsline{toc}{chapter}{Conclusions}

Let us briefly summarize what we achieved so far and make an attempt to identify possible directions to be explored in the future. 

Concerning the extension of the asymptotic analysis to higher dimensions, we were able to reproduce several aspects of the discussion of memory effects in the context of even-dimensional Maxwell and Yang-Mills theories, encompassing ordinary/linear, null/nonlinear, phase and color memories. 

While it is fair to say that the existence of memory in even dimensions is by now no longer disputed, still, much remains to be done in this respect. Kick, phase and color memory effects are just a few of the many more types of memory that were put forward in the four-dimensional setup. Therefore, it would be natural to understand whether also different aspects of memory formulas admit a generalization to arbitrary even spacetime dimensions: for instance, the distinction between electric and magnetic memories \cite{Mao_em}, spin memory \cite{spinmemory}, or refraction memory \cite{CompereRefraction}. Of course, an even more interesting possibility is that in higher dimensions memory effects may possess a richer zoology and may allow to define \emph{new} types of observable phenomena. For instance, while the Coulombic order $\mathcal O(1/r^{D-3})$ is, with hindsight, a natural place where to look for memory effects, since it is associated with DC phenomena, it is conceivable that there may exist effects that scale as $\mathcal O(1/r^k)$ with $\tfrac{D-2}{2}\le k<D-3$, provided that the test probe is chosen appropriately and that suitable observables are identified. Independently of the dimension of spacetime, it would also be interesting to better understand whether proper analogs of color and phase memories in the gravitational case, \emph{i.e.} memory effects encoded in the gravitational quantum numbers of a test particle, comprise qualitatively new phenomena or just reduce to a quantum rephrasing of the standard displacement memory (see \emph{e.g.} \cite{Afshar:2018sbq} for a discussion of the quantum kick memory effect).

The status of memory effects in odd spacetime dimensions is much less clear, as a consequence of the dispersion phenomena occurring in this case. These manifest themselves as the failure of the Huygens principle and, in the asymptotic expansion near null infinity, as non-analytic terms of the form $1/r^{k/2}$ with $k$ odd. While a proof of concept has been provided here for the existence in odd dimensions of null memory, which overcomes the blurring induced by dispersion since the emitting source effectively reaches the probe near null infinity, we still feel that the phenomenology of this effect is much less studied and understood compared to its odd-dimensional counterpart. 

The non-analyticities in $1/r$, which are also at the basis of the failure of the conformal approach to null infinity for solutions containing gravitational waves in odd dimensions, are tightly related to difficulties encountered in the discussion of soft theorems and asymptotic symmetries which were resolved in \cite{He-Mitra-photond+2, He-Mitra-gauged+2,He-Mitra-magneticd+2}. The solution proposed highlights a major drawback of the standard perturbative expansion in powers of $1/r$. In odd dimensions, the perturbative treatment of the field equations  is not sufficient in order to identify constraints that link  radiation $\mathcal O(1/r^{\frac{D-2}{2}})$ and Coulombic $\mathcal O(1/r^{D-3})$ components, since the corresponding recursion relations only involve perturbative orders that differ by an integer number. Such constraints only arise nonperturbatively, \emph{e.g.} by considering exact solutions to the field equations, and are crucial for soft theorems: they are needed in order to connect the information on the propagating soft photon modes, stored in the radiation components, to the soft charges involved in the Ward identities, which are instead expressed in terms of Coulombic components. 

While these works focused on the discussion of field strengths and relied on the structure of soft charges for the identification of the underlying asymptotic symmetries, thus working ``backwards'' compared to the logic we employed in our discussion, an explicit realization of these symmetries is precluded unless one introduces the gauge fields and defines their dynamics by means of  a gauge-fixing. 
Asymptotic symmetries can be then evaluated in any dimension according to the following proposed strategy. After gauge fixing, one first considers the set of solutions characterized by radiation falloffs, whose behavior at infinity allows one to have control over physical observable properties, but which are too restrictive and effectively forbid infinite-dimensional asymptotic symmetries. Then, one acts on this set of solutions by gauge symmetries that preserve the gauge fixing and have finite and nonzero asymptotic charge, but which do not necessarily leave radiation falloffs invariant. The wider solution space so-obtained still retains all the desirable physical properties of radiation solutions, such as finite energy and charge fluxes, but it admits nontrivial asymptotic symmetries by construction. 

Asymptotic surface charges, on the other hand, may first be safely evaluated on $\mathscr I^+$ (resp. $\mathscr I^-$) in regions where radiation is absent, \emph{i.e.} where the solutions are stationary. In particular one may first consider the limit $\mathscr I^+_-$ (resp. $\mathscr I^-_+$), where, under standard assumptions, only Coulombic contributions matter. The evolution of the charges \emph{along} $\mathscr I$ can then be defined, even in the presence of radiation, by means of the equations of motion.

Combining this observation with the previous one, we may formulate the following systematic criterion to detect the existence of nontrivial asymptotic symmetries in higher dimensions: it is sufficient to find the residual subset of transformations preserving  the assigned gauge-fixing and giving rise to a finite asymptotic charge when evaluated on Coulombic solutions.

However, at this point, one may still wonder to which extent the nature of the asymptotic symmetry group is gauge-independent, or, if it does depend on the gauge, whether there exists a maximal asymptotic symmetry group and a systematic way in which this group allows for an embedding of the smaller ones. These are, to our knowledge, open questions.

Moving to the case raised by the soft scalars, we have been able to provide a possible resolution of the purported paradox related to the existence of soft scalar charges, in the four dimensional case, by appealing to a two-form gauge symmetry giving rise to dual soft charges. Nevertheless, the analysis of this point leaves open two interesting aspects. First, again in connection with the possible gauge dependence of the asymptotic analysis, there is the problem of knowing, if possible \emph{a priori}, when the chosen gauge fixing results in too stringent a condition and actually hides the presence of asymptotic symmetries. This seems to be the case, in this context, for the radial gauge, which trivializes the asymptotic charges, while the Lorenz gauge appears to be more flexible and allows one to overcome this difficulty. Second, the example of the scalar and the two-form in four dimensions paves the way to the possibility of discussing asymptotic structures by exploiting dualities in more general contexts, although one should bear in mind that such a possibility is essentially restricted to the free/asymptotic level.
 
On the higher-spin side, we have seen how a strategy similar to the one employed in the context of linearized gravity allows us, upon introducing a suitable Bondi-like gauge, to perform a meaningful discussion of asymptotic symmetries for all integer spins in any dimension, for free fields.
The calculation of the associated charges has also been carried out explicitly in four dimensions and,  at the level of global symmetries, in any dimension. We have verified the connection of such symmetries with Weinberg's soft theorem in four dimensions, where the infinite dimensional asymptotic symmetry enhancement occurs more naturally, but it is reasonable to envision that, once the strategy proposed above is employed in order to overcome the difficulties induced by the imposition of radiation falloffs, such a link should hold independently of the spacetime dimension as in the case of lower spins.

Moving to issues less directly related to the ones we have discussed in detail, it appears natural, also in connection with cosmological observations, to wonder how far the discussion of asymptotic symmetries and their observable consequences can be carried in the presence of a cosmological constant. This issue has already attracted attention in the literature on the gravitational setting with promising results (see \emph{e.g.} \cite{BieridS, CompereLambda, Chumemory} for recent developments in this direction), especially for de Sitter spacetime, but the fact that many of the above results on flat spacetime hold regardless of spin strongly motivates analogous investigations also for Maxwell, Yang-Mills and higher spins. Identifying similarities and differences between flat spacetime and (A)dS in this respect could prove challenging, due to their different causal structures, but it could also provide fruitful insights into the problem of flat space holography.

Also the analysis of higher-spin asymptotic symmetries on Anti de Sitter spacetime may be regarded as promising, since AdS offers the only arena in which non-Abelian higher-spin gauge algebras are explicitly known so far \cite{BCIValgebras}. A tightly related issue is the possibility of uncovering a non-Abelian algebra that may underly the infinite-dimensional family of (commuting) asymptotic symmetries that has been put forward in this thesis for any spin on flat space. A first step in this direction would be the introduction of cubic vertices in the analysis of higher-spin asymptotic symmetries.

Finally, let us mention again the tempting analogy between our higher-spin symmetry enhancement and enhanced symmetries exhibited by string amplitudes in the high-energy regime \cite{Gross}, whose relation, if any, would provide further insight into the role played by asymptotic symmetries in the structure of string theory. A related issue would be the understanding of the role, if any, played by these symmetries of massless higher-spin states in the tensionless limit of string theory and in the mechanism of symmetry breaking that gives rise to a mass for such states.

\appendix

\chapter{Coordinate Conventions}\label{app:Laplaciano}

In $D$-dimensional Minkowski spacetime, retarded Bondi coordinates are defined as follows in terms of the standard Cartesian coordinates $x^\mu= (t, \mathbf x)=(x^0,x^I)$, with $I=1$, $2$,$\ldots$, $D-1$:   retarded time $u=t-|\mathbf x|$, a radial coordinate $r=|\mathbf x|$, and angular coordinates $x^i$ on the Euclidean unit $(D-2)$-sphere, with metric $\gamma_{ij}$.  

The Minkowski metric, in such coordinates, reads
$$
ds^2 = -dt^2 + d\mathbf x^2= -du^2-2 du dr + r^2 \gamma_{ij} dx^i dx^j\,,
$$
while the nonvanishing Christoffel symbols are
$$
\Gamma\indices{^i_{rj}}=\frac{1}{r}\delta\indices{^i_j}\, ,\qquad
\Gamma\indices{^u_{ij}}=-\Gamma\indices{^r_{ij}}=r\gamma_{ij}\, ,\qquad
\Gamma\indices{^i_{jk}}=\frac{1}{2}\gamma^{il}(\partial_j \gamma_{lk}+\partial_k \gamma_{jl}-\partial_l \gamma_{jk})\,.
$$
With $D_i$ we denote the covariant derivative on the sphere associated to $\gamma_{ij}$ and $\Delta= D_iD^i$ is the corresponding Laplace-Beltrami operator. In particular, the d'Alembert operator $\Box$ acting on a scalar $\varphi$ takes the explicit form
$$
\Box \varphi  = -\left(2\partial_r + \tfrac{D-2}{r}\right)\partial_u \varphi + \left(\partial_r^2 + \tfrac{D-2}{r}\partial_r + \tfrac{1}{r^2}\Delta\right)\varphi\,.
$$

In terms of a given parametrization $\mathbf n = \mathbf n(x^i)$ of unit vectors, $|\mathbf n|=1$, identifying points on the sphere, one has $\mathbf x=r\,\mathbf n$ and $\gamma_{ij}=\partial_i\mathbf n \cdot \partial_j \mathbf n$.

The Cartesian components of the radial vector $\mathbf x$ satisfy $\nabla_a \nabla_b \mathbf x=0$, since in Cartesian coordinates this reduces to $\partial_I \partial_J \mathbf x=0$. In spherical coordinates this identity gives, 
$$D_iD_j\mathbf n = -\gamma_{ij}\mathbf n\,,$$
which in its turn implies
$$\left(D_i D_j-\tfrac{1}{D-2}\gamma_{ij}\Delta\right)\mathbf n=0\qquad\text{ and }\qquad (\Delta+D-2) \mathbf n =0\,.$$
More generally, for any homogeneous polynomial $P_l(\mathbf x)=r^l Q_l(\mathbf n)$ of degree $l\ge0$, we have that
$$
\delta^{IJ}\nabla_I \nabla_J P_l= r^{2-D}\partial_r(r^{D-2}\partial_r P_l)+r^{-2}\Delta P_l
$$ 
vanishes if and only if $[\Delta + l(D+l-3)]Q_l=0$. This shows that $\Delta$ has eigenvalues $-l(D+l-3)$ for $l=0$, $1$, $2\ldots$, each with multiplicity $g_l$ equal to the dimension of the space of harmonic polynomials of degree $l$. In particular $g_0=1$, associated with the constant polynomial, and $g_1=D-1$, corresponding to the monomials given by the Cartesian components of $\mathbf x$. For $l\ge2$, we have \cite{libroRusso} $$g_l=\tbinom{D+l}{l}-\tbinom{D-2+l}{l}\,.$$

\chapter{Spin-Three Symmetries at $\mathscr I^+$}
\section{Spin-three Asymptotic Symmetries}\label{app:spin3symm}
In this appendix we explicitly calculate the asymptotic symmetries that preserve the following conditions for a spin-three field in $D=n+2$ dimensions:
\be\label{r+traces}
\varphi_{rab}=0\,,\qquad \gamma^{ij}\varphi_{ija}=0\,,
\ee
together with the requirement that
\be\label{eqWeakest}
\varphi_{uuu}\,,\qquad \frac{\varphi_{uui}}{r}\,,\qquad \frac{\varphi_{uij}}{r^2}\,,\qquad \frac{\varphi_{ijk}}{r^3}
\ee
scale as $\mathcal O(r^{-1})$ near future null infinity. The significance of this set of conditions is that, when $D=4$, they coincide with the standard Bondi-like gauge conditions discussed in Chapter \ref{chap:HSP}, while, for generic dimensions,  \eqref{eqWeakest} provides the strongest falloffs still allowing for the presence of infinite-dimensional higher-spins supertranslations, in complete analogy with the gravitational case.

Let us start from 
\begin{align}
\label{rrr}
\delta_\epsilon\varphi_{rrr}&=3\partial_r\epsilon_{rr}=0\,,\\
\label{rru}
\delta_\epsilon\varphi_{rru}&=2\partial_r\epsilon_{ur}+\partial_u\epsilon_{rr}=0\,,\\
\label{ruu}
\delta_\epsilon\varphi_{ruu}&=2\partial_u\epsilon_{ur}+\partial_r\epsilon_{uu}=0\,,\\
\label{uuu} \delta_\epsilon\varphi_{uuu}&=3\partial_u\epsilon_{uu}=\mathcal O(r^{-1})\,.
\end{align}
From \eqref{rrr} we get 
\be
\epsilon_{rr}=F(u,\mathbf n)\,,
\ee
for some $r$-independent function $F$. Then, by \eqref{rru}, 
\be
\epsilon_{ur}=-\frac{r}{2}\partial_uF(u,\mathbf n)+G(u,\mathbf n)\,,
\ee
for an arbitrary $r$-independent function $G$. Hence, \eqref{ruu} implies
\be
\epsilon_{uu}=\frac{r^2}{2}\partial_u^2F(u,\mathbf n)-2r\partial_uG(u,\mathbf n)+H(u,\mathbf n)\,,
\ee
with $H(u,\mathbf n)$ a third arbitrary function of $u$ and the angles. Imposing \eqref{uuu}, we have
\be\label{uuuvariation}
\frac{r^2}{2}\partial_u^3F - 2r\partial_u^2G + \partial_u H = \mathcal O(r^{-1})\,,
\ee
where we see that the left-hand side must actually vanish, so that
\be\begin{aligned}\label{uexpFGH}
F(u,\mathbf n)&=u^2 F_2(\mathbf n)+u F_1(\mathbf n)+T(\mathbf n)\,,\\
G(u,\mathbf n)&=u G_1(\mathbf n)+S(\mathbf n)\,,\\
H(u,\mathbf n)&=H(\mathbf n)\,,
\end{aligned}\ee
where we have introduced a number of arbitrary angular functions.

Now, let us analyze
\begin{align}
\label{rri}
\delta_\epsilon \varphi_{rri}&=2\left(\partial_r-\tfrac{2}{r}\right)\epsilon_{ri}+D_i\epsilon_{rr}=0\,,\\
\label{rui}
\delta_\epsilon \varphi_{rui}&=\left(\partial_r-\tfrac{2}{r}\right)\epsilon_{ui}+\partial_u\epsilon_{ri}+D_i\epsilon_{ur}=0\,,\\
\label{uui}
\delta_\epsilon \varphi_{uui}&=2\partial_u\epsilon_{ui}+D_i\epsilon_{uu}=0\,.
\end{align}
Eq. \eqref{rri} requires
\be
\epsilon_{ri}=r^2 I_i(u,\mathbf n)+\frac{r}{2}D_i F(u,\mathbf n)\,,
\ee
while, from \eqref{rui}, 
\be
\epsilon_{ui}=-r^3 \partial_u I_i(u,\mathbf n)+r^2 J_i(u,\mathbf n)+r D_i G(u, \mathbf n)\,.
\ee
In its turn, by \eqref{uui},
\be\label{uuivariation}
-2r^3 \partial_u^2 I_i+\frac{r^2}{2}(D_i\partial_u^2 F + 4\partial_u J_i)+D_iH=\mathcal O(1)\,,
\ee
which requires that the coefficients of $r^3$ and $r$ vanish:
\be\begin{aligned}\label{uexpIJ}
I_i(u,\mathbf n)&= u r_i(\mathbf n)+\rho_i(\mathbf n)\,,\\
J_i(u,\mathbf n)&= - \frac{u}{2}D_i F_2(\mathbf n)+\sigma_i(\mathbf n)\,.
\end{aligned}\ee

Let us now turn to the variation
\be
\delta_\epsilon\varphi_{rij}=\left(\partial_r-\tfrac{4}{r}\right)\epsilon_{ij}+D_{(i}\epsilon_{j)r}-2r\gamma_{ij}(\epsilon_{ur}-\epsilon_{rr})=0
\ee
which gives
\be
\epsilon_{ij}=r^4 K_{ij}+r^{3}\left[D_{(i}I_{j)}+\gamma_{ij}\partial_uF\right]+\frac{r^2}{2}\left[D_iD_jF+2\gamma_{ij}(F-G)\right]\,,
\ee
where $K_{ij}(u,\mathbf n)$ is an arbitrary symmetric tensor,
and to the trace constraint that must be obeyed by the gauge parameter,
\be
\eta^{ab}\epsilon_{ab}=\epsilon_{rr}-2\epsilon_{ur}+\frac{1}{r^2}\gamma^{ij}\epsilon_{ij}=0\,,
\ee
from which we obtain three conditions,
\begin{align}
\label{traceK}
\gamma^{ij}K_{ij}&=0\,,\\
\label{2DdotJ}
2D\cdot I + (n+1)\partial_u F &=0\,,\\
\label{Delta+2(n+1)}
[\Delta + 2(n+1)]F-2(n+2)G&=0\,.
\end{align}
We thus learn that $K_{ij}$ is traceless, while \eqref{2DdotJ} and \eqref{Delta+2(n+1)} using the expansions \eqref{uexpFGH}, \eqref{uexpIJ} give the constraint
\be\label{constrF2}
[\Delta+2(n+1)]F_2=0
\ee
and allow to solve for the following quantities,
\be\label{interrelF1F2}
F_2=-\frac{1}{n+1}D\cdot r\,,\qquad F_1=-\frac{2}{n+1}D\cdot \rho\,,
\ee
and
\be\label{interrelG1S}
G_1=\frac{1}{2(n+2)}[\Delta+2(n+1)]F_1\,,\qquad S=\frac{1}{2(n+2)}[\Delta+2(n+1)]T\,.
\ee

The variation
\be
\delta_\epsilon \varphi_{uij}=\partial_u\epsilon_{ij}+D_{(i}\epsilon_{j)u}-2r \gamma_{ij}(\epsilon_{uu}-\epsilon_{ur})=\mathcal O(r)\,,
\ee
namely
\be\label{uijvariation}
r^4 \partial_u K_{ij}+
r^2\left[\tfrac{1}{2}D_iD_j\partial_uF+\gamma_{ij}\partial_u G+D_{(i}J_{j)}\right]
+2r[D_iD_jG-\gamma_{ij}(G-H)]=\mathcal O(r)\,,
\ee
imposes
\be\label{duKij}
\partial_u K_{ij}=0\,,\qquad \frac{1}{2}D_iD_j\partial_uF+\gamma_{ij}\partial_u G+D_{(i}J_{j)}=0\,,
\ee
because the coefficients of $r^4$ and $r^2$ in \eqref{uijvariation} must vanish.
Furthermore the coefficient of $r$, while allowed to be nonzero, must be traceless, so that
\be\label{DeltaG+n}
\Delta G+n(G-H)\,.
\ee
More explicitly, we obtain from \eqref{duKij} that $K_{ij}$ is $u$-independent and, using \eqref{uexpFGH}, \eqref{uexpIJ},
\be
\frac{1}{2}D_iD_j\partial_uF_1+\gamma_{ij}G_1+D_{(i}\sigma_{j)}=0\,,
\ee
while \eqref{DeltaG+n} yields
\be\label{interrelH}
H=\frac{1}{n}(\Delta+n)S\,,
\ee
together with the constraint equation
\be\label{constrG1}
(\Delta+n)G_1=0\,.
\ee

The last variation to be taken into account is 
\be
\delta_{\epsilon} \varphi_{ijk}=D_{(i}\epsilon_{jk)}-2r\gamma_{(ij}[\epsilon_{k)u}-\epsilon_{k)r}]=\mathcal O(r^{2})\,,
\ee
namely
\be\begin{aligned}\label{ijkvariation}
	&r^4\left[D_{(i}K_{jk)}+2\partial_u I_{(i}\gamma_{jk)}\right]
	+r^3\left\{\D_{(i}D_j I_{k)}+\gamma_{(ij}[D_{k)}\partial_u F+2I_{k)}-2J_{k)}]\right\}\\
	&+r^2\left[\tfrac{1}{2}D_{(i}D_jD_{k)}F+\gamma_{(ij}D_{k)}(2F-3G)\right]=\mathcal O(r^2)\,,
\end{aligned}\ee
which, imposing that the coefficients of $r^4$ and $r^3$ be zero, affords
\begin{align}
\label{protoKilling}
D_{(i}K_{jk)}+2\partial_u I_{(i}\gamma_{jk)}&=0\,,\\
\label{DiDjIk}
\D_{(i}D_j I_{k)}+\gamma_{(ij}[D_{k)}\partial_u F+2I_{k)}-2J_{k)}]&=0\,.
\end{align}
Furthermore, the coefficient of $r^2$ in \eqref{ijkvariation} must be traceless:
\be\label{tracelessr2}
\frac{1}{2}D_i\Delta F+\Delta D_i F + (n+2)D_i(2F-3G)=0\,.
\ee
The trace of \eqref{protoKilling}, recalling that $\gamma^{ij}K_{ij}=0$ by \eqref{traceK} and using \eqref{uexpIJ},  gives
\be\label{interrelri}
r_i = -\frac{1}{n+2}D\cdot K_i\,,
\ee
and substituting back into \eqref{protoKilling} we see that $K_{ij}$ must satisfy the conformal Killing equation
\be\label{Killingtens}
D_{(i}K_{jk)}-\frac{2}{n+2}\gamma_{(ij}D\cdot K_{k)}=0\,.
\ee
Recalling \eqref{uexpFGH}, \eqref{uexpIJ} we obtain from \eqref{DiDjIk}
\be
D_{(i}D_j r_{k)}+\gamma_{(ij}\left[2r_{k)}+3D_{k)}F_2\right]=0
\ee
and
\be
D_{(i}D_j \rho_{k)} + \gamma_{(ij}\left[D_{k)}F_1-2\sigma_{k)}+2\rho_{k)}\right]=0\,.
\ee
Recalling \eqref{interrelF1F2}, \eqref{interrelG1S}, \eqref{interrelri} and taking the trace of the last equation also gives
\begin{align}
\label{constrD3K}
D_{(i}D_j D\cdot K_{k)}+2\gamma_{(ij}D\cdot K_{k)}+\frac{3}{n+1}\gamma_{(ij}D_{k)}D\cdot D\cdot K&=0\,,\\
\label{interrelsigmai}
\rho_i+\frac{1}{n+2}(\Delta \rho_i + \{D_i, D_j\}\rho^{j})-\frac{1}{n+1}D_i D\cdot \rho&=\sigma_i\,,\\
\label{Killingrho}
D_{(i}D_j \rho_{k)}-\frac{2}{n+2}\gamma_{(ij}\left[
\Delta \rho_{k)}+D\cdot D_{k)}\rho + D_{k)}D\cdot \rho
\right]&=0\,.
\end{align}
Finally, from \eqref{tracelessr2} we obtain three constraints
\be\begin{aligned}\label{constrFGTS}
\left(\tfrac{1}{2}D_i\Delta+\Delta D_i\right)F_2+2(n+2)D_i F_2&=0\,,\\
\left(\tfrac{1}{2}D_i\Delta+\Delta D_i\right)F_1+2(n+2)D_i F_1-3(n+2)D_iG_1&=0\,,\\
\left(\tfrac{1}{2}D_i\Delta+\Delta D_i\right)T+2(n+2)D_i T-3(n+2)D_iS&=0\,.
\end{aligned}\ee

Before proceeding further, let us recall our conventions for the Riemann tensor on the Euclidean sphere, 
\be
[D_i, D_j]\rho^k = R\indices{^k_{lij}}\rho^l\,,\qquad\qquad R_{ijkl}=\gamma_{ik}\gamma_{jl}-\gamma_{il}\gamma_{jk}\,. 
\ee
Repeatedly use of this relation allows one to derive a number of useful identities, which we will employ extensively in the following, for generic functions $T$, vectors $\rho^i$ and symmetric, traceless tensors $K^{ij}$ on the $n$-sphere:
\be\begin{aligned}\label{Ridentities}
	[\Delta, D_i]T&=(n-1)D_iT\,,
	\\
	[D_l,D_{(i}]K\indices{^l_{j)}}&=2n K_{ij}\,,
	\\
	[\Delta,D_{(i}]K_{jk)} &= (n+3)D_{(i}K_{jk)}-4\gamma_{(ij}D\cdot K_{k)}\,,
	\\
	[D_l, D_i]\rho^l &=(n-1)\rho_i\,,
	\\
	[D_i, \Delta]\rho^i &=(n-1)D\cdot \rho\,,
	\\
	D^k[D_k, D_{(i}]\rho_{j)}&=D_{(i}\rho_{j)}-2\gamma_{ij}D\cdot \rho\,,
	\\
	[\Delta, D_{(i}]\rho_{j)} &=(n+1)D_{(i}\rho_{j)}-4\gamma_{ij}D\cdot\rho\,,
	\\
	[D_l, D_{(i}D_{j)}]\rho^l &=(2n-1)D_{(i}\rho_{j)}-2\gamma_{ij}D\cdot\rho\,,
	\\
	[D_i, \Delta^2]\rho^i &=2(n-1)\Delta D\cdot\rho+(n-1)^2 D\cdot\rho\,,
	\\
	[D^j, \Delta]D_{(i}\rho_{j)}&= (n+1)\Delta \rho_i + (n-3)D_i D\cdot \rho+(n^2-1)\rho_i\,,\\
	[\Delta, D_{(i}]D_j \rho_{k)} &=
	(n+3)D_{(i}D_j\rho_{k)} - 4\gamma_{(ij}[D_{k)}D\cdot\rho+\Delta\rho_{k)}+(n-1)\rho_{k)}]\,.
\end{aligned}\ee
We begin the analysis of the equations we obtained by noting that \eqref{interrelF1F2}, \eqref{interrelG1S}, \eqref{interrelH}, \eqref{interrelri} and \eqref{interrelsigmai} allow us to determine all integration functions in terms of $T(\mathbf n)$, $\rho_i(\mathbf n)$ and $K_{ij}(\mathbf n)$, namely
\be
F_2 = \frac{D\cdot D\cdot K}{(n+1)(n+2)}\,,\qquad F_1=-\frac{2D\cdot\rho}{n+1}\,,\qquad G_1=-\frac{\Delta+2(n+1)}{(n+1)(n+2)}D\cdot\rho\,, 
\ee 
while
\be
S=\frac{\Delta+2(n+1)}{2(n+2)}T\,,\qquad H=\frac{(\Delta+n)[\Delta+2(n+1)]}{2n(n+2)}T\,,
\ee
and 
\be
r_i = -\frac{1}{n+2}D\cdot K_i\,,
\qquad
\sigma_i= \frac{n}{(n+1)(n+2)}D_iD\cdot\rho + \frac{\Delta+2n+1}{n+2}\rho_i\,.
\ee
On the other hand, $T$,  $\rho_i$ and $K_{ij}$ must satisfy \eqref{traceK}, \eqref{Killingtens} and \eqref{Killingrho}, namely
\begin{align}\label{equationsK}
\mathcal K_{ijk}&= D_{(i}K_{jk)}-\frac{2}{n+2}\gamma_{(ij}D\cdot K_{k)}=0\,,\qquad \gamma^{ij}K_{ij}=0\,,\\
\label{equationrho}
\mathcal R_{ijk}&=D_{(i}D_j \rho_{k)}-\frac{2}{n+2}\gamma_{(ij}\left[
2D_{k)}D\cdot \rho + (\Delta+n-1)\rho_{k)}
\right]=0\,,
\end{align}
plus a number of consistency conditions \eqref{constrF2}, \eqref{constrG1}, \eqref{constrD3K}, \eqref{constrFGTS}, which take the following form: the one involving $T$ reads
\be
[\Delta, D_i]T-(n-1)D_iT=0\,,
\ee
which is actually identically satisfied on account of \eqref{Ridentities}, the constraints on $\rho_i$ are
\be\begin{aligned}\label{rhoconstr}
[\Delta, D_i]D\cdot \rho-(n-1)D_iD\cdot \rho&=0\,,\\
(\Delta+n)[\Delta + 2(n+1)]D\cdot\rho&=0\,,\\
(\Delta+n)D_{(i}\rho_{j)}+\frac{n-2}{n+1}D_i D_j D\cdot\rho -\frac{3\gamma_{ij}}{n+1}\Delta D\cdot \rho-2\gamma_{ij} D\cdot\rho&=0\,,
\end{aligned}\ee
of which the first is again identically true by \eqref{Ridentities}, while the consistency conditions involving $K_{ij}$ are
\be\begin{aligned}\label{Kconstr}
[\Delta + 2(n+1)]D\cdot D\cdot K&=0\,,\\
D_i[\Delta + 2(n+1)]D\cdot D\cdot K&=0\,,\\
D_{(i}D_jD\cdot K_{k)}+2\gamma_{(ij}D\cdot K_{k)}-\frac{3}{n+1}\gamma_{(ij}D_{k)}D\cdot D\cdot K&=0\,.
\end{aligned}\ee
We will now prove that all these equations are actually identically satisfied, provided that $\rho_i$ and $K_{ij}$ obey \eqref{equationsK} and \eqref{equationrho}. To see this, we take divergences of $\mathcal K_{ijk}$ and $\mathcal R_{ijk}$, which, making extensive use of \eqref{Ridentities}, read
\be\begin{aligned}\label{rhodivergences}
	D\cdot \mathcal R_{ij}&=\frac{2(n+1)}{n+2}\left[
	(\Delta+n)D_{(i}\rho_{j)}+\frac{n-2}{n+1}D_i D_j D\cdot \rho\right.\\
	& \left.- \frac{3\gamma_{ij}}{n+1}\Delta D\cdot\rho - 2\gamma_{ij}D\cdot\rho
	\right],\\
	D\cdot D\cdot \mathcal R_i &= \frac{2(n+1)}{n+2} \left[
	\Delta^2\rho_i + 3n \Delta\rho_i + \frac{2(n-2)}{n+1}D_i\Delta D\cdot \rho \right. \\
	& \left.+ \frac{2(2n+1)(n-2)}{n+1}D_i D\cdot\rho+(2n+1)(n-1)\rho_i
	\right],\\
	D\cdot D\cdot D\cdot \mathcal R &= \frac{6(n-1)}{n+2}(\Delta+n)[\Delta+2(n+1)]D\cdot\rho\,,
\end{aligned}\ee
and 
\be\begin{aligned}\label{kappadivergences}
	D\cdot \mathcal K_{ij}&=
	\frac{n}{n+2}\left[D_{(i} D\cdot K_{j)}-\frac{2}{n}\gamma_{ij}D\cdot D\cdot K\right]
	+
	(\Delta+2n)K_{ij}\,,\\
	D\cdot D\cdot \mathcal K_i &= \frac{2(n+1)}{n+2} \left[ \frac{n-2}{2(n+1)} D_i D\cdot D\cdot K+
	\Delta D\cdot K_i + (2n+1)D\cdot K_i
	\right]\,,\\
	D\cdot D\cdot D\cdot \mathcal K &= \frac{3n}{n+2}[\Delta+2(n+1)]D\cdot D\cdot K\,.
\end{aligned}\ee
Equations \eqref{rhodivergences} clearly show that all remaining constraints \eqref{rhoconstr} involving $\rho_i$ are actually proportional to divergences of equation \eqref{equationrho}, namely $D\cdot\mathcal R_{ij}$ and $D\cdot D\cdot D\cdot \mathcal R$. Furthermore, equation \eqref{kappadivergences} reduces the first two constraints in \eqref{Kconstr} to $D\cdot D\cdot D\cdot \mathcal K$ and $D_iD\cdot D\cdot D\cdot \mathcal K$. 

To see that also the last equation in \eqref{Kconstr} is identically satisfied by \eqref{equationsK}, we first take its trace and note that it is proportional to  $D\cdot D\cdot \mathcal K_i$. The traceless projection instead reads
\be\label{DiDjDdotcalK}
D_{(i}D_jD\cdot K_{k)}-\frac{2}{n+2}\gamma_{(ij}\left[2D_{k)}D\cdot D\cdot K + (\Delta + n -1)D\cdot K_{k)}\right]=0\,.
\ee
We then consider
\be
D_{(i}D\cdot \mathcal K_{jk)}=
\frac{n}{n+2}D_{(i}D_jD\cdot K_{k)}
+(\Delta+n-3)D_{(i}K_{jk)}+4\gamma_{(ij}D\cdot K_{k)}-\frac{2}{n+2}\gamma_{(ij}D_{k)}D\cdot D\cdot K\,,
\ee
which, upon substituting $\mathcal K_{ijk}=0$, yields
\be\label{DiDdotKjk}
D_{(i}D\cdot \mathcal K_{jk)}=\frac{n}{n+2}\left\{
D_{(i}D_jD\cdot K_{k)}-\frac{2}{n}\gamma_{(ij}\left[
D_{k)}D\cdot D\cdot K - (\Delta + 3n + 1)D\cdot K_{k)}
\right]
\right\}=0\,.
\ee
The traceless projection of this equation is precisely \eqref{DiDjDdotcalK}, which proves that it is identically satisfied.

To reiterate, the gauge transformations that preserve the conditions \eqref{r+traces} and the ``weak'', dimension-independent falloffs \eqref{eqWeakest} are parametrized by a conformal Killing tensor $K_{ij}(x^k)$ on the sphere, by a vector $\rho_i(x^k)$ satisfying \eqref{equationrho} and by an arbitrary function $T(x^i)$.

Making explicit the dependence of the gauge parameters on these function affords
\be\begin{aligned}
\epsilon_{rr}&=
\frac{u^2D\cdot D\cdot K}{(n+1)(n+2)} 
-\frac{2u}{n+1} D\cdot \rho
+T\,,\\
\epsilon_{ur}&=
-\frac{ur\, D\cdot D\cdot K}{(n+1)(n+2)} 
+ \frac{r}{n+1}D\cdot \rho
-\frac{u[\Delta+2(n+1)]}{(n+1)(n+2)} D\cdot \rho 
+ \frac{\Delta+2(n+1)}{2(n+2)}T\,,\\
\epsilon_{uu}&=
\frac{r^2D\cdot D\cdot K}{(n+1)(n+2)}
+\frac{2r\,[\Delta+2(n+1)]}{(n+1)(n+2)} D\cdot \rho
+\frac{(\Delta+n)[\Delta+2(n+1)]}{2n(n+2)} T\,,\\
\epsilon_{ri}&=
-\frac{ur^2D\cdot K_i}{n+2}
+r^2\rho_i
+\frac{u^2r\, D_iD\cdot D\cdot K}{2(n+1)(n+2)}
-\frac{ur}{n+1}D_iD\cdot\rho
+\frac{r}{2}D_iT\,,\\
\epsilon_{ui}&=
\frac{r^3}{n+2}D\cdot K_i
-\frac{ur^2 D_iD\cdot D\cdot K}{2(n+1)(n+2)}
+\frac{r^2}{n+2}\left[\frac{n}{n+1}D_iD\cdot\rho+(\Delta+2n+1)\rho_i\right]\\
&-\frac{ur}{(n+1)(n+2)}D_i[\Delta+2(n+1)]D\cdot\rho
+\frac{r}{2(n+2)}[\Delta+2(n+1)]T\,,\\
\epsilon_{ij}&=
r^4 K_{ij}
-\frac{ur^3}{n+2}\left[
D_{(i}D\cdot K_{j)}-\frac{2\gamma_{ij}}{n+1}D\cdot D\cdot K
\right]
+r^3 \left[
D_{(i}\rho_{j)}-\frac{2\gamma_{ij}}{n+1}D\cdot \rho
\right]\\
&+\frac{u^2r^2}{2(n+1)(n+2)}(D_iD_j+2\gamma_{ij})D\cdot D\cdot K
\\
&-\frac{ur^2}{n+1}\left[
D_iD_j - \frac{\gamma_{ij}}{n+2}(\Delta-2)
\right]D\cdot \rho
+\frac{r^2}{2}\left[
D_iD_j - \frac{\gamma_{ij}}{n+2}(\Delta-2)
\right]T\,.
\end{aligned}\ee
while the corresponding nonvanishing symmetry variations read
\be\begin{aligned}\label{spin3variations}
\delta_\epsilon\varphi_{uui}&=
\frac{1}{2n(n+2)}D_i(\Delta+n)[\Delta+2(n+1)]T\,,\\
\delta_\epsilon\varphi_{uij}&=
-\frac{2ur}{(n+2)(n+2)}(D_iD_j+\gamma_{ij})[\Delta+2(n+1)]D\cdot\rho
\\
&+\frac{r}{n+2}\left(
D_iD_j-\tfrac{1}{n}\gamma_{ij}\Delta
\right)
[\Delta+2(n+1)]T\,,\\
\delta_\epsilon\varphi_{ijk}&=
\frac{u^2r^2}{2(n+1)(n+2)}\left[D_{(i}D_j D_{k)}+4\gamma_{(ij}D_{k)}\right]D\cdot D\cdot K
\\
&-\frac{ur^2}{d+1}\left[
D_{(i}D_jD_{k)}-\frac{1}{n+2}\gamma_{(ij}D_{k)}[3\Delta+2(n-1)]
\right]D\cdot\rho\\
&+\frac{r^2}{2}\left[
D_{(i}D_jD_{k)}-\frac{1}{n+2}\gamma_{(ij}D_{k)}[3\Delta+2(n-1)]
\right]T\,.
\end{aligned}\ee
 
\section{Spin-three Global Symmetries}\label{app:spin3exact}
We now specialize the calculation of spin-three residual symmetries in $D=n+2$ performed above to the more conservative radiation falloff conditions. We thus require that 
\be
\delta_\epsilon \varphi_{uui}\,,\qquad 
\frac{\delta_\epsilon \varphi_{uij}}{r}\,,\qquad 
\frac{\delta_\epsilon \varphi_{ijk}}{r^2}
\ee 
scale as $\mathcal O(r^{-\frac{n}{2}})$.
This choice will actually force us to consider, for any dimension greater than four, exact Killing tensors, namely the solutions of $\delta_\epsilon \varphi_{abc}=0$, due to the occurrence of the following additional constraints. 

First, $D_iH=0$ from the $\mathcal O(1)$ term in \eqref{uuivariation}, namely 
\be\label{addconstr1}
H=\frac{(\Delta+n)[\Delta+2(n+1)]}{2n(n+2)}T
\ee 
would be a constant for any dimension greater than four. Second, the coefficient of $r$ in \eqref{uijvariation} would not just be traceless, but would be actually set to zero, 
\be\label{addconstr2}
D_iD_jG-\gamma_{ij}(G-H)=0\,.
\ee 
Third, we would have the vanishing of the coefficient of $r^2$ in \eqref{ijkvariation},
\be\label{addconstr3}
\frac{1}{2}D_{(i}D_jD_{k)}F+\gamma_{(ij}D_{k)}(2F-3G)=0\,.
\ee
As already remarked, equations \eqref{addconstr1}, \eqref{addconstr2} and \eqref{addconstr3} are equivalent to the vanishing of \eqref{spin3variations}.
We now express the additional constraints \eqref{addconstr1}, \eqref{addconstr2} and \eqref{addconstr3}, or more precisely their traceless projections, since the vanishing of their traces has already been imposed above, in terms of $K_{ij}$, $\rho_i$ and $T$, thereby obtaining
\begin{align}
\label{DiDjDlDDK}
D_{(i}D_jD_{l)}D\cdot D\cdot K
-\frac{1}{n+2} \gamma_{(ij}\left[2\Delta D_{l)}D\cdot D\cdot K+ D_{l)}\Delta D\cdot D\cdot K\right]&=0\,,\\
\label{DiDjDrho}
\left(D_iD_j - \tfrac{1}{n}\gamma_{ij}\Delta
\right)[\Delta+2(n+1)]D\cdot \rho&=0\,,\\
\label{DiDjDlDrho}
D_{(i}D_jD_{l)}D\cdot \rho
-\frac{1}{n+2} \gamma_{(ij}\left[2\Delta D_{l)}D\cdot \rho+ D_{l)}\Delta D\cdot \rho\right]&=0\,,\\
\label{DiDeltaT}
D_i(\Delta+n)[\Delta+2(n+1)]T&=0\,,\\
\label{DiDjT}
\left(D_iD_j - \tfrac{1}{n}\gamma_{ij}\Delta
\right)[\Delta+2(n+1)]T&=0\,,
\end{align}
and
\be\label{constrT}
\mathcal T_{ijk}=D_{(i}D_jD_{k)}T
-\frac{1}{n+2} \gamma_{(ij}\left[2\Delta D_{k)}T+ D_{k)}\Delta T\right]=0\,.\\
\ee
We will now prove that equation \eqref{constrT} is the only genuinely new condition, while all other constraint actually follow from $\mathcal K_{ijk}=0$, $\mathcal R_{ijk}=0$ and $\mathcal T_{ijk}=0$. 

The conditions \eqref{DiDeltaT} and \eqref{DiDjT} are proportional to divergences of $\mathcal T_{ijk}=0$: employing \eqref{Ridentities},
\begin{align}
D\cdot \mathcal T_{ij} &= \frac{3n}{n+2}
\left(D_iD_j - \tfrac{1}{n}\gamma_{ij}\Delta
\right)[\Delta+2(n+1)]T\\
D\cdot D\cdot \mathcal T_{i} &= 
\frac{3n(n-1)}{n+2}D_i(\Delta+n)[\Delta+2(n+1)]T\,.
\end{align}
Equation \eqref{DiDjDrho} then follows in the same way from \eqref{DiDjDlDrho}. To see that the latter holds, we can start from equation \eqref{rhodivergences} and consider $D_{(i}D\cdot \mathcal R_{jk)}$, which takes the form
\be
D_{(i}D\cdot \mathcal R_{jk)}=\frac{2(n+1)}{n+2}\left\{
(\Delta-3)D_{(i}D_j\rho_{k)} 
+\frac{n-2}{n+1} D_{(i}D_jD_{k)}D\cdot \rho+\gamma_{(ij}[\cdots]_{k)}
\right\}\,,
\ee
where dots in the square bracket denote a number of terms obtained from $\rho_i$ and its derivatives, with one free index.
Using $\mathcal R_{ijk}=0$, given in \eqref{equationrho}, we can eliminate the first term on the right-hand side to cast this equation in the final form
\be
D_{(i}D\cdot \mathcal R_{jk)}=\frac{2(n-2)}{n+2}\left\{
D_{(i}D_jD_{k)}D\cdot \rho+\gamma_{(ij}[\cdots]_{k)}
\right\}=0\,,
\ee
whose traceless projection is \eqref{DiDjDlDrho}.

Equation \eqref{DiDjDlDDK} can be seen to hold adopting a similar strategy. Starting from equation \eqref{kappadivergences}, we can consider $D_{(i}D_j D\cdot D\cdot \mathcal K_{k)}$, which has the form
\be\begin{aligned}
D_{(i} D_j D\cdot D\cdot \mathcal K_{k)}&=\frac{2(n+1)}{n+2}\left\{
(\Delta-3)D_{(i}D_jD\cdot K_{k)} 
\right.\\
& \left.
+\frac{n-2}{n+1} D_{(i}D_jD_{k)}D\cdot D\cdot K
+\gamma_{(ij}[\cdots]_{k)}
\right\}\,,
\end{aligned}\ee
where the square brackets now contain a suitable combination of $K_{ij}$ and its derivatives. Employing the expression for $D_{(i}D\cdot \mathcal K_{jk)}=0$ given in \eqref{DiDdotKjk}, we can substitute the first term on the right-hand side and obtain
\be
D_{(i} D_j D\cdot D\cdot \mathcal K_{k)}=\frac{2(n-2)}{n+2}\left\{
D_{(i}D_jD_{k)}D\cdot D\cdot K+\gamma_{(ij}[\cdots]_{k)}
\right\}=0\,,
\ee
which, upon taking the traceless part, yields \eqref{DiDjDlDDK} as desired.

Thus, to summarize, the only new condition that is introduced by enforcing the exact Killing equation is \eqref{constrT}, whose effect is that of reducing the allowed choices for the previously arbitrary function $T(x^i)$ to (linear combinations of) spherical harmonics with $l=0$, $1$, $2$, namely to a solution space with finite dimension $1+(n+1)+\left[\tfrac{(n+1)(n+2)}{2}-1\right]=\frac{(n+1)(n+4)}{2}$.

%\printbibliography
%\bibliography{MyBib}

\end{document}